# A FAULT-TOLERANCE LINGUISTIC STRUCTURE FOR DISTRIBUTED APPLICATIONS

Promoter:
Prof. Dr. ir. R. Lauwereins

Dissertation submitted in partial
fulfilment of the requirements
for the degree of "doctor in de
toegepaste wetenschappen"
by

**Vincenzo DE FLORIO**

October 2000

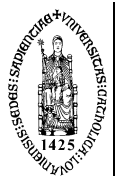

**KATHOLIEKE UNIVERSITEIT LEUVEN**
FACULTEIT TOEGEPASTE WETENSCHAPPEN
DEPARTEMENT ELEKTROTECHNIEK – ESAT
ONDERZOEKSEENHEID ACCA
Kardinaal Mercierlaan 94, B-3001 Leuven — België

# A FAULT-TOLERANCE LINGUISTIC STRUCTURE FOR DISTRIBUTED APPLICATIONS

Jury:
Prof. Dr. ir. J. Berlamont, chairman
Prof. Dr. ir. R. Lauwereins, promoter
Prof. Dr. ir. Y. Berbers
Prof. Dr. ir. B. Preneel
Prof. Dr. ir. G. Deconinck
Prof. Dr. ir. A. M. Tyrrell (University of York)

U.D.C. 681.3*B23

October 2000

**Copyright**

© Katholieke Universiteit Leuven – Faculteit Toegepaste Wetenschappen – Departement Elektrotechniek ESAT – Onderzoekseenheid ACCA, Kardinaal Mercierlaan 94, B-3001 Leuven (Belgium)

# ACKNOWLEDGEMENTS

This work would not have been possible without the contributions of a number of people who have supported me throughout these years.

I thank very much my promoter, Prof. R. Lauwereins, whose availability, guidance, and support allowed me to develop my doctoral studies. Many thanks also for his important remarks on the contents of this thesis.

Special thanks to Prof. G. Deconinck, whose continuous support, guidance and insightful comments during these years significantly contributed to the contents of this thesis. I am also grateful to him for getting me started in the field of dependability and in particular of fault-tolerance.

Many thanks to Prof. Y. Berbers, member of the reading committee, for her profound reading of the manuscript and for her valuable comments and suggestions. Many thanks to Prof. B. Preneel, member of the reading committee, for his important suggestions, including better formulations and shorter proofs for some of the theorems in Appendix A. I am convinced that the numerous contributions of the members of the reading committee greatly improved the overall structure and the contents of this work. Many thanks to Prof. J. Berlamont for being chairman of the Jury.

I am very grateful to Prof. A. Tyrrell, Head of the Department of Electronics of the University of York, for the great honor he so kindly conceded me by accepting to come to Leuven and be a member of the Jury.

Many thanks to Prof. J. Peperstraete, who encouraged and supported me at the beginning of my doctoral studies.

It has been a sheer pleasure for me to work in collaboration with several other people during these last four years. I wish to thank the many friends I had the great pleasure to work with in the ESPRIT projects EFTOS and TIRAN. I would like to thank also the reviewers of those two projects, whose recommendations, suggestions, and enthusiasm, considerably and positively influenced their development.

Many thanks are also due to the people from the ESAT Department of the K. U. Leuven and, in particular, to those from the ACCA Division.

I like to acknowledge the support of Dr. D. Laforenza, who introduced me to the domain of parallel computing, and whose friendly enthusiastic encouragement and support helped me to begin this exciting research experience.

My love and gratitude go to my dearest wife Tiziana, to whom I dedicate this work.

Leuven, October 2000

Vincenzo De Florio

# ABSTRACT

The structures for the expression of fault-tolerance provisions into the application software are the central topic of this dissertation.

Structuring techniques provide means to control complexity, the latter being a relevant factor for the introduction of design faults. This fact and the ever increasing complexity of today's distributed software justify the need for simple, coherent, and effective structures for the expression of fault-tolerance in the application software. A first contribution of this dissertation is the definition of a base of structural attributes with which application-level fault-tolerance structures can be qualitatively assessed and compared with each other and with respect to the above mentioned need. This result is then used to provide an elaborated survey of the state-of-the-art of software fault-tolerance structures.

The key contribution of this work is a novel structuring technique for the expression of the fault-tolerance design concerns in the application layer of those distributed software systems that are characterised by soft real-time requirements and with a number of processing nodes known at compile-time. The main thesis of this dissertation is that this new structuring technique is capable of exhibiting satisfactory values of the structural attributes in the domain of soft real-time, distributed and parallel applications. Following this novel approach, beside the conventional programming language addressing the functional design concerns, a special-purpose linguistic structure (the so-called "recovery language") is available to address error recovery and reconfiguration. This recovery language comes into play as soon as an error is detected by an underlying error detection layer, or when some erroneous condition is signalled by the application processes. Error recovery and reconfiguration are specified as a set of guarded actions, i.e., actions that require a pre-condition to be fulfilled in order to be executed. Recovery actions deal with coarse-grained entities of the application and pre-conditions query the current state of those entities.

An important added value of this so-called "recovery language approach" is that the executable code is structured so that the portion addressing fault-tolerance is distinct and separated from the rest of the code. This allows for division of complexity into distinct blocks that can be tackled independently of each other.

This dissertation also describes a prototype of a compliant architecture that has been developed in the framework of two ESPRIT projects. The approach is illustrated via a few case studies. Some preliminary steps towards an overall analysis and assessment of the novel approach are contributed by means of reliability models, discrete mathematics, and simulations.

Finally, it is described how the recovery language approach may serve as a harness with which to trade optimally the complexity of failure mode against number and type of faults being tolerated. This would provide dynamic adaptation of the application to the variations in the fault model of the environment.

# List of abbreviations

| Abbreviation | Meaning | Section | Page |
|---|---|---|---|
| A | Adaptability | 1.1.2 | 7 |
| ALFT | Application-Level Fault-Tolerance | 1.1 | 1 |
| AMS | Algorithm of Mutual Suspicion | 5.2.4 | 80 |
| AOP | Aspect-Oriented Programming | 3.4 | 42 |
| APL | ENEL application component | 6.3.2 | 126 |
| AS | Alarm Scheduler | B.1 | 177 |
| AT | Alarm Thread | B.1 | 177 |
| AW | Algorithmic Worker Processes | 6.1.3 | 113 |
| BB | TIRAN Backbone | 4.2.1 | 52 |
| BSL | TIRAN Basic Services Library | 4.2.1 | 52 |
| BSW | ENEL "basic software" component | 6.3.2 | 126 |
| BT | TIRAN Basic Tool | 4.2.1 | 52 |
| COTS | Commercial-Off-The-Shelf | 4.1.1 | 50 |
| CSP | Communicating Sequential Processes | 3.5 | 45 |
| DB | TIRAN Database | 4.2.1 | 52 |
| DIR net | EFTOS Detection, Isolation and Recovery network | 3.1.1 | 27 |
| DM | TIRAN Dependable Mechanism | 5.1 | 71 |
| DV | TIRAN Distributed Voting Tool | 5.2.3 | 74 |
| EE | $N$-Version Executive | 3.1.2.2 | 33 |
| EFTOS | Embedded Fault-Tolerant Supercomputing, ESPRIT project 21012 | 3.1.1 | 26 |
| EM | ENEL Exchange Memory component | 6.3.2 | 127 |
| EMI | Electro-Magnetic Interference | 1.1.2 | 6 |
| FSM | Fail-Stop Modules | 3.3.1.4 | 39 |
| FTAG | Fault-Tolerant Attribute Grammars | 3.3.2.3 | 40 |
| GLIMP | Siemens Gray Level IMage Processing package | 6.1.1 | 112 |
| ID | Image Dispatcher | 6.1.3 | 113 |
| IMP | Siemens Integrated Mail Processor | 6.1 | 110 |
| LCL | Local Control Level | 6.3.1 | 126 |
| MC | Machine Control | 6.1.1 | 112 |
| MOP | Metaobject Protocol | 3.2 | 35 |
| MV | Multiple-Version software fault-tolerance | 3.1.2 | 29 |
| NMR | $N$-Modular Redundancy | 5.2.3 | 74 |

# List of the symbols used and of their definitions

| Symbol | | Definition | Section | Page |
|---|---|---|---|---|
| $A(t)$ | = | Availability (probability that a service is operating correctly and is available to perform its functions at time $t$) | 2.1.2.3 | 14 |
| $A_{ss}$ | = | $\frac{\text{MTTF}}{\text{MTTF+MTTR}}$ | 2.1.2.3 | 14 |
| | = | Steady-state availability | | |
| $b(f)$ | = | Failure behaviours of failure class $f$ | 2.1.3.1 | 16 |
| $C$ | = | Error recovery coverage | 7.1.1 | 137 |
| correct $x$ | = | Event "the system provides its service during time interval $x$" | 2.1.2.1 | 13 |
| $\delta_{i,t}$ | = | 1 if $U(t,i)$ is true, 0 otherwise | 7.2.1.1 | 145 |
| $\varepsilon_N$ | = | Efficiency | 7.2.1.1 | 145 |
| | = | Percentage of slots used during a run | | |
| fail$(t)$ | = | $\frac{dQ(t)}{dt}$ | 2.1.2.1 | 13 |
| | = | Failure density function | | |
| $I_m$ | = | Set $\{0, \ldots, m-1\}$ $(m > 1)$ | 4.1.1 | 50 |
| $\lambda$ | = | Failure rate | 2.1.2.1 | 13 |
| $\lambda_N$ | = | Length of a run with $N+1$ processors | 7.2.1.1 | 145 |
| $M(t)$ | = | $1 - \exp^{-\mu t}$ (if $\mu$ is constant) | 2.1.2.4 | 14 |
| | = | Maintainability (probability that a failed system will be repaired in a time less than or equal to $t$) | | |
| MTBF | = | MTTF + MTTR | 2.1.2.2 | 14 |
| | = | Mean Time Between Failure | | |
| MTTF | = | $\int_0^\infty R(t)\,dt$ | 2.1.2.2 | 13 |
| | = | Mean Time To Failure | | |
| MTTR | = | Mean Time To Repair | 2.1.2.2 | 14 |
| | = | Average time required to repair a system | | |
| $\mu$ | = | Repair rate | 2.1.2.2 | 13 |
| $\mu_N$ | = | Average slot utilisation | 7.2.1.1 | 145 |
| $\nu_t$ | = | $\sum_{i=0}^{N} \delta_{i,t}$ | 7.2.1.1 | 145 |
| | = | Number of slots used during time step t | | |
| $\vec{\nu}$ | = | $[\nu_1, \nu_2, \ldots, \nu_\lambda]$ | 7.2.1.1 | 145 |
| | = | Utilisation string (number of slots used during each time step) | | |
| $n\text{T}/m\text{H}/p\text{S}$ | = | $n$ executions, on $m$ hardware channels, of $p$ programs | 3.1 | 26 |
| | = | Avižienis' classification of fault-tolerance approaches | | |

| Symbol | | Definition | Section | Page |
|---|---|---|---|---|
| $\mathcal{P}$ | $=$ | $\mathcal{P}_1, \ldots, \mathcal{P}_N$ | 7.2.1.1 | 145 |
| | $=$ | Permutation of the $N$ integers $[0, N] - \{i\}$, $N > 0, 0 \leq i \leq N$ | | |
| $Q(t)$ | $=$ | $1 - R(t)$ | 2.1.2.1 | 13 |
| | $=$ | Unreliability | | |
| $R$ | $=$ | Probability of successful recovery for a component affected by a transient fault | 7.1.2 | 138 |
| $R(t_0, t)$ | $=$ | $P\{\text{correct}[t_0, t] \mid \text{correct}[t_0, t_0]\}$ | 2.1.2.1 | 13 |
| | $=$ | Reliability in interval $[t_0, t]$ | | |
| $R(t)$ | $=$ | $R(0, t)$ | 2.1.2.1 | 13 |
| | $=$ | Reliability | | |
| $R^{(1)}(C, t)$ | $=$ | $(-3C^2 + 6C) \times [R(t)(1 - R(t))]^2 + R^{(0)}(t)$ | 7.1.1 | 138 |
| | $=$ | Reliability of a TMR-and-a-spare system | | |
| $R^{(M)}(C, t)$ | $=$ | Reliability of a TMR-and-$M$-spare system | 7.1.1 | 138 |
| $R_{\text{TMR}}^{\alpha}(t)$ | $=$ | $3\exp^{-2(1-RT)\lambda t} - 2\exp^{-3(1-RT)\lambda t}$ | 7.1.2 | 138 |
| | $=$ | Reliability of a TMR system exploiting the $\alpha$-count technique | | |
| $i\,R^t j$ | $=$ | At time $t$, processor $i$ is receiving a message from processor $j$ | 7.2.1.1 | 144 |
| $i\,S^t j$ | $=$ | At time $t$, processor $i$ is sending a message to processor $j$ | 7.2.1.1 | 144 |
| $\sigma(N)$ | $=$ | $\sigma(N) = (N+1)\lambda_N$ | 7.2.1.1 | 145 |
| | $=$ | Number of slots available within a run of $N + 1$ processors | | |
| $T$ | $=$ | Probability that the current fault is a transient one | 7.1.2 | 138 |
| $T_{\text{cycle}}$ | $=$ | Cycle time for the ENEL pilot application | 6.3.2 | 126 |
| $U(N)$ | $=$ | Number of used slots in a run of $N$ processors | A.3.4 | 175 |
| $U(t, i)$ | $=$ | Predicate "$\exists j \; (i\,S^t j \vee i\,R^t j)$" | 7.2.1.1 | 144 |
| $[x \; is \; odd]$ | $=$ | 1 when $x$ is odd, 0 otherwise | 7.2.1.2 | 147 |
| $\prec$ | $=$ | Failure semantics partial-order relation | 2.1.3.1 | 16 |

# Contents

# Chapter 1

# Introduction

The central topic of this dissertation is the system structure for expressing fault-tolerance provisions in the application layer of a computer program. The lack of a simple and coherent system structure for software fault-tolerance engineering, capable to offer the designer effective support towards fulfilling goals such as maintainability, re-usability, and service portability of the fault-tolerant software, has been the main motivation for starting this work. It addresses the class of distributed applications, written in a procedural language, to be executed on distributed or parallel computers consisting of a set of processing nodes known at compile time.

This chapter presents the scenario of application-level fault-tolerance engineering. A summary of the main contributions and the structure of this work are finally presented.

## 1.1 Rationale and Thesis

This section first justifies the need of application-level fault-tolerance (ALFT) as a consequence of the crucial role and of the complexity of modern computing systems and of software in particular. The requirements, impairments, and key properties of ALFT are then introduced. Finally, the thesis of this dissertation is enunciated.

### 1.1.1 The Need for Application-Level Fault-Tolerance

#### 1.1.1.1 A Need for Fault-Tolerance

No man conceived tool in human history has ever permeated as many aspects of human life as the computer has done during the last half a century. An outstanding aspect of this success is certainly given by the overwhelming increase in its performance, though probably the most important one is the growth in complexity and crucial character of the roles nowadays assigned to computers—human society more and more expects and relies on good quality of complex services supplied by computers. More and more these services become vital, in the sense that lack of timely delivery ever more often can have immediate consequences on capitals, the environment, and even human lives.

In other words, if in the early days of modern computing it was to some extent acceptable that outages and wrong results occurred rather often[1], being the main role of computers basically that of a fast solver of numerical problems, the criticality associated with many tasks nowadays appointed to computers does require high values for properties such as availability and data integrity. This happens because the magnitude of consequences associated with a failure has constantly increased, to the point that now, with computers controlling nuclear plants, airborne equipment, health care systems and so forth, a computer failure may likely turn into a catastrophe, in the sense that its associated penalty can be incalculable [HLKC99].

As a consequence, with the growth in complexity and criticality, it has become evident how important it is to adopt techniques for assessing and enhancing, in a justifiable way, the reliance to be placed on the services provided by computer systems, together with those techniques aiming at avoiding the risks associated with computer failures, or, at least, at bounding the extent of their consequences. The work reported herein deals in particular with **fault-tolerance**, that is, on how to ensure a service up to fulfilling the system's function even in the presence of "faults" [Lap98], namely, events having their origin inside or outside the system boundaries and possibly introducing unexpected deviations of the system state.

### 1.1.1.2   A Need for Software Fault-Tolerance

Research in fault-tolerance concentrated for many years on *hardware* fault-tolerance, i.e., on devising a number of effective and ingenious hardware structures to cope with faults [Joh89]. For some time this approach was considered as the only one needed in order to reach the requirements of availability and data integrity demanded by nowadays complex computer services. Probably the first researcher who realized that this was far from being true was B. Randell who in 1975 [Ran75] questioned hardware fault-tolerance as the only approach to pursue—in the cited paper he states:

> "Hardware component failures are only *one* source of unreliability in computing systems, decreasing in significance as component reliability improves, while software faults have become increasingly prevalent with the steadily increasing size and complexity of software systems."

Indeed most of the complexity supplied by modern computing services lies in their software rather than in the hardware layer [Lyu98a, Lyu98b, HK95, Wie93, Ran75]. This state of things could only be reached by exploiting a powerful conceptual tool for managing complexity in a flexible and effective way, i.e., devising hierarchies of sophisticated abstract machines [Tan90]. This translates in implementing software with high-level computer languages lying on top of other software strata—the device drivers layers, the basic services kernel, the operating system, the run-time support of the involved programming languages, and so forth.

---

[1]This excerpt from a report on the ENIAC activity [Wei61] gives an idea of how dependable computers were in 1947: "power line fluctuations and power failures made continuous operation directly off transformer mains an impossibility [...] down times were long; error-free running periods were short [...]". After many considerable improvements, still "trouble-free operating time remained at about 100 hours a week during the last 6 years of the ENIAC's use", i.e., an availability of about 60%!

Partitioning the complexity into stacks of software layers allowed the implementor to focus exclusively on the high-level aspects of their problems, and hence it allowed to manage a larger and larger degree of complexity. But *though made transparent, still this complexity is part of the overall system* being developed. A number of complex algorithms are executed by the hardware at the same time, resulting in the simultaneous progress of many system states—under the hypothesis that no involved abstract machine nor the actual hardware be affected by faults. Unfortunately, as in real life faults do occur, the corresponding deviations are likely to jeopardise the system's function, also propagating from one layer to the other, unless appropriate means are taken to avoid in the first place, or to remove, or to tolerate these faults. In particular, faults may also occur in the **application layer**, that is, in the abstract machine on top of the software hierarchy[2]. These faults, possibly having their origin at design time, or during operation, or while interacting with the environment, *are not different in the extent of their consequences from those faults originating, e.g., in the hardware or the operating system.*

An efficacious argument to bring evidence to the above statement is the case of the so-called "millennium bug", i.e., the most popular class of design faults that ever showed up in the history of computing technologies, also known as "the year 2000 problem", or as "Y2K": most of the software still in use today was developed using a standard where dates are coded in a 6-digit format. According to this standard, two digits were considered as enough to represent the year. Unfortunately this translates into the impossibility to distinguish, e.g., year 2000 from year 1900, which was recently recognised as the possible cause of an unpredictably large number of failures when calculating time elapsed between two calendar dates, as for instance year 1900 was not a leap year while year 2000 is. The adoption of the above mentioned standard for representing dates resulted in a hidden, forgotten design fault, never considered nor tested by application programmers. As society got closer and closer to the year 2000, the unaware presence of this design fault became a nightmare that seemed to jeopardise many crucial functions of our society appointed to programs manipulating calendar dates, such us utilities, transportation, health care, communication, public administration. Luckily the expected many and possibly crucial system failures due to this one application-level fault [HLKC99] were not so many and not that crucial, though probably for the first time the whole society became aware of the extent of the relevance of dependability in software.

These facts and the above reasoning suggest that, the higher the level of abstraction, the higher the complexity of the algorithms into play and the consequent error proneness of the involved (real or abstract) machines. As a conclusion, full tolerance of faults and complete fulfilment of the dependability design goals of a complex software application *must include means to avoid, remove, or tolerate faults working at all levels, including the application layer*.

### 1.1.1.3 Software Fault-Tolerance in the Application Layer

The need of software fault-tolerance provisions, located in the application layer, is supported by studies that showed that the majority of failures experienced by nowadays computer systems are due to *software faults*, including those located in the application layer [Lyu98a, Lyu98b, Lap98];

---

[2]In what follows, the application layer is to be intended as the programming and execution context in which a complete, self-contained program that performs a specific function directly for the user is expressed or is running.

for instance, NRC reported that 81% of the total number of outages of US switching systems in 1992 were due to software faults [NRC93]. Moreover, nowadays application software systems are increasingly networked and distributed. Such systems, e.g., client-server applications, are often characterised by a loosely coupled architecture whose global structure is in general more prone to failures[3]. Due to the complex and temporal nature of interleaving of messages and computations in distributed software systems, no amount of verification, validation and testing can eliminate all faults in an application and give complete confidence in the availability and data consistency of applications of this kind [HK95]. Under these assumptions, *the only alternative (and effective) means for increasing software reliability is that of incorporating in the application software provisions of software fault-tolerance* [Ran75].

Another argument that justifies the addition of software fault-tolerance means in the application layer is given by the widespread adoption of object orientation. Many object-oriented applications are indeed built from *reusable components* the sources of which are unknown to the application developers. The object abstraction fostered the capability of dealing with higher levels of complexity in software and at the same time eased and therefore encouraged software reuse. This has a big, positive impact on development costs though translates the application in a sort of collection of reused, pre-existing objects made by third parties. The reliability of these components and therefore their impact on the overall reliability of the user application is often unknown, to the point that Grey defines as "art" creating reliable applications using off-the-shelf software components [Gre97]. The case of the Ariane 501 flight is a well-known example that shows how improper reuse of software may have severe consequences[4] [Inq96].

Though probably the most convincing reasoning for not excluding the application layer from a fault-tolerance strategy is the so-called "end-to-end argument", a system design principle introduced by Saltzer, Reed and Clark [SRC84]. This principle states that, rather often, functions such as reliable file transfer, can be *completely* and *correctly* implemented only with the knowledge and help of the application standing at the endpoints of the underlying system (for instance, the communication network).

This does not mean that everything should be done at the application level—fault-tolerance strategies in the underlying hardware and operating system can have a strong impact on the system's performance. However, an extraordinarily reliable communication system, that guarantees that no packet is lost, duplicated, or corrupted, nor delivered to the wrong addressee, does not reduce the burden of the application program to ensure reliability: for instance, for reliable file transfer, the application programs that perform the transfer must still supply a file-transfer-specific, end-to-end reliability guarantee.

---

[3]As Leslie Lamport efficaciously synthesised in his quotation, "a distributed system is one in which I cannot get something done because a machine I've never heard of is down".

[4]Within the Ariane 5 programme, it was decided to reuse the long-tested software used in the Ariane 4 programme. Such software had been thoroughly tested and was compliant to Ariane 4 specifications. Unfortunately, specifications for Ariane 5 were different—in particular, they stated that the launcher had a trajectory characterised by considerably higher horizontal velocity values. A dormant design fault, related to an overflow while casting a 64-bit floating point value representing horizontal velocity to a 16-bit signed integer was never unravelled simply because, given the moderate horizontal velocity values of Ariane 4, such an overflow would have never occurred. Reusing this same software, including the routine affected by this dormant design fault, in both the primary and the active backup Inertial Reference Systems that were in use in the Ariane 5 flight 501 the morning of 4 June 1996, triggered almost at the same time the failure of two Inertial Reference Systems. Diagnostic information produced by that component was then interpreted as correct flight information, making the launcher veer off its flight path, then initiate a self-destruction procedure, and eventually explode. It is worth mentioning that the faulty routine served an Ariane 4 specific functionality, that was of no use in Ariane 5. It had been nevertheless included in order to reuse exactly the same software adopted in the previous programme. Such a software was indeed considered design fault free. This failure entailed a loss of about 1.9 billion French francs (approximately 0.37 billion Euros) and a delay of about one year for the Ariane 5 programme [Le 96].

Hence one can conclude that:

> Pure hardware-based or operating system-based solutions to fault-tolerance, though often characterised by a higher degree of transparency, are not *fully* capable of providing complete end-to-end tolerance to faults in the user application. Furthermore, relying solely on the hardware and the operating system develops only partially satisfying solutions; requires a large amount of extra resources and costs; and is often characterised by poor service portability [SRC84, SS92].

## 1.1.2 Strategies, Problems, and Key Properties

The above conclusions justify the strong need for ALFT; as a consequence of this need, several approaches to ALFT have been devised during the last three decades (see Chapter 3 for a brief survey). Such a long research period hints at the complexity of the design problems underlying ALFT engineering, which include:

1. How to incorporate fault-tolerance in the application layer of a computer program.

2. Which fault-tolerance provisions to support.

3. How to manage the fault-tolerance code.

Problem 1 is also known as the problem of the **system structure to software fault-tolerance**, first proposed by B. Randell in 1975 [Ran75]. It states the need of appropriate structuring techniques such that the incorporation of a set of fault-tolerance provisions in the application software might be performed in a simple, coherent, and well structured way. Indeed, poor solutions to this problem result in a huge degree of **code intrusion**: in such cases, the application code that addresses the functional requirements and the application code that addresses the fault-tolerance requirements are mixed up into one large and complex application software.

- This greatly complicates the task of the developer and requires expertise in both the application domain and in fault-tolerance. Negative repercussions on the development times and costs are to be expected.

- The maintenance of the resulting code, *both for the functional part and for the fault-tolerance provisions*, is more complex, costly, and error prone.

- Furthermore, the overall complexity of the software product is increased—which is detrimental to its resilience to faults.

One can conclude that, with respect to the first problem, an ideal system structure should guarantee an adequate **Separation between the functional and the fault-tolerance Concerns** (SC).

Moreover, the design choice of *which fault-tolerance provisions to support* can be conditioned by the adequacy of the syntactical structure at "hosting" the various provisions. The well-known quotation by B. L. Whorf efficaciously captures this concept:

"Language shapes the way we think, and determines what we can think about":

Indeed, as explained in Sect. 2.2, a non-optimal answer to Problem 2 may

- require a high degree of redundancy, and

- rapidly consume large amounts of the available redundancy,

which at the same time would increase the costs and reduce the reliability. One can conclude that, devising a syntactical structure offering *straightforward support* to *a large set of fault-tolerance provisions*, can be an important aspect of an ideal system structure for ALFT. In the following this property will be called **Syntactical Adequacy** (SA).

Finally, one can observe that another important aspect of an ALFT architecture is *the way the fault-tolerance code is managed*, at compile time as well as at run time. Evidence to this statement can be brought by observing the following facts:

- A number of important choices pertaining to the adopted fault-tolerance provisions, such as the parameters of a temporal redundancy scheme, are a consequence of an analysis of the environment in which the application is to be deployed and is to run[5]. In other words, depending on the target environments, the set of (external) impairments that might affect the application can vary considerably (see Chapter 2). Now, while it may be in principle straightforward to port an existing code to an other computer system, **porting the service** supplied by that code may require a proper adjustment of the above mentioned choices, namely the parameters of the adopted provisions. Effective support towards the management of the parametrisation of the fault-tolerance code and, in general, of its maintenance, could guarantee **fault-tolerance software reuse**.

- The *dynamic* (run-time) adaptation of the fault-tolerance code and of its parameters would allow the application to stand also unexpected changes in the environment, such as an increase of temperature triggered by a failure of the cooling system. These events may trigger unexpected faults. Furthermore, the ever increasing diffusion of *mobile software components*, coupled with the increasing need for dependability, will require more and more the capability to guarantee an agreed quality of service even when the environment changes during the run-time because of mobility.

---

[5]For instance, if an application is to be moved from a domestic environment to another one characterised by an higher electro-magnetic interference (EMI), it is reasonable to assume that, e.g., the number of replicas of some protected resource should be increased accordingly.

Therefore, the off-line as well as on-line (dynamic) management of the fault-tolerance provisions and of their parameters may be an important requirement for any satisfactory solution of Problem 3. As further motivated in Sect. 2.2, ideally, the fault-tolerance code should *adapt* itself to the current environment. Furthermore, any satisfactory management scheme should not overly increase the complexity of the application—which would be detrimental to dependability. Let us call this property **Adaptability** (A).

Let us refer collectively to properties SC, SA and A as to the *structural attributes* of ALFT.

The various approaches to ALFT surveyed in Chapter 3 provide different system structures to solve the above mentioned problems. The three structural attributes are used in that chapter in order to provide a qualitative assessment with respect to various application requirements. The structural attributes constitute, in a sense, a *base* with whom to perform this assessment. One of the major conclusions of that survey is that *none* of the surveyed approaches is capable to provide the best combination of values of the three structural attributes in *every* application domain. For specific domains, such as object-oriented distributed applications, satisfactory solutions have been devised at least for SC and SA, while only partial solutions exist, for instance, when dealing with the class of distributed or parallel applications *not based on the object model*. An example of the applications in this class is given by GLIMP (gray level image processing package), embedded in the Siemens Integrated Mail Processor, and discussed in Sect. 6.1.

The above matter of facts has been efficaciously captured by Lyu, who calls this situation "*the software bottleneck*" of system development [Lyu98b]: in other words, there is evidence of an urgent need for *systematic approaches to assure software reliability within a system* [Lyu98b] while effectively addressing the above problems. In the cited paper, Lyu remarks how "developing the required techniques for software reliability engineering is a major challenge to computer engineers, software engineers and engineers of related disciplines".

### 1.1.3   Thesis of the Dissertation

As in a formal proof the theoretician finds it a useful conceptual tool to separate the special case from the normal one (following the *divide et impera* design principle), likewise it is reasonable to assume that the designer of a fault-tolerant software system could take great advantage from dividing the non-faulty case from the specification and the management of the strategy to adopt when faults do occur. The above observation brings us to the thesis of this dissertation:

> *It is possible to catch up satisfactory values for the structural attributes, at least in the domain of soft real-time, distributed and parallel applications, by means of a fault-tolerance linguistic structure that realises a distinct, secondary, programming and processing context, directly addressable from the application layer.*

This statement is proved by describing and assessing a software system developed by the author of this dissertation in the framework of his participation to the ESPRIT-IV project 21012 "EFTOS" (*Embedded Fault-TO*lerant *S*upercomputing) [DDFLV97, EVM+98, DVB+97] and to the ESPRIT-IV project 28620 "TIRAN" (*TaI*lorable fault-tole*RAN*ce frameworks for embedded applications) [BDFD+99, BDFD+00]. Using this system, the bulk of the strategy to cope with errors needs to be expressed by the user in a "recovery script", *conceptually as well physically* distinct from the functional application layer. Such script is to be written in a so-called "recovery language", i.e., a specialised linguistic structure devoted to the management of the fault-tolerance strategies, which allows to express scenarios of isolation, reconfiguration, and recovery. These scenarios take place by referencing symbolic representations of the processing nodes, of the application tasks, and of user-defined groups of tasks.

Basic characteristics of this system are the separation of the functional code from the code addressing error recovery strategies, both at design time and during the execution of the application. The developer can modify the parameters of the fault-tolerance provisions, or even set up diverse fault-tolerance strategies with no modifications in the application part, and vice-versa, which allows to tackle more easily and effectively any of the two fronts. This can result in a better maintainability of the target fault-tolerant application and in support for reaching portability of the service when moving the application to other unfavourable environments. Dynamic management of the fault-tolerance code can also be attained, thanks to the strict separation between the functional and the fault-tolerance aspects.

## 1.2    Structure and Contributions

Section 1.1 provides the elements of ALFT and introduces the concept of structural attributes of ALFT—an original contribution of the author. The rest of this dissertation is structured as follows:

**Chapter 2** is an introduction to the subject of computer dependability. The aim of this chapter is to give the reader a concise but formal definition of concepts and key words already used, informally, in this section. For this sake, the Laprie model of dependability is briefly presented. In particular, dependability is characterised as a collective property:

- described by a set of sub-properties or basic attributes,

- affected by a set of impairments, and

- improved by the adoption of a number of techniques.

The chapter also states the problem of managing the trade off between the benefits of adaptable (dynamic) dependable strategies and the additional complexity required by the adoption of these strategies, which is inherently detrimental to dependability. The formulation of this problem is a contribution of the author.

**Chapter 3** reviews six categories of design approaches for software fault-tolerance in the application layer, namely, *single-version* and *multiple-version software fault-tolerance*; the adoption of *metaobject protocols*; the embedding of fault-tolerance into a high-level programming language (*language approach*); a structuring technique called *aspect-oriented programming*; and an approach called *recovery meta-program*. Each category is qualitatively assessed, with respect to the structural attributes introduced in Sect. 1.1.2, for various application domains. Positive and negative aspects of the "evolutionary" approaches with respect to the "revolutionary" ones are remarked.

Two conjectures are proposed in this chapter. The first one is that the coexistence of two separate though cooperating layers for the functional and the fault-tolerance aspects may allow to get the best of both the evolutionary and the revolutionary classes of approaches. The second conjecture is that a satisfactory solution to the design problem of the management of the fault-tolerance code may translate into an optimal management of the fault-tolerance provisions with respect to varying environmental conditions.

Survey, conjectures and ideas in this chapter are original contributions of the author of this dissertation.

**Chapter 4** describes the elements of a novel system structure for ALFT, which has been called "the recovery language approach". Main contributions in this chapter are:

- The system, application, and fault models.

- The basic ideas of the novel approach.

- An *abstract* architecture compliant to the approach.

This chapter also draws a workflow diagram of the approach and reports on the *specific differences* between the recovery language approach and the ALFT approaches discussed in Chapter 3. Finally, the *specific limitations* of the novel approach are discussed.

The conception of this new approach and the other ideas presented in this chapter are original contributions of the author of this dissertation.

**Chapter 5** presents a prototype implemented in the framework of the ESPRIT projects 21012 "EFTOS" and 28620 "TIRAN". This software system is one of the main results of the above mentioned ESPRIT projects, the latter of which is also briefly presented in this chapter. It also introduces the ARIEL recovery and configuration language and describes its relations with the architectural entities outlined in Chapter 4 and with the system structures described in Chapter 3. Once again, the basis for these comparisons are the structural attributes of ALFT.

The main contributions of the author while taking part in these two projects have been the concept of recovery and configuration language and the design and development of this prototypic implementation and of most of of its core components, such as, e.g., the distributed application serving as a backbone and coordinating the activities of a set of basic tools for error detection, isolation, and recovery. The work of the author in this context includes also the design and development of ARIEL. This translated in:

- The choice of a number of fault-tolerance provisions.

- The design and implementation of these provisions.

- The design of a linguistic structure to host the above mechanisms in a coherent and effective way.

- The development of the above structure and its inclusion in the grammar of ARIEL.

- The development of a run-time executive for the execution of ARIEL programs.

**Chapter 6**  provides an example of how to make use of the recovery language approach in order to develop or to enhance dependable mechanisms with minimal code intrusion. Four cases are described: the GLIMP software, embedded into the Siemens integrated mail processor; the redundant watchdog timer designed at ENEL; an $N$-version programming executive developed using ARIEL as both a configuration tool and a recovery language and using the TIRAN framework. Finally, the current use of the recovery language approach at ENEL is explained.

The case studies described in this chapter have been co-designed by the author.

**Chapter 7**  describes a number of results obtained via reliability and performance analysis and via measurements during simulations.  Key properties evaluated in this chapter are *reliability* (assessed via Markov models), *performance* (analysed within a formal, discrete time model), and *costs* (both development and maintenance costs; the limited code intrusion as well as the compact size of the related ARIEL script are used as arguments to show the higher degree of maintainability and service portability offered by the approach).

All the models, analyses, and simulations in this chapter have been devised and carried out by the author of this dissertation.

**Chapter 8**  concludes this dissertation. The conjectures drawn in the preceeding chapters are reviewed and justified in the light of the facts discussed throughout the dissertation. A number of possible evolutions and improvements are presented.  In particular, a structure for the adaptable management of the fault-tolerance code is sketched. It is also suggested how, making use of that structure, dynamic management of libraries of fault-tolerance "applets" may be used to set up movable components adapting themselves to varying environmental conditions with no need to recompile the code.  A detailed listing of the contributions of the author concludes this work.

# Chapter 2

# The Laprie Model of Dependability

This chapter briefly introduces the basic concepts and terminology adopted in this dissertation.

The central topic herein presented is **dependability**, defined as the trustworthiness of a computer system such that *reliance* can *justifiably* be placed on the service it delivers. In this context,

**service**  means the behaviour of that system as perceived by its users;

**user**  means another system, e.g., a human being, or a physical device, or a computer application, interacting with the former one.

The concept of dependability as described herein was first introduced by J.-C. Laprie [Lap85] as a contribution to an effort by IFIP Working Group 10.4 (Dependability and Fault-Tolerance) aiming at the establishment of a standard framework and terminology for discussing reliable and fault-tolerant systems. The cited paper and other works by Laprie are the main sources for this survey—in particular [Lap92], later revised as [Lap95] and eventually as [Lap98].

## 2.1   The Dependability Tree

A precise and complete characterisation of dependability is given

1. by enumerating its basic properties or *attributes*,

2. by explaining what phenomena constitute potential *impairments* to it, and

3. by reviewing the scientific disciplines and the techniques that can be adopted as *means* for improving dependability.

Attributes, impairments, and means can be globally represented into one picture as a tree, traditionally called the dependability tree [Lap95] (see Fig. 2.1).

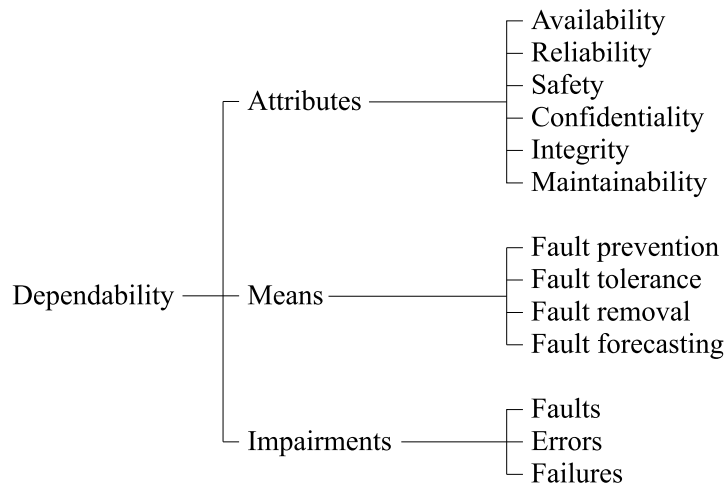

**Figure 2.1: The dependability tree**

## 2.1.1   Attributes of Dependability

As already mentioned, dependability is a general concept that embraces a number of different properties. These properties correspond to different viewpoints from which the user perceives the quality of the offered service—in other words, for different users there will be in general different key properties corresponding to a positive assessment for the service:

- the property that addresses the readiness for usage, that has been termed as **availability**,

- the property that measures the continuity of service delivery, that has been termed **reliability**,

- the property expressing the reliance on the non-occurrence of events with catastrophic consequences on the environment, known as **safety**,

- the property that measures the reliance on the non-occurrence of unauthorised disclosure of information, i.e., **confidentiality**,

- the property that measures the reliance on the non-occurrence of improper alterations of information, that has been called **integrity**,

- the property that expresses the ability to undergo repairs and upgrades, that has been called **maintainability**.

These properties qualify dependability, and therefore are known as its **attributes** [Lap95]. A special combination of these attributes has been termed as **security** and defined as the conjoint requirement for integrity, availability, and confidentiality.

A more formal definition exists for most of the above properties. Section 2.1.2 formally defines those attributes and quantities that are more pertaining to the rest of this dissertation, including availability, reliability, and maintainability.

### 2.1.2 Formal Definitions

This section defines a number of important measures of the quality of service of a system, including three of the attributes presented in 2.1.1 that are most relevant in what follows.

#### 2.1.2.1 Reliability

Reliability is defined as follows:

$$R(t_0, t) = P\{\text{correct}[t_0, t] \mid \text{correct}[t_0, t_0]\},$$

where "correct $x$" represents the event "the system provides its service during time interval $x$", and $R(t_0, t)$ is the conditional probability that the system will perform correctly throughout the interval $[t_0, t]$, given that the system was performing correctly at time $t_0$ [Joh89]. Time $t_0$ is usually omitted and taken as the current time. The general notation for reliability is therefore $R(t)$.

The negative counterpart of reliability, **unreliability**, is defined as $Q(t) = 1 - R(t)$, and represents the conditional probability that the system will perform incorrectly during the interval $[t_0, t]$, given that the system was performing *correctly* at time $t_0$. Unreliability is also known as the **probability of failure**, and quantity

$$\text{fail}(t) = \frac{dQ(t)}{dt}$$

is the **failure density function**.

If the failure rate $\lambda$ of a system is assumed to be fixed, i.e., if the expected number of failures per a given period of time can be considered as constant, as it happens in the "useful life period" of electronic equipment, then it is possible to show [Joh89] that

$$R(t) = \exp^{-\lambda t}.$$

The above equation is known as the exponential failure law.

#### 2.1.2.2 Mean Time To Failure, Mean Time To Repair, and Mean Time Between Failures

Mean Time to Failure (MTTF) is defined as the expected time that a system will operate before the occurrence of its first failure. More formally, it is the expected value of the time of failure, or, from probability theory,

$$\text{MTTF} = \int_{-\infty}^{\infty} t \, \text{fail}(t) \, dt = \int_{0}^{\infty} t \frac{dQ(t)}{dt} \, dt = \int_{0}^{\infty} R(t) \, dt.$$

If reliability obeys the exponential failure law with constant failure rate $\lambda$, then

$$\text{MTTF} = \int_{0}^{\infty} \exp^{-\lambda t} \, dt = \frac{1}{\lambda}.$$

Mean Time to Repair (MTTR) is defined as the average time required to repair a system. It is often specified by means of a repair rate $\mu$, namely the average number of repairs that occur per time unit. If $\mu$ is a constant, MTTR is the inverse of repair rate.

Mean Time Between Failures (MTBF) is the average time between any two consecutive failures of a system. This is slightly different from MTTF which concerns a system's very first failure. The following relation holds:

$$\text{MTBF} = \text{MTTF} + \text{MTTR}.$$

As it is usually true that MTTR is a small fraction of MTTF, it is usually allowed to assume that $\text{MTBF} \approx \text{MTTF}$.

#### 2.1.2.3  Availability

Availability is defined as a function of time representing the probability that a service is operating correctly and is available to perform its functions at the instant of time $t$ [Joh89]. It is usually represented as function $A(t)$. Availability represents a property at a given *point* in time, whereas reliability concerns time *intervals*. These two properties are not to be mistaken with each other— a system might exhibit a good degree of availability and yet be rather unreliable, e.g., when inoperability is pointwise or rather short.

Availability can be approximated as the total time that a system has been capable of supplying its intended service divided by the elapsed time that system has been in operation, i.e., the percentage of time that the system is available to perform its expected functions. The steady-state availability can be proven [Joh89] to be

$$A_{\text{ss}} = \frac{\text{MTTF}}{\text{MTTF} + \text{MTTR}}.$$

#### 2.1.2.4  Maintainability

Maintainability is a function of time representing the probability that a failed system will be repaired in a time less than or equal to $t$. It can be estimated as

$$M(t) = 1 - \exp^{-\mu t},$$

$\mu$ being the repair rate, assumed to be constant (see Sect. 2.1.2.2).

### 2.1.3  Impairments to Dependability

Hardware and software systems must conform to certain specifications, i.e., agreed descriptions of the system response for any initial system state and input, as well as the time interval within which the response should occur. This includes a description of the functional behaviour of the system—basically, what the system is supposed to do, or in other words, a description of its service—and possibly a description of other, non-functional requirements. Some of these requirements may concern the dependability of the service.

In real life, any system is subject to internal or external events that can affect in different ways the quality of its service. These events have been partitioned into three classes by their cause-effect relationship: depending on this, an impairment can be classified as a **fault**, an **error**, or a **failure**. When the delivered service of a system deviates from its specification, the user of the system experiences a **failure**. Such failure is due to a deviation from the correct state of the system, known as an **error**. That deviation is due to a given cause, for instance related to the physical state of the system, or to bad system design. This cause is called a **fault**. A failure of a system could give rise to an event that is perceived as a fault by the user of that system, bringing to a concatenation of cause-and-effects events known as the "fundamental chain" [Lap85]:

$$\ldots \text{fault} \rightarrow \text{error} \rightarrow \text{failure} \rightarrow \text{fault} \rightarrow \text{error} \rightarrow \text{failure} \rightarrow \ldots$$

(symbol "→" can be read as "brings to"). Attributes defined in Sect. 2.1.1 can be negatively affected by faults, errors, and failures. For this reason, failures, errors, and faults have been collectively termed as the "impairments" of dependability. They are characterised in the following three paragraphs.

### 2.1.3.1 Failures

System failures occur when the system does not behave as agreed in the system specifications. This can happen in many different ways [Cri91]:

**omission** failures occur when an agreed reply to a well defined request is missing. The request appears to be ignored;

**timing** failures occur when the service is supplied, though outside the real-time interval agreed upon in the specifications. This may occur when the service is supplied *too soon* (early timing failure), or *too late* (late timing failure, also known as **performance** failure);

**response** failures happen either when the system supplies an incorrect output (in which case the failure is said to be a **value** failure), or when the system executes an incorrect state transition (**state transition** failure);

**crash** failure is when a system continuously exhibits omission failures until that system is restarted. In particular, a **pause-crash** failure occurs when the system restarts in the state it had right before its crash, while a **halting-crash** occurs when the system simply never restarts. When a restarted system re-initialises itself wiping out the state it had before its crash, that system is said to have experienced an **amnesia crash**. It may also be possible that some part of a system's state is re-initialised while the rest is restored to its value before the occurrence of the crash—this is called a **partial-amnesia crash**.

Defining the above failure classes allows to extend a system's specification—that is, the set of its failure-free behaviours—with failure semantics, i.e., with the failure behaviour that system is likely to exhibit upon failures. This is important when programming strategies for recovery after failure [Cri91]. For instance, if the service supplied by a communication system may delay transmitted messages but never lose or corrupt them, then that system is said to have *performance* failure semantics. If that system can delay and also lose them, then it is said to have *omission/performance* failure semantics.

In general, if the failure semantics of a system $s$ allows it to exhibit a behaviour in the union of two failure classes $F$ and $G$, then $s$ is said to have $F/G$ failure semantics. In other words, the "slash" symbol can be read as the union operator among sets. For any given $s$ it is possible to count the possible failure behaviours in a failure class. Let us call $b$ this function from the set of failure classes to integers. Then, given failure classes $F$ and $G$,

$$b(F/G) = b(F \cup G) = b(F) + b(G).$$

Failure semantics can be partially ordered by means of function $b$: given any two failure semantics $F$ and $G$, then $F$ is said to exhibit a weaker (less restrictive) failure semantics than $G$:

$$F \prec G \quad \Leftrightarrow \quad b(F) > b(G).$$

In particular $F/G \prec F$. Therefore, the union of all possible failure classes represents the weakest failure semantics possible. If system $s$ exhibits such semantics, $s$ is said to have **arbitrary failure semantics**, i.e., $s$ can exhibit *any* failure behaviour, without any restriction. By its definition, arbitrary failure semantics is also weaker than arbitrary *value* failure semantics. This latter is also known as **Byzantine failure semantics** [LSP82].

In the case of stateless systems, pause-crash and halting-crash behaviours are subsets of omission failure behaviours [Cri91], so omission failure semantics is in this case weaker than pause-crash and halting-crash failure semantics.

As clearly stated in [Cri91], it is the responsibility of a system designer to ensure that it properly implements a specified failure semantics. For instance, in order to implement a processing service with crash failure semantics, one can use duplication with comparison: two physically independent processors executing in parallel the same sequence of instructions and comparing their results after the execution of each instruction. As soon as a disagreement occurs, the system is shut down [Pow97]. Another possibility is to use self-checking capabilities. Anyway, given any failure semantics $F$, it is up to the system designer to decide how to implement it, also depending on the designer's other requirements, e.g., those concerning costs and expected performance. In general, *the weaker the failure semantics, the more expensive and complex to implement it*. Moreover, a weak failure semantics imply *higher costs in terms of redundancy exhaustion* (see Sect. 2.2) and, often, higher performance penalties. For this reason, the designer may leave the ultimate choice to the user—for instance, the designer of the Motorola C compiler for the PowerPC allows the user to choose between two different modes of compilation—the fastest mode does not guarantee that the state of the system pipeline be restored on return from interrupts [Sun96]. This translates into behaviours belonging to the partial-amnesia crash semantics. The other mode guarantees the non-occurrence of these behaviours at the price of a lower performance for the service supplied by that system—programs compiled with this mode run slower.

Failures can also be characterised according to the classification in Fig. 2.2 [Lap95], corresponding to the different viewpoints of

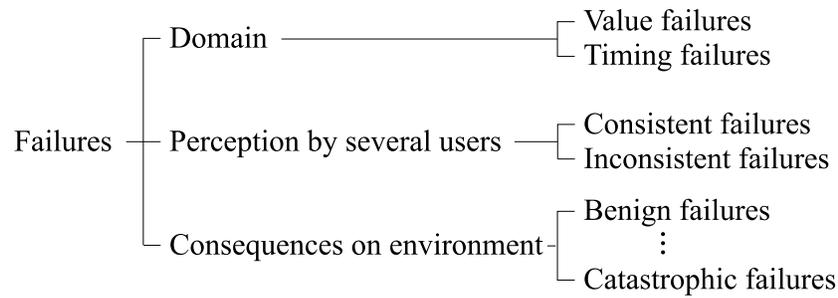

**Figure 2.2: Failure classes.**

- failure **domain** (i.e., whether the failure manifests itself in the time or value domain),

- failure **perception** (i.e., whether any two users perceive the failure in the same way, in which case the failure is said to be *consistent*, or differently, in which the failure is said to be *inconsistent*),

- and **consequences on the environment**. In particular a failure is said to be *benign* when consequences are of the same order as the benefits provided by normal system operation, while it is said *catastrophic* when consequences are incommensurably more relevant than the benefits of normal operation [Lap95].

Systems that provide a given failure semantics are often said to exhibit a "failure mode". For instance, systems having arbitrary failure semantics (in both time and value domains) are called **fail-uncontrolled** systems, while those only affected by benign failures are said to be **fail-safe** systems; likewise, systems with halt-failure semantics are referred to as **fail-halt** systems. These terms are also used to express the behaviour a system should have when dealing with multiple failures—for instance, a "fail-op, fail-op, fail-safe" system is one such that is able to withstand two failures and then behaves as a fail-safe system [Rus94] (fail-op stands for "after failure, the system goes back to operational state"). Finally, it is worth mentioning the **fail-time-bounded** failure mode, introduced in [Cuy95], which assumes that all errors are detected within a pre-defined, bounded period after the fault has occurred.

### 2.1.3.2 Errors

An error is the manifestation of a fault [Joh89] in terms of a deviation from accuracy or correctness of the system state. An error can be either **latent**, i.e., when its presence in the system has not been yet perceived, or **detected**, otherwise. Error latency is the length of time between the occurrence of an error and the appearance of the corresponding failure or its detection.

### 2.1.3.3 Faults

A fault is a defect, or an imperfection, or a lack in a system's hardware or software component. It is generically defined as the adjudged or hypothesised cause of an error. Faults can have their origin within the system boundaries (*internal faults*) or outside, i.e., in the environment (*external faults*). In particular, an internal fault is said to be *active* when it produces an error, and *dormant* (or *latent*) when it does not. A dormant fault becomes an active fault when it is *activated* by the computation process or the environment. Fault latency is defined as either the length of time between the occurrence of a fault and the appearance of the corresponding error, or the length of time between the occurrence of a fault and its removal.

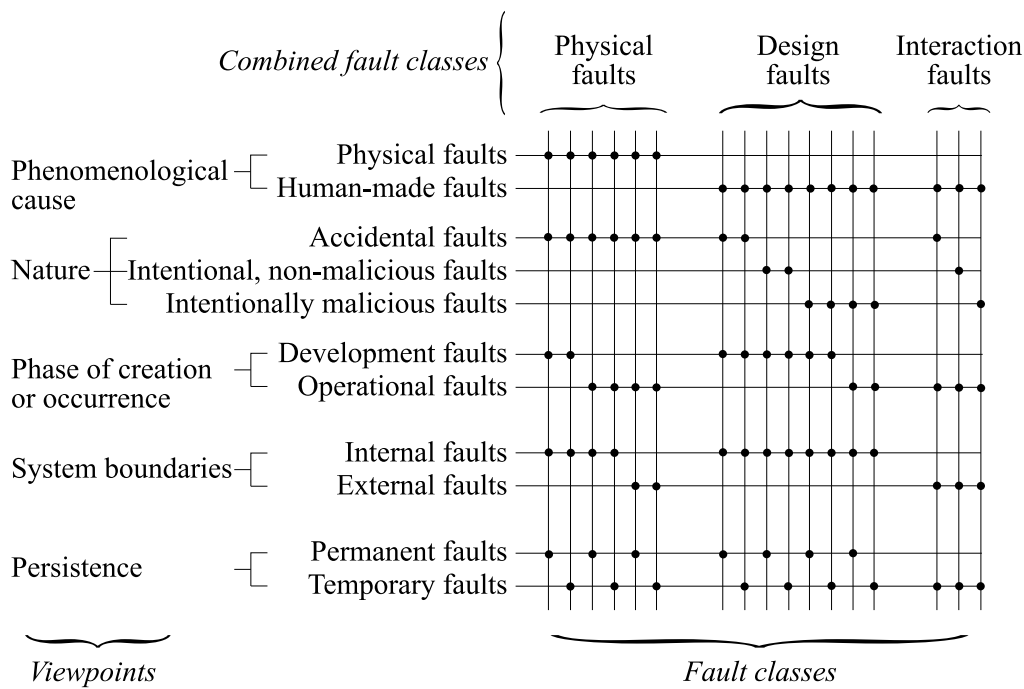

**Figure 2.3: Laprie's fault classification scheme.**

Faults can be classified according to five viewpoints [Lap92, Lap95, Lap98]—phenomenological cause, nature, phase of creation or occurrence, situation with respect to system boundaries, persistence. Not all combinations can give rise to a fault class—this process only defines 17 *fault classes*, summarised in Fig. 2.3. These classes have been further partitioned into three "groups", known as combined fault classes.

The combined fault classes that are more relevant in the rest of the dissertation are now briefly characterised:

**Physical faults:**

- Permanent, internal, physical faults. This class concerns those faults that have their origin within hardware components and are continuously active. A typical example is given by the fault corresponding to a worn out component.

- Temporary, internal, physical faults (also known as *intermittent faults*) [BCDGG97]. These are typically internal, physical defects that become active depending on a particular pointwise condition.

- Permanent, external, physical faults. These are faults induced on the system by the physical environment.

- Temporary, external, physical faults (also known as *transient faults*) [BCDGG97]. These are faults induced by environmental phenomena, e.g., EMI.

**Design faults:**

- Intentional, though not malicious, permanent / temporary design faults. These are basically trade-offs introduced at design time. A typical example is insufficient dimensioning (underestimations of the size of a given field in a communication protocol[1], and so forth).

- Accidental, permanent, design faults (also called systematic faults, or Bohrbugs): flawed algorithms that systematically turn into the same errors in the presence of the same input conditions and initial states—for instance, an unchecked divisor that can result in a division-by-zero error.

- Accidental, temporary design faults (known as Heisenbugs, for "bugs of Heisenberg", after their elusive character): while systematic faults have an evident, deterministic behaviour, these bugs depend on subtle combinations of the system state and environment.

**Interaction faults:**

- Temporary, external, operational, human-made, accidental faults. These include operator faults, in which an operator does not correctly perform his or her role in system operation.

- Temporary, external, operational, human-made, non-malicious faults: "neglect, interaction, or incorrect use problems" [Sib98]. Examples include poorly chosen passwords and bad system parameter setting.

- Temporary, external, operational, human-made, malicious faults. This class includes the so-called malicious replication faults, i.e., faults that occur when replicated information in a system becomes inconsistent, either because replicates that are supposed to provide identical results no longer do so, or because the aggregate of the data from the various replicates is no longer consistent with system specifications.

---

[1] A noteworthy example is given by the bad dimensioning of IP addresses. Currently, an IP address consists of four sections separated by periods. Each section contains an 8-bit value, for a total of 32 bits per address. Normally this would allow for more than 4 billion possible IP addresses—a rather acceptable value. Unfortunately, due to a lavish method for assigning IP address space, IP addresses are rapidly running out. A new protocol, IPv6 [HD95], is going to fix this problem through larger data fields (128-bit addresses) and a more flexible allocation algorithm.

### 2.1.4 Means for Dependability

Developing a dependable service, i.e., a service on which reliance can be placed justifiably, calls for the combined utilisation of a set of methods and techniques globally referred to as the "means for dependability" [Lap98]:

**fault prevention** aims at preventing the occurrence or introduction of faults. Techniques in this category include, e.g., quality assurance and design methodologies;

**fault-tolerance** groups methods and techniques to set up services capable of fulfilling their function in spite of faults;

**fault removal** methods try to reduce the number, incidence, and consequences of faults. Fault removal is composed of three steps: verification, diagnosis and correction. Verification checks whether the system adheres to certain properties—the verification conditions—during the design, development, production or operation phase; if it does not, the fault(s) preventing these conditions to be fulfilled must be diagnosed, and the necessary corrections (corrective maintenance) must be made;

**fault forecasting** investigates how to estimate the present number, the future incidence and the consequences of faults. Fault forecasting is conducted by evaluating the system behaviour with respect to fault occurrence or activation. Qualitatively, it aims at identifying, classifying and ordering failure modes or at identifying event combinations leading to undesired effects. Quantitatively, it aims at evaluating (in terms of probabilities) some of the attributes of dependability.

Of the above mentioned methods, fault-tolerance represents the core tool for the techniques and tools presented in this dissertation. Therefore, it is discussed in more detail in Sect. 2.1.4.1.

#### 2.1.4.1 Fault-Tolerance

Fault-tolerance methods come into play the moment a fault enters the system boundaries. Its core objective is "preserving the delivery of expected services despite the presence of fault-caused errors within the system itself" [Avi85]. Fault-tolerance has its roots in hardware systems, where the assumption of *random* component failures is substantiated by the physical characteristics of the adopted devices [Rus94].

According to [AL81], fault-tolerance can be decomposed into two sub-techniques—error processing and fault treatment.

**Error processing** aims at removing errors from the computational state (if possible, before failure occurrence). It can be based on the following primitives [Lap95]:

  **Error detection** which aims to detect the presence in the system of latent errors before they are activated. This can be done, e.g., by means of built-in self-tests or by comparison with redundant computations [Rus94].

**Error diagnosis** i.e., assessing the damages caused by the detected errors or by errors propagated before detection.

**Error recovery** i.e., the replacement of an erroneous state by an error-free state. This replacement takes one of the following forms:

1. Compensation, which means reverting the erroneous state into an error-free one exploiting information redundancy available in the erroneous state, predisposed, e.g., through the adoption of error correcting codes [Joh89].

2. Forward recovery, which finds a new state from which the system can operate (frequently in degraded mode). This method only allows to recover from errors of which the damage can be *anticipated*[2]—therefore, this method is system dependent [LA90]. The main tool for forward error recovery is exception handling [Cri95].

3. Backward recovery, which substitutes an erroneous state by an error-free state prior to the error occurrence. As a consequence, the method requires that, at different points in time (known as *recovery points*), the current state of the system be saved in some stable storage means. If a system state saved in a recovery point is error-free, it can be used to restore the system to that state, thus wiping out the effects of transient faults. For the same reason, this technique allows also to recover from errors of which the damage cannot or has not been anticipated. The need for backward error recovery tools and techniques stems from their ability to prevent the occurrence of failures originated by transient faults, which are many times more frequent than permanent faults [Rus94]. The main tools for backward error recovery are based on checkpoint-and-rollback [Dec96] and recovery blocks [Ran75] (see Sect. 3.1.2.1).

**Fault treatment** aims at preventing faults from being re-activated. It can be based on the following primitives [Lap95]:

**Fault diagnosis** i.e., identifying the cause(s) of the error(s), in location and nature, i.e. determining the fault classes to which the faults belong. This is different from error diagnosis; besides, different faults can lead to the same error.

**Fault passivation** i.e., preventing the re-activation of the fault. This step is not necessary if the error recovery step removes the fault, or if the likelihood of re-activation of the fault is low enough.

**Reconfiguration** updates the structure of the system so that non-failed components fulfil the system function, possibly at a degraded level, even though some other components have failed.

---

[2]In general, program specifications are not *complete*: there exist input states for which the behaviour of the corresponding program $P$ has been left unspecified. No forward recovery technique can be applied to deal with errors resulting from executing $P$ on these input states. On the contrary, if a given specification is complete, that is, if each input state is covered in the set $G$ of all the standard and exceptional specifications for $P$, and if $P$ is *totally correct*, i.e. fully consistent with what prescribed in $G$, then $P$ is said to be *robust* [Cri95]. In this case forward recovery can be used as an effective tool for error recovery.

## 2.2   Fault-Tolerance, Redundancy, and Complexity

A well-known result by Shannon [SWS93] tells us that, from any unreliable channel, it is possible
to set up a more reliable channel by increasing the degree of information redundancy. This
means that *it is possible to trade off reliability and redundancy* of a channel. The author of this
dissertation observes that the same can be said for a fault-tolerant system, because fault-tolerance
is in general the result of some strategy effectively exploiting some form of redundancy—time,
information, and/or hardware redundancy [Joh89]. This redundancy has a cost penalty attached,
though. Addressing a weak failure semantics, able to span many failure behaviours, effectively
translates in higher reliability—nevertheless,

1. it **requires** large amounts of extra resources, and therefore implies a high cost penalty, and

2. it **consumes** large amounts of extra resources, which translates into the rapid exhaustion
   of the extra resources.

For instance, a well-known result by Lamport *et al.* [LSP82] sets the minimum level of redun-
dancy required for tolerating Byzantine failures to a value that is greater than the one required
for tolerating, e.g., value failures. Using the simplest of the algorithms described in the cited
paper, a 4-modular-redundant (4-MR) system can only withstand any *single Byzantine failure*,
while the same system may exploit its redundancy to withstand up to *three crash faults—though
no other kind of fault* [Pow97]. In other words:

> After the occurrence of a crash fault, a 4-MR system with strict Byzantine fail-
> ure semantics has exhausted its redundancy and is no more dependable than a non-
> redundant system supplying the same service, while the crash failure semantics sys-
> tem is able to survive to the occurrence of that and two other crash faults. On the
> other hand, the latter system, subject to just one Byzantine fault, would fail regard-
> less its redundancy.

Therefore, for any given level of redundancy, *trading complexity of failure mode against
number and type of faults tolerated* may be considered as an important capability for an effec-
tive fault-tolerant structure. Dynamic adaptability to different environmental conditions[3] may
provide a satisfactory answer to this need, especially when the additional complexity does not
burden (and jeopardise) the application. Ideally, such complexity should be part of a custom
architecture and not of the application. On the contrary, the embedding in the application of
complex failure semantics, covering many failure modes, implicitly promotes complexity, as it
may require the implementation of many recovery mechanisms. This complexity is detrimental
to the dependability of the system, as it is in itself a significant source of design faults. Fur-
thermore, the isolation of that complexity outside the user application may allow cost-effective
verification, validation and testing. These processes may be unfeasible at application level.

---

[3]The following quote by J. Horning [Hor98] captures very well how relevant may be the role of the environment
with respect to achieving the required quality of service: "What is the most often overlooked risk in software
engineering? That the environment will do something the designer never anticipated".

The author of this dissertation conjectures that a satisfactory solution to the design problem of the management of the fault-tolerance code (presented in Sect. 1.1.2) may translate in an optimal management of the failure semantics (with respect to the involved penalties). The fault-tolerance linguistic structure proposed in Chapter 4 and following ones allows to solve the above problems exploiting its *adaptability* (A).

## 2.3  Conclusions

This chapter has introduced the reader to Laprie's model of dependability describing its attributes, impairments, and means. The central topic of this dissertation, fault-tolerance, has also been briefly discussed. The complex relation between the management of fault-tolerance, of redundancy, and of complexity, has been pointed out. In particular, a link has been conjectured between attribute A and the dynamic ability to trade off the complexity of failure mode against number and type of faults being tolerated.

# Chapter 3

# Current Approaches for Application-Level Fault-Tolerance

One of the conclusions drawn in Chapter 1 is that the system to be made fault-tolerant must include provisions for fault-tolerance also in the application layer of a computer program. In that context, the problem of which system structure to use for ALFT has been proposed. This chapter provides a critical survey of the state-of-the-art on embedding fault-tolerance means in the application layer.

According to the literature, at least six classes of methods, or approaches, can be used for embedding fault-tolerance provisions in the application layer of a computer program. This chapter describes these approaches and points out positive and negative aspects of each of them *with respect to the structural attributes defined in* Sect. 1.1.2 *and to various application domains*. A non-exhaustive list of the systems and projects implementing these approaches is also given. Conclusions are drawn in Sect. 3.6, where the need for more effective approaches is recognised.

Two of the above mentioned approaches derive from well-established research in software fault-tolerance—Lyu [Lyu98b, Lyu96, Lyu95] refers to them as single-version and multiple-version software fault-tolerance. They are dealt with in Sect. 3.1. A third approach, described in Sect. 3.2, is based on metaobject protocols. It is derived from the domain of object-oriented design and can also be used for embedding services other than those related to fault-tolerance. A fourth approach translates into developing new custom high-level distributed programming languages or enhancing pre-existent languages of that kind. It is described in Sect. 3.3. A structuring technique called aspect-oriented programming, is reported in Sect. 3.4. Finally, Sect. 3.5 describes an approach, based on a special recovery task monitoring the execution of the user task.

## 3.1 Single- and Multiple-Version Software Fault-Tolerance

A key requirement for the development of fault-tolerant systems is the availability of **replicated resources**, in hardware or software. A fundamental method employed to attain fault-tolerance is **multiple computation**, i.e., $N$-fold ($N \geq 2$) replications in three domains:

**time** i.e., repetition of computations,

**space**  i.e., the adoption of multiple hardware channels (also called "lanes"), and

**information**  i.e., the adoption of multiple versions of software.

Following Avižienis [Avi85], it is possible to characterise at least some of the approaches towards fault-tolerance by means of a notation resembling the one used to classify queueing systems models [Kle75]:

$$n\text{T}/m\text{H}/p\text{S},$$

the meaning of which is "$n$ executions, on $m$ hardware channels, of $p$ programs". The non-fault-tolerant system, or 1T/1H/1S, is called *simplex* in the cited paper.

### 3.1.1   Single-Version Software Fault-Tolerance

Single-version software fault-tolerance (SV) is basically the embedding into the user application of a simplex system of error detection or recovery features, e.g., atomic actions [JC85], checkpoint-and-rollback [Dec96], or exception handling [Cri95]. The adoption of SV in the application layer requires the designer to concentrate in one physical location, namely, the source code of the application, both the specification of what to do in order to carry on some user computation and the strategy such that faults are tolerated when they occur. As a result, the size of the problem addressed is increased. A fortiori, this translates into increasing the size of the user application. This induces loss of transparency, maintainability, and portability while increasing development times and costs.

   A partial solution to this loss in portability and these higher costs is given by the development of libraries and frameworks created under strict software engineering processes. In the following, two examples of this approach are presented—the EFTOS library and the SwIFT system. Special emphasis is reserved in particular to the first system, where the author of this work provided a number of contributions.

**The EFTOS library.**  EFTOS [DDFLV97, DVB+97] (the acronym stands for "embedded, fault-tolerant supercomputing") is the name of ESPRIT-IV project 21012. The aims of this project were to integrate fault-tolerance into embedded distributed high-performance applications in a flexible and effective way. The EFTOS library has been first implemented on a Parsytec CC system [Par96b], a distributed-memory MIMD supercomputer consisting of processing nodes based on PowerPC 604 microprocessors at 133MHz, dedicated high-speed links, I/O modules, and routers. As part of the project, this library has been then ported to a Microsoft Windows NT / Intel PentiumPro platform and to a TEX / DEC Alpha platform [TXT97, DEC97] in order to fulfil the requirements of the EFTOS application partners. The main characteristics of the CC system are the adoption of the thread processing model and of the message passing communication model: communicating threads exchange messages through a proprietary message passing library called EPX [Par96a]. The porting of the EFTOS library was achieved by porting EPX on the various target platforms and developing suitable adaptation layers.

Through the adoption of the EFTOS library, the target embedded parallel application is plugged into a hierarchical, layered system whose structure and basic components (depicted in Fig. 3.1) are:

- At the base level, a distributed net of "servers" whose main task is mimicking possibly missing (with respect to the POSIX standards) operating system functionalities, such as remote thread creation;

- One level upward (detection tool layer), a set of parametrisable functions managing error detection, referred to as "Dtools". These basic components are plugged into the embedded application to make it more dependable. EFTOS supplies a number of these Dtools, e.g., a watchdog timer thread and a trap-handling mechanism, plus an API for incorporating user-defined EFTOS-compliant tools;

- At the third level (control layer), a distributed application called "DIR net" (its name stands for "detection, isolation, and recovery network") is used to coherently combine the Dtools, to ensure consistent fault-tolerance strategies throughout the system, and to play the role of a backbone handling information to and from the fault-tolerance elements [DTDF$^+$99]. The DIR net can be regarded as a fault-tolerant network of crash-failure detectors, connected to other peripheral error detectors. Each node of the DIR net is "guarded" by an  thread that requires the local component to send periodically "heartbeats" (signs of life). A special component, called RINT, manages error recovery by interpreting a custom language called RL—the latter being a sort of ancestor of the language described in this thesis (see Sect. 5.3);

- At the fourth level (application layer), the Dtools and the components of the DIR net are combined into dependable mechanisms i.e., methods to guarantee fault-tolerant communication [EVM$^+$98], tools implementing a virtual Stable Memory [DBC$^+$98], a distributed voting mechanism called "voting farm" [DF97b, DFDL98a, DFDL98c], and so forth;

- The highest level (presentation layer) is given by a hypermedia distributed application based on standard World-Wide Web (WWW) [BLCGP92] technology, which monitors the structure and the state of the user application [DFDT$^+$98]. This application is based on a special CGI script [Kim96], called DIR Daemon, which continuously takes its inputs from the DIR net, translates them in HTML [BLC95], and remotely controls a WWW browser [Zaw94] so that it renders these HTML data.

The author of this dissertation contributed to this project designing and developing a number of basic tools, e.g., the distributed voting system, the monitoring tool, the RL language and its run-time system, that is, the task responsible for the management of error recovery [DFDL98b, DFDL98c]. Furthermore, he took part in the design and development of the first version of the DIR net. He also designed and developed a second version of the DIR net [DF98].

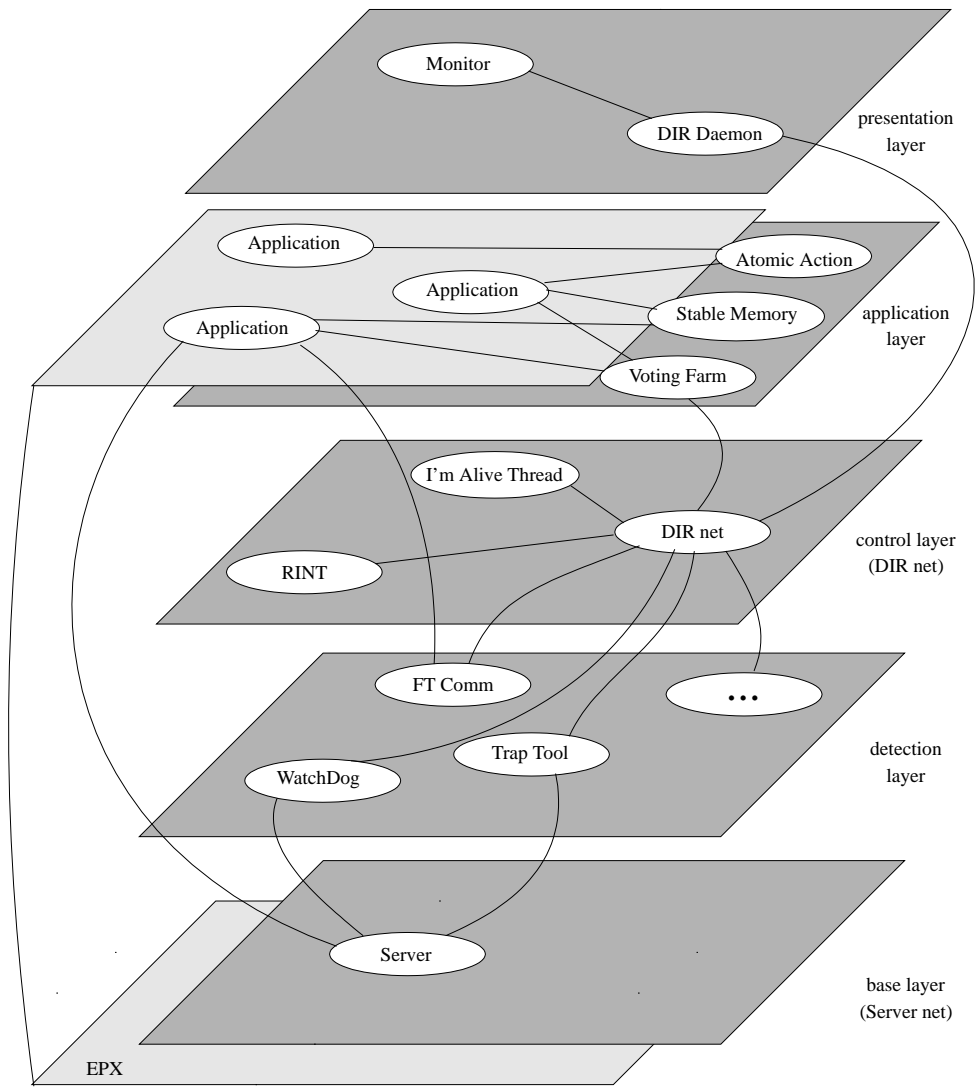

**Figure 3.1: The structure of the EFTOS library. Light gray has been used for the operating system and the user application, while dark gray layers pertain EFTOS.**

**The SwIFT System.**  SwIFT [HKBW96] stands for Software Implemented Fault-Tolerance, a system including a set of reusable software components (`watchd`, a general-purpose UNIX daemon watchdog timer; `libft`, a library of fault-tolerance methods, including single-version implementation of recovery blocks and $N$-version programming (see Sect. 3.1.2); `libckp`, i.e., a user-transparent checkpoint-and-rollback library; a file replication mechanism called `REPL`; and `addrejuv`, a special "reactive" feature of `watchd` [HKKF95] that allows for software rejuvenation[1].  The system derives from the HATS system [HK95] developed at AT&T. Both have been successfully used and proved to be efficient and economical means to increase the level of fault-tolerance in a software system where residual faults are present and their toleration is less costly than their full elimination [Lyu98b]. A relatively small overhead is introduced in most cases [HK95].

---

[1]Software rejuvenation [HKKF95] offers tools for periodical and graceful termination of an application with immediate restart, so that possible erroneous internal states, due to transient faults, be wiped out before they turn into a failure.

**Conclusions.**    Two SV systems have been described. One can observe that for pure SV systems *full transparency is not reached*, because the user needs anyway to manage the location and inter-dependency of such tools within one source code. In other words, SV requires the application developer to be an expert in fault-tolerance as well, because he / she has to integrate in the application a number of fault-tolerance provisions among those available in a set of ready-made basic tools, and has the responsibility for doing it in a coherent, effective, and efficient way. As it has been observed in Sect. 1.1.2, the resulting code is a mixture of functional code and of custom error-management code that does not always offer an acceptable degree of *portability* and *maintainability*. The functional and non-functional design concerns are not kept apart with SV, hence one can conclude that (qualitatively) *SV exhibits poor separation of concerns (*SC*)*.

Moreover, as the fault-tolerance provisions are offered to the user via an interface based on a general-purpose language like C or C++, very limited syntactical adequacy (SA) can be offered by SV as a system structure for ALFT.

Furthermore, no support is provided for off-line and on-line configuration of the fault-tolerance provisions, therefore also adaptability (A) can be qualitatively assessed as insufficient.

Note also that, even just in order to reach code portability, one relies on the portability and availability of the related libraries and products on the target platform, and that in some cases these libraries are only available in the form of object code.

On the other hand, tools in these libraries and systems give the user the ability to deal with fault-tolerance "atoms" without having to worry about their actual implementation and with a good ratio of costs over improvements of the dependability attributes, sometimes introducing a relatively small overhead. Using these toolsets the designer can re-use existing, long tested, sophisticated pieces of software without having each time "to re-invent the wheel". The EFTOS experience with the Siemens GLIMP application, described in Sect. 6.1, brings evidence to this statement.

Finally, it is important to remark that, in principle, SV poses no restrictions on the class of applications that may be tackled with it.

### 3.1.2    Multiple-Version Software Fault-Tolerance

This section describes multiple-version software fault-tolerance (MV), an approach which requires $N$ ($N \geq 2$) independently designed versions of software. MV systems are therefore $x$T/$y$H/$N$S systems. In MV, a same service or functionality is supplied by $N$ pieces of code that have been designed and developed by different, independent software teams[2]. The aim of this approach is to reduce the effects of design faults due to human mistakes committed at design time. The most used configurations are $N$T/1H/$N$S, i.e., $N$ sequentially applicable alternate programs running on the same hardware channel, and 1T/$N$H/$N$S, based on the parallel execution of the alternate programs on $N$, possibly diverse, hardware channels.

---

[2]This requirement is well explained by Randell [Ran75]: "All fault-tolerance must be based on the provision of useful redundancy, both for error detection and error recovery. In software the redundancy required is not simple replication of programs but *redundancy of design*." It is noteworthy to remark how, in the infamous Ariane 5 flight

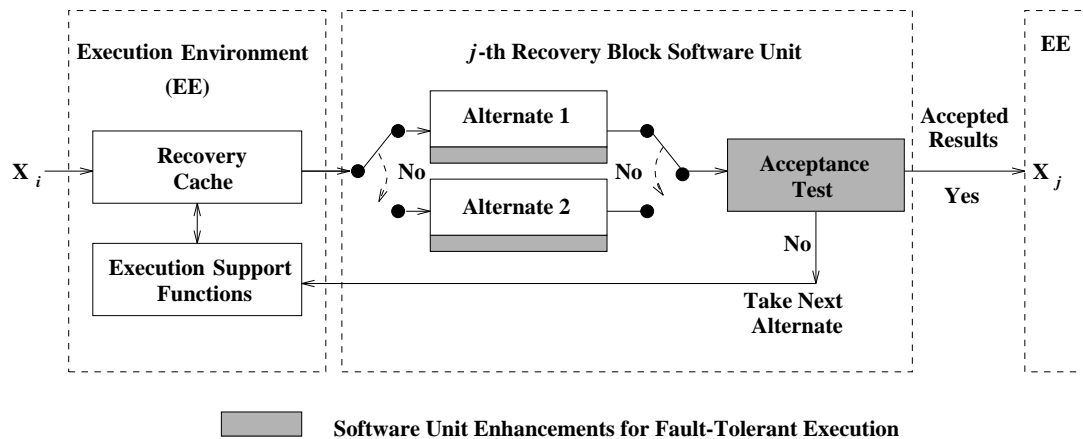

**Figure 3.2: The recovery block model with two alternates. The execution environment is charged with the management of the recovery cache and the execution support functions (used to restore the state of the application when the acceptance test is not passed), while the user is responsible for supplying both alternates and the acceptance test (picture from [Avi95]).**

Two major approaches exist: the first one is known as recovery block [Ran75, RX95], and is dealt with in Sect. 3.1.2.1. The second approach is the so-called $N$-version programming [Avi85, Avi95]. It is described in Sect. 3.1.2.2.

### 3.1.2.1   The Recovery Block Technique

Recovery Blocks (RB) are usually implemented as $N$T/1H/$N$S systems. The technique addresses residual software design faults. It aims at providing fault-tolerant functional components which may be nested within a sequential program. Other versions of the approach, implemented as 1T/$N$H/$N$S systems, are suited for parallel or distributed programs [SGM85, RX95].

RB is similar to the hardware fault-tolerance scheme known as "stand-by sparing" and described, e.g., in [Joh89]. The scheme is summarised in Fig. 3.2: on entry to a recovery block, the current state of the system is checkpointed. A primary alternate is executed. When it ends, an acceptance test checks whether the primary alternate successfully accomplished its objectives. If not, a backward recovery scheme reverts the system state back to its original value and a secondary alternate takes over the task of the primary alternate. When the secondary alternate ends, the acceptance test is executed again. The scheme goes on until either an alternate fulfils its tasks or all alternates are executed without success. In such a case, an error routine is executed.

---

501, two special components were operating in parallel, *with identical hardware and software* [Inq96]. The failure of that mission has been attributed—among other causes—to the nearly consecutive failures of these two components, triggered by the activation of the same design fault, hidden in the same, replicated program (see cited paper).

Recovery blocks can be nested—in this case the error routine invokes recovery in the enclosing block [RX95]. An exception triggered within an alternate is managed as a failed acceptance test. A possible syntax for recovery blocks is as follows:

```
ensure          acceptance test
by              primary alternate
else by         alternate 2
       .
       .
else by         alternate N
else error
```

Note how this syntax does not explicitly show the recovery step that should be carried out transparently by a run-time executive.

The effectiveness of RB rests to a great extent on the coverage of the error detection mechanisms adopted, the most crucial component of which is the acceptance test. A failure of the acceptance test is a failure of the whole RB scheme. For this reason, the acceptance test must be simple, must not introduce huge run-time overheads, must not retain data locally, and so forth. It must be regarded as the *ultimate* means for detecting errors, though not the *exclusive* one. Assertions and run-time checks, possibly supported by underlying layers, need to buttress the scheme and reduce the probability of an acceptance test failure. Another possible failure condition for the RB scheme is given by an alternate failing to terminate. This may be detected by a time-out mechanism that could be added to the RB scheme. This addition obviously further increases the complexity.

**Some systems supporting RB and their assessment.** The SwIFT library that has been described in Sect. 3.1.1 (p. 28) implements recovery blocks in the C language as follows:

```
#include <ftmacros.h>
...
ENSURE(acceptance-test) {
            primary alternate;
} ELSEBY {
   alternate 2;
} ELSEBY {
   alternate 3;
}
...
ENSURE;
```

Unfortunately this scheme does not cover any of the above mentioned requirements for enhancing the error detection coverage of the acceptance test. This would clearly require a run-time executive that is not part of this scheme. Other solutions, based on enhancing the grammar of pre-existing programming languages like Pascal [Shr78] and Coral [ABHM85], have some impact on portability. In both cases, code intrusion is not avoided. This translates in difficulties when trying to modify or maintain the application program without interfering "much" with the recovery structure, and vice-versa, when trying to modify or maintain the recovery structure without interfering "much" with the application program. Hence one can conclude that the RB is characterised by unsatisfactory values of the structural attribute SC. Furthermore, a system structure for ALFT based exclusively on the RB scheme does not satisfy attribute SA[3]. Finally, what attribute A is concerned, one can observe that RB is a rigid scheme that does not allow off-line configuration nor (a fortiori) code adaptability.

On the other hand, the RB scheme has been successfully adopted throughout 25 years in many different application fields. It has been successfully validated by a number of statistical experiments and through mathematical modelling [RX95]. Its adoption as the sole fault-tolerance means, while developing complex applications, resulted in some cases [ABHM85] in a failure coverage of over 70%, with acceptable overheads in memory space and CPU time.

### 3.1.2.2   N-Version Programming

$N$-Version Programming (NVP) systems are built from generic architectures based on redundancy and consensus. Such systems usually belongs to class $1\mathrm{T}/N\mathrm{H}/N\mathrm{S}$, less often to class $N\mathrm{T}/1\mathrm{H}/N\mathrm{S}$. NVP is defined by its author [Avi85] as "the independent generation of $N \geq 2$ functionally equivalent programs from the same initial specification." These $N$ programs, called versions, are developed for being executed in parallel. This system constitutes a fault-tolerant software unit that depends on a generic decision algorithm to determine a consensus or majority result from the individual outputs of two or more versions of the unit.

Such a scheme (depicted in Fig. 3.3) has been developed under the fundamental conjecture that independent designs translate in *random component failures*—i.e., statistical independence. Such a result would guarantee that correlated failures do not translate in immediate exhaustion of the available redundancy, as it would happen, e.g., using $N$ copies of the same version. Replicating software would also mean replicating any dormant software fault in the source version—see, e.g., the accidents with the Therac-25 linear accelerator [Lev95] or the Ariane 5 flight 502 [Inq96]. According to Avižienis, independent generation of the versions significantly reduces the probability of correlated failures. Unfortunately a number of experiments [ECK+91] and theoretical studies [EL85] have shown that this assumption is not always correct.

The main differences between RB and NVP are:

- RB (in its original form) is a sequential scheme whereas NVP allows concurrent execution;

- RB requires the user to provide a fault-free, application-specific, effective acceptance test, while NVP adopts a *generic* consensus or majority voting algorithm that can be provided by the execution environment (EE);

---

[3]Randell himself states that, given the ever increasing complexity of modern computing, there is still an urgent need for "richer forms of structuring for error recovery and for design diversity" [RX95].

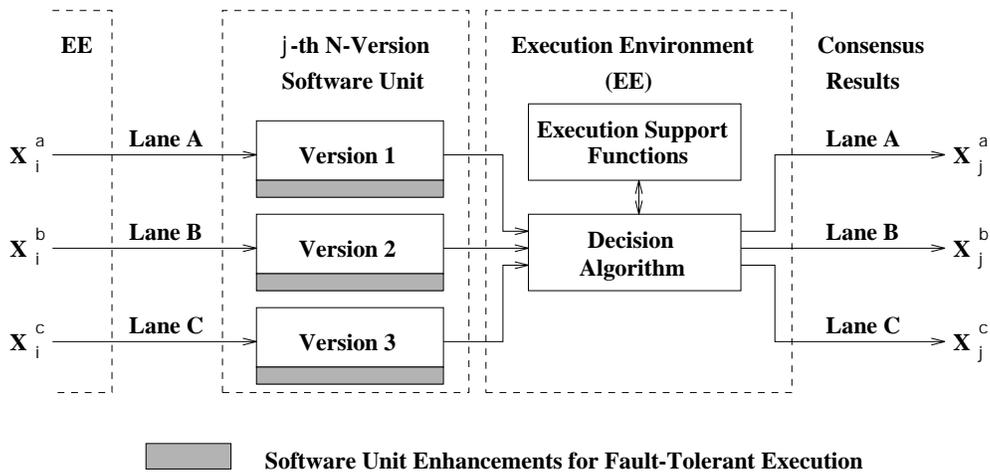

Figure 3.3: The $N$-Version Software model when $N = 3$. The execution environment is charged with the management of the decision algorithm and with the execution support functions. The user is responsible for supplying the $N$ versions. Note how the Decision Algorithm box takes care also of multiplexing its output onto the three hardware channels—also called "lanes" (picture from [Avi95]).

• RB allows different correct outputs from the alternates, while the general-purpose character of the consensus algorithm of NVP calls for a single correct output[4].

The two models collapse when the acceptance test of RB is done as in NVP, i.e., when the acceptance test is consensus on the basis of the outputs of the different alternates.

**Conclusions.** As RB, also NVP has been successfully adopted for many years in various application fields, including safety-critical airborne and spaceborne applications. The generic NVP architecture, based on redundancy and consensus, addresses parallel and distributed applications written according any programming paradigm. A generic, parametrisable architecture for real-time systems that supports straightforwardly the NVP scheme is GUARDS [PABD+99].

It is noteworthy to remark that the EE (also known as $N$-Version Executive) is a *complex* component that needs to manage a number of basic functions, for instance the execution of the decision algorithm, the assurance of input consistency for all versions, the inter-version communication, the version synchronisation and the enforcement of timing constraints [Avi95]. On the other hand, *this complexity is not part of the application software*—the $N$ versions—that *does not need to be aware of the fault-tolerance scheme*. An excellent degree of transparency can be reached, thus guaranteeing a good value for attribute SC. Furthermore, as mentioned in Sect. 1.1.2, costs and times required by a thorough verification, validation, and testing of this architectural complexity may be worth pursuing, while charging them to each application component is not.

---

[4]This weakness of NVP can be narrowed, if not solved, adopting the approach used in the so-called "voting farm" [DFDL98c, DFDL98a, DF97b], a generic voting tool designed by the author of this dissertation in the framework of his participation to the ESPRIT project "EFTOS" (see Sect. 3.1.1 at page 26): such a tool works with opaque objects that are compared by means of a user-defined function. This function returns an integer value representing a "distance" between any two objects to be voted. The user may choose between a set of predefined distance functions or may develop an application-specific distance function. Doing the latter, a distance may be endowed with the ability to assess that bitwise different objects are semantically equivalent. Of course, the user is still responsible for supplying a bug-free distance function—though is assisted in this simpler task by a number of template functions supplied with that tool.

What attribute SA is concerned, the same considerations provided when describing the RB scheme holds for NVP: the latter only covers one fault-tolerance scheme as well. One can conclude that NVP reaches unsatisfactory values of SA.

Off-line adaptability to "worse" environments may be reached increasing the value of $N$—though this requires developing new versions—a costly activity for both times and costs. Furthermore, the architecture does not allow any dynamic management of the fault-tolerance provisions. Attribute A can be then assessed as being poorly addressed by NVP.

Portability is restricted by the portability of the EE and of each of the $N$ versions. Maintainability actions may also be problematic, as they need to be replicated and validated $N$ times—as well as performed according to the NVP paradigm, so not to impact negatively on statistical independence of failures. These latter considerations apply to RB as well of course. In other words, the adoption of multiple-version software fault-tolerance provisions always implies a penalty on maintainability and portability.

Limited NVP support has been developed for "conventional" programming languages such as C. For instance, `libft` (see Sect. 3.1.1, p. 28) implements NVP as follows:

```
#include <ftmacros.h>
...
NVP
VERSION{
            block 1;
            SENDVOTE(v_pointer, v_size);
}
VERSION{
            block 2;
            SENDVOTE(v_pointer, v_size);
}
...
ENDVERSION(timeout, v_size);
if (!agreeon(v_pointer)) error_handler();
ENDNVP;
```

Note that this particular implementation extinguishes the potential transparency that in general characterises NVP, as it requires some non-functional code to be included. This translates in an unsatisfactory value for attribute SC. Note also that the execution of each block is in this case carried out sequentially.

It is important to remark how the adoption of NVP as a system structure for ALFT requires a substantial increase in development and maintenance costs. The author of the scheme claims that such costs are well payed back by the gain in trustworthiness. This is certainly true when dealing with systems possibly subjected to catastrophic failures—let us recall once more the case of the Ariane 5 flight 501 [Inq96]. Nevertheless, the risks related to the chances of rapid exhaustion of redundancy, due to a burst of correlated failures caused by a single or a few design faults, justifies and calls for the adoption of other fault-tolerance provisions within and around the NVP unit, in order to deal with the case of a failed NVP unit.

## 3.2 Metaobject Protocols and Reflection

Some of the negative aspects pointed out while describing single and multiple version software approaches can be in some cases weakened, if not solved, by means of a generic structuring technique which allows to reach in some cases an adequate degree of flexibility, transparency, and separation of design concerns: the adoption of *metaobject protocols* (MOPs) [KdRB91]. The idea is to "open" the implementation of the run-time executive of an object-oriented language like C++ or Java so that the developer can adopt and program different, custom semantics, adjusting the language to the needs of the user and to the requirements of the environment. Using MOPs, the programmer can modify the behaviour of fundamental features like methods invocation, object creation and destruction, and member access. The transparent management of spatial and temporal redundancy [TMB80] is a context where MOPs seem particularly adequate.

The key concept behind MOPs is that of *computational reflection*, or the causal connection between a system and a meta-level description representing structural and computational aspects of that system [Mae87]. MOPs offer the metalevel programmer a representation of a system as a set of *metaobjects*, i.e., objects that represent and reflect properties of "real" objects, i.e., those objects that constitute the functional part of the user application. Metaobjects can for instance represent the structure of a class, or object interaction, or the code of an operation. This mapping process is called *reification* [Rob99].

The causality relation of MOPs could also be extended to allow for a dynamical reorganisation of the structure and the operation of a system, e.g., to perform reconfiguration and error recovery. The basic object-oriented feature of inheritance can be used to enhance the reusability of the FT mechanisms developed with this approach.

**Project FRIENDS.** An architecture supporting this approach is the one developed in the framework of project FRIENDS [FP96, FP98]. Name FRIENDS stands for "flexible and reusable implementation environment for your next dependable system". This project aims at implementing a number of fault-tolerance provisions (e.g., replication, group-based communication, synchronisation, voting... [VA97]) at meta-level. In FRIENDS a distributed application is a set of objects interacting via the proxy model, a proxy being a local intermediary between each object and any other (possibly replicated) object. FRIENDS uses the metaobject protocol provided by Open C++, a C++ preprocessor that provides control over instance creation and deletion, state access, and invocation of methods.

Other ALFT architectures, exploiting the concept of metaobject protocols within custom programming languages, are reported in Sect. 3.3.

**Conclusions.** MOPs are indeed a promising system structure for embedding different non-functional concerns in the application-level of a computer program. MOPs work at *language* level, providing means to modify the semantics of basic object-oriented language building blocks like object creation and deletion, calling and termination of class methods, and so forth. This appears to match perfectly to a proper subset of the possible fault-tolerance provisions, especially those such as transparent object redundancy, which can be straightforwardly managed with the metaobject approach. When dealing with these fault-tolerance provisions, MOPs provide a perfect separation of the design concerns, i.e., optimal SC. Some other techniques, specifically those who might be described as "the most coarse-grained ones", such as distributed recovery blocks [KW89], appear to be less suited for being efficiently implemented via MOPs. These techniques work at distributed, i.e., macroscopic, level.

The above situation reminds the author of this dissertation of another one, regarding the "quest" for a novel computational paradigm for parallel processing which would be capable of dealing effectively with the widest class of problems, like the Von Neumann paradigm does for sequential processing, though with the highest degree of efficiency and the least amount of changes in the original (sequential) user code. In that context, the concept of computational *grain* came up—some techniques were inherently looking at the problem "with coarse-grained glasses," i.e., at macroscopic level, others were considering the problem exclusively at microscopical level. One can conclude that MOPs offer an elegant system structure to embed a set of non-functional services (including fault-tolerance provisions) in an object-oriented program. It is still unclear whether this set is general enough to host, *efficaciously*, *many* forms of fault-tolerance, as is remarked for instance in [RX95, LVL00]. It is therefore difficult to establish a qualitative assessment of attribute SA for MOPs.

The run-time management of libraries of MOPs may be used to reach satisfactory values for attribute A. To the knowledge of the author of this dissertation, this feature is not present in any language supporting MOPs.

As evident, the target application domain is the one of object-oriented applications written with languages extended with a MOP, such as Open C++.

## 3.3  Enhancing or Developing Fault-Tolerance Languages

Another approach is given by working at language level enhancing a pre-existing programming language or developing an ad hoc distributed programming language so that it hosts specific fault-tolerance provisions. The following two sections cover these topics.

### 3.3.1  Enhancing Pre-existing Programming Languages

Enhancing a pre-existing programming language means upgrading the grammar of a wide-spread language such as C or Pascal so that it directly supports features that can be used to enhance the dependability of its programs, e.g., recovery blocks [Shr78].

In the following, four classes of systems based on this approach are presented: Arjuna, Sina, Linda, and FT-SR.

#### 3.3.1.1 The Arjuna Distributed Programming System

Arjuna is an object-oriented system for portable distributed programming in C++ [Shr95]. It can be considered as a clever blending of useful and widespread tools, techniques, and ideas—as such, it is a good example of the evolutionary approach towards application-level software fault-tolerance. It exploits remote procedure calls [BN84] and UNIX daemons [HS87]. On each node of the system an *object server* connects client objects to objects supplying services. The object server also takes care of spawning objects when they are not yet running (in this case they are referred to as "passive objects"). Arjuna also exploits a "naming service", by means of which client objects request a service "by name". This transparency effectively supports object migration and replication. Arjuna offers the programmer means for dealing with atomic actions (via the two-phase commit protocol) and persistent objects. Unfortunately, it requires the programmers to explicitly deal with tools to save and restore the state, to manage locks, and to declare in their applications instances of the class for managing atomic actions. As its authors state, in many respects Arjuna asks the programmer to be aware of several complexities—as such, it is prejudicial to transparency and separation of design concerns. On the other hand, its good design choices result in an effective, portable environment.

#### 3.3.1.2 The SINA Extensions

The SINA [ADT91] object-oriented language implements the so-called composition filters object model, a modular extension of the object model. In SINA, each object is equipped with a set of "filters". Messages sent to any object are trapped by the filters of that object. These filters possibly manipulate the message before passing it to the object. SINA is a language for composing such filters—its authors refer to it as a "composition filter language". It also supports meta-level programming through the reification of messages. The concept of composition filters allows to implement several different "behaviours" corresponding to different non-functional concerns. SINA has been designed for being attached to existing languages: its first implementation, SINA/st, was for Smalltalk. It has been also implemented for C++ [Gla95]—the extended language has been called C++/CF. A preprocessor is used to translate a C++/CF source into standard C++ code.

#### 3.3.1.3 Fault-Tolerant Linda Systems

The Linda [CG89b, CG89a] approach adopts a special model of communication, known as *generative communication* [Gel85]. According to this model, communication is still carried out through messages, though messages are not sent to one or more addressees, and eventually read by these latter—on the contrary, messages are included in a distributed (virtual) shared memory, called tuple space, where every Linda process has equal read/write access rights. A tuple space is some sort of a shared relational database for storing and withdrawing special data objects called tuples, sent by the Linda processes. Tuples are basically lists of objects identified by their contents, cardinality and type. Two tuples match if they have the same number of objects, if the objects are pairwise equal for what concerns their types, and if the memory cells associated to the objects are bitwise equal. A Linda process inserts, reads, and withdraws tuples via blocking or non-blocking primitives. Reads can be performed supplying a template tuple—a prototype tuple consisting of constant fields and of fields that can assume any value. A process trying to access a missing tuple via a blocking primitive enters a wait state that continues until any tuple matching its template tuple is added to the tuple space. This allows processes to synchronise. When more than one tuple matches a template, the choice of which actual tuple to address is done in a non-deterministic way. Concurrent execution of processes is supported through the concept of "live data structures": tuples requiring the execution of one or more functions can be evaluated on different processors—in a sense, they become active, or "alive". Once the evaluation has finished, a (no more active, or passive) output tuple is entered in the tuple space.

Parallelism is implicit in Linda—there is no explicit notion of network, number and location of the system processors, though Linda has been successfully employed in many different hardware architectures and many applicative domains, resulting in a powerful programming tool that sometimes achieves excellent speedups without affecting portability issues. Unfortunately the model does not cover the possibility of failures—for instance, the semantics of its primitives are not well defined in the case of a processor crash, and no fault-tolerance means are part of the model. Moreover, in its original form, Linda only offers single-op atomicity [BS95], i.e., atomic execution for only a single tuple space operation. With single-op atomicity it is not possible to solve problems arising in two common Linda programming paradigms when faults occur: both the distributed variable and the replicated-worker paradigms can fail [BS95]. As a consequence, a number of possible improvements have been investigated to support fault-tolerant parallel programming in Linda. Apart from design choices and development issues, many of them implement stability of the tuple space (via replicated state machines [Sch90] kept consistent via ordered atomic multicast [BSS91]) [BS95, XL89, PTHR93], while others aim at combining multiple tuple-space operations into atomic transactions [BS95, AS91, CD92]. Other techniques have also been used, e.g., tuple space checkpoint-and-rollback [Kam91]. The author of this dissertation also proposed an augmented Linda model for solving inconsistencies related to failures occurring in a replicated-worker environment and an algorithm for implementing a resilient replicated worker scheme for message-passing farmer-worker applications. This algorithm can masking failures affecting a proper subset of the set of workers [DFDL99].

Linda can be described as an extension that can be added to an existing programming language. The greater part of these extensions requires a preprocessor translating the extension in the host language. This is the case, e.g., for FT-Linda [BS95], PvmLinda [DFMS94], C-Linda [Ber89], and MOM [AS91]. A counterexample is, e.g., the POSYBL system [Sch91], which implements Linda primitives with remote procedure calls, and requires the user to supply the ancillary information for distinguishing tuples.

#### 3.3.1.4 FT-SR

FT-SR [ST95] is basically an attempt to augment the SR [AO93] distributed programming language with mechanisms to facilitate fault-tolerance. FT-SR is based on the concept of fail-stop modules (FSM). A FSM is defined as an abstract unit of encapsulation. It consists of a number of threads that export a number of operations to other FSMs. The execution of operations is atomic. FSM can be composed so to give rise to complex FSMs. For instance it is possible to replicate a module $n > 1$ times and set up a complex FSM that can survive to $n - 1$ failures. Whenever a failure exhausts the redundancy of a FSM, be that a simple or complex FSM, a failure notification is automatically sent to a number of other FSMs so to trigger proper recovery actions. This feature explains the name of FSM: as in fail-stop processors, either the system is correct or a notification is sent and the system stops its functions. This means that the computing model of FT-SR guarantees, to some extent, that in the absence of explicit failure notification, commands can be assumed to have been processed correctly. This greatly simplifies program development because it masks the occurrence of faults, offers guarantees that no erroneous results are produced, and encourages the design of complex, possibly *dynamic* failure semantics (see Sect. 2.2) based on failure notifications. Of course this strategy is fully effective only under the hypothesis of perfect failure detection coverage—an assumption that sometimes may be found to be false.

**Conclusions.** It is worth remarking how the approach of designing fault-tolerance enhancements for a pre-existing programming language implies an *explicit code intrusion*: the extensions are designed purposely in order to explicitly host a number of fault-tolerance provisions within the single source code. Being explicit, this code intrusion is such that the fault-tolerance code be generally easy to locate and distinguish from the functional code of the application. Hence, attribute SC may be positively assessed for systems belonging to this category.

On the contrary, the problem of hosting an adequate structure for ALFT can be complicated by the syntax constraints in the hosting language. This may prevent to incorporate a wide set of fault-tolerance provisions within the same syntactical structure. One can conclude that attribute SA does not reach satisfactory values—at least in the cases reviewed in this section.

Enhancing a pre-existing language is an *evolutionary approach*: doing so, portability problems are weakened—especially when the extended grammar is translated into the plain grammar, e.g., via a preprocessor—and can be characterised by good execution efficiency [ABHM85, ST95].

The approach is generally applicable, though the application must be written (or rewritten) using the enhanced language. Its adaptability (attribute A) is unsatisfactory, because at run-time the fault-tolerance code is indistinguishable from the functional code.

### 3.3.2 Developing Novel Fault-Tolerance Programming Languages

The adoption of a custom-made language especially conceived to write fault-tolerant distributed software is discussed in the rest of this subsection.

### 3.3.2.1   ARGUS

Argus [Lis88] is a distributed object-oriented programming language and operating system. Argus was designed to support application programs like banking systems. To capture the object-oriented nature of such programs, it provides a special kind of objects, called guardians, which perform user-definable actions in response to remote requests. To solve the problems of concurrency and failures, Argus allows computations to run as atomic transactions. Argus' target application domain is the one of transaction processing.

### 3.3.2.2   The Correlate Language

The Correlate object-oriented language [Rob99] adopts the concept of *active object*, defined as an object that has control over the synchronisation of incoming requests from other objects. Objects are active in the sense that they do not process immediately their requests—they may decide to delay a request until it is accepted, i.e., until a given precondition (a guard) is met—for instance, a mailbox object may refuse a new message in its buffer until an entry becomes available in it. The precondition is a function of the state of the object and the invocation parameters—it does not imply interaction with other objects and has no side effects. If a request cannot be served according to an object's precondition, it is saved into a buffer until it becomes servable, or until the object is destroyed. Conditions like an overflow in the request buffer are not dealt with in [Rob99]. If more than a single request becomes servable by an object, the choice is made non-deterministically. Correlate uses a communication model called "pattern-based group communication"—communication goes from an "advertising object" to those objects that declare their "interest" in the advertised subject. This is similar to Linda's model of generative communication, introduced in Sect. 3.3.1.3. Objects in Correlate are autonomous, in the sense that they may not only react to external stimuli but also give rise to autonomous operations motivated by an internal "goal". When invoking a method, the programmer can choose to block until the method is fully executed (this is called synchronous interaction), or to execute it "in the background" (asynchronous interaction). Correlate supports MOPs. It has been effectively used to offer transparent support for transaction, replication, and checkpoint-and-rollback. The first implementation of Correlate consists of a translator to plain Java plus an execution environment, also written in Java.

### 3.3.2.3   Fault-Tolerance Attribute Grammars

The system models for application-level software fault-tolerance encountered so far all have their basis in an imperative language. A different research trend exists, which is based on the use of functional languages. This choice translates in a program structure that allows a straightforward inclusion of fault-tolerance means, with high degrees of transparency and flexibility. Functional models that appear particularly interesting as system structures for software fault-tolerance are those based on the concept of *attribute grammars* [Paa95]. This paragraph briefly introduces the model known as FTAG (fault-tolerant attribute grammars) [SKS96], which offers the designer a large set of fault-tolerance mechanisms. A noteworthy aspect of FTAG is that its authors explicitly address the problem of providing a syntactical model for the widest possible set of fault-tolerance provisions and paradigms, developing coherent abstractions of those mechanisms while maintaining the linguistic integrity of the adopted notation. This means that optimising the value of attribute SA is one of the design goals of FTAG.

FTAG regards a computation as a collection of pure mathematical functions known as *modules*. Each module has a set of input values, called inherited attributes, and of output variables, called synthesised attributes. Modules may refer to other modules. When modules do not refer any other module, they can be performed immediately. Such modules are called primitive modules. On the other hand, non-primitive modules require other modules to be performed first—as a consequence, an FTAG program is executed by decomposing a "root" module into its basic sub-modules and then applying recursively this decomposition process to each of the submodules. This process goes on until all primitive modules are encountered and executed. The execution graph is clearly a tree called *computation tree*. This scheme presents many benefits, e.g., as the order in which modules are decomposed is exclusively determined by attribute dependencies among submodules, a computation tree can be mapped onto a parallel processing means straightforwardly.

The linguistic structure of FTAG allows the integration of a number of useful fault-tolerance features that address the whole range of faults—design, physical, and interaction faults. One of this features is called *redoing*. Redoing replaces a portion of the computation tree with a new computation. This is useful for instance to eliminate the effects of a portion of the computation tree that has generated an incorrect result, or whose executor has crashed. It can be used to implement easily "retry blocks" and recovery blocks by adding ancillary modules that test whether the original module behaved consistently with its specification and, if not, give rise to a "redoing", a recursive call to the original module.

Another relevant feature of FTAG is its support for *replication*, a concept that in FTAG translates into a decomposition of a module into $N$ identical submodules implementing the function to replicate. The scheme is known as *replicated decomposition*, while involved submodules are called *replicas*. Replicas are executed according to the usual rules of decomposition, though only one of the generated results is used as the output of the original module. Depending on the chosen fault-tolerance strategy, this output can be, e.g., the first valid output or the output of a demultiplexing function, e.g., a voter. It is worth remarking that no syntactical changes are needed, only a subtle extension of the interpretation so to allow the involved submodules to have the same set of inherited attributes and to generate a collated set of synthesised attributes.

FTAG stores its attributes in a stable object base or in primary memory depending on their criticality—critical attributes can then be transparently retrieved from the stable object base after a failure. Object versioning is also used, a concept that facilitates the development of checkpoint-and-rollback strategies.

FTAG provides a unified linguistic structure that effectively supports the development of fault-tolerant software. Conscious of the importance of supporting the widest possible set of fault-tolerance means, its authors report in the cited paper how they are investigating the inclusion of other fault-tolerance features and trying to synthesise new expressive syntactical structures for FTAG—thus further improving attribute SA.

Unfortunately, the widespread adoption of this valuable tool is conditioned by the limited acceptance and spread of the functional programming paradigm outside the academia.

**Conclusions.**   The ad hoc development of a fault-tolerance programming language allows in some cases to reach optimal values for attribute SA. The explicit, controlled intrusion of fault-tolerance code *explicitly encourages* the adoption of high-level fault-tolerance provisions and *requires dependability-aware design processes*, which translates in a positive assessment for attribute SC.  On the contrary, with the same reasoning of Sect. 3.3.1, attribute A can be in general assessed as unsatisfactory[5].

The target application domain for this approach is restricted by the characteristics of the hosting language and of its programming model.  Obviously this also requires the application to be (re-)written using the hosting language.

## 3.4   Aspect-oriented Programming Languages

Aspect-oriented programming [KLM+97] (AOP) is a programming methodology and a structuring technique that explicitly addresses, at system-wide level, the problem of the best code structure to express different, possibly conflicting design goals such as high performance, optimal memory usage, and dependability.

Indeed, when coding a non-functional service within an application—for instance a system-wide error handling protocol—using either a procedural or an object-oriented programming language, one is required to decompose the original goal, in this case a certain degree of dependability, into a multiplicity of fragments scattered among a number of procedures or objects. This happens because those programming languages only provide abstraction and composition mechanisms to cleanly support the *functional* concerns. In other words, specific non-functional goals, such as high performance, cannot be easily captured into a single unit of functionality among those offered by a procedural or object-oriented language, and must be *fragmented* and *intruded* into the available units of functionality.  As already observed, this code intrusion is detrimental to maintainability and portability of both functional and non-functional services (the latter called "aspects" in AOP terms).  These aspects tend to crosscut the system's class and module structure rather than staying, well localised, within one of these unit of functionality, e.g., a class.  This increases the complexity of the resulting systems.

The main idea of AOP is to use:

1. A "conventional" language (that is, a procedural, object-oriented, or functional programming language) to code the basic functionality. The resulting program is called *component program*. The program's basic functional units are called *components*.

2. A so-called *aspect-oriented language* to implement given aspects by defining specific interconnections ("aspect programs" in AOP lingo) among the components in order to address various systemic concerns.

---

[5]In the case of FTAG, though, one could design a run-time interpreter that could dynamically "decide" the "best" values for the parameters of the fault-tolerance provisions being executed—where "best" refers to the current environmental conditions.  This same idea, which is an original contribution of the author of this dissertation, is further detailed and discussed in Chapter 8, where an architecture for adaptable ALFT is also sketched.

3. An *aspect weaver*, that takes as input both the aspect and the component programs and produces with those ("weaves") an output program ("tangled code") that addresses specific aspects.

The weaver first generates a data flow graph from the component program. In this graph, nodes represent components, and edges represent data flowing from one component to another. Next, it executes the aspect programs. These programs edit the graph according to specific goals, collapsing nodes together and adjusting the corresponding code accordingly. Finally, a code generator takes the graph resulting from the previous step as its input and translates it into an actual software package written, e.g., for a procedural language such as C. This package is only meant to be compiled and produce the ultimate executable code fulfilling a specific aspect like, e.g., higher dependability.

In a sense, AOP systematically automatises and supports the process to adapt an existing code so that it fulfils specific aspects. AOP may be defined as a software engineering methodology supporting those adaptations in such a way that they do not destroy the original design and do not increase complexity. The original idea of AOP is a clever blending and generalisation of the ideas that are at the basis, for instance, of optimising compilers, program transformation systems, MOPs, and of literate programming [Knu84].

**AspectJ** is an example of aspect-oriented language [Kic00, LVL00]. Developed as a Xerox PARC project, AspectJ can be defined as an aspect-oriented extension to the Java programming language. AspectJ provides its users with the concept of a "join points", i.e., relevant points in a program's dynamic call graph. Join points are those that mark the code regions that can be manipulated by an aspect weaver (see above). In AspectJ, these points can be

- method executions,

- constructor calls,

- constructor executions,

- field accesses, and

- exception handlers.

Another extension to Java is AspectJ's support of the Design by Contract methodology [Mey97], where *contracts* [Hoa69] define a set of pre-conditions, post-conditions, and invariants, that *determine how to use* and *what to expect* from a computational entity.

A study has been carried out on the capability of AspectJ as an AOP language supporting exception detection and handling [LVL00]. It has been shown how AspectJ can be used to develop so-called "plug-and-play" exception handlers: libraries of exception handlers that can be plugged into many different applications. This translates into better support for managing different configurations *at compile-time*. This addresses one of the aspects of attribute A defined in Sect. 1.1.2.

**Conclusions.**    AOP is a recent approach to software development. AOP can in principle address any application domain and can use a procedural, functional or object-oriented programming language as component language. The isolation and coding of aspects requires extra work and expertise that may be well payed back by the capability of addressing new aspects while keeping a single unmodified and general design.

For the time being it is not yet possible to tell whether AOP will spread out as a programming paradigm among academia and industry the way object-oriented programming has done since the Eighties. The many qualities of AOP are currently being quantitatively assessed, both with theoretical studies and with practical experience, and results seem encouraging. From the point of view of the dependability aspect, one can observe that AOP exhibits optimal SC ("by construction", in a sense), and that attribute A can be in principle optimised, e.g., by means of specific run-time weaving. The adequacy at fulfilling attribute SA is indeed debatable also because, to date, no fault-tolerance aspect languages have been devised[6]—which may possibly be an interesting research domain.

## 3.5   The Recovery Meta-Program

The Recovery Meta-Program (RMP) [ADG+90] is a mechanism that alternates the execution of two cooperating processing contexts. The concept behind its architecture can be captured by means of the idea of a debugger, or a monitor, which:

- is scheduled when the application is stopped at some *breakpoints*,

- executes some sort of a *program*, written in a specific language,

- and finally returns the control to the application context, until the next breakpoint is encountered.

Breakpoints outline portions of code relevant to specific fault-tolerance strategies—for instance, breakpoints can be used to specify alternate blocks or acceptance tests of recovery blocks (see Sect. 3.1.2.1)—while programs are implementations of those strategies, e.g., of recovery blocks or $N$-version programming. The main benefit of RMP is in the fact that, while breakpoints require a (minimal) intervention of the functional-concerned programmer, RMP scripts can be designed and implemented without the intervention and even the awareness of the developer. In other words, RMP guarantees a good separation of design concerns. As an example, recovery blocks are implemented, from the point of view of the functionally concerned designer, specifying alternates and acceptance tests, while the execution goes like in Fig. 3.4:

- When the system encounters a breakpoint corresponding to the entrance of a recovery block, control flows to the RMP, which saves the application program environment and starts the first alternate.

---

[6]AspectJ only addresses exception error detection and handling. Remarkably enough, the authors of a recent study on AspectJ and its support to this field conclude [LVL00] that "whether the properties of AspectJ [documented in this paper] lead to programs with fewer implementation errors and that can be changed easier, is still an open research topic that will require serious usability studies as AOP matures".

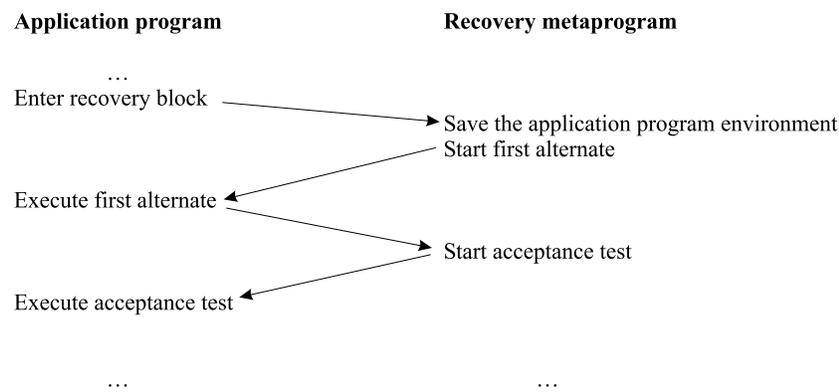

**Figure 3.4: Control flow between the application program and RMP while executing a fault-tolerance strategy based on recovery blocks.**

- The execution of the first alternate goes on until its end, marked by another breakpoint. The latter returns the control to RMP, this time in order to execute the acceptance test.

- Should the test succeed, the recovery block is exited, otherwise control goes to the second alternate, and so forth.

In RMP, the language to express the meta-programs is Hoare's Communicating Sequential Processes language [Hoa78] (CSP).

**Conclusions.** In the RMP approach, all the technicalities related to the management of the fault-tolerance provisions are coded in a separate programming context. Even the language to code the provisions may be different from the one used to express the functional aspects of the application. One can conclude that RMP is characterised by optimal SC.

The design choice of using CSP to code the meta-programs influences negatively attribute SA. Choosing a pre-existent formalism clearly presents many practical advantages, though it means adopting a fixed, immutable syntactical structure to express the fault-tolerance strategies. The choice of a pre-existing general-purpose distributed programming language as CSP is therefore questionable, as it appears to be rather difficult or at least cumbersome to use it to express at least some of the fault-tolerance provisions. For instance, RMP proves to be an effective linguistic structure to express strategies such as recovery blocks and $N$-version programming, where the main components are coarse grain processes to be arranged into complex fault-tolerance structures. Because of the choice of a pre-existing language like CSP, RMP appears not to be the best choice for representing provisions such as, e.g., atomic actions [JC85]. This translates in very limited SA.

Our conjecture is that the coexistence of two separate layers for the functional and the non-functional aspects could have been better exploited to reach the best of the two approaches: using a widespread programming language such as, e.g., C, for expressing the functional aspect, while devising a custom language for dealing with non-functional requirements, e.g., a language especially designed to express error recovery strategies. This design choice has been taken in the approach described in this dissertation.

| Approach | SC | SA | A |
|----------|----|----|---|
| SV | 1 | 2 | 1 |
| MV (RB) | 1 | 1 | 1 |
| MV (NVP) | 4 | 1 | 1 |
| MOP | 5 | 3 | 3 |
| EL | 3 | 1 | 1 |
| DL | 3 | 5 | 1 |
| AOP | 5 | 3 | 4 |
| RMP | 5 | 2 | 1 |

Table 3.1: A quantitative interpretation of the qualitative assessments proposed in Chapter 3. The mapping between adjectives and numbers is as follows: "poor" and "unsatisfactory" map to 1, "very limited" is 2, "medium" and "positive" are 3, "good" is 4, and "optimal" is 5. MV has been differentiated into RB and NVP. EL stands for the approach in Sect. 3.3.1, DL is the approach in Sect. 3.3.2.

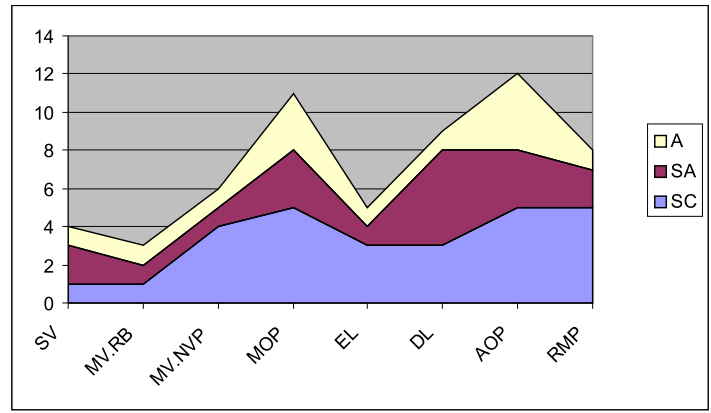

Figure 3.5: The data in Table 3.1 are here represented as a stacked area.

Satisfactory values for attribute A cannot be reached in the only implementation available of RMP [ADG⁺90], because no dynamic management of the executable code has been foreseen.

The RMP approach appears to be characterised by a large overhead due to frequent context switching between the main application and the RMP [RX95]. Run-time requirements may be jeopardised by these large overheads, especially when it is difficult to establish time bounds for their extent. No other restrictions appear to be posed by RMP on the target application domain.

## 3.6 Conclusions

Six approaches to ALFT have been described and critically reviewed, qualitatively, with respect to the structural attributes SC, SA, and A. Table 3.1 and Fig. 3.5 summarise this survey providing a comparison of the six approaches. This chapter has also remarked the positive and negative issues of the evolutionary solution—using a pre-existing language—with respect to a "revolutionary" approach—based on devising a custom made, ad hoc language. It has been conjectured how, using an approach based on two languages, one covering the functional concerns and the other covering the fault-tolerance concerns, it may be possible to address, within one efficacious linguistic structure, the widest set of fault-tolerance provisions, reaching optimal values for SA and SC.

# Chapter 4

# The Recovery Language Approach

This chapter introduces a novel approach towards software engineering of dependable systems that has been called the "recovery language approach" ($\mathcal{R}\mathcal{L}$). The elements of $\mathcal{R}\mathcal{L}$ are introduced in general terms, coupling each concept to the technical foundations buttressing it. The chapter casts the basis of a general approach in abstract terms, while a particular instance of the herein presented concepts is described in Chapter 5 as a prototypic distributed architecture supporting a fault-tolerance linguistic structure for ALFT. System, application and fault models are drawn. The approach is also reviewed with respect to the structural attributes (SC, SA and A) and to the approaches presented in Chapter 3. All the concepts presented in this chapter are original ideas of the author of this dissertation.

The structure of this chapter is as follows:

- Models are introduced in Sect. 4.1.

- Key ideas, concepts, and technical foundations are described in Sect. 4.2.

- Section 4.3 shows the workflow corresponding to using $\mathcal{R}\mathcal{L}$.

- Section 4.4 points out the *specific differences* between $\mathcal{R}\mathcal{L}$ and other system structures for ALFT.

- Sect. 4.5 closes this chapter summarising the positive values of the structural attributes SA, SC, and A for $\mathcal{R}\mathcal{L}$.

## 4.1 System, Application and Fault Models

This section introduces the system and application models that will be assumed in the rest of this dissertation.

### 4.1.1 System Assumptions

In the following, the target system is assumed to be a distributed or parallel system. Basic components are nodes, tasks, and the network.

- A node can be, e.g., a workstation in a networked cluster or a processor in a MIMD parallel computer.

- Tasks are independent threads of execution running on the nodes.

- The network system allows tasks on different nodes to communicate with each other.

Nodes can be commercial-off-the-shelf (COTS) hardware components with no special provisions for hardware fault-tolerance. It is not mandatory to have memory management units and secondary storage devices.

A general-purpose operating system (OS) is required on each node. No special purpose, distributed, or fault-tolerant OS is required.

The number $N$ of nodes is assumed to be known ahead of the run-time. Nodes are addressable by the integers in $\{0, \ldots, N-1\}$. For any integer $m > 0$ let us call the set of integers $\{0, \ldots, m-1\}$ as $I_m$. Let us furthermore refer to the node addressed by integer $i$ as $n_i, i \in I_N$.

Tasks are pre-defined at compile-time: in particular for each $i \in I_N$, it is known that node $n_i$ is to run $t_i$ tasks, up to a given node-specific limit. No special constraints are posed on the task scheduling policy.

On each node, say node $i$, tasks are identified by user-defined unique local labels—integers greater than or equal to zero. Let us call $I_{n_i}$ the set of labels for tasks to be run on node $n_i$, $i \in I_N$. The task with local label $j$ on node $i$ will be also referred to as $n_i[j]$.

The system obeys the *timed asynchronous distributed system model* [CF99]:

- Tasks communicate through the network via a datagram service with omission/performance failure semantics (see Sect. 2.1.3.1). Low level software in the nodes and the network implements this service.

- Services are timed: specifications prescribe not only the outputs and state transitions that should occur in response to inputs, but also the time intervals within which a client task can expect these outputs and transitions to occur.

- Tasks (including those related to the OS and the network) have crash/performance failure semantics.

- Tasks have access to a node-local hardware clock. If more than one node is present, clocks on different nodes have a bounded drift rate.

- A "time-out" service is available at application-level: tasks can schedule the execution of events so that they occur at a given future point in time, as measured by their local clock.

In particular, this model allows a straightforward modelling of system partitioning—as a consequence of sufficiently many omission or performance communication failures, correct nodes may be temporarily disconnected from the rest of the system during so-called periods of instability [CF99]. Moreover it is assumed that, at reset, tasks or nodes restart from a well-defined, initial state—partial-amnesia crashes (defined in Sect. 2.1.3.1) are not considered.

A message passing library is assumed to be available, built on the datagram service. Such library offers asynchronous, non-blocking multicast primitives.

As clearly explained in [CF99], the above hypotheses match well to nowadays distributed systems based on networked workstations—as such, they represent a general model with no practical restriction.

### 4.1.2 Application-specific Assumptions

The following assumptions characterise the user application:

- (When $N > 1$ nodes are available): the target application is distributed on the system nodes.

- It is written or is to be written in a procedural language such as, e.g., C.

- Inter-process communication takes place by means of the functions in the above mentioned message passing library. Higher-level communication services, if available, must be using the message passing library as well.

The reason behind the third assumption is that, forcing communication through a single virtual provision, namely the functions for sending and for receiving messages, allows a straightforward implementation of provisions for task isolation. This concept is explained in more detail in Sect. 5.3.3.

### 4.1.3 Fault Model

As suggested in the conclusion of Sect. 1.1.1.2, any effective design including dependability goals requires provisions, *located at all levels*, to avoid, remove, or tolerate faults. Hence, as an *application-level* structure, $\mathcal{REL}$ is complementary to other approaches addressing fault-tolerance at *system level*, i.e., hardware-level and OS-level fault-tolerance. In particular, a system-level architecture such as GUARDS [PABD+99], that is based on redundancy and hardware and OS provisions for systematic management of consensus, appears to be particularly appropriate for being coupled with $\mathcal{REL}$ which offers application-level provisions for NVP and replication (see later on).

The main classes of faults addressed by $\mathcal{REL}$ are those of accidental, permanent or temporary design faults, and temporary, external, physical faults. Both value and timing failures are considered. The architecture addresses one fault at a time: the system is ready to deal with new faults only after having recovered from the present one.

## 4.2 Key Ideas and Technical Foundations

Key concepts of $\mathcal{REL}$'s design are:

- Adopting a fault-tolerance toolset.

- Separating the configuration of the toolset from the specification of the functional service.

- Separating the system structure for the specification of the functional service from that for error recovery and reconfiguration.

These concepts and their technical foundations are illustrated in the rest of this section.

## 4.2.1   Adoption of a Fault-Tolerance Toolset

A requirement of $\mathcal{R}\mathcal{L}$ is the availability of a fault-tolerance **toolset**, to be interpreted herein as the conjoint adoption of:

- A set of fault-tolerance tools addressing error detection, localisation, containment and recovery, such as the ones in SwIFT [HKBW96] or EFTOS [DDFLV97, DVB+97]. Fault-tolerance services provided by the toolset include, e.g., watchdog timers and voting. These are called **basic tools** (BT) hereafter.

- A "**basic services library**" (BSL) is assumed to be present, providing functions for:

    - intra-node and remote communication;

    - task management;

    - access to the local clock;

    - application-level assertions;

    - functions to reboot or shut down a node.

    This library is required to be available in source code so that it can be instrumented, e.g., with code to forward information to some collector (described below) transparently. Information may include, for instance, the notification of a successful task creation or any failure of this kind.  If supported, MOPs may also be used to implement the library and its instrumentation. It is also suggested that the functions for sending messages work with opaque objects that reference either single tasks or groups of tasks.  In the first case, the function would perform a plain "send", while in the second case it would perform a multi-cast. This would increase the degree of transparency.

- A distributed component serving as a sort of backbone controlling and monitoring the toolset and the user application.  Let us call this application "the **backbone**" (BB). It is assumed that the BB has a component *on each node of the system* and that, through some software (and, possibly, hardware) fault-tolerance provisions, *it can tolerate crash failures of up to all but one node or component*.  An application such as the EFTOS DIR net [DTDF+99] may be used for this.

    Notifications from the BSL and from the BTs are assumed to be collected and maintained by the BB into a data structure called "the **database**" (DB). The DB therefore holds data related to the current *structure* and *state* of *the system, of the user application, and of the*

*BB*. A special section of the DB is devoted to keeping track of *error notifications*, such as, for instance, "divide-by-zero exception caught while executing task 11" sent by a trap handling tool. If possible, error detection support at hardware or kernel level may be also instrumented in order to provide the BB with similar notifications. The DB is assumed to be stored in a reliable storage device, e.g., a stable storage device, or replicated and protected against corruption or other unwanted modifications.

- Following the hypothesis of the timed asynchronous distributed system model [CF99], a time-out management system is also assumed to be available. This allows an application to define *time-outs*, namely, to schedule an event to be generated a given amount of "clock ticks" in the future [CS95]. Let us call this component the "**time-out manager**" (TOM).

A prototype of a $\mathcal{REL}$-compliant toolset has been developed within the ESPRIT project "TIRAN". Section 5.2 describes its main components.

## 4.2.2 Configuration Support Tool

The second key component of $\mathcal{REL}$ is a tool to support *fault-tolerance configuration*, defined herein as the deployment of customised instances of fault-tolerance tools and strategies. $\mathcal{REL}$ envisages a **translator** to help the user configure the toolset and his / her application. The translator has to support a custom **configuration language** especially conceived to facilitate configuration and therefore to reduce the probability of fault-tolerance design faults—the main cause of failure for fault-tolerant software systems [Lap98, Lyu98b].

As an output, the translator could issue, e.g., C or C++ header files defining configured objects and symbolic constants to refer easily to the configured objects. Recompilation of the target application is therefore required after each execution of the translator.

Configuration can group a number of activities, including:

- configuration of system and application entities,

- configuration of the basic tools,

- configuration of replicated tasks,

- configuration for retry blocks,

- configuration for multiple-version software fault-tolerance,

The above configuration activities are now briefly described.

### 4.2.2.1 Configuration of System and Application Entities

One of the tasks of a configuration language is to declare the key entities of the system and to define a global naming scheme in order to refer to them. Key entities are nodes, tasks, and groups of tasks.

For each node $n_i, 0 \leq i < N$ and for each task $n_i[j], 0 \leq j < I_{n_i}$, a unique, global-scope identifier must be defined by the user. Let us call this identifier the task's *unique-id*. This can be done by editing a configuration script with rules of the form

$$\text{task}_t \equiv n_i[j], \tag{4.1}$$

which assigns unique-id $t$ to $n_i[j]$, that is, task number $j$ on node $i$. Similarly one could define groups of task with rules of the form

$$\text{group}_g \equiv \{\vec{u}\}, \tag{4.2}$$

where $\vec{u}$ is a list of comma-separated integers representing unique-ids. Such rule defines then a group named $g$ made of the tasks corresponding to the mentioned unique-ids. The translator would then turn a configuration script containing rules of these kinds into a header file to be compiled with the target application and, if necessary, in configuration files expected by the toolset.

### 4.2.2.2   Configuration of the Fault-Tolerance Tools in the Toolset

Specific instances of the tools in the toolset can be statically configured by means of the translator. For instance, in the case of a watchdog timer, configuration can specify:

- The unique-id of the watching task as well as that of the watched task.

- The initial expected frequency of "heartbeats" to be sent from the watched task to the watchdog.

- The actions to be taken when an expected heartbeat is not received in time,

and so forth. Ideally, the output of the translator should be a configuration file for the BSL to associate transparently the creation of the configured instance of the watchdog to the creation of the watched task. Doing like this, the only non-functional code to be intruded in the watched task can be the function call corresponding to sending the heartbeat to the watchdog. A generic "HEARTBEAT" method, with no arguments, can be used. This can be a symbolic name properly defined in a header file written by the translator and automatically included when compiling the user application. Even when instrumenting is not possible, the watched task can start the watchdog by means of a symbolic name properly defined in the above mentioned header file.

These kind of translations can be applied to most of the tools of a library such as SwIFT or EFTOS (see Sect. 3.1.1). Note how the minimal or absent code intrusion provides an optimal SC. The adoption of a custom, ad hoc language for the expression of the configuration concerns can be used to reach high values for SA and compile-time A.

### 4.2.2.3   Configuration of Replicated Tasks

The translator and the BSL may be used to implement replicated tasks, i.e., multiple instances of the same task that perform like a more dependable entity. The goal of the translator is to mask this choice and any other fault-tolerance technicalities, including, in this specific case, replication. This can be obtained, e.g., by solving separately the following sub-problems, both at syntactical and at semantical level:

1. replication and forwarding of the input value,

2. execution support of a fault-tolerance strategy,

3. output management.

Problem 1 can be solved by defining a group-of-tasks object that, once passed to the BSL function for sending messages, triggers a multicast of the same message to the whole group, as suggested in Sect. 4.2.2.1.

Problem 2 can be solved in various ways—for instance, through a temporal redundancy scheme, executing the involved tasks one after the other, on the same node, or via spatial redundancy, executing tasks in parallel. Increased dependability may be obtained via a number of software techniques, implementing schemes such as active replication or primary-backup replication [GS97]. As noticed in the cited paper, each scheme has both positive and negative aspects and requires solving specific problems. All these problems are low-level design issues that can be made transparent to the user of an $\mathcal{REL}$ system. Other choices and options, e.g., which type and degree of replication and which redundancy scheme to adopt, that would result in non-functional code intrusion, can be also made transparent to the user by means of the translator (see Sect. 4.2.2.5), hence increasing configurability.

Problem 3, depending on the adopted scheme, can be as simple as sending a message (when in primary-backups mode) or could require special processing. For instance, in the case of active replication, two sub-problems would call for specific treatment:

3.1. Routing the outputs produced by the base tasks.

3.2. De-multiplexing, i.e., production of a unique output value from the values routed in sub-problem 3.1.

Sub-problem 3.1 can be solved, e.g., by a proper combination of pipelining and redirection, two basic programming tools that originated in UNIX environments. Other approaches may be used when the OS does not support the above tools. De-multiplexing, i.e., solution of sub-problem 3.2, can be for instance the result of a *voting procedure* performed among the outputs produced by the base tasks. Also in this case, the availability of a translator and an appropriate syntax rule could guarantee an almost complete separation of design concerns.

The configuration of a replicated task requires the specification of:

- The unique-id of a task globally representing a set of replicas.

- The unique-ids of the replicas.

- The replication method and its parameters.

- The actions to be taken when an output is produced by the replicated task.

• The actions to be taken when an error occurs.

In the case of replicated tasks, the translator is also responsible for the set up of a proper
run-time executive. The latter would then be responsible, at run-time, for the orchestration of
the services required by the replicated tasks—task management, distributed voting, and so forth.
Proper calls to the BSL and to instances of the basic tools may be used for this. A voting tool such
as the EFTOS voting farm [DFDL98a, DFDL98c] appear to be a natural choice, as it addresses
many of the required issues.

It is worth noting how, also in this case, full transparency is reached: a client of a replicated
task would have no way to tell whether its server be simple or replicated—possibly apart from
some performance penalty and a higher quality of service. This translates into optimal SC. The
same applies to SA and A for the reasons mentioned in Sect. 4.2.2.2.

### 4.2.2.4   Configuration for Retry Blocks

Transparent support for redoing (see Sect. 3.3.2.3), another important fault-tolerance provisions,
can be provided via "retry blocks". Again, a proper run-time executive is to be produced by the
translator. The problems to be solved at this level include

1. reversibility of a failed task and

2. input replication.

The first problem can be solved by implementing some "recovery cache" (as in the RB tech-
nique; see Sect. 3.1.2.2), that is, a mechanism to checkpoint the state of the calling task before
entering the retry block and to roll it back to its original value, in case the acceptance test fails.
This may be done transparently or with the intervention of the user. In the latter case, one re-
stricts the size of the recovery cache and reduces the corresponding overheads, at the same time
increasing the code intrusion.

The second problem could be solved via an "input cache", i.e., a mechanism that:

• Intercepts the original input message.

• Stores the original message into some stable means.

• And forwards the saved original message to each new retry instance.

Transparent adoption of an input cache can also be realized, e.g., by means of pipelining and
redirection (when the OS supports these).

The configuration of a retry block requires the specification of:

• The unique-id of a task to be retried in case of errors.

• An acceptance test, in the form of the name of a function returning a Boolean value or of
a task that, upon termination, returns a Boolean value[1].

---

[1]For instance, via the `exit` function call from the C standard libraries.

- A threshold $r$ representing the maximum number of retries.

- The actions to be taken when the base task fails for $r$ times in a row.

### 4.2.2.5 Configuration for Multiple-Version Software Fault-Tolerance

Compile-time support towards multiple-version software fault-tolerance can be provided by the translator through a syntax and techniques similar to those described for task replication.

The configuration of a provision for MV requires the specification of:

- The unique-id of a task representing the provision.

- The unique-id's of the tasks running the versions.

- A set of thresholds representing time-outs on the execution of the version tasks[2].

- The name of the user-specified function to be executed by each version task.

- A method to de-multiplex the multiple outputs produced by the versions into a single output value.

- Possible arguments to the de-multiplexing method.

- The unique-id of a task to be notified each time an execution cycle is successfully completed.

- The unique-id of a task to be notified each time an execution cycle fails.

Support towards consensus recovery blocks [SGM85] may be provided in a similar way. Acceptance tests should be specified as described in Sect. 4.2.2.4.

### 4.2.2.6 Example Scenario and Conclusions

A possible compile-time and run-time scenario is now described for the case of the configuration of multiple-version software fault-tolerance. This is done in order to provide the reader with a more concrete view of the kind of support supplied by a configuration language.

It is assumed that the OS supports pipelining and stream redirection. It is also assumed that, by agreement, the user tasks that are going to be used as NVP versions forward a single output value onto the standard output stream. Finally, user tasks are assumed to be side effect-free.

Once fed with a configuration script, the translator writes a number of source files for the tasks corresponding to the employed versions. Each of these source files, set up from some template file, specify how to:

- Set up a configured instance of a distributed voting tool (for instance, the EFTOS voting farm).

---

[2]Note how, according to the hypothesis of adherence to the timed-asynchronous distributed system model, such thresholds are known because all services are timed.

- Redirect standard output streams.

- Execute one of the version tasks.

During the execution, when a client needs to access a service supplied by the provisions, it simply sends a request message to the corresponding task. The client does not need to know that the latter is actually an "NVP task", that is, a group. Through the BSL, this sending turns into a multicast to the version tasks. These tasks, which in the meanwhile have transparently set up a distributing voting tool,

- get their input,

- compute some user-specified function,

- produce an output,

- and (by the above agreement) write that output into their standard output stream.

This output, which was already redirected through a piped stream to the template task, is fed into the voting system. The latter eventually produces an output that, in case of success, is sent to some output task with the notification of successful completion of a processing cycle.

Note that the client task of such an "NVP task" is completely unaware of the context in which it is running, with full transparency and separation of design concern. One can conclude that the adoption of a configuration strategy like the one just sketched can lead to an optimal SC. The variety of fault-tolerance provisions that can be supported by configuration and the adoption of a linguistic environment separated from the functional application layer can be exploited by programming language designers in order to attain optimal values for SA and A as well.

### 4.2.3   Recovery Languages

Two of the three key concepts of $\mathcal{R}\mathcal{L}$ have been described, namely the adoption of a fault-tolerance toolset and that of a configuration language. This section now introduces the third component of $\mathcal{R}\mathcal{L}$. This component supports two application layers, namely:

- the traditional user application layer, i.e., the one devoted to the fulfilment of the functional requirements and to the specification of an intended user service, supported by the run-time modules of conventional programming languages such as, e.g., C; and

- an ancillary application layer, specifically devoted to error processing and fault identification, to be switched in either *asynchronously*, when errors are detected in the system, or when the user *synchronously* signals particular run-time conditions, such as a failed assertion, or when the control flow runs into user-defined breakpoints.

Let us call **service language** the language constituting the front layer and **recovery language** the one related to error processing.

Note how a recovery language can be specified in a separate script. The latter can then be translated into pseudo-code to be interpreted at run-time, or, e.g., into plain C to be compiled with the user application. In the first case, the functional code and the recovery code are fully separated, also at run-time, which can be exploited to reach optimal A as explained in Chapter 8. The second case eliminates the overhead of interpreting the pseudo-code. The same **translator** used for fault-tolerance configuration may be used to generate the pseudo-code or the C source code.

The general strategy is as follows:

- During the lifetime of the application, *in the absence of errors*, the front layer controls the progress of the service supply while the BB collects and maintains in the DB the data concerning the state of each component of the application, the state of each node of the system, the state and progress of public and private resources, and so forth.

- *As soon as an error is detected* by a BT, such as a watchdog timer, the latter transparently forwards a notification of this event to the BB, which awakes the ancillary layer by enabling a module to interpret or execute the recovery code. Let us call this module **RINT**, for recovery interpreter.

A possible syntactical structure for the recovery language is that of a list of **guarded actions**, i.e., statements in the form

$$g : a,$$

where $g$ is a **guard**, i.e., a Boolean expression on the contents of the DB, and $a$ is one or more **actions** (to be specified further on). Hence, adequate paradigms for the recovery language could be that of *procedural* or *logic programming languages*. The following two paragraphs describe guards and actions in more detail.

### 4.2.3.1 Guards

Guards represent conditions that require recovery. As just said, they are Boolean expressions made of basic queries called **atoms**. Possible atoms may express conditions such as:

- task $t$ has been detected as faulty;

- task $t$ has been detected as faulty by error detection tool $d$;

- for $m$ times in a row, task $t$ has been detected as faulty;

- a time-out concerning task $t$ has expired;

- an $N$-version task has signalled that no full consensus has been reached, or in more detail:

  - for $m$ times in a row, the same version, say task $t'$, has been found in minority with respect to the other versions;

  - for $m$ times in a row, version $t'$ did not produce any output within its deadline;

- – and so forth.

- • node $n$ is down, or, in more detail:

    - – node $n$ has crashed;

    - – node $n$ is unreachable;

    - – no sign of life from node $n$ in the last $s$ seconds, as measured by the local clock;

    - – and so forth.

- • task $t$ could not be restarted;

- • some of the tasks in group $g$ are faulty.

and so forth. Most of the atoms in a Boolean clause will require a DB query or proper actions on DB fields—for instance, conditions like "task $t$ is affected by a transient fault" require the adoption, within the DB, of a thresholding statistical technique such as $\alpha$-count [PABD$^+$99, BCDGG97], which is capable of distinguishing between transient and permanent/intermittent faults. In particular, $\alpha$-count (described in Sect. 5.2.4.2) can be straightforwardly "wired" into the DB management system within the BB, because that mechanism is based on adjusting some counters according to the contents in a stream of error messages forwarded by a set of error detectors.

#### 4.2.3.2   Actions

Actions are local or remote commands to be executed when their guard is evaluated as true. Actions can specify, for instance, recovery or reconfiguration services. A special case of action can be also another guarded action—this allows to have nested guarded actions that can be represented as a tree. The execution of an inner guard is, again, its evaluation. When a parent guard is evaluated as false, all its actions are skipped, including its child guards.

Actions can include, for instance:

- • switching tasks in and out of a fault-tolerance structure such as, e.g., an "NVP task" (see Sect. 4.2.2.6);

- • synchronising groups of tasks;

- • isolating[3] groups of tasks;

- • instructions to roll the execution of a task or a group of tasks back to a previously saved checkpoint,

---

[3]"Virtual" isolation of a task can be obtained, when the task obeys the third assumption of Sect. 4.1.2, "disactivating" the corresponding BSL communication descriptors.

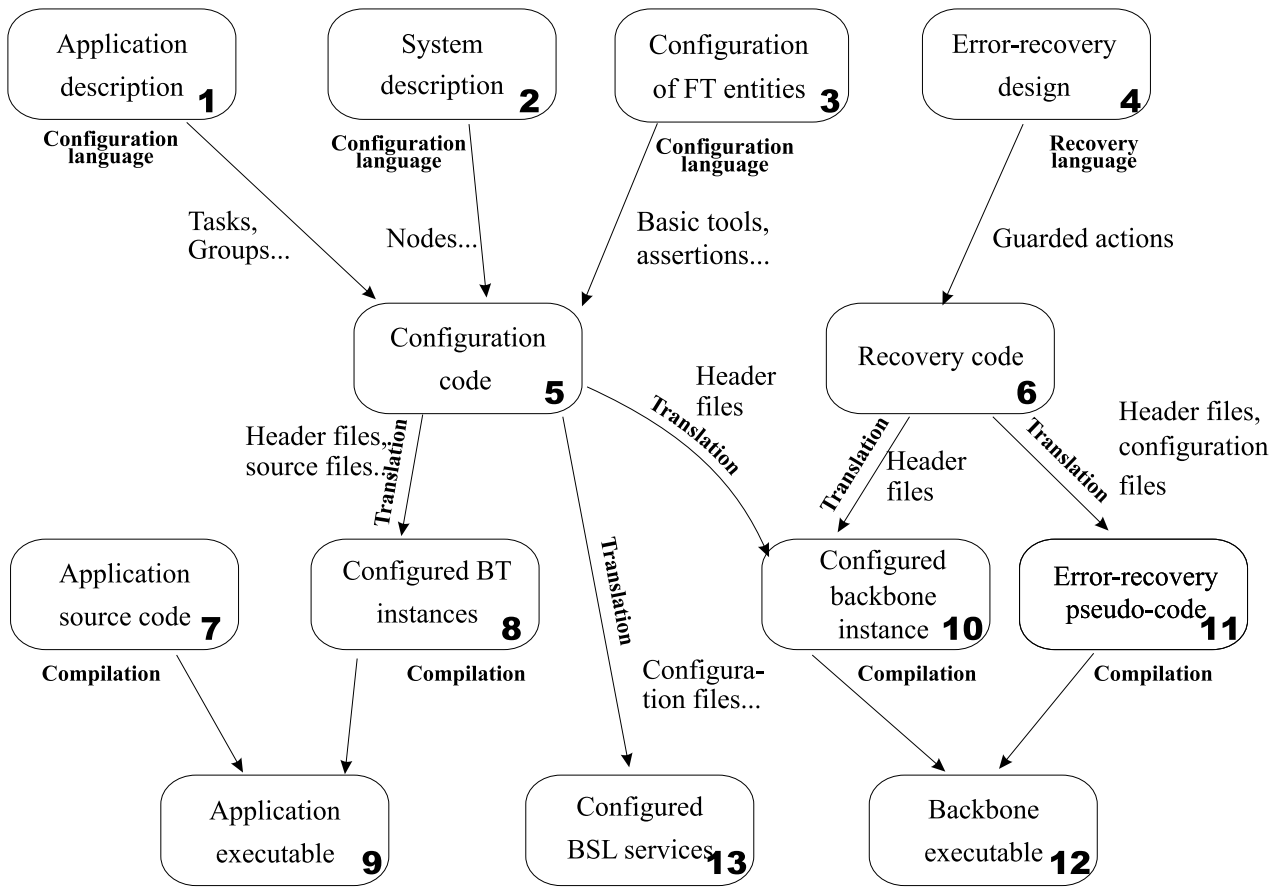

**Figure 4.1: A workflow diagram for $\mathcal{REL}$. Labels refer to usage steps and are described in Sect. 4.3.**

and so forth. Actions have a system-wide scope and are executed on any processing node of the system where a non-faulty component of the BB is running. They may include commands to send control signals to specific components of the BB or to user tasks. Basic tools such as a distributed voting tool like the EFTOS voting farm [DFDL98c, DFDL98a] can be instructed so that they respond to that signal supplying transparent support towards graceful degradation, task switching, and task migration, as described for instance in [DFDL98c]. Furthermore, the ability to distinguish between transient and permanent/intermittent faults can be exploited, in order to avoid unnecessary reconfigurations or other costly or redundancy-consuming actions, attaching different strategies to these two cases.

## 4.3   Workflow

This section describes the workflow corresponding to the adoption of the $\mathcal{REL}$ approach.

Figure 4.1 summarises the workflow. The following basic steps have been foreseen:

- In the first steps (labels 1 and 2 in the cited figure), the designer describes the key applica-

tion and system entities, such as tasks, groups of tasks, and nodes. The main tool for this phase is the configuration language.

- Next (step 3), the designer configures the basic tools and the fault-tolerance provisions he / she has decided to use. The configuration language is used for this. The output of steps 1–3 is the configuration code.

- Next (step 4), the designer defines what conditions need to be caught, and what actions should follow each caught condition. The resulting list is coded as a number of guarded action via a recovery language.

- The configuration code and the recovery code are then converted via the translator into a set of C header files, C fragments, and system-specific configuration files (steps 5 and 6). These files represent: configured instances of the basic tools, of the system and of the application; initialisation files for the communication management functions; user preferences for the backbone; and the recovery pseudo-code.

- On steps 7–9, the application source code and the configured BT instances are compiled in order to produce the executable code of the application.

- Next, the backbone is compiled on steps 10–12.

- Finally, on step 13, the communication management services of the BSL are configured in order to allow the proper management of multicasting and other communication services.

## 4.4 Relation with Other ALFT Approaches and Specific Limitations

This section draws a qualitative comparison between $\mathcal{REL}$ and the system structures for ALFT described in Chapter 3. The structural attributes SC, SA and A are the key properties being assessed and compared. Finally, the specific limitations of $\mathcal{REL}$ are summarised.

First, one can observe that the concept of recovery languages shows some similarities with respect to the recovery meta-program approach (RMP, defined in Sect. 3.5) and to aspect-oriented programming languages (AOP, see Sect. 3.4). A recovery language is indeed *a language to express application-level error recovery concerns*, as it is for the RMP language and as it can be for custom aspect languages. Nevertheless, a number of differences can be pointed out:

- RMP is invoked synchronously, under the control of the user, who is responsible for placing specific "breakpoints" in his/her application. Control is transferred to the meta-program as soon as a breakpoint is encountered. On the contrary, using the $\mathcal{REL}$, a recovery language executor (RINT) is awoken, asynchronously, each time a new error is entered in the DB, or synchronously, when the user signals some condition. Less code intrusion is required by $\mathcal{REL}$, which thus reaches better SC.

- Furthermore, a recovery language is incorporated in a wider $\mathcal{R\&L}$ scheme that includes an off-line configuration tool—the translator, by means of which the user can specify in a separate programming environment issues such as, for instance, replication.

- The adoption of a special grammar for the recovery language allows the fault-tolerance designer to experiment with the system structure for software fault-tolerance in order to increase the ability of the language at expressing a wide variety of fault-tolerance provisions in an adequate way. This allows, e.g., to set up a refinement process that can be used to adjust incrementally the syntax of a recovery language. On the contrary, RMP opted for a pre-existing, standard language (CSP). One can conclude that $\mathcal{R\&L}$ may be better than RMP with respect to attribute SA.

- To date, RMP is only available on a prototypic system [ADG$^+$90]. On the other hand, a $\mathcal{R\&L}$ prototypic architecture has been developed on a number of hardware platforms (see Sect. 5.1).

Furthermore, with respect to AOP, one can first observe that:

- Both approaches reach optimal separation of the design concerns (attribute SC).

- Very limited research has been conducted so far on the design of *fault-tolerance aspect languages*—as far as the author of this dissertation could find in the literature—and on the adequacy of AOP as a syntactical structure to host efficaciously a set of fault-tolerance provisions. The many fault-tolerance provisions supported by the configuration language and a proper syntax for the recovery language can guarantee adequate degrees of SA for the $\mathcal{R\&L}$.

- Designing the aspect programs, and in particular those related to fault-tolerance, may be more complex than writing a set of guarded actions as required by $\mathcal{R\&L}$. The same may be true when considering verification issues: the logic structure of guarded actions may allow the automatic translation of a recovery language program into the form of a set of logic clauses with whom a verification tool may be fed.

- The separation between the executable code addressing fault-tolerance and the one addressing the functional objectives can be exploited to reach optimal A with $\mathcal{R\&L}$ (see Chapter 8). A similar result can be reached also with AOP by executing the weaving process at run-time. Note how this calls for the interpretation of the whole resulting "tangled" program, and leads to some performance penalty that only regards error recovery in $\mathcal{R\&L}$.

Secondly, it is worth observing how *some of the ALFT approaches*, such as SV, MV, and MOPs, *are not complementary with respect to* $\mathcal{R\&L}$; on the contrary, as it has been shown for the configuration language, most of the SV and MV provisions can be supported and hosted by $\mathcal{R\&L}$. Furthermore, the adoption within a $\mathcal{R\&L}$-based architecture of fault-tolerance provisions based on MOPs would result in a straightforward improvement of $\mathcal{R\&L}$ which would further increase its degrees of SA and SC. One can conclude that, using $\mathcal{R\&L}$, the user can take advantage of a wider set of provisions, and can express error recovery strategies into one coherent syntactical structure, where "coarse-grained" techniques such as NVP are used next to "fine-grained" techniques, such as those based on transparent object redundancy. This way the user is set free to choose the provisions which match better with the problems to be addressed.

Finally, one can note that the language-based approaches differ considerably from $\mathcal{REL}$. Despite their many positive characteristics, their very axioms—using a custom syntax for both the functional and the fault-tolerance concerns—may restrict significantly their usability. The lack of standards is further detrimental to their diffusion. On the contrary, the choice to support a standard language for the functional aspects and an ad-hoc syntactical structure for the fault-tolerance aspects allows to reach optimal SA with no impact on the usability.

### 4.4.1   Limitations of $\mathcal{REL}$

As it is remarked in Chapter 6, a complete and detailed evaluation of any ALFT approach requires strong usability studies in order to assess its effectiveness as a structuring technique for hosting the fault-tolerance aspects into the application-layer of a computer program [LVL00, RX95]. In this context, relevant properties other than the structural attributes SC, SA and A play an important role. Among them, the performance penalties in terms of time and memory size may be crucial in order to determine whether a $\mathcal{REL}$ architecture may be hosted, for instance, on a given streamlined embedded system.

At the moment, some steps have been conducted towards a process of overall evaluation (see Chapter 6 and Chapter 7). Promising results have been observed for the single prototypic $\mathcal{REL}$ architecture that is available to date (see Chapter 5), though no complete assessment of $\mathcal{REL}$ is available yet, mainly because of its novelty.

Even from the so far limited experience, a few facts can be observed, which can be used to clarify the usability domain of $\mathcal{REL}$:

- The $\mathcal{REL}$ *does imply* a performance and memory overhead. A rough idea of the magnitude of these figures can come from the penalties of a prototypic $\mathcal{REL}$ system reported in Sect. 6.1.5.

- The system is assumed to be "static", in the sense that it must be known completely at compile-time. This means that, to date, the system cannot grow dynamically beyond the boundaries that have been set at compile-time. This assumption may match well with some parallel or distributed embedded system but does not allow the adoption of unlimited distributed systems, such as Internet-based wide-area metacomputing systems [Buy99].

- Specific support for checkpoint-and-rollback has been designed but not implemented in the current prototypic architecture described in next chapter. This means that the current implementation of $\mathcal{REL}$ requires external support to checkpoint-and-rollback. Coupling the current structure of $\mathcal{REL}$ with the system described in [Dec96] is a straightforward way to solve this limitation.

- The current implementation of $\mathcal{R}\!\mathcal{E}\!\mathcal{L}$ supports spare task management and a fault-identification mechanism. The adoption of these features does impact on the real-time behaviour of the system. This may turn into the unsuitability to fulfilling hard real-time constraints.

- The crash/performance failure semantics assumption may require specific support.

It is worth remarking how some of these limitations pertain to specific *implementation* aspects, and that the overheads measured so far derive from a prototypic, non-optimised implementation.

## 4.5 Conclusions

A novel approach to ALFT has been described. The approach provides its users with a single linguistic structure to express application-level error recovery concerns and to configure a number of fault-tolerance provisions *outside the functional application layer*. The latter can be provided by any procedural language.

The $\mathcal{R}\!\mathcal{E}\!\mathcal{L}$ approach addresses dependability goals for distributed or parallel applications written in any procedural language. Its target hardware platforms include distributed and parallel computers consisting of a set of processing nodes known at compile time. Within this application and system domain, the novel approach can be qualitatively assessed as reaching optimal values of the structural attributes. In particular, with respect to the ALFT approaches reviewed in Chapter 3:

- Error recovery is expressed in a separate programming context.

- The executable code is separable—that is, at run-time, the portion of the executable code devoted to the fault-tolerance aspects is clearly distinct from the functional code.

- A large number of well-known fault-tolerance strategies are supported in a straightforward way.

- The $\mathcal{R}\!\mathcal{E}\!\mathcal{L}$ addresses a wide class of applications: that of distributed applications, with non-strict real-time requirements, written in a procedural language, to be executed on distributed or parallel computers consisting of a set of processing nodes known at compile time.

- It reaches both the benefits of the evolutionary approaches, which base themselves on standards, and those of "revolutionary" approaches, exploiting ad-hoc, non-standard solutions.

- It can host other provisions and approaches.

- As shown in Sect. 6.1, in same cases $\mathcal{R}\!\mathcal{E}\!\mathcal{L}$ can be adopted with minimal programming effort and minimal adaptation of a pre-existing, non fault-tolerant application.

Table 4.1 and Fig. 4.2 extend respectively Table 3.1 and Fig. 3.5 with the value of the structural attributes for $\mathcal{R}\!\mathcal{E}\!\mathcal{L}$. Next chapter describes a prototype architecture based on the $\mathcal{R}\!\mathcal{E}\!\mathcal{L}$ approach.

| Approach | SC | SA | A |
|----------|----|----|----|
| SV | 1 | 2 | 1 |
| MV (RB) | 1 | 1 | 1 |
| MV (NVP) | 4 | 1 | 1 |
| MOP | 5 | 3 | 3 |
| EL | 3 | 1 | 1 |
| DL | 3 | 5 | 1 |
| AOP | 5 | 3 | 4 |
| RMP | 5 | 2 | 1 |
| $\mathcal{REL}$ | 5 | 5 | 5 |

**Table 4.1:** Comparison between $\mathcal{REL}$ and the approaches proposed in Chapter 3 and summarised in Table 3.1.

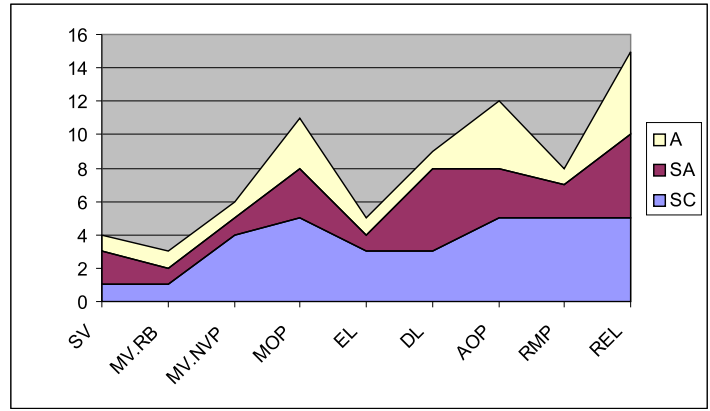

**Figure 4.2:** The data in Table 4.1 are here represented as a stacked area.

# Chapter 5

# A Distributed Architecture Based on the Recovery Language Approach

This chapter describes a prototypic architecture based on $\mathcal{R\mathcal{L}}$ that has been designed in the context of the European ESPRIT project TIRAN.

The structure of the chapter is as follows: first, project TIRAN is described in Sect. 5.1. The main components of the TIRAN architecture are covered in Sect. 5.2. Section 5.3 describes the TIRAN configuration and recovery language. Section 5.4 closes the chapter and draws a few conclusions.

Throughout the chapter, the contributions of the author of this dissertation are pointed out.

## 5.1 The ESPRIT Project TIRAN

TIRAN is the name of the European ESPRIT project 28620. Its name stands for "tailorable fault-tolerance frameworks for embedded applications" [BDFD+99]. The main objective of TIRAN is to develop a software *framework*[1] that provides fault-tolerant capabilities to automation systems. Application-level support to fault-tolerance is provided by means of a $\mathcal{R\mathcal{L}}$ architecture, which is described in the rest of this chapter. To date, version 2 of the framework has been issued and the final version is due by the end of the project (October 2000). The framework provides a library of software fault-tolerance provisions that are parametric and support an easy configuration process. Using the framework, application developers are allowed to select, configure and integrate provisions for fault masking, error detection, isolation and recovery among those offered by the library. Goal of the project is to provide a tool that significantly reduces the development times and costs of a new dependable system. The target market segment concerns non-safety-critical[2] distributed real-time embedded systems [BDFD+00].

---

[1]To be intended herein as a set of software libraries, distributed components, and formal techniques.

[2]Safety-critical systems, i.e., computer, electronic or electromechanical systems whose failure may cause injury or death to human beings, such as an aircraft or nuclear power station control system [FOL00], are not covered within TIRAN. This allows the crash failure semantics assumption (see Sect. 4.1.2) to be satisfied with less strict coverage, which translates into lower development costs for the framework.

TIRAN explicitly adopts formal techniques to support requirement specification and predictive evaluation. This, together with the intensive testing on pilot applications, is exploited in order to:

- Guarantee the correctness of the framework.

- Quantify the fulfilment of real-time, dependability and cost requirements.

- Provide guidelines to the configuration process of the users.

The project partners are:

- The ACCA Division of the ESAT Department of the University of Leuven.

- The Computer Science department of the University of Turin (UniTo, Italy).

- The Belgian company Eonic Systems (supplier of high performance, application-specific light-weight RTOS for DSP processors and embedded ASIC cores).

- The Italian company TXT Ingegneria Informatica (provider of systems and services in many fields of real-time automation).

- CESI, a research centre of ENEL (the main electricity supplier in Italy, the third largest world-wide).

- Siemens (the leading German company in the field of electrical engineering and electronics, one of the largest world-wide).

The partners are globally referred to as "the consortium".

Within the consortium, ACCA provided expertise in fault-tolerance design and developed the $\mathcal{REL}$ infrastructure; Siemens and CESI provided two industrial pilot applications; UniTo developed specific Petri net models for the validation and the verification of the framework and to facilitate the user in the process of validating its overall application design; Eonic Systems provided expertise in hard real-time systems design and developed some of the TIRAN BTs; TXT managed the design of the TIRAN case studies, co-designed and developed the TIRAN BSL and most of the TIRAN BTs. Furthermore, TXT, Siemens, and Eonic Systems ported the original BSL, developed for Windows NT, to a number of hardware and operating system combinations— which are described in what follows. The project management was carried out by CESI, which also authored a fault-tolerance software methodology [DB00] spanning the whole software life-cycle, from requirement specification up to product development.

Most of this framework has been designed for being platform independent. A single version of the framework has been written in the C programming language making use of a BSL designed by the consortium and initially developed by TXT for the development platform (Windows NT). Within TIRAN, a number of target platforms and systems compliant to the model sketched in Sect. 4.1 have been selected. Porting TIRAN to these platforms is mainly obtained through the porting of the BSL. The project's target platforms include:

- A set of Mosaic020[3] boards running the Virtuoso microkernel [Sys98, Mos98].

- A set of VME boards based on PowerPC processors and running the VxWorks operating system [Win99].

- A set of DEC Alpha boards running the TEX kernel.

- Clusters of Windows-CE personal computers (PCs).

The final version of the TIRAN framework, to run on all the above systems, is demonstrated in November 2000 at the final TIRAN review meeting and at the TIRAN workshop [Thi00].

Some elements of the framework are also available on Parsytec CC and Parsytec Xplorer systems based on the PowerPC processor and the EPX kernels [Par96a, Par96b] and on a proprietary hardware board based on the DEC Alpha chip and running the TEX kernel [DEC97, TXT97].

The project results, driven by industrial users' requirements and market demand, will be integrated into the Virtuoso RTOS at Eonic Systems and are being tested and adopted by ENEL and SIEMENS within their application fields.

The main components of the TIRAN framework are:

1. The TIRAN toolset, the components of which are discussed in Sect. 5.2.

2. The TIRAN configuration and recovery language ARIEL, described in Sect. 5.3. In particular:

   - Its translator is dealt with in Sect. 5.3.3.1.

   - The run-time executive of ARIEL, which is sketched in Sect. 5.3.3.2.

TIRAN builds on top of the ESPRIT project EFTOS that has been described in Sect. 3.1.1. In particular, the author of this dissertation has conceived, designed and developed a number of components while taking part in these two projects within the ACCA Division. They include:

- The TIRAN BB, including its fault-tolerance distributed algorithm AMS [DFDL00a] (see Sect. 5.2.4.1) and the DB management system.

- The ARIEL configuration and recovery language [DFDL98b, DDFB01].

- The ARIEL translator, `art`.

- The RINT module.

- The TOM module [DFDL00b].

---

[3]Mosaic020 boards have been developed in the framework of the ESPRIT project Dipsap II [Dip97] by DASA/DSS and Eonic Systems. They are based on the Analog Devices ADSP-21020 DSP and the SMCS communication chip. The SMCS chip—co-designed by the ACCA Division —complies with the IEEE 1355 standard [IEE95] and has hardware support for detecting transmission and connection errors and for higher level system protocols such as reset-at-runtime.

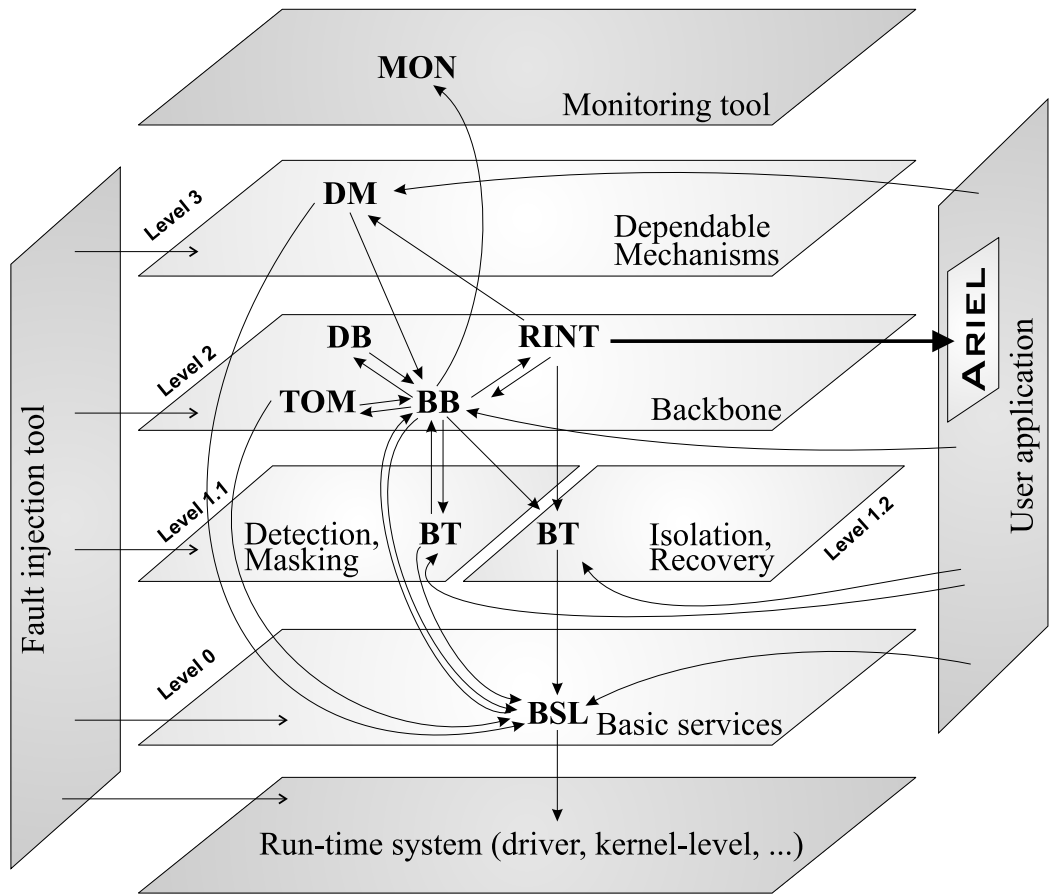

**Figure 5.1: A representation of the TIRAN elements. The central, whiter layers constitute the TIRAN framework. This same structure is replicated on each processing node of the system.**

- The MON module, that is, a monitoring tool based on WWW technology [DFDT$^+$98].

He also wrote and adapted the EFTOS voting farm [DFDL98c] in order to function as a distributed voting system in the TIRAN framework.

Figure 5.1 draws the TIRAN architecture and positions its main components into it. Two types of edges are used in the picture:

- The thinner, directed edge represents relation "<Sends>": if such an edge goes from entity $a$ to entity $b$ then it means that $a$ sends a *data* message to $b$ or requests a service provided by $b$ through a *control* message or a method invocation.

- The thicker edge is only used between the RINT and ARIEL entities and means that *the former implements the latter* through the process of interpreting the r-codes (see Sect. 5.3.3.2).

The central, whiter layers represent the TIRAN framework. In particular:

- The level 0 hosts the TIRAN BSL, which gives system-independent access to the services provided by the underlying run-time system (see Sect. 4.2.1). The user application, the BTs, the BB, the TOM, and the dependable mechanisms (DMs, see Sect. 5.2.3), all make use of the BSL, e.g., to create tasks or to send messages through a mailbox.

- Level 1 services are provided by the BTs for error detection and fault masking (level 1.1) and by those for error isolation, recovery and reconfiguration (level 1.2). Both services are supplied by node-local provisions. Section 5.2.1 describes in more detail the BTs. The edge from BT to BB represents the sending of error detection or diagnostic messages. The edge from BB to BT represents control messages, such as, for instance, a request to modify a parameter of a watchdog, or a request to reboot the local node (see Fig. 5.7).

- Level 2 hosts the TIRAN BB, including the TOM component, the DB management functions, and the recovery interpreter, RINT. In Fig. 5.1, the thicker edge connecting RINT to ARIEL means that RINT actually implements (executes) the ARIEL programs. Note the control and data messages that flow from BB to TOM, DB, and RINT. RINT also sends control messages to the isolation and recovery BTs. Data messages flow also from BB to the monitoring tool.

- Dependable mechanisms, i.e., high-level, distributed fault-tolerance tools exploiting the services of the BB and of the BTs, are located at level 3. These tools include the distributed voting tool, the distributed synchronisation tool, the stable memory tool, and the distributed memory tool, described in Sect. 5.2.3. DMs receive notifications from RINT in order to execute reconfigurations such as introducing a spare task to take over the role of a failed task (see Table 5.4).

The layers around the TIRAN framework in Fig. 5.1 represent (from the layer at the bottom and proceeding clockwise):

- The run-time system.

- A provision to inject software faults at all levels of the framework and in the run-time system.

- The monitoring tool, for hypermedia rendering of the current state of the system within a WWW browser windows.

- The functional application layer and the recovery language application layer (the latter is represented as the box labelled ARIEL).

Figure 5.2 locates the key elements of the TIRAN architecture within the workflow diagram in Fig. 4.1.

The following sections describe these components in detail.

## 5.2 Some Components of the TIRAN Framework

This section describes some key components of the TIRAN framework, namely:

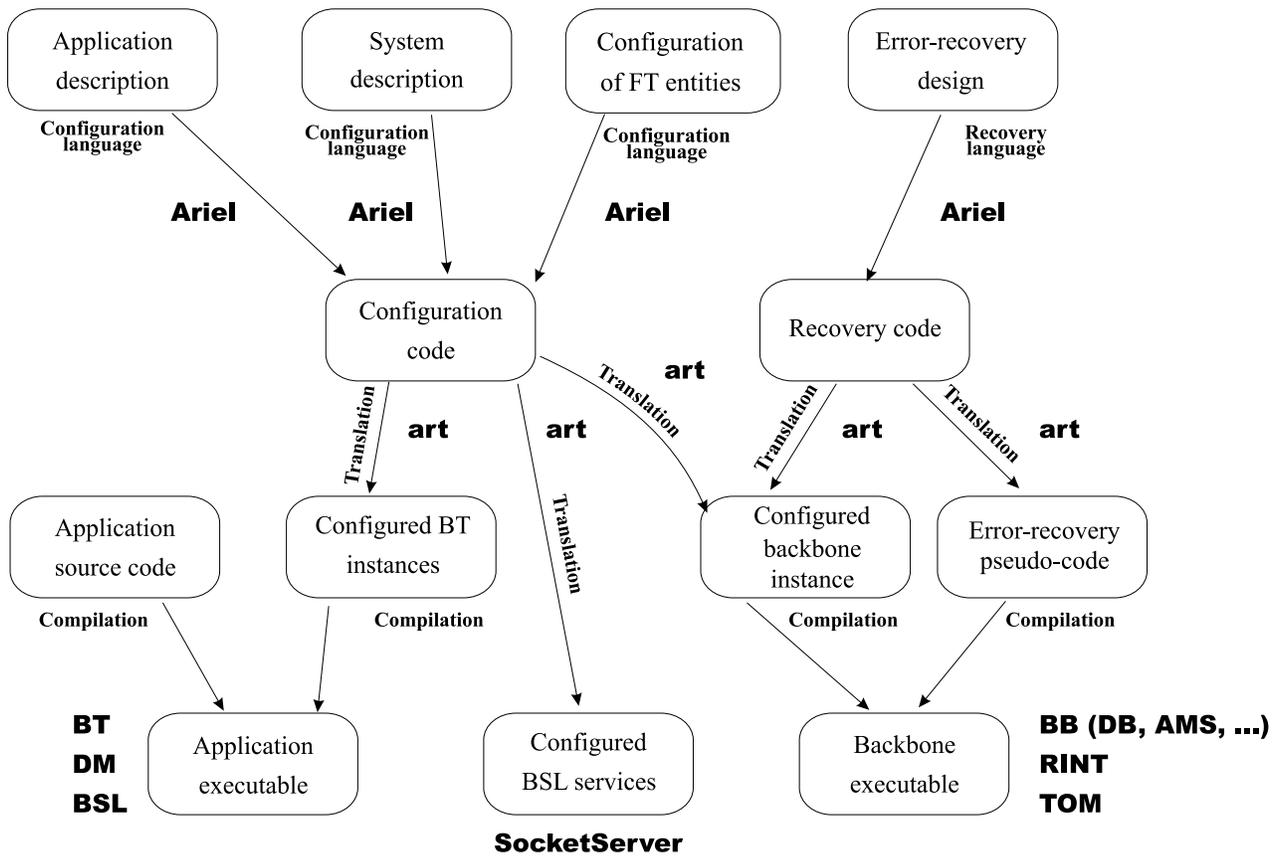

**Figure 5.2:** The main components of the $\mathcal{REL}$ architecture developed in TIRAN are located within the workflow diagram of Fig. 4.1. A so-called "SocketServer" has been also depicted, i.e., a BSL component managing message transmission and dispatching.

- The TIRAN toolset, i.e., the BTs, the BSL, the BB, and the TOM service.

- A level-3 mechanism, i.e., an application-level provision for distributed voting.

## 5.2.1   The TIRAN Basic Tools

The TIRAN BTs represent a layer of node-local services for error detection, fault masking, error recovery, and isolation. The TIRAN BTs include:

1. A watchdog timer.

2. A node-local voting system.

3. An input replicator, to be used with the voting system.

4. An output collector, to be used with the voting system.

5. A tool responsible for node shutdowns.

6. A tool responsible for node reboots.

7. A tool responsible for isolating a task.

Some of the BTs (such as the level 1.1 BTs with numbers 1–4) are application-level software tools, others (for instance, the level-1.2 BTs number 5–7) provide hooks to lower-level services. Most of these tools have been designed and developed by Eonic Systems and have been especially conceived in order to meet hard real-time requirements. A deeper description of some of these tools can be found in [DBC+98, DTDF+99]. As a side effect of using these tools, the TIRAN BB (see Sect. 5.2.4) is transparently notified of a number of events related to fault-tolerance.

Notifications generally describe the state of a given component and have the form of a 4-tuple of integers

$$(c, t, u, p),$$

where $c$ identifies a specific condition, $t$ is a label that specifies the type of BT, $u$ is the unique-id of either a task, or a group of tasks, or a node, and $p$ is a possibly empty list of optional arguments. Within TIRAN, function `RaiseEvent` is used to forward notifications. Function `RaiseEvent` has the following prototype:

```
int RaiseEvent(int condition, int actor, TIRAN_TASK_ID uniqueId,
                int nargs, ...).
```

## 5.2.2 The TIRAN Basic Services Library

The TIRAN BSL offers specific services including communication, task creation and management, access to the local hardware clock, management of semaphores, and so forth. This BSL supports multicasting: messages are sent to so-called "*logicals*," i.e., groups of tasks. It has been purposely designed by TXT and the TIRAN Consortium and developed by TXT, Siemens and Eonic Systems for the target platforms and operating systems. Specific adaptation layers may be designed for mapping existing communication libraries, such as MPI/RT [KSR98] or PVM [GBD+94, DFM92], to the TIRAN BSL.

As a side effect of using some of the BSL functions, the TIRAN BB (see Sect. 5.2.4) can be transparently notified of events such as a successful spawning of a task, or the failure state of a communication primitive. BSL notifications are similar to BT notifications—4-tuples of integers

$$(c, t, u, p),$$

where $c$ and $p$ are as in Sect. 5.2.1, while in this context $t$ is a label that specifies the class of BSL services and $u$ is the unique-id of the task that experienced condition $c$.

A full description of the TIRAN BSL can be found in [CSV+99].

## 5.2.3 The TIRAN Distributed Voting Tool

As already mentioned, the TIRAN DMs are distributed, high-level fault-tolerance provisions that can be used to facilitate the development of dependable services. The TIRAN DMs include:

- A tool to enhance data integrity by means of a stabilising procedure (data is only allowed to pass through the system boundaries when it is confirmed a user-defined number of times). This is known as Stable Memory Tool and is described in [DBC+98, DFDLG01].

- A Distributed Synchronisation Tool, i.e., a software tool for synchronising a set of tasks. A description of this tool can be found in [CSV+99].

- A so-called Distributed Memory Tool, that creates and manages distributed replicas of memory cells.

- A "Redundant Watchdog", i.e., a distributed system exploiting multiple instances of the watchdog BT in order to enhance various dependability properties (see Sect. 6.4).

- The TIRAN distributed voting tool (DV), an adaptation of the EFTOS voting farm, designed and developed by the author of this dissertation.

This section describes one of the above tools, namely the DV.

The TIRAN DV is a $\mathcal{REL}$-compliant software component that can be used to implement restoring organs i.e., $N$-modular redundancy systems ($N$MR) with $N$-replicated voters [Joh89] (see Fig. 5.3). The basic goals of the tools include fault transparency but also replication transparency, a high degree of flexibility and ease-of-use, and good performance. Restoring organs allow to overcome the shortcoming of having one voter, the failure of which leads to the failure of the whole system even when each and every other module is still running correctly. From the point of view of software engineering, such systems though are characterised by two major drawbacks:

- Each module in the $N$MR must be aware of and responsible for interacting with the whole set of voters;

- The complexity of these interactions, which is a function that increases quadratically with $N$ (the cardinality of the set of voters), burdens each module in the $N$MR.

To overcome these drawbacks, DV adopts a different scheme, as described in Fig. 5.4: in this new scheme, each module only has to interact with, and be aware of *one* voter, regardless of the value of $N$. Moreover, the complexity of such a task is fully shifted to the voter.

The key component of the DV architecture is indeed the *voter* (see Fig. 5.5), defined as a local software module connected to *one* user module and to a set of peer-levels connected with each other. The attribute "local" means that both user module and voter run on the same processing node. As a consequence, the user module has no other interlocutor than its voter, whose tasks are completely transparent to the user module. It is therefore possible to model the whole system as a simple client-server application: on each user module the same client protocol applies while the same server protocol is executed on every instance of the voter. The following paragraphs briefly describe these two protocols.

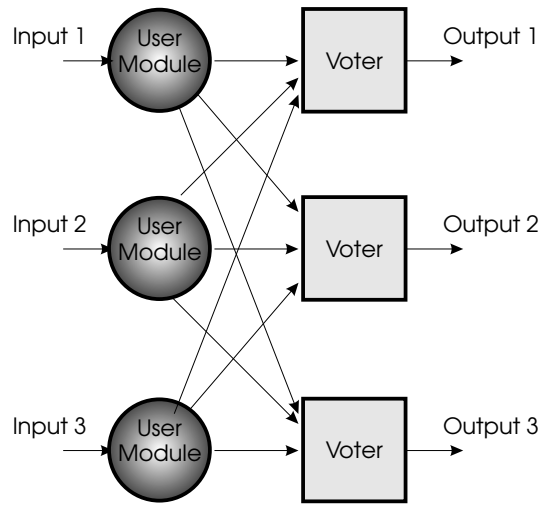

**Figure 5.3: A restoring organ [Joh89], i.e., a** $N$**-modular redundant system with** $N$ **voters, when** $N = 3$**. Note that a de-multiplexer is required to produce the single final output.**

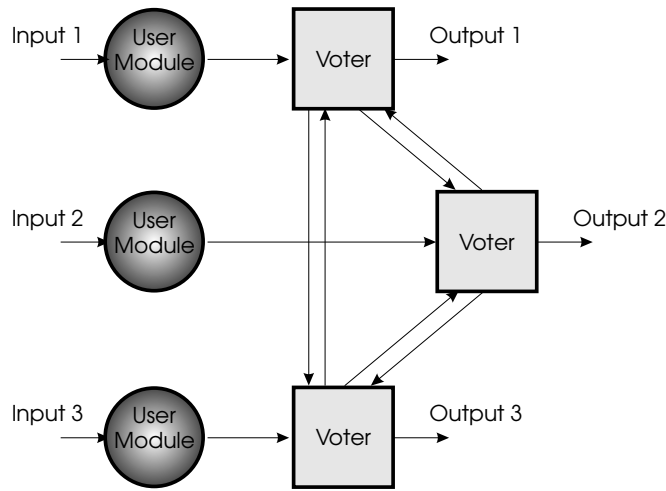

**Figure 5.4: Structure of the TIRAN DV for** $N = 3$**.**

#### 5.2.3.1 Client-Side Protocol

Table 5.1 gives an example of the client-side protocol, i.e., the sequence of actions that need to be executed in each user module in order to make proper use of the DV. Six "methods" have been set up in order to declare, define, describe, activate, control, and query the DV. In particular, *describing* an instance of DV stands for creating a static map of the allocation of its components; *activating* an instance means spawning the local voter (this is described in next paragraph); *controlling* an instance means requesting its service by means of control and data messages (this includes starting a voting session); finally, an instance of DV can also be *queried* about its state, the current voted value, and so forth.

As already mentioned, the above steps have to be carried out in the same way on each user module: this coherency is transparently supported by the TIRAN configuration language (see Sect. 5.3).

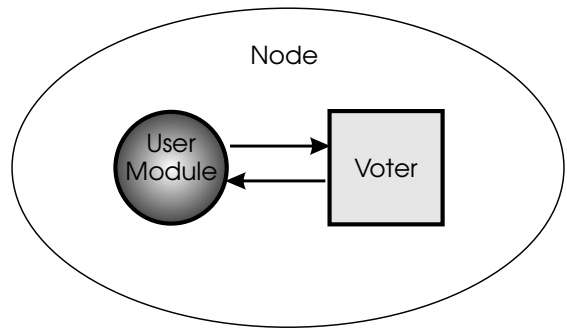

**Figure 5.5:  A user module and its voter.  The latter is the only voter the user module needs to be aware of:  control and data messages only flow between these two ends.  This has been designed in order to minimise the burden of the user module and to keep it free to continue undisturbed as much as possible.**

This protocol is available to the user as a class-like collection of functions dealing with opaque objects referenced through pointers.

#### 5.2.3.2   Server-Side Protocol

The local voter process executes the server-side of DV. After the set up of the static description of the local instance (Table 5.1, line 3) in the form of an ordered list of process identifiers (integers greater than 0), the server-side of the application is launched by the user by means of the `TIRAN_VotingRun` function. This turns the static representation of the DV instance into an "alive" (running) object, the voter process.

Each voter then connects to its user module via a pipeline and to the rest of the farm via sockets. From then on, in the absence of faults, it reacts to the arrival of the user messages as a finite-state automaton: in particular, the arrival of input messages triggers a number of broadcasts among the voters—as shown in Fig. 5.6.

Besides the input value, which represents a request for voting, the user module may send its voter a number of other requests. In particular, the user can choose to adopt a voting algorithm from the following ones:

- Formalised majority voter technique,

- Generalised median voter technique,

- Formalised plurality voter technique,

- Weighted averaging technique,

- Consensus.

```
1   int main(void) {
2        /* declarations */
         TIRAN_Voting_t *dv;
         int objcmp(const void*, const void*);
3        /* definition */
         dv ← TIRAN_VotingOpen(objcmp);
4        /* description */
         ∀i ∈ {1,...,N} :
             TIRAN_VotingDescribe(dv, task_i, version_i, ident_i);
5        /* activation */
         TIRAN_VotingRun(dv);
6        /* control */
         TIRAN_VotingOption(dv,
             TIRAN_Voting_input(obj, sizeof(VFobj_t)),
             TIRAN_Voting_output(link),
             TIRAN_Voting_algorithm (VFA_WEIGHTED_AVERAGE),
             TIRAN_Voting_timeout (TIMEOUT),
             TIRAN_Voting_scaling_factor(1.0) );
7        /* query */
         do { MsgWait(); } while (TIRAN_Voting_error ≡ DV_NONE
             ∧ TIRAN_VotingGet(dv) ≠ DV_DONE);
8        /* deactivation */
         TIRAN_VotingClose(dv);
9   }
```

**Table 5.1: An example of usage of the TIRAN DV. A static description of a DV instance is completed in lines 2–4. In particular, each call of** `TIRAN_VotingDescribe` **describes an input module, specifying its task identifier, its version number (if the tool is used within an NVP scheme), and a Boolean flag that must be true only for the current input module. At line 5 the instance is activated. After this step, the system can be controlled by sending messages, e.g., to select a voting algorithm or to set a timeout (line 6). The state of the system can be queried by reading incoming state information (line 7). Finally, the system can be deactivated as in line 8.**

The first four items are the voting techniques that have been generalised in [LCE89] to "arbitrary $N$-version systems with arbitrary output types using a metric space framework". To use these algorithms, a metric function can be supplied by the user when he or she "opens" the DV instance (Table 5.1, step 3). Other requests include the setting of some algorithmic parameters and the removal of the voting farm (function `TIRAN_VotingClose`).

The voters' replies to the incoming requests are straightforward. In particular, a `DV_DONE` message is sent to the user module when a voting session has been performed (see line 7 in Table 5.1).

Note how a function like `TIRAN_VotingGet` simply sets the caller in a waiting state from which it exits either on a message arrival or on the expiration of a time-out.

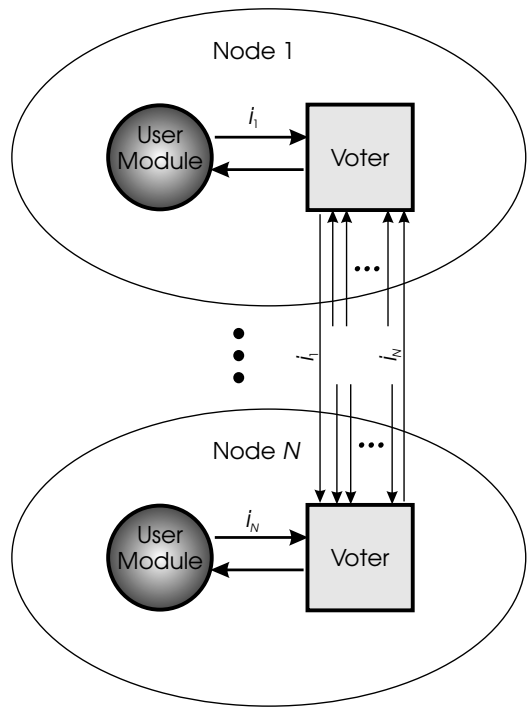

**Figure 5.6: The "local" input value has to be broadcast to $N-1$ fellows, and $N-1$ "remote" input values have to be collected from each of the fellows. The voting algorithm takes place as soon as a complete set of values is available.**

The author of this dissertation designed and developed DV and also performed a theoretical study on the performance of the core algorithm of the TIRAN DV. This algorithm realizes the process such that every voter communicates its own input value to all the other voters and receives the input value owned by all the other voters. The study is reported in Sect. 7.2.1 and in [DFDL00c].

## 5.2.4   The TIRAN Backbone

The TIRAN BB is the core component of the fault-tolerance distributed architecture described in this dissertation. The main objectives of the TIRAN BB include:

1. Gathering and maintaining error detection information produced by TIRAN BTs and BSL.

2. Using this information at error recovery time.

In particular, as described in Sect. 5.2.1, the TIRAN BTs related to error detection and fault masking forward their deductions to the BB. The BB maintains these data in the TIRAN DB, replicated on multiple nodes of the system. Incoming data are also fed into an $\alpha$-count [PABD+99] filter (see Sect. 5.2.4.2). This mechanism allows to identify statistically the nature of faults—whether a given fault is transient or whether it is permanent or intermittent. A set of private functions to query the current state of the DB is available within the BB. Other private functions request remote services such as, for instance, rebooting node $n_2$ or starting task number 3 on node $n_1$.

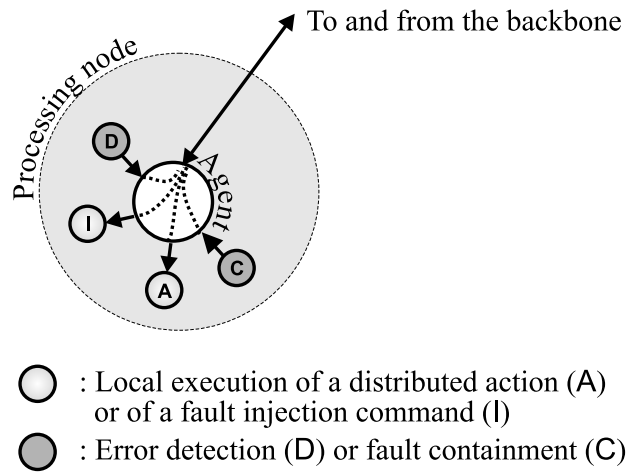

Figure 5.7: **On each processing node exactly one component of the BB is located. This component is the intermediary of the BB on that node. In particular, it gathers information from error detection and fault containment BTs (grey circles) and forwards requests to error containment and recovery BTs (light grey circles). These latter execute recovery actions and possibly, at test time, fault injection requests.**

When the underlying architecture is built around a host computer controlling a number of target boards by means of a custom OS, the latter may execute remote commands on the boards. On the contrary, when the OS is general-purpose and node-local, remote services can be executed on any node of the system by sending command messages from one component to the other of the BB—for instance, a "reboot" message may be sent from the component of the BB on node 0 to the component on node 2. On receipt, the latter may execute the reboot command by forwarding a local request to a basic tool addressing error isolation or recovery, e.g., a tool managing the reboot service. Figure 5.7 represents this scheme. The above remote services can be the basis for more complex, system-wide recovery and reconfiguration actions.

The TIRAN BB consists of two core components. In the absence of system partitioning, within the system there is exactly one **manager**, holding the main copy of the DB and with recovery management responsibilities. On other nodes, BB components called **assistants** deal with DB replicas, forward local deductions to the manager and can take over the role of the manager should the latter or its node fail.

A key point in the effectiveness of this approach is *guaranteeing that the BB itself tolerates internal and external faults*. A custom distributed algorithm has been devised by the author of this dissertation in order to tolerate crash failures triggered by faults affecting at most all but one of the components of the BB or at most all but one of the nodes of the system. The same algorithm also tolerates system partitioning during the "periods of instability" (see the timed-asynchronous distributed system model [CF99]). This procedure, described in [DFDL00a], is one of the contributions of the author of this dissertation to the state-of-the-art of fault-tolerance distributed algorithms. It has been called the "algorithm of mutual suspicion" (AMS) since each component of the BB continuously questions the correctness of all the other valid components. It is briefly described in the following section.

### 5.2.4.1   The Algorithm of Mutual Suspicion

The AMS is built on a twofold error detection strategy:

- On a local basis, on each node of the system, a special watchdog called " Task" (IAT) guards the local component of the BB. This component has to reset periodically an " flag", which is checked and set by the IAT. A faulty or slowed down component can be detected at the end of each period if the flag is found to be still set.

- On a remote basis, each assistant expects a MIA ("Manager Is Alive") message from the manager at most every $p_m$ seconds, while the manager expects a TAIA ("This Assistant Is Alive") at most every $p_a$ seconds.

In the normal, non-faulty case, the IAT and the guarded BB component would simply keep on setting and clearing the  flag, while the manager and its assistants would keep on receiving and sending MIA's and TAIA's (possibly piggybacking them, i.e., together with data related to the DB).

When a crash failure affects, e.g., a guarded BB component, this results in a violation of both the local and the remote strategy.  Locally, this is detected by the guarding IAT, which reacts to this by broadcasting a TEIF ("This Entity Is Faulty") to the rest of the BB. Remotely, this appears as a lack of a MIA or TAIA message from the crashed component, and within a period of $p_i$ seconds, the arrival of a TEIF message from the remote IAT.

When a crash failure affects, e.g., the node where a BB component was running, this is detected (remotely) by a lack of a MIA or TAIA message for $p_m$ (resp. $p_a$) seconds, and by a further lack of a TEIF message after another $p_i$ seconds.

This scheme allows to detect a crash failure of a node and of a component of the BB. Figures 5.8, 5.9, and 5.10 supply a graphical representation of this algorithm.

Each component of the backbone and the IAT need to perform a set of periodical tasks, like clearing or setting the  flag, or other tasks logically to follow the expiration of time periods $p_m$, $p_a$ or $p_i$. These tasks are carried out by means of lists of time-outs, and managed by the TIRAN TOM, described in next section.

One of the partners in the consortium (UniTo) has developed a model of the AMS using stochastic Petri Nets [CSV+99]. Simulations of their model revealed the absence of deadlocks. Testing has been also performed via fault injection. Figures 5.11–5.14 show a scenario in which a fault is injected on node 0 of a system consisting of four nodes.  A Netscape World-Wide Web browser is remotely controlled by the MON component (a non-parsing-header CGI application [Kim96, DF96b] conceived and developed by the author of this dissertation) in order to render dynamically the shape and state of the system [DFDT+98]. Node 0 hosts the manager of the BB and, as a result of the fault injection, the role of manager is appointed to backup node 1, formerly running as a BB assistant. No node recovery is performed in this case, therefore node 0 is soon detected to be inactive and removed from the backbone.

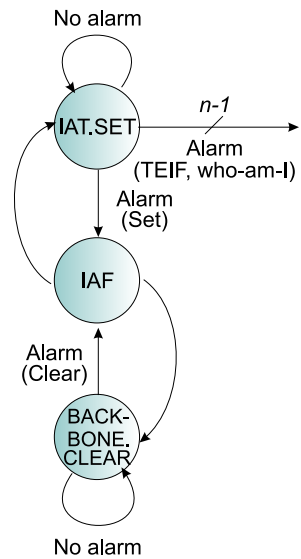

**Figure 5.8: A representation of the conjoint action of the IAT and of the BB component on the  flag.**

It is worth remarking how AMS is compliant to the timed asynchronous distributed system model described in Sect. 4.1.1. In particular, it tolerates system partitioning, during which correct nodes are temporarily disconnected from the rest of the system. When such an event occurs, for instance because of transient faults affecting the communication subsystem, the BB splits up into separate blocks, in each of which a manager is elected according to a prefixed scheme (currently, the node with the highest label is elected as manager). Blocks are then automatically merged back into one coherent system when the erroneous conditions that brought the system to system partitioning disappear. As a special case of this feature, a crashed node that is restarted is automatically accepted back into the BB.

The above mentioned features have been successfully demonstrated at a review meeting of project TIRAN. In that occasion, faults were injected by removing temporarily the Ethernet cable connecting two PCs. The backbone reacted to this event partitioning itself into two blocks. In the block hosting the manager, the event was followed by graceful degradation. In the block hosting no manager, a manager was elected. Reconnecting the network cable, the two blocks merged back with one of the manager being demoted to the role of assistant. The reviewers assessed this demonstration and the others which were shown in that occasion as impressing [Mor00].

### 5.2.4.2 The $\alpha$-Count Fault Identification Mechanism

It is well known that reconfiguration is a costly activity, in the sense that it always results in redundancy consumption. This translates in a possibly graceful, though actual degradation of the quality of service of a system.  This may have drastic consequences, especially in systems where service degradations are static and irreversible.  For instance, in satellites and space probes, rapid exhaustion of redundancy may severely affect the duration of the useful mission hours and therefore reliability [Inq96].  Two important issues towards solving the problem just stated are:

1. Understanding the nature of faults, and in particular identifying faults as permanent (and thus actually requiring reconfiguration) with respect to transient ones.

2. Tolerating transient faults with less redundancy consuming techniques, possibly not based on reconfiguration.

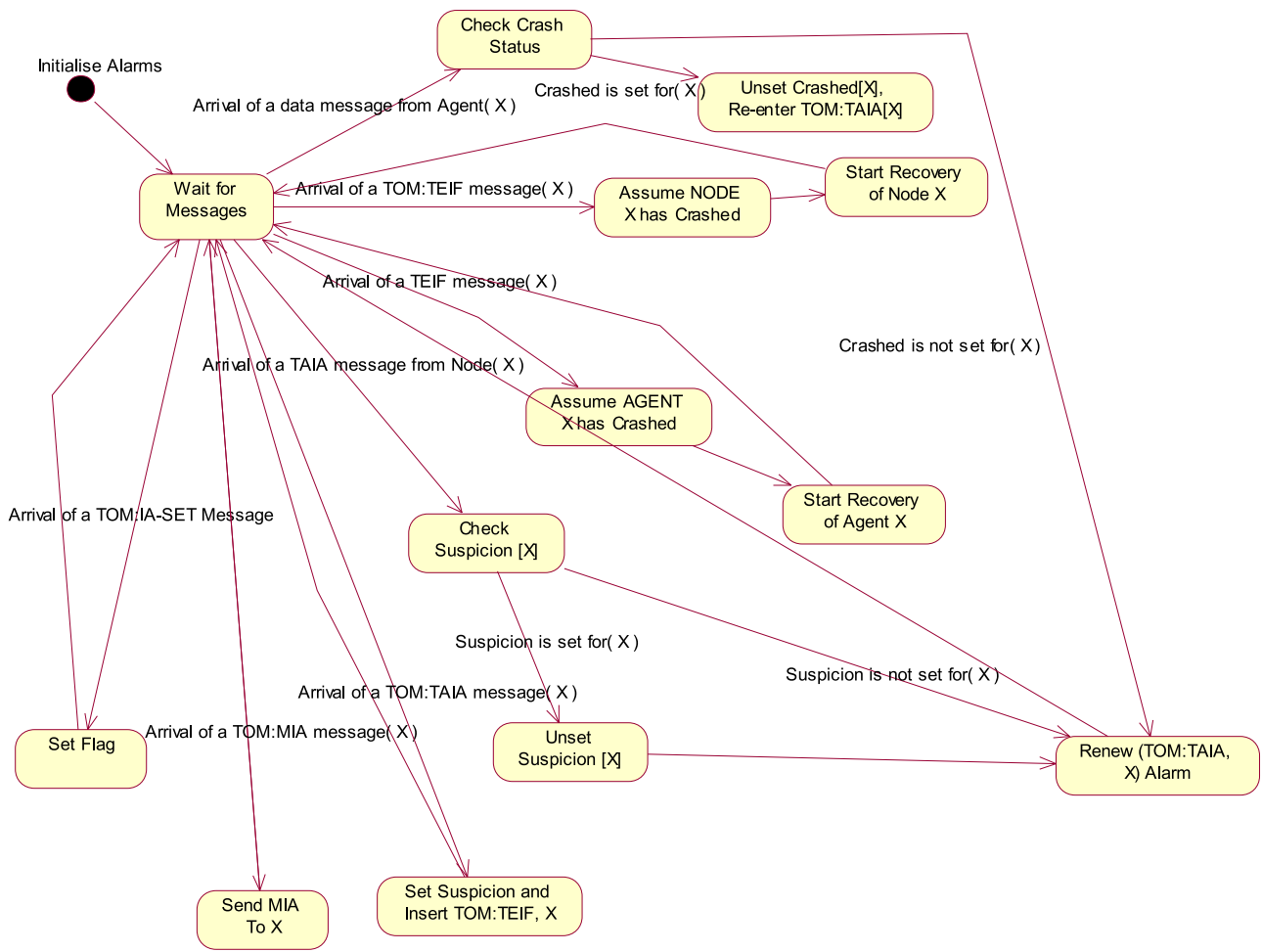

**Figure 5.9:  A state diagram of the algorithm of the manager.**

Issue 1 means that the adopted fault-tolerance mechanism is required not only to locate a component subject to an error, but also to assess the nature of the originating fault.  This implies processing additional information and unfortunately translates also in a larger delay in fault diagnosis.  Despite this larger delay, in same cases the benefits of techniques based on the above issues may be greater than its penalties in performance and latency [BCDGG97].

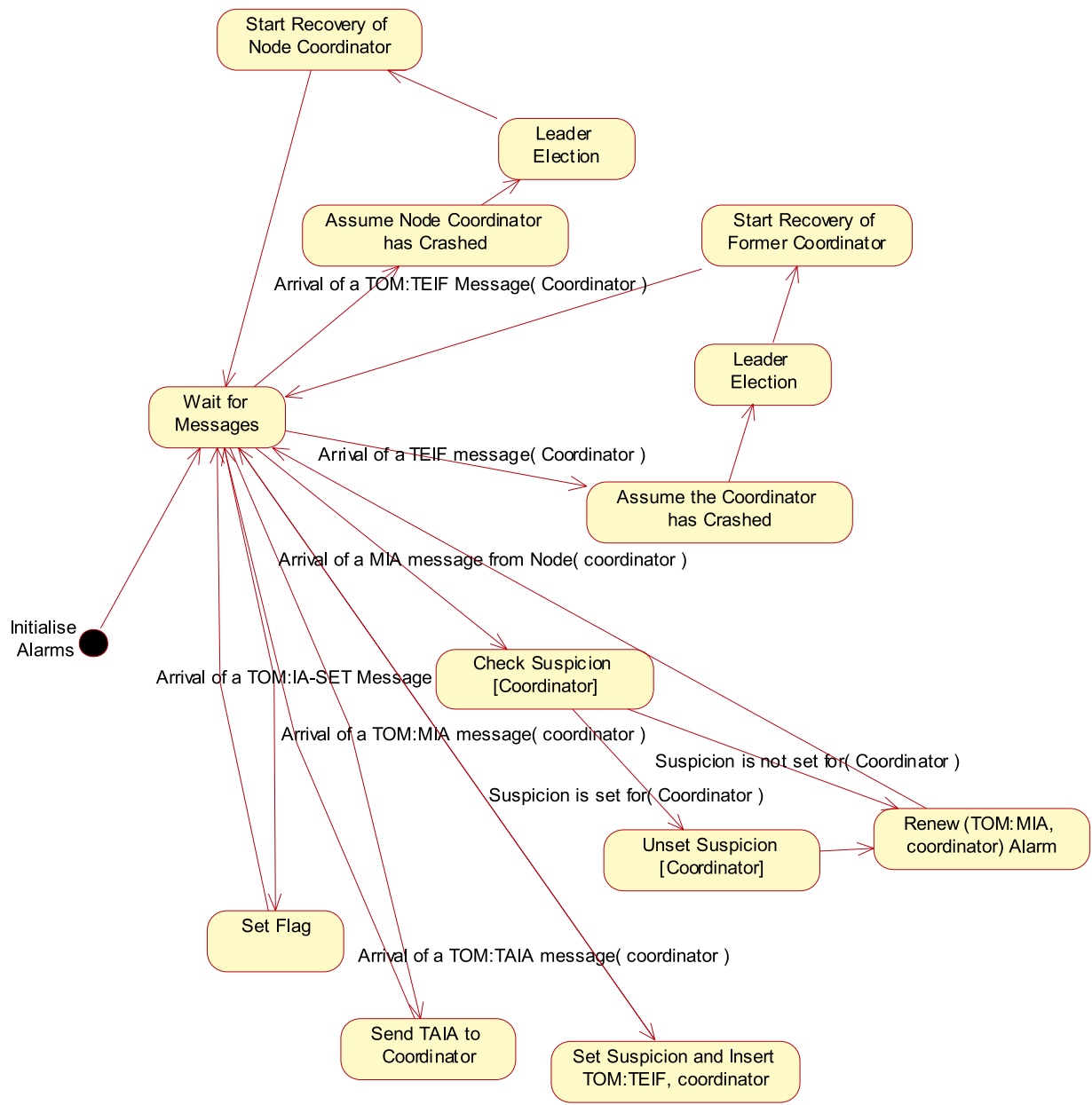

**Figure 5.10: A state diagram of the algorithm of the assistant.**

A number of techniques have been devised to assess the character of faults—some of them are based on tracking the occurrences and the frequency of faults and adopting thresholds of the kind "a device is diagnosed as affected by a permanent fault when four failures occurs within 24 hours". This and other similar heuristics are described in [LS90]. A novel fault identification mechanism, called $\alpha$-count, has been described in [PABD$^{+}$99] and recently generalised in [BCDGG00]. $\alpha$-count is also based on thresholds. The basic idea is that each system component to be assessed is to be "guarded" by an error detection device. This device should forward its deductions, in the form of a binary digit, to a device called $\alpha$-count filter. For each incoming judgement, the filter would then perform the $\alpha$-count technique and issue an assessment about the nature of the fault leading to the detected error.

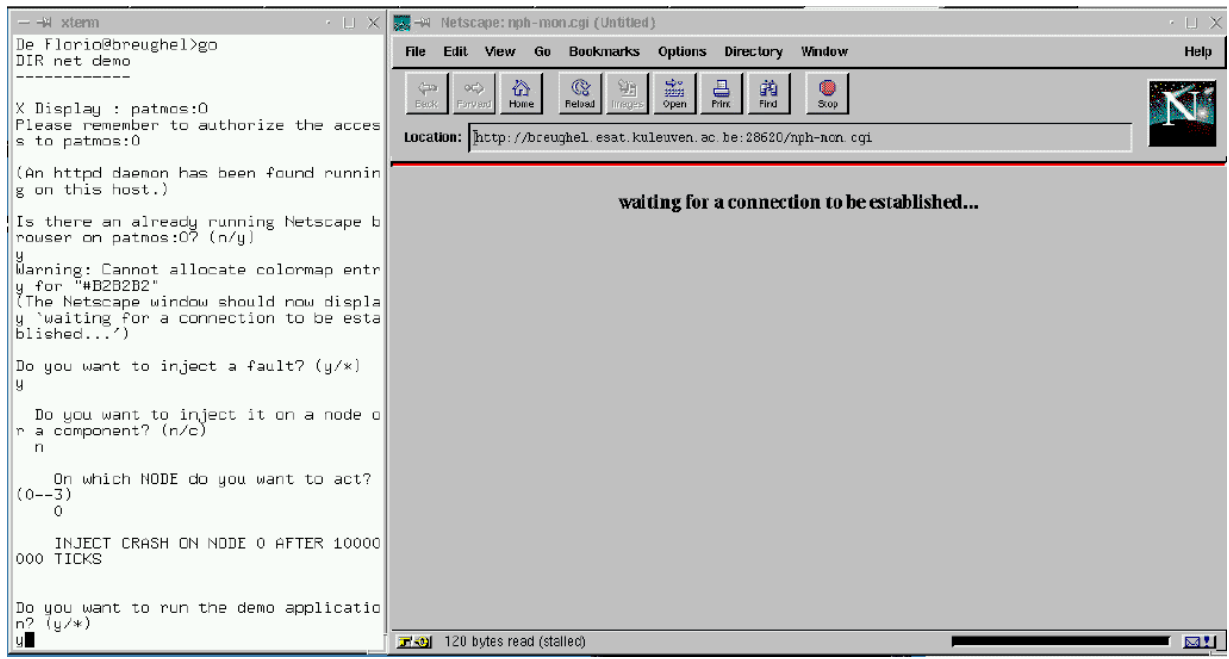

**Figure 5.11:  A fault is injected on node 0 of a system of four nodes. The Node hosts the manager of the BB. The user selects the fault to be injected and connects to a Netscape browser remotely controlled by the MON application.**

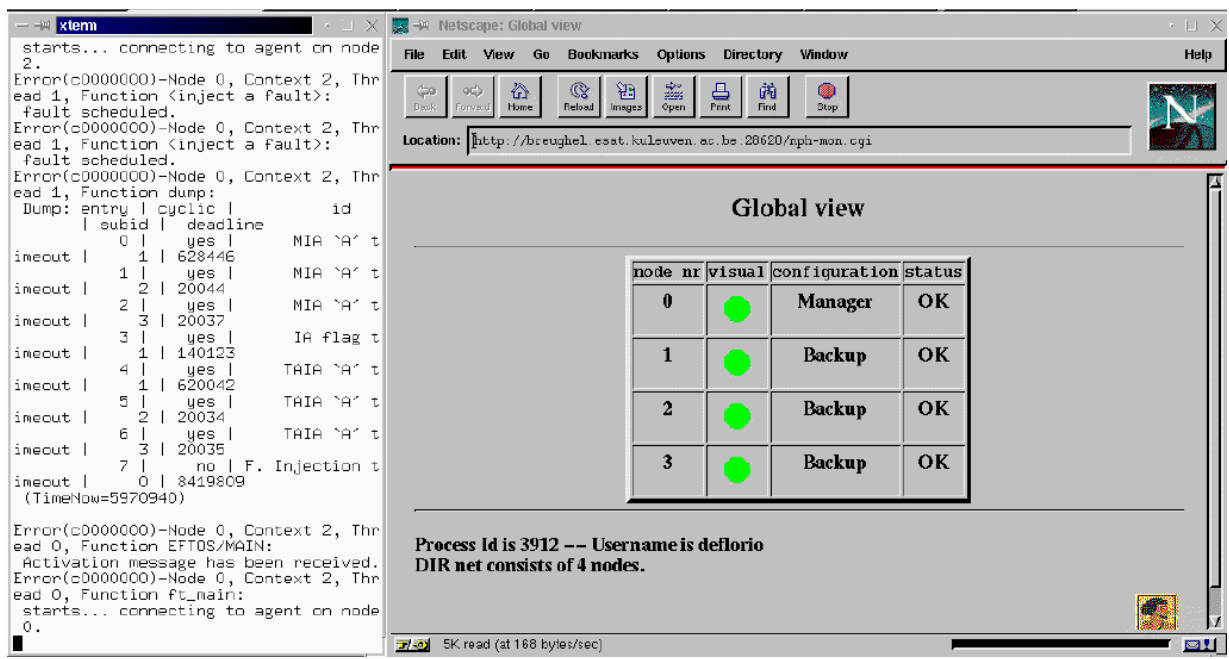

**Figure 5.12:  The browser acts as a renderer of the shape and state of the system.  The textual window reports the current state of the list of time-outs used by the TOM component on node 0.**

**Figure 5.13: The crash of node 0 has been detected and a new manager has been elected.**

**Figure 5.14: On election, the manager expects node 0 to be back in operation. In this case this is not true (that node has been shut down), so node 0 is detected as inactive and labelled as "KILLED".**

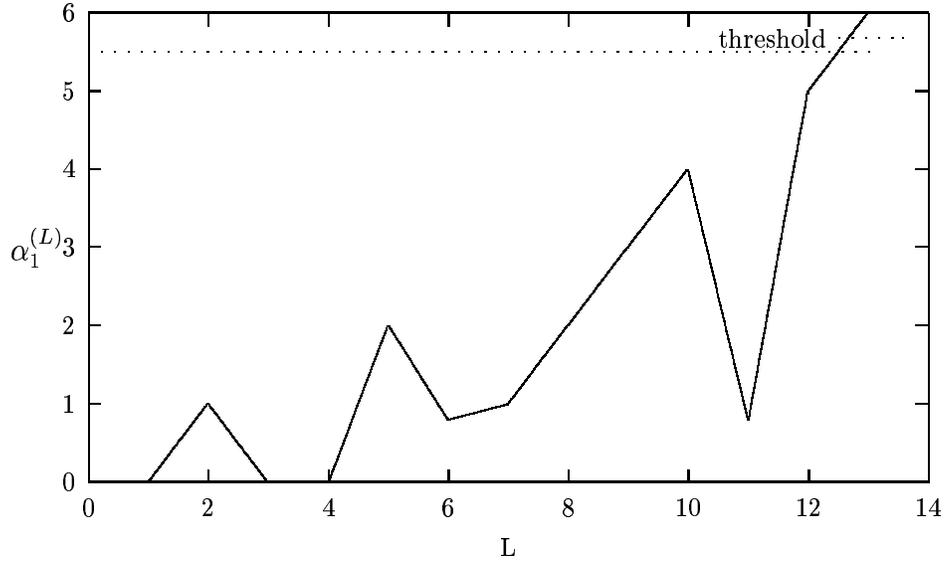

**Figure 5.15:** The picture shows how an $\alpha$-count filter, $\alpha_1^{(L)}$, is updated when it receives 14 error detection notifications ($0 \leq L \leq 13$) from the BB. Notifications are in this case randomly-chosen. Note that, on the last notification ($L = 13$), a threshold (in this case, 5.5) is reached and the corresponding fault is assessed as permanent or intermittent.

More formally, given $n$ "guarded" components $u_i, 0 \leq i < n$, the authors of the strategy call $J_i^{(L)}$ the $L$-th judgement on $u_i$. Judgement 0 means success ($u_i$ is "healthy"), 1 means failure ($u_i$ is faulty). A score vector, $\alpha_i, 0 \leq i < n$, is initially set to zero and updated, for each judgement $L > 1$, as follows:

$$\alpha_i^{(L)} = \begin{cases} \alpha_i^{(L-1)}K & \text{if } J_i^{(L)} = 0 \\ \alpha_i^{(L-1)} + 1 & \text{if } J_i^{(L)} = 1, \end{cases} \tag{5.1}$$

being $0 \leq K \leq 1$. When $\alpha_i^{(L)}$ becomes greater than a certain threshold $\alpha_T$, $u_i$ is diagnosed as affected by a permanent or intermittent fault. This event may be signalled to a reconfigurator or another fault passivation mechanism. The authors of $\alpha$-count show that their mechanism is asymptotically able to identify all components affected by permanent or intermittent faults—if the threshold $\alpha_T$ is set to any finite positive integer $A$, and if the component is indeed affected by a permanent or intermittent fault, they prove that $\alpha_i^{(L)}$ will eventually become greater than or equal to $A$ [BCDGG97]. The authors also prove that similar results can be reached with some variants of formula (5.1). Figure 5.15 shows how that counter evolves when random-valued notifications (either positive or negative assessments) are sent to the BB.

It is worth noting how the mechanism described above requires an approach slightly similar to the one of the $\mathcal{REL}$ toolset: in both cases, a stream of detection data has to flow from a periphery of detection tools to a collector. This collector is the $\alpha$-count filter in the one case and the TIRAN BB and its DB in the other one. This makes it straightforward adopting $\alpha$-count in TIRAN: the filter can simply be fed with data coming from the detection BTs just before this data in inserted into the DB. The added value of this approach is that it is possible to set up a function that returns the current estimation of the nature of a fault (see Sect. 5.3.3). Clearly, a requirement of this technique is that each detection tool not only forward notifications of erroneous activity, but also confirmations of normal behaviour. This may have negative consequences on the performance of the approach, as it may result in many communication requests. In the prototype version developed by the author, each notification from any detection tool to the backbone is tagged with an integer label starting at 0 and incremented by 1 at each new deduction. This way, only negative assessments can be sent—in fact, if, for instance, the backbone receives from the same detection tool two consecutive negative assessments labelled respectively, e.g., 24 and 28, then this implicitly means that there have been three positive assessments in between, and the $\alpha$-counter can be updated accordingly—though possibly with some delay. This strategy has been conceived by the author of this dissertation.

Figure 5.16 shows the $\alpha$-counter in action—a task has reached its threshold and has been declared as affected by a permanent-or-intermittent fault.

The design choice to support $\alpha$-count within the strategy of the TIRAN BB was taken by the author of this dissertation.

## 5.2.5 The TIRAN Time-Out Manager

As already mentioned, the main assumption of $\mathcal{REL}$ is the adoption of the timed asynchronous distributed system model [CF99], a recently developed promising model for solving problems such as dynamic membership [CS95] in distributed systems. The availability of a class of functions for managing time-outs is a fundamental requirement for that model. The TIRAN TOM [DFDL00b] fulfils this need—it is basically a mechanism for managing lists of *time-outs*, defined as objects that postpone a function call for a certain number of local clock units. Time-outs are ordered by clock-units-to-expiration, and time-out lists are managed in such a way that only the head of the list needs to be checked for expiration. When the specified amount of time elapses for the head of the list, its function—let us call it "alarm"—is executed and the object is either thrown out of the list or renewed (in this second case, a time-out is said to be "cyclic"). A special thread monitors and manages one or more of such lists, checking for the expiration of the entries in the time-out list.

Within the strategy of the TIRAN BB, a TOM task is available on each node of the system, spawned at initialisation time by the local BB component. TOM is used in this context to translate time-related clauses, such as, "$p_a$ seconds have elapsed", into message arrivals. In other words, each TOM instance of the BB may be represented as a function

$$a : C \rightarrow M$$

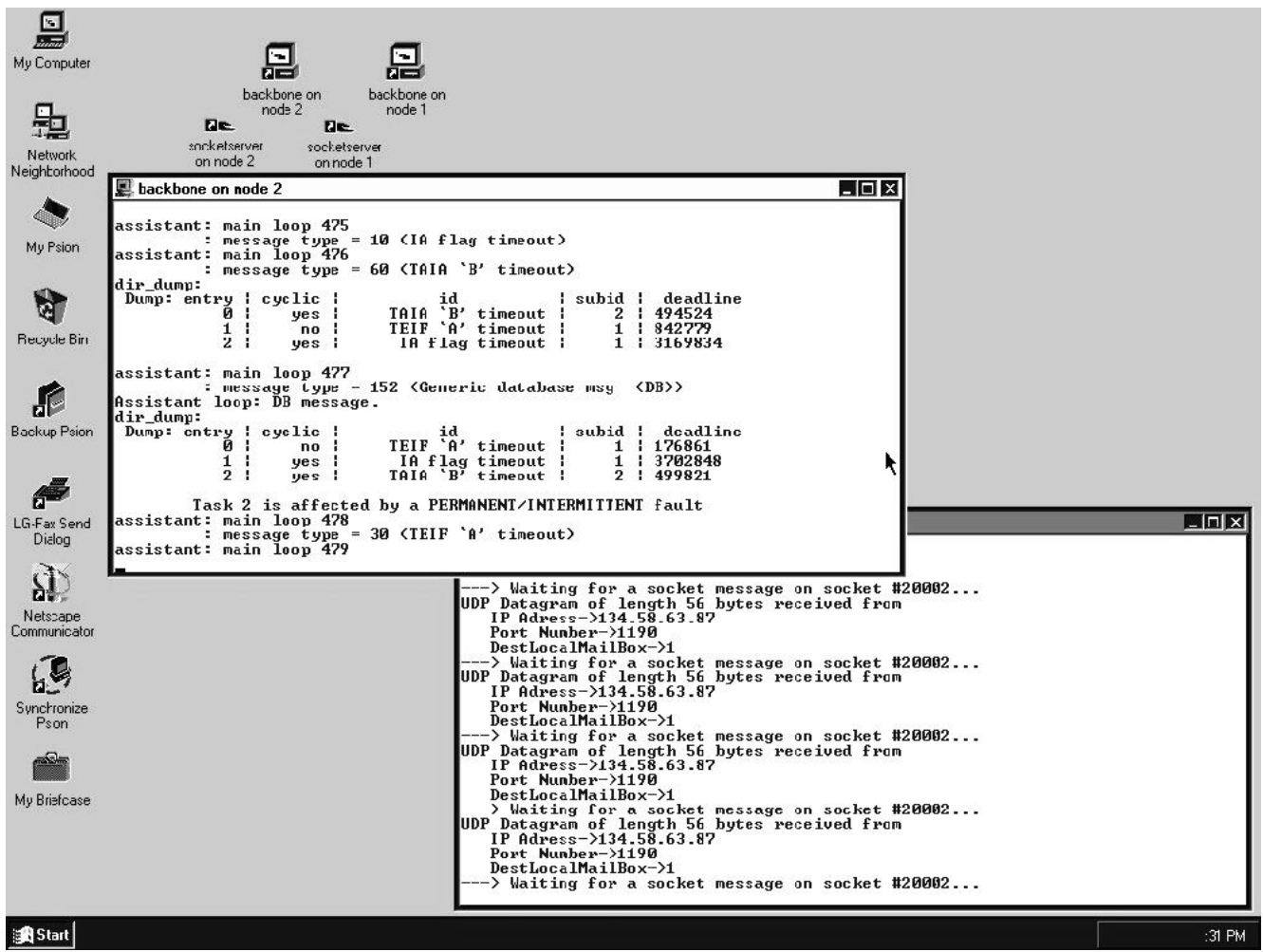

**Figure 5.16:** An assistant running on a node of a cluster of PCs.  The upper window hosts the assistant, the lower one its SocketServer (see Fig. 5.2).  In the former one, note how the $\alpha$-counter detects that task 2 is affected by a permanent-or-intermittent fault.

such that, for any time-related clause $c \in C$:

$$a(c) = \text{message “clause } c \text{ has elapsed”} \in M.$$

This homomorphism is useful because it allows to deal with the entire set of possible events—both messages and time-related events—as *the arrival of one set of messages*.  Hence, one multiple-selection statement such as the C language `switch` can be used, which translates into a simpler and more straightforward implementation for error detection protocols such as the AMS.

TOM uses a well-known algorithm for managing its time-outs [Tan96].  Once the first time-out is entered, TOM creates a linked-list of time-outs and polls the top of the list.  For each new time-out to be inserted, a proper position in the list is found and the list is modified accordingly, as described in Fig. 5.17.  If the top entry expires, a user-defined alarm function is invoked.  This is a general mechanism that allows to associate any event with the expiring of a time-out. In the case of the backbone, the TOM component on node $k$ sends a message to BB component on the same node—the same result may also be achieved by sending that component a UNIX signal [HS87]. Special time-outs are defined as "cyclic", i.e., they are automatically renewed at each new expiration, after invoking their alarm function. A special function renews a time-out, i.e., it deletes and re-enters that entry. It is also possible to temporarily suspend[4] a time-out and re-enable it afterwards.

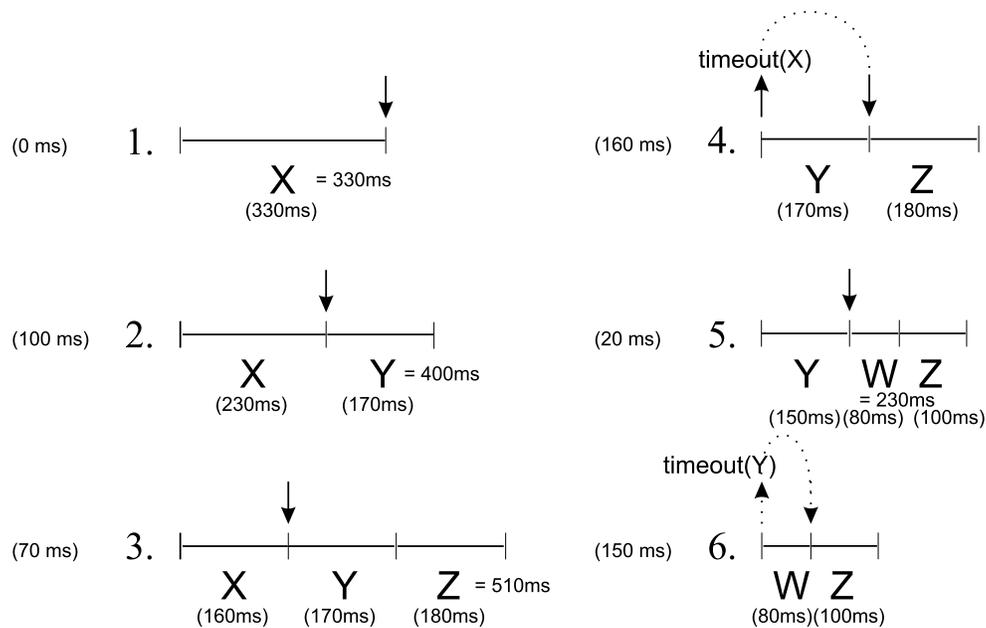

**Figure 5.17:** How the alarm manager works: in 1., a 330ms alarm called X is inserted in the list. In 2., after 100ms, X has been reduced to 230ms and a 400ms alarm, called Y, is inserted (its value is 170, i.e., 400-230). Another 70ms have passed in 3., so X has been reduced to 160ms. At that point, a 510ms alarm, Z is inserted–it goes at the third position. In 4., after 160ms, alarm X occurs–Y becomes then the top of the list; its decrement starts. In 5. another 20ms have passed and Y is at 150ms–at that point a 230ms alarm, called W is inserted. Its position is in between Y and Z, therefore the latter is adjusted. In 6., after 150ms, Y expires and W goes on top.

TOM exploits multiple alarm execution threads in order to reduce the congestion that is due to concurrent execution of alarms and the consequent run-time violations. A description of this strategy can be found in [DFDL00b] and in Appendix B.

The design and the implementation of TOM and of the above mentioned strategy are an original contribution of the author of this dissertation.

---

[4]A time-out is said to be suspended when, on expiration, no alarm is executed. The corresponding entry stays in the time-out list and obeys its rules—in particular, if the time-out was cyclic, on expiration the entry is renewed.

## 5.3    The ARIEL Configuration and Recovery Language

ARIEL  My master through his art foresees the danger
That you, his friend, are in; and sends me forth—
For else his project dies—to keep them living.
(Shakespeare, *The Tempest*, Act II, Scene I)

Within TIRAN, a single syntactical structure—provided by the ARIEL language [DFDL01, DDFB01]—has been devised by the author of this dissertation as both a configuration and a recovery language. This chapter describes this language and its compile-time and run-time tools. The structure of the section is as follows:

- The general characteristics of ARIEL are described in Sect. 5.3.1.

- ARIEL as a configuration language is introduced in Sect. 5.3.2.

- Section 5.3.3 is on ARIEL as a recovery language.

ARIEL *bears its name after the character with the same name of Shakespeare's last Comedy,* The Tempest. *In fact, in the Tempest, spirit Ariel is a magic creature, the invisible assistant of Prospero, the Duke of Milano. While Prospero plays his role in the foreground of the Comedy—orchestrating a strategy to regain possession of his dukedom, which had been usurped by Prospero's brother Antonio—Ariel faithfully serves Prospero performing his magic in the background, exploiting his powers to execute Prospero's commands when his "Master through his art foresees the danger" (see above quote). In a sense, the role of the* ARIEL *language is somewhat similar to the one of Prospero's ally. Its run-time support is given by the agent that is sent forth when the BB, through its nervous terminals—the error detection basic tools—senses a potentially dangerous condition. The author of this dissertation thinks that the name Ariel captures this similarity. This is also the reason that led the author of this dissertation to the choice of "$\mathcal{REL}$" as an abbreviation for "Recovery Language": indeed, the spelling of that word is "[aː*]-[iː]-[el]".*

### 5.3.1    General Characteristics of ARIEL

ARIEL is a declarative language with a syntax somewhat similar to that of the UNIX shells. One instruction per line is allowed. Comments are like in the C shell ("#" starts a comment which ends at next new line or at end-of-file). Names are not case-sensitive. ARIEL deals with five basic types: "*nodes*", "*tasks*", "*logicals*", integers, and real numbers. A node is a uniquely identifiable processing node of the system, e.g., a processor of a MIMD supercomputer. A task is a uniquely identifiable process or thread in the system. A logical is a uniquely identifiable collection of tasks, possibly running on different nodes. Nodes, tasks, and logicals are generically called *entities*. Entities are uniquely identified via non-negative integers; for instance, NODE3 or N3 refer to processing node currently configured as number 3.

Integer symbolic constants can be "imported" from C language header files through the statement `INCLUDE`. For instance, if the C language header file `"vf.h"` contains a define statement such as:

$$\texttt{\#define PROC\_NUM 4,}$$

then it is possible to use that symbolic constant wherever an integer is expected in the language. To de-reference a symbolic constant imported via `INCLUDE`s a "brace-operator" has been defined—for instance, under the above assumptions the following valid ARIEL statement:

$$\texttt{NPROCS = \{PROC\_NUM\}}$$

(described later on) is equivalent to

$$\texttt{NPROCS = 4.}$$

An ARIEL script basically consists of two parts:

- A part dedicated to configuration. This is described in Sect. 5.3.2.

- A part containing the guarded actions which constitute the user-defined error recovery strategy. They are described in Sect. 5.3.3.

## 5.3.2 ARIEL **as a Configuration Language**

Special linguistic support has been designed by the author of this dissertation while taking part in the TIRAN project. Aim of this linguistic support is to facilitate the configuration of the instances of the framework elements, of the system and application parameters, and of the fault-tolerance provisions. Let us call these elements a *framework instance*.

Once the user has configured a framework instance with ARIEL, the TIRAN ARIEL translator "`art`" must be used to translate these high level specifications into the actual C language calls that set up configured tasks such as, for instance, user-configured watchdogs. The output of the translator is a set of C files that need to be compiled with the user application.

A subset of the fault-tolerance provisions described in Chapter 4 are supported by the version of ARIEL described in this dissertation. The rest of this section describes the process of configuring a framework instance with ARIEL.

### 5.3.2.1 System and Application Parameters

The ARIEL configuration language can be used to define and configure the target system and application entities, e.g., nodes, tasks, and group of tasks. The rules defined in Sect. 4.2.2.1 are coded as follows:

- Rule (4.1), as for instance in
$$\text{task}_3 \equiv n_0[8],$$
  is coded as

```
TASK 3 = "TMR.EXE" IS NODE 0, TASKID 8.
```

In other words, the above statement declares that task 8, local to node 0, is to be globally referred to as "task 3". String `"TMR.EXE"` may also be used to refer symbolically[5] to $task_3$. More complex rules are possible—for instance,

```
TASK [1,3] = "Triple" IS NODE 0, TASKID [6,8]
```

is equivalent to

```
TASK 1 = "Triple1" IS NODE 0, TASKID 6,
TASK 2 = "Triple2" IS NODE 0, TASKID 7,
TASK 3 = "Triple3" IS NODE 0, TASKID 8.
```

- Rule (4.2), as in
$$\text{group}_{10} \equiv \{5, 6, 7\},$$

  is coded as

```
LOGICAL 10 = "TMR" IS TASK 5, TASK 6, TASK 7 END LOGICAL.
```

(String `"TMR"` may also be used to refer symbolically to $\text{group}_{10}$). The example defines a logical to be globally referred to as "logical 10", symbolically known as `TMR`, and corresponding to the three tasks whose unique-id are 5, 6, and 7.

Let us assume the following lines have been written in a file called "`test.ariel`":

```
TASK 5 = "TMR_LEADER" IS NODE 0, TASKID 8
TASK 6 = "TMR2" IS NODE 1, TASKID 8
TASK 7 = "TMR3" IS NODE 2, TASKID 8
LOGICAL 10 = "TMR" IS   TASK 5, TASK 6, TASK 7   END LOGICAL
```

File `test.ariel` can be translated by executing `art` as follows:

```
bash-2.02$ art -i test.ariel
Ariel translator, v2.0g 03-Aug-2000, (c) K.U.Leuven 1999, 2000.
Parsing file test.ariel...
...done (4 lines in 0.01 CPU secs, or 400 lines per CPU sec.)
Output written in file .rcode.
Logicals written in file LogicalTable.csv.
Tasks written in file TaskTable.csv.
Alpha-count parameters written in file alphacount.h.
```

---

[5]An associative array (see, for instance, [DF96a]) may then be used to de-references an entity through its symbolic name.

The bold typefaced string is the command line. What follows is the output printed to the user screen. The italics-highlighted strings are the names of two configuration files that are written by `art` on successful execution. These two files declare tasks and logicals in the format expected by the BSL and by other ancillary components (e.g., on the version for Windows NT, these tables are also used by a "SocketServer[6]").

The user can also define a number of other parameters of the TIRAN world like, for instance:

- $N$, i.e., the total number of processing nodes in the system.

- $t_i$, i.e., the number of tasks running on node $n_i$.

- Task-specific parameters of the $\alpha$-count fault identification mechanism supported by the TIRAN BB.

As an example, the following lines declare a system consisting of two nodes, each of which has to host 10 user tasks, and define the $\alpha$-count parameters of task 5:

```
NPROCS = 2
NUMTASKS 1 = 10
NUMTASKS 2 = 10
ALPHA-COUNT 2 IS  threshold = 3.0, factor = 0.4 END ALPHA-COUNT.
```

The output produced by the `art` translator is given in this case by a number of header files:

```
bash-2.02$ art -i test.ariel -s
Ariel translator, v2.0g 03-Aug-2000, (c) K.U.Leuven 1999, 2000.
Parsing file test.ariel...
...done (8 lines in 0.01 CPU secs, or 800 lines per CPU sec.)
Output written in file .rcode.
Logicals written in file LogicalTable.csv.
Tasks written in file TaskTable.csv.
static version
Preloaded r-codes written in file trl.h.
Time-outs written in file timeouts.h.
Identifiers written in file identifiers.h.
Alpha-count parameters written in file alphacount.h.
```

Again, the command has been given in bold typeface and relevant lines have been highlighted using the italics typeface. The "-s" option, for "static", requests the writing of a number of header files. The produced header files contain definitions like the following one, from file "timeouts.h":

---

[6]In TIRAN lingo, a SocketServer is a task run on each node of the system, which is used by the TIRAN BSL for managing off node communication (via UDP sockets) and for local dispatching of remote messages. This is a well-known technical solution which is used, e.g., in PVM, where a single component, `pvmd` (PVM daemon), is launched on each node of a PVM cluster to manage global tasks [GBD+94].

```
/* Number of available nodes
 */
#define MAX_PROCS                 2
```

These are the first few lines of the output file "alphacount.h":

```
/**********************************************************************
 *                                                                    *
 * Header file alphacount.h                                           *
 *                                                                    *
 * This file contains the parameters of the alphacount filter        *
 * (factor and threshold)                                             *
 * Written by art (v2.0g 03-Aug-2000) on Wed Aug 16 15:46:40  2000 *
 * (c) Katholieke Universiteit Leuven / ESAT / ACCA - 2000.          *
 *                                                                    *
 **********************************************************************/

#ifndef      __ALPHA_COUNT__
#define      __ALPHA_COUNT__

#include "DB.h"

alphacount_t alphas[] = {
                { 0.000000, 0.000000, 0 }, /* entry 0 */
                { 0.000000, 0.000000, 0 }, /* entry 1 */
                { 0.400000, 3.000000, 1 }, /* entry 2 */
                { 0.000000, 0.000000, 0 }, /* entry 3 */
```

(Note how, at entry 2, the threshold and factor of the $\alpha$-count related to task 2 have been entered in an array. The latter will then be used by the TIRAN backbone when updating the $\alpha$-count filters—see Sect. 5.2.4.2).

### 5.3.2.2   Backbone Parameters

ARIEL can be used to configure the BB. In particular, the initial role of each BB component must be specified through the DEFINE statement. For instance, the following two lines configure a system consisting of four nodes and place the BB manager on node 0 and three assistants on the other nodes:

```
    DEFINE 0 = MANAGER
    DEFINE 1-3 = ASSISTANTS
```

A number of backbone-specific parameters can also be specified via the ARIEL configuration language. These parameters include, for instance, the frequency of setting and checking the  flag of the backbone. Values are specified in microseconds. A complete example can be seen in Table 5.2. Again, the art translator changes the above specifications into the appropriate C language settings as expected by the TIRAN BB.

```
    # Specification of a strategy in the recovery language Ariel
    # Include files
1   INCLUDE "phases.h"
2   INCLUDE "vf.h"

    # Definitions
    #     After keyword 'DEFINE', the user can specify
    #     an integer, an interval, or a list, followed by
    #     the equal sign and a backbone role, that may be
    #     ASSISTANT(s) or MANAGER
3   NPROCS = 4
4   Define 0 = MANAGER
5   Define 1-3 = ASSISTANTS

    # Time-out values for the BB and the  mechanism
6   MIA_SEND_TIMEOUT = 800000 # Manager Is Alive -- manager side
7   TAIA_RECV_TIMEOUT = 1500000 # This Assistant Is Alive -- manager side

8   MIA_RECV_TIMEOUT = 1500000 # Manager Is Alive -- backup side
9   TAIA_SEND_TIMEOUT = 1000000 # This Assistant Is Alive -- backup side

10  TEIF_TIMEOUT = 1800000 # After this time a suspected node is assumed
    # to have crashed.

11  I'M_ALIVE_CLEAR_TIMEOUT = 900000 #  timeout -- clear IA flag
12  I'M_ALIVE_SET_TIMEOUT = 1400000 #  timeout -- set IA flag

    # Number of tasks
13  NUMTASKS 0 = 11 # node 0 is to be loaded with 11 tasks
14  NUMTASKS 1 = 10
15  NUMTASKS 2 = 10
16  NUMTASKS 3 = 10

17  TASK [0,10] IS NODE 0, TASKID [0,10]
18  TASK [11,20] IS NODE 1, TASKID [1,10]
19  TASK [21,30] IS NODE 2, TASKID [1,10]
20  TASK [31,40] IS NODE 3, TASKID [1,10]

21  LOGICAL 1 IS TASK 1, TASK 2, TASK 3 END LOGICAL
```

**Table 5.2: An excerpt from an** ARIEL **script: configuration part. Line numbers have been added for the sake of clarity.**

#### 5.3.2.3   Basic Tools

ARIEL can be used to configure statically the TIRAN tools.  The current prototypic version can configure only one tool, the TIRAN watchdog.  The following syntax is recognised by `art` to configure it:

```
WATCHDOG 10 WATCHES TASK 14
  HEARTBEATS EVERY 100 MS
  ON ERROR WARN TASK 18
END WATCHDOG.
```

The output in this case is a C file that corresponds to a configured instance of a watchdog. The application developer needs only to send heartbeats to that instance, which can be done as follows:

```
HEARTBEAT 10.
```

#### 5.3.2.4   Configuring Multiple-Version Software Fault-Tolerance

As described in Sect. 4.2.2.5, it is possible to design a syntax to support the configuration of the software fault-tolerance provisions described in Sect. 3.1.2.  This section describes the solution provided by ARIEL in order to support $N$-version programming [Avi85], and a possible syntax to support consensus recovery blocks [SGM85].

**$N$-version programming.**    The following is an example that shows how it is possible to define an "$N$-version task" with ARIEL:

```
#include "my_nvp.h"
N-VERSION LOGICAL {NVP_LOGICAL}
  VERSION 1 IS TASK{VERSION1}  TIMEOUT {VERSION_TIMEOUT}
  VERSION 2 IS TASK{VERSION2}  TIMEOUT {VERSION_TIMEOUT}
  VERSION 3 IS TASK{VERSION3}  TIMEOUT {VERSION_TIMEOUT}
METRIC "nvp_comp"
ON SUCCESS TASK{NVP_OUTPUT}
ON ERROR   TASK{NVP_ERROR} {NVP_LOGICAL}
VOTING ALGORITHM IS MAJORITY
END N-VERSION
```

The `art` translator, once fed with the above lines, produces three source files for tasks the unique-id of which is {VERSION1}, {VERSION2} and {VERSION3}.  These source files consist of code that

- sets up the TIRAN distributed voting tool (described in Sect. 5.2.3) using metric function

  `int nvp_comp(const void*, const void*)`, setting the voting algorithm to majority voting, and so forth;

- redirects standard output streams;

- executes a user task, e.g., task {VERSION3}.

By agreement, each user task (i.e., each version) has to write its output onto the standard output stream.

During run-time, when the user needs to access a service supplied by an NVP logical, it simply sends a message to entity {NVP_LOGICAL}. This translates into a multicast to tasks {VERSION1}, {VERSION2} and {VERSION3}. These tasks, which in the meanwhile have transparently set up a distributing voting tool,

- get their input,

- compute a generic function,

- produce an output

- and (by the above stated agreement) they write the output onto their standard output stream.

This, which had been already redirected through a piped stream to the template task, is fed into the voting system. This latter eventually produces an output that goes to task {NVP_OUTPUT}.

A time-out can also be set up so to produce an error notification when no output is sent by a version within a certain deadline.

Table 5.3 shows one of the three files produced by the ARIEL translator when it parses the script of Sect. 5.3.2.4. Note how this file basically configures an instance of the TIRAN DV tool described in Sect. 5.2.3. Note also how all technicalities concerning:

- the API of the tool,

- input replication,

- the adopted voting strategy,

- output communication,

and so forth are fully made transparent to the designer, who needs only be concerned with the functional service. This allows the fault-tolerance designer to modify all the above mentioned technicalities with no repercussions on the tasks of the application designer, and even to deploy different strategies depending on the particular target platform. This can be exploited in order to pursue performance design goals.

**Consensus recovery blocks.** Support towards consensus recovery block may be provided in a similar way, e.g., as follows:

```
#include "TIRAN.h"
/* Task 101 of NVersion Task 20
    Version 1 / 3
 */
int TIRAN_task_101(void) {
    TIRAN_Voting *dv; size_t size;
    double task20_cmp(const void*, const void*);

    dv = TIRAN_VotingOpen(task20_cmp);
    if (dv == NULL) {
        RaiseEvent(TIRAN_ERROR_VOTING_CANTOPEN,TIRAN_DV,101,0);
        TIRAN_exit(TIRAN_ERROR_VOTING_CANTOPEN);
    }

    /* voting task description: which tasks and which versions */
    /* constitute the n-version task */
    TIRAN_VotingDescribe(dv, 101, 1, 1);
    TIRAN_VotingDescribe(dv, 102, 2, 0);
    TIRAN_VotingDescribe(dv, 103, 3, 0);

    TIRAN_VotingRun(dv);

    /* output should be sent to task 40 */
    TIRAN_VotingOutput(dv, 40);
    TIRAN_VotingOption(dv, TIRAN_VOTING_IS_MAJORITY);

    /* redirect stdout into a pipe input stream */
    TIRAN_pipework();

    /* execute the version */
    task_101();

    size = read(0, buff, MAX_BUFF);
    if (size > 0) {
        /* forward the input buffer to the local voter of this version */
        TIRAN_VotingInput(dv, buff, size);
    } else {
        /* signal there's no input */
        TIRAN_VotingInput(dv, NULL, 0);
        RaiseEvent(TIRAN_ERROR_VOTING_NOINPUT,TIRAN_DV,101,0);
        TIRAN_NotifyTask(60, TIRAN_ERROR_VOTING_NOINPUT);
    }
}
/* EOF file TIRAN_task_101.c */
```

**Table 5.3: Translation of the N-Version Task defined in Sect. 5.3.2.4.**

```
#include "my_crb.h"
CONSENSUS RECOVERY BLOCK LOGICAL {CRB_LOGICAL}
   VARIANT 1 IS TASK{VARIANT1}  TIMEOUT {VARIANT_TIMEOUT}
     ACCEPTANCE TEST TASK{ACCEPT1}
   VARIANT 2 IS TASK{VARIANT2}  TIMEOUT {VARIANT_TIMEOUT}
     ACCEPTANCE TEST TASK{ACCEPT2}
   VARIANT 3 IS TASK{VARIANT3}  TIMEOUT {VARIANT_TIMEOUT}
     ACCEPTANCE TEST TASK{ACCEPT3}
METRIC "crb_comp"
ON SUCCESS TASK{CRB_OUTPUT}
ON ERROR    TASK{CRB_ERROR} {CRB_LOGICAL}
VOTING ALGORITHM IS MAJORITY
END CRB
```

This is syntactically similar to the previous example, but the user is asked to supply tasks for the execution of acceptance tests. Other possibilities might also be considered, e.g., supplying a function name corresponding to the acceptance test, in order to avoid the overhead of spawning a task for that purpose.

### 5.3.3    ARIEL **as a Recovery Language**

The same linguistic structure that realises the TIRAN configuration language is used also to host the structure in which the user defines his / her error recovery strategies. Recovery strategies are collections of *sections* with the following syntax[7]:

```
section :  if elif else fi ;

if      :  IF '[' guard ']' THEN actions ;

elif    :
        |  ELIF '[' guard ']'
           THEN actions elif ;

else    :
        |  ELSE actions ;

fi      :  FI ;
```

where non-terminals `guard` and `actions` are the syntactical terms defined respectively in Sect. 4.2.3.1 and Sect. 4.2.3.2.

---

[7]Here and in the following, context-free grammars are used in order to describe syntax rules. The syntax used for describing those rules is that of the yacc [Joh75] parser generator—so, for instance, symbol "|" supplies an alternative definition of a non-terminal symbol. Terminal symbols like `GT` are in capital letters. They are considered as intuitive and their definition (in this case, string ">") in general will not be supplied.

**ARIEL's guards.**   An excerpt of the context-free grammar rules for guards follows:

```
status   :FAULTY | RUNNING | REBOOTED | STARTED | ISOLATED
          | RESTARTED | TRANSIENT ;

entity   :GROUP | NODE | TASK ;

expr     :status entity
         |'(' expr ')'
         |expr AND expr
         |expr OR expr
         |NOT expr
         |ERRN '(' entity ')' comp NUMBER
         |PHASE '(' entity ')' comp NUMBER ;

comp     :EQ | NEQ | GT | GE | LT | LE ;
```

The following conditions and values have been foreseen:

**Faulty.**  This is true when an error notification related to a processing node, a group of tasks, or a single task, can be found in the TIRAN DB.

**Running.**  True when the corresponding entity is active and no error has been detected that regards it.

**Rebooted**  (only applicable to nodes).  This means that a node has been rebooted at least once during the run-time of the application.

**Started**  (not applicable to nodes). This checks whether a waiting task or group of task has been started.

**Isolated.**  This clause is true when its argument has been isolated from the rest of the application through a deliberate action.

**Phase**  (only applicable to tasks). It returns the current value of an attribute set by any task via the public function `RaiseEvent`. This value is forwarded to the BB to represent its current "phase" or state (e.g., an identifier referring to its current algorithmic step, or the outcome of a test or of an assertion).  For instance, a voting task could inform the BB that it has completed a given algorithmic step by setting a given integer value after each step (this approach is transparently adopted in the EFTOS voting tool and is described in more detail in [DFDL98c]).  Recovery block tests can take advantage of this facility to switch back and try an alternate task when a primary one sets a "failure" phase or when a guarding watchdog expires because a watched task sent it no signs of life. This condition returns an integer symbol that can be compared via C-like arithmetic operators.

**Restarted**  (not applicable to nodes). This returns the number of times a given task or group has been restarted.  It implies **started**.

**Transient** is true when an entity has been detected as faulty and the current assessment of the
$\alpha$-count fault identification mechanism (see Sect. 5.2.4.2) is "transient". It implies **faulty**.

Furthermore, it is possible to query the number of errors that have been detected and pertain
to a given entity. Complex guards can be built via the standard logic connectives and parentheses.
As an example, the following guard:

```
FAULTY TASK{MASTER} AND ERRN(TASK{MASTER}) > 10 AND
              RESTARTED TASK{MASTER}.
```

checks whether the three conditions:

- the task, the unique-id of which is the value of the symbolic constant `MASTER`, has been
  detected as faulty;

- more than 10 errors have been associated to that task;

- that task has been restarted,

are all true.

**ARIEL's actions.** "Actions" can be attached to the `THEN` or `ELSE` parts of a section. In the
current implementation of the language, these actions allow to start, isolate, restart, terminate a
task or a group of tasks, to isolate or reboot a node, to invoke a local function. Moreover, it is
possible to multicast messages to groups of tasks and to purge events from the DB.

An excerpt of the context-free grammar for ARIEL's actions follows:

```
actions    :
           |  actions action ;

action     :
           |  section
           |  recovery_action ;

recovery_action
           :  STOP entity
           |  ISOLATE entity
           |  START entity
           |  REBOOT entity
           |  RESTART entity
           |  ENABLE entity
           |  SEND NUMBER TASK
           |  SEND NUMBER GROUP
           |  WARN entity ( condition )
           |  REMOVE PHASE entity FROM ERRORLIST
           |  REMOVE ANY entity FROM ERRORLIST ;
```

```
        |   CALL NUMBER
        |   CALL NUMBER '(' list ')'

condition :  ERR NUMBER entity ;
```

As suggested in Sect. 4.2.3.2, a special case of action is a section, i.e., another guarded action. This allows to specify hierarchies (trees) of sections such that, during the run-time evaluation of the recovery strategies, a branch is only visited when its parent clause has been evaluated as true.

In the current, prototypic implementation, the following actions have been foreseen:

**Stop**  terminates a task or a group of tasks, or initiates the shutdown procedure of a node[8].

**Isolate**  prevents an entity to communicate with the rest of the system[9].

**Reboot**  reboots a node (via the `TIRAN_Reboot_Node` BT).

**Start**  spawns (or, in static environments, awakes) a task or a group.

**Restart**  is reverting a task or group of tasks to their initial state or, if no other means are available, stopping that entity and spawning a clone of it.

**Enable**  awakes a task or group, or boots a node.

**Send**  multicasts (or sends) signals to groups of tasks (or single tasks).

**Warn**  informs a task or group of tasks that an error regarding an entity has been detected.  Action "`WARN` $x$" is equivalent to action "`SEND {WARN}` $x$"

**Remove**  purges records from the section of the DB collecting the errors or the phases of a given entity.

Custom actions and conditions may be easily added to the grammar of ARIEL[10].

When actions are specified, it is possible to use meta-characters to refer implicitly to a subset of the entities involved in the query. For instance, when the first atom specifies a group of tasks,

---

[8]These services are obtained via specific function calls to the level-1.2 BTs (see the edge from RINT to those BTs in Fig. 5.1). Such BTs, in turn, can either execute, through the BSL, a kernel-level function for stopping processes— if available—or send a termination signal to the involved processes.  The actual choice is taken transparently, and RINT only calls one or more times either a `TIRAN_Stop_Task` or a `TIRAN_Stop_Node` function.

[9]This service is obtained as described in the previous footnote. Depending on the characteristics of the adopted platform, isolation can be reached either through support at the communication driver or kernel level, or as follows: when a task opens a connection, a reference to a given object describing the connection is returned to both the user code and the local component of the BB. Action `ISOLATE` simply substitutes this object with another one, the methods of which prevent that task to communicate.  This explains the third application-specific assumption of Sect. 4.1.2.

[10]For instance, condition "`DEADLOCKED`" and action "`CALL`" (see Appendix C) were added to test the inclusion in ARIEL, respectively, of a provision for checking whether two tasks are in a deadlock (see [EVM+98] for a description of this provision) and of a hook to the function call invocation method. These two provisions were easily included in the grammar of ARIEL.

```
IF [ FAULTY (GROUP{My_Group}) AND NOT TRANSIENT (GROUP{My_Group}) ]
THEN
       STOP TASK@
       SEND {DEGRADE} TASK~
FI
```

**Table 5.4: This section queries the state of group {My_Group} and, if any of its tasks have been detected as affected by a permanent or intermittent fault, it stops those tasks and sends a control message to those considered as being correct so that they reconfigure themselves gracefully degrading their overall service.**

STOP TASK@1 means "terminate those tasks, belonging to the group mentioned in the first atom of the guard, that fulfil that condition", while WARN TASK~2 means "warn those tasks, belonging to the group mentioned in the second atom of the guard, that *do not fulfil* the second condition". If one does not specify any number, as in STOP TASK@, then the involved group of tasks is the one that fulfils the whole clause. Table 5.4 shows an example of usage of this feature. Metacharacter "star" (*) can be used to refer to any task, group, or processing node in the system. Actions like STOP TASK* or STOP GROUP* are trapped by the translator and are not allowed. Metacharacter "dollar" ($) can be used to refer in a section to an entity mentioned in an atom. For instance, STOP GROUP$2 means "stop the group mentioned in the second atom of the clause".

The whole context-free grammar of ARIEL is provided in Appendix C.

### 5.3.3.1 Compile-time Support for Error-Recovery

Once fed with a recovery script, the art translator produces a binary pseudo-code, called the **r-code**. In the current version, this r-code is written in a binary file and in a C header file as a statically defined C array, as in Table 5.5. As can be seen in that table, the r-code is made of a set of "triplets" of integers, given by an opcode and two operands. These are called "r-codes".

This header file needs to be compiled with the application. Run-time error recovery is carried out by the RINT module, which basically is an r-code interpreter. This module and its recovery algorithm are described in the following section. The rest of this section describes how to translate an ARIEL script into the r-code. Within this section and the following one the simple script of Table 5.6 will be used as an example.

The following scenario is assumed: a triple modular redundancy (TMR) system consisting of three voting tasks, identified by integers {VOTER1}, {VOTER2}, and {VOTER3} is operating. A fourth task, identified as T{SPARE}, is available and waiting. It is ready to take over one of the voting tasks should the latter fail. The failed voter signals its state to the backbone by entering phase HAS_FAILED through some self-diagnostic module (e.g., assertions or control-flow monitoring). The spare is enabled when it receives a {WAKEUP} message and it requires the identity of the voter it has to take over. Finally, it is assumed that once a voter receives a control message with the identity of the spare, it has to initiate a reconfiguration of the TMR

```
/************************************************************
 *                                                          *
 *  Header file trl.h                                       *
 *                                                          *
 *  Hardwired set of r-codes for Ariel file         ariel  *
 *  Written by art (v2.0g 03-Aug-2000) on   Fri Aug 18 2000 *
 *  (c) Katholieke Universiteit Leuven 1999, 2000.          *
 *                                                          *
 ************************************************************/
#ifndef _T_R_L__H_
#define _T_R_L__H_

#include "rcode.h"
#define RCODE_CARD 15 /* total number of r-codes */

rcode_t rcodes[] = {
/*line#*/        /* opcode */     /* operand 1 */  /* operand 2 */
/*0*/            { R_INC_NEST,            -1,            -1 },
/*1*/            { R_STRPHASE,             0,            -1 },
/*2*/            { R_COMPARE,              1,          9999 },
/*3*/            { R_FALSE,               10,            -1 },
/*4*/            { R_STOP,                18,             0 },
/*5*/            { R_PUSH,                18,            -1 },
/*6*/            { R_SEND,                18,             3 },
/*7*/            { R_PUSH,                 0,            -1 },
/*8*/            { R_SEND,                18,             3 },
/*9*/            { R_PUSH,                 3,            -1 },
/*10*/           { R_SEND,                18,             1 },
/*11*/           { R_PUSH,                 3,            -1 },
/*12*/           { R_SEND,                18,             2 },
/*13*/           { R_DEC_NEST,            -1,            -1 },
/*14*/           { R_OANEW,                1,            -1 },
/*15*/           { R_STOP,                -1,            -1 },
};
```

**Table 5.5: The beginning of header file `trl.h`, produced by `art` specifying option "`-s`". Array `rcodes` is statically initialised with the r-code translation of the recovery strategy in the ARIEL script.**

such that the failed voter is switched out of and the spare is switched in the system.

Table 5.6 shows a recovery section that specifies what to do when task {`VOTER1`} fails. The user needs to supply a section like the one in lines 8–12 for each voting task. Another, more concise method based on defining a group of tasks (as in line 23 of Table 5.2) is discussed in Sect. 7.2.2.4.

```
1   INCLUDE "my_definitions.h"

2   TASK {VOTER1} IS NODE {NODE1}, TASKID {VOTER1}
3   TASK {VOTER2} IS NODE {NODE2}, TASKID {VOTER2}
4   TASK {VOTER3} IS NODE {NODE3}, TASKID {VOTER3}
5   TASK {SPARE} IS NODE {NODE4}, TASKID {SPARE}

6   IF [ PHASE (T{VOTER1}) == {HAS_FAILED} ]
7   THEN
8           STOP T{VOTER1}

9           SEND {WAKEUP} T{SPARE}
10          SEND {VOTER1} T{SPARE}

11          SEND {SPARE} T{VOTER2}
12          SEND {SPARE} T{VOTER3}
13  FI
```

**Table 5.6: Another excerpt from a recovery script: after the declarative part, a number of "sections" like the one portrayed in here can be supplied by the user.**

Once the specification has been completed, the user can translate it, by means of the `art` program, into a pseudo-code whose basic blocks are the r-codes (see Table 5.7). A textual representation of the r-codes is also produced (see Table 5.8).

Other than syntax errors, `art` catches a number of semantical inconsistencies which are reported to the user—as an example, a non-sense request, such as asking the phase of a node, gives rise to the following response:

```
bash-2.02$ art -i .ariel -s
Ariel translator, v2.0f 25-Jul-2000, (c) K.U.Leuven 1999, 2000.
Parsing file .ariel...
        Line 76: semantical error: Can only use PHASE with tasks
        if-then-else: ok
...done (85 lines.)
1 error detected --- output rejected.
```

### 5.3.3.2  The ARIEL Recovery Interpreter

This section briefly describes the ARIEL recovery interpreter, RINT. Basically RINT is a virtual machine executing r-codes. Its algorithm is straightforward: each time a new error or burst of errors is detected,

- it executes the r-codes one triplet at a time;

- if the current r-code requires accessing the DB, a query is executed and the state of the entities mentioned in the arguments of the r-code is checked;

```
bash-2.02$ art -i ariel -v -s
Ariel translator, v2.0g 03-Aug-2000, (c) K.U.Leuven 1999, 2000.
Parsing file ariel...
[ Including file 'my_definitions.h' ...9 associations stored. ]
substituting {VOTER1} with 0
substituting {NODE1} with 1
substituting {VOTER2} with 1
substituting {NODE2} with 2
substituting {VOTER3} with 2
substituting {NODE3} with 3
substituting {SPARE} with 3
substituting {NODE4} with 4
substituting T{VOTER1} with T0
substituting {HAS_FAILED} with 9999
substituting {WAKEUP} with 18
substituting T{SPARE} with T3
substituting T{VOTER2} with T1
substituting T{VOTER3} with T2
if-then-else: ok
...done (17 lines in 0.02 CPU secs, or 850.000 lines per CPU sec.)
Output written in file .rcode.
Tasks written in file TaskTable.csv.
Preloaded r-codes written in file trl.h.
Time-outs written in file timeouts.h.
Identifiers written in file identifiers.h.
```

**Table 5.7: The `art` program translates the section mentioned in Table 5.6 into r-codes. The "`-i`" option is used to specify the input filename, "`-v`" sets the verbose mode, while "`-s`" allows to create three header files containing, among other things, an array of pre-loaded r-codes (see Table 5.5).**

- if the current r-codes requires executing actions, a request for execution is sent to the BB.

RINT plays an important role within the $\mathcal{REL}$ architecture—its main objective is establishing and managing a *causal connection* between the entities of the ARIEL language (identifiers of nodes, tasks, and groups of tasks) and the corresponding components of the system and of the target application. This causal connection is supported by the BB and its DB. In particular, each atom regarding one or more entities is translated at run-time into one or more DB queries. Under the hypothesis that the DB reflects—with a small delay—the actual state of the system, the truth value of the clauses on the entities of the language will have a large probability to tally with the truth value of the assertions on the corresponding components of the system. Furthermore, by means of RINT, symbolic actions on the entities are translated into actual commands on the components. These commands are then managed by the BB as described in Fig. 5.7.

```
Art translated Ariel strategy file: .... ariel
into rcode object file : ............... .rcode

 line              rcode      opn1    opn2
----------------------------------------------
00000          SET_ROLE        0    Manager
00001          SET_ROLE        1    Assistant
00002          SET_ROLE        2    Assistant
00003          SET_ROLE        3    Assistant
00004               IF
00005      STORE_PHASE...    Thread       0
00006         ...COMPARE        ==    9999
00007             FALSE       10
00008             STOP     Thread       0
00009            PUSH...      18
00010          ...SEND     Thread       3
00011            PUSH...       0
00012          ...SEND     Thread       3
00013            PUSH...       3
00014          ...SEND     Thread       1
00015            PUSH...       3
00016          ...SEND     Thread       2
00017               FI
00018      ANEW_OA_OBJECTS      1
00019             STOP
```

**Table 5.8: A textual representation of the r-code produced when translating the recovery section in Table 5.6.**

The RINT task is available and disabled on each BB assistant while it is enabled on the BB manager. Only one execution process is allowed. RINT has the architecture of a stack-based machine—a run-time stack is used during the evaluation of clauses. In a future release of RINT, the run-time stack shall also be used as a means for communicating information between actions. For any r-code being executed, a message will be broadcast to the non-faulty assistants. Next r-code will only be executed when all the non-faulty assistants acknowledge the receipt of this message and update their stack accordingly. This allows to maintain a consistent copy of the current status of the run-time stack on each assistant. Should the manager fail while executing recovery, the new manager would then be able to continue recovery seamlessly, starting from the last r-code executed by the previous manager.

**Chameleon.**    It is worth noting how some of the assumptions as well as some of the architectural solutions adopted in the TIRAN prototype and in its EFTOS predecessor have been recently adopted in a system called Chameleon [KIBW99]. This system uses a set of basic tools (called armors) and a distributed component, hierarchically structured into managers and non-managers. These architectural solutions were adopted independently. No configuration nor any linguistic support to error recovery is currently part of Chameleon.

## 5.4   Conclusions

The feasibility of an architecture supporting the recovery language approach for dependable-software engineering has been proved by describing the TIRAN framework, and in particular the TIRAN backbone and the configuration and recovery language ARIEL. This language and its run-time system provide the user with a fault-tolerance linguistic structure that appears to the user as a sort of second application-layer especially conceived and devoted to address the error recovery concerns. The strategies and the algorithms of the TIRAN framework have been also described. Next chapters complement this one by focusing on the effectiveness of this architecture.

# Chapter 6

# Using the Recovery Language Approach

This chapter focuses on some examples of practical use of the $\mathcal{R}\mathcal{E}\mathcal{L}$ architecture described in Chapter 5. Next chapter complements this one by focusing on the assessment of some of the elements of that architecture.

As a matter of facts, an exhaustive analysis and assessment of an ALFT approach is a research activity that requires both serious usability studies *as the approach matures* and the analysis of *many diverse real-life case studies*—as it has been observed by the authors of the AOL approach what concerns the systems based on their approach [LVL00]. Hence, a *complete* and *detailed evaluation* for a large and complex architecture such as the TIRAN one, in its current prototypic state, is out of the scope of this dissertation. The case studies described in this chapter, as well as the architectural components evaluated in the next one, provide the reader with an idea of the class of applications that $\mathcal{R}\mathcal{E}\mathcal{L}$ may tackle well and of the reasons behind these results, and as such it may be considered as a first effort towards a complete characterisation and evaluation of the $\mathcal{R}\mathcal{E}\mathcal{L}$ approach. Further experimentation will also be carried out in the framework of a new ESPRIT project that is currently being evaluated for acceptance and financing[1].

A first case study is given by one of the pilot applications of the EFTOS project: the GLIMP software, embedded into the Siemens Integrated Mail Processor. It is described in Sect. 6.1, and focuses on the SC attribute and on *performance* issues.

Section 6.2 describes how $\mathcal{R}\mathcal{E}\mathcal{L}$ may be used to further improve the dependability of the class of applications that adopt NVP as their main fault-tolerance provision. It is shown how it is possible to exploit $\mathcal{R}\mathcal{E}\mathcal{L}$ in order to enhance the TIRAN distributed voting tool described in Sect. 5.2.3 and attain both *location transparency* and *replication transparency*.

Section 6.3 reports on the current use that is being made of ARIEL in an industrial application at ENEL in the framework of the TIRAN project.

Section 6.4 shows how $\mathcal{R}\mathcal{E}\mathcal{L}$ can be used to enhance the capabilities of the TIRAN basic tools in a straightforward way. In particular, a strategy is described, devised at CESI and ENEL, which exploits the availability of a number of processing nodes in order to obtain high-safety or high-availability watchdogs from multiple instances of the simple, node-local TIRAN watchdog.

---

[1]A proposal for a project called DepAuDE has recently passed the first steps towards official acceptance and financing. DepAuDE stands for "*Dep*endability for embedded *Au*tomation systems in *D*ynamic *E*nvironments with intra-site and inter-site distribution aspects". Aim of the project is to develop an architecture and methodology to ensure dependability for non-safety critical, distributed, embedded automation systems with both IP (inter-site) and dedicated (intra-site) connections. Dynamic behaviour of the systems is also required to accommodate for remote diagnosis, maintenance or upgrades.

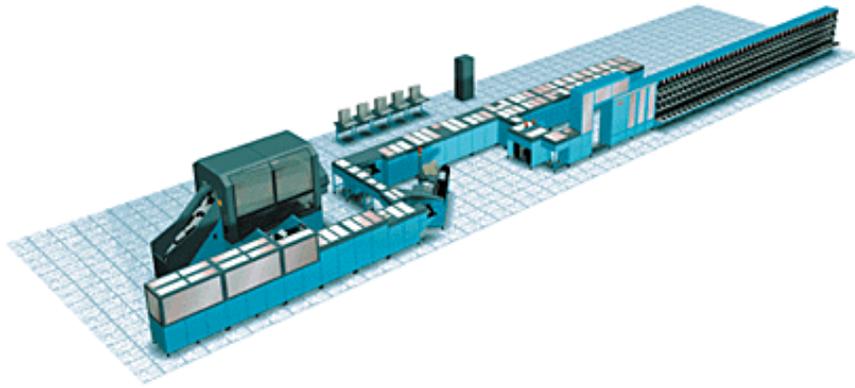

**Figure 6.1: The IMP system.**

Section 6.5 concludes the chapter summarising the positive aspects of $\mathcal{REL}$ in the reported case studies.

## 6.1    A Case Study: The Siemens Integrated Mail Processor

A first case study deals with an application in the industrial automation domain: an automatic mail sorting system developed by Siemens, the Integrated Mail Processor (IMP) [Ele00], is described. Figure 6.1 draws a typical configuration of the IMP. Such system may be defined as an integrated system for automated letter sorting designed to operate at country-level—according to Siemens, "the latest innovation in the automation of letter sorting", combining the processing steps of several individual machines, from pre-processing, to reading and encoding, and up to sorting. Conceived and developed in close cooperation with one of the largest European postal services, the IMP is designed to avoid any manual handling and to ensure that no machinable mail is rejected under normal operating conditions.

This section:

- First introduces the class of problems addressed by IMP (in Sect. 6.1.1).

- Then, in Sect. 6.1.2 and 6.1.3, it describes the hardware and software structures of a version of IMP, based on the Parsytec CC system, which has been subject of investigation during project EFTOS.

- Next, Sect. 6.1.4 describes the main design requirements of IMP related to both real-time and dependability.

- Section 6.1.5 finally describes how the $\mathcal{REL}$ approach was used in this case study and what results were obtained.

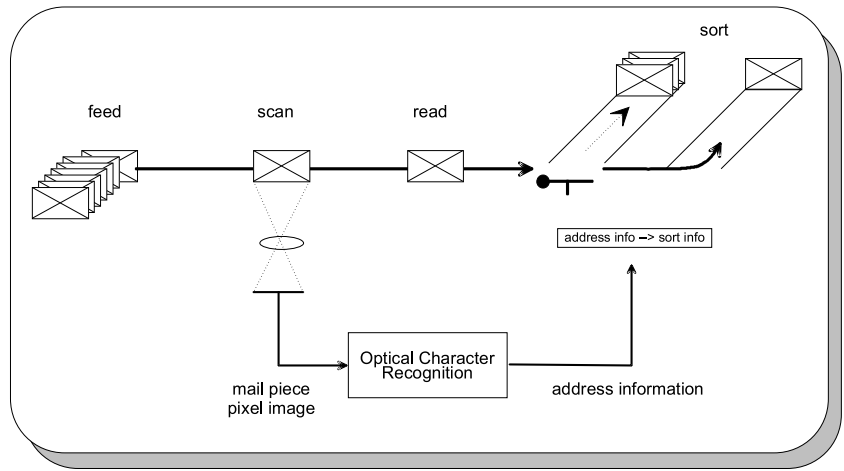

**Figure 6.2: General Structure of Automatic Letter Sorting.**

The main reference for this section is [BGB$^+$96]. In particular, all figures related to real-time, throughput, and dependability requirements for the described application are from the cited document.

## 6.1.1 Automatic Mail Sorting Machines

Automatic Mail Sorting machines are used by many postal services all over the world to optimise the delivery of letters, flats, and packets. Typically, they must be able to transport and sort up to 36000 letters per hour. The main components of an automatic mail sorting machine are the transport machine, the scanner, the address reader (or OCR, i.e., Optical Character Recognition module), and the mechanical sorter. Today OCRs are able to recognise more than 90% of machine printed addresses with error rates less than 1%. The general structure of a typical sorting automation is illustrated in Figure 6.2.

The OCR has to extract the address information from the mail piece image. A strong requirement for the OCR is that the address information must arrive at the sorter *within the time interval given by the mail transport between scanner and sorter*, otherwise the mail piece would be rejected. The required performance calls for a solution based on high-performance computing technology. The actual hardware structure adopting this technology is described in 6.1.2.

Two main parts can be distinguished in the OCR software: a *control module* and the *algorithmic modules*.

- The control module is responsible for the correct collaboration of all algorithmic modules. Part of its task is to detect anomalous behaviour and take appropriate actions in such a case. As an example of this, it supervises the deadlines imposed by the rate at which letters are provided given by the physical transport characteristics of the system. In case of breaking of deadlines, corrective actions are required.

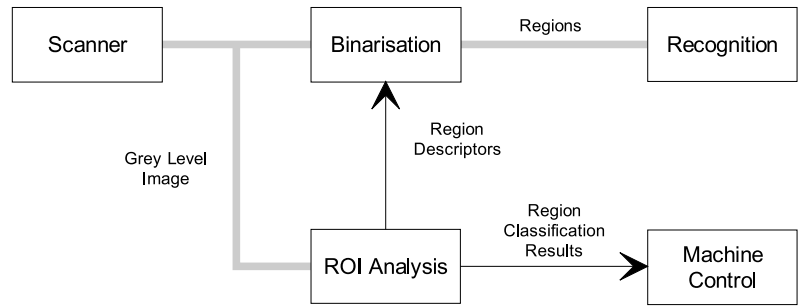

**Figure 6.3: Schematic Data Flow of Image Processing.**

- The algorithmic modules are responsible for the recognition task. They are *purely data-flow driven* and are not affected by modifications of the control software (and vice-versa).

During the EFTOS project, a specific functional part of the recognition module has been subject to investigation—the GLIMP (Gray Level IMage Processing) package. The purpose of GLIMP is to analyse the images of the scanned letters according to their various regions of interest (ROIs). From the mail sorting point of view, the most important ROI is clearly the address, however other ROIs located on the envelope of a letter are very important to the postal services (for example, for purposes of revenue protection and fraud detection, it is required to find and classify the stamp or the meter mark of the letter). Other labels indicate registered mail or foreign mail. The task of this ROI-analysis sub-system is performed in parallel in order to meet the real-time constraints of the sorting system.

The global data flow can be described by the scheme of Fig. 6.3. The raw image of the scanner is fed into two image paths, one processing the image towards binarisation, the other one providing the ROI-analysis of the image, that execute in parallel. The results of this analysis is then used to extract the relevant parts of the image from the binarised image. Only these parts are further transmitted to the character recognition system. The second output of the ROI-analysis is the classification result, e.g. the value of a recognised stamp, or the presence and position of certain labels. These results are transmitted to the machine control (MC) in order to initiate special processing of the respective mail piece in the sorting system.

## 6.1.2   Hardware Structure

The ROI-analysis of the mail piece images is performed in the original scanned grey level image. It is realised on a hardware platform physically separated from the recognition system. The implementation addressed in EFTOS and reported herein is based on the Parsytec CC system. As mentioned before, the CC is a MIMD supercomputer based on the 133-MHz PowerPC processor. This system hosts two types of nodes—the so-called *I/O nodes*, that communicate with external devices via a PCI bus, and the so-called *compute nodes*, purely devoted to processing. Both nodes communicate with each other via HS-links (high-speed links). The interfaces between the CC and the rest of the system is realised by PCI plug-in boards and via an Ethernet connection. A simple representation of the hardware architecture can be seen in Fig. 6.4.

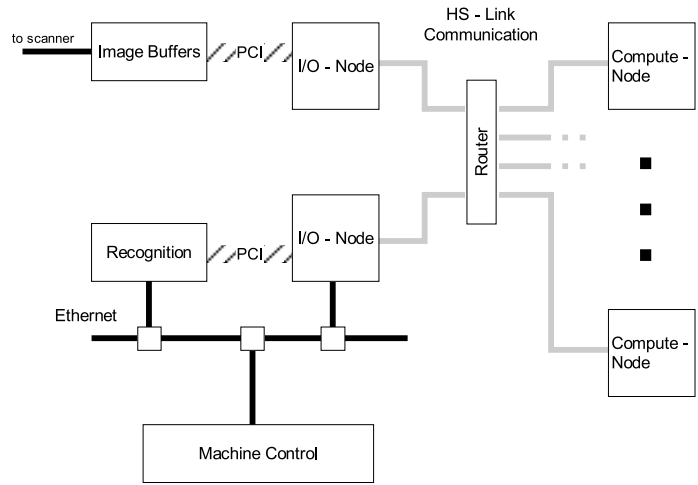

**Figure 6.4: Hardware Architecture of Image Processing Sub-System.**

The images generated in the scanner are stored in an appropriate number of image buffers. The PCI bus interface of the I/O-node allows the images to be read into the system. Further distribution of parts of the images towards the compute node is realised via the HS-link connections. The compute nodes execute the algorithms for image processing and classification of the region of interest. The results of the analysis are passed on to the recognition system across the second I/O node and its PCI interface. The overall machine control interacts with the ROI-analysis system via an Ethernet connection.

## 6.1.3 Software Structure

The overall software structure is completely *data driven*. An initial process, the Image Dispatcher (ID), receives parts of the images (regions) from the image buffer on request. Along with the image data comes a Region Descriptor (RD) indicating the type of the region. Both information is then distributed to a farm of algorithmic worker processes (AW) doing the image processing and classification algorithms. The distribution is controlled by the ID process according to some appropriate load balancing heuristics. The results of the AWs are collected in a Result Manager (RM) process. Once all relevant results belonging to a mail piece are available at the RM, it forwards them to the MC (machine control) and releases the images at the ID. The process graph of the software is depicted in Fig. 6.5.

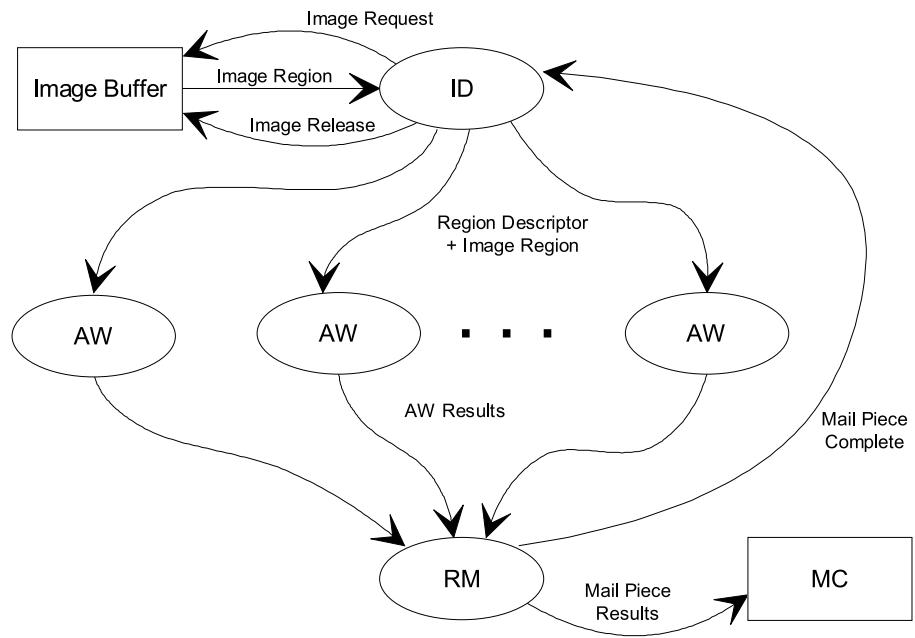

**Figure 6.5: Software Process Graph.**

## 6.1.4  Design Requirements

The overall performance characteristics of the mail sorting system vary according to the different customer requirements. The peak throughput requirements that had to be met at the time were in the order of 40000 mail pieces per hour. This means that the machine had to process about 11 mail pieces each second on the average. However, the mechanical transport of the mail pieces through the sorting machine is setting the upper limit for throughput. Transport speeds of several metres per second were achieved; higher speeds increase the probability of damaging the mail pieces.

Beside the requirements for throughput, further boundary conditions influence the real-time behaviour of the system. In on-line mail sorting systems, the mail pieces are continuously moving through the system from the feeder to the final sorter bin[2]. This means that each processing stage has to produce its results within some given amount of time. For example, the address recognition result needs to be available before the mail piece physically arrives at the sorter entry. The time each processing stage needs requires for an expensive mechanical storage line. The total latency of a mail piece in the complete sorting system is in the order of 10–20 seconds. For the ROI-analysis module investigated in EFTOS, i.e., GLIMP, the allowed latency time is in the order of one second.

Dependability requirement for such a system include:

- minimising the failures of the system;

- confining the effects of errors through graceful degradation, avoiding as much as possible

---

[2]Or to a so-called "unsorted bin", in case of occasional failures of the classification algorithms.

total system failures and allowing operability at reduced throughput.

- In the event of total system failures due to unforeseen situations, i.e., a power failure, the recovery times must be kept to a minimum.

In GLIMP, the algorithms of the image analysis are implemented as distributed software processes communicating with each other and with the rest of the system. The input to GLIMP is originating from a piece of hardware (image buffer, filled by the scanner, triggered by the mechanical letter movement) with strict real-time requirements, while the output is sent down a TCP/IP Ethernet connection, which is not real-time. The most frequent problems with the realisation of the ROI-analysis are caused by *unstable behaviour of the software processes*. Dealing with real-life images means processing of unknown data with unpredictable properties. As mentioned in Sect. 1.1.1.3, given the amount of software complexity and the requirement of reduced development cycle times of nowadays' software, the incremental process of software testing is an activity that can never be comprehensive at justifiable costs. Hence, data induced malfunctioning of the software may cause a software process to hang up[3]. One possible consequence of this is that a real-time deadline is missed. However, another consequence could be that *a second process waiting for communication with this faulty process also gets stuck*. In this way, the error condition can *propagate* through the whole system. Another effect of a data induced error may be the corruption of memory on the processor by writing outside of the processes address space. In this way other processes running on the same processor might be also corrupted by the faulting process. The occurrence of data induced software errors cannot be completely avoided. In the operational environment of the mail sorting, it is acceptable to lose the processing of one mail piece, but it is not acceptable that *one faulty mail piece causes subsequent faults* resulting in the loss of more mail pieces. Hence, it is important to detect errors as they occur, to confine them in a limited area, and to recover from the error as quick as possible.

## 6.1.5   Results Obtained and Lessons Learned

The author of this dissertation and two colleagues of the ACCA Division had the possibility to work with the actual GLIMP libraries and to embed it into the so-called EFTOS Vertical Integration Test Application (VITA) —a farmer worker application developed by the colleagues of the National Technical University of Athens and used within EFTOS in order to exercise and test different demonstrations. This test application played the role of the control module, while the actual GLIMP libraries were used for the algorithmic modules. The VITA/GLIMP application was successfully demonstrated at one of the EFTOS review meetings.

The enhanced version was simply the combination of GLIMP with the VITA, which in turn used a watchdog process, the prototypic backbone component developed in EFTOS and a few other ancillary codes. The watchdog informed the backbone of errors concerning the GLIMP processes. On receipt of such messages, the backbone automatically stopped and restarted the watched processes. In case of node crashes affecting both watched tasks and watchdogs, it was the backbone that detected such event by means of a variant of the AMS algorithm described in

---

[3]The GLIMP workers have performance/crash failure semantics.

Sect. 5.2.4.1[4]. The performance overhead for the overall system with respect to the unmodified Siemens application was of about 15%. The increase of the binary size was of approximately 500 kilobytes. This quantity includes the codes of: the backbone, the watchdog, the VITA, and an ancillary code for communication management.

The code intrusion in the VITA was limited to the code to configure and manage the watchdog (no configuration language was available in EFTOS). Error recovery produced no code intrusion, because a simple recovery action had been hardwired in the EFTOS backbone, hence outside the user application (the VITA). *No amount of code was intruded* in the GLIMP libraries, which were only available in the form of object code. Fast reboot and task restart were obtained through specific kernel-level provisions made available by Parsytec on their CC system. From the point of view of the structural attributes defined in Sect. 1.1.2, one can conclude that it was possible to obtain an optimal degree of SC and SA.

Obtained results [Tea98] have been encouraging: the resulting application, run on a 2-node system, was able to tolerate process failures and node crash failures triggered by software fault injection, but also *to revert to full throughput mode* at the end of error recovery. Another version of the VITA, mimicking the original OCR control module but exploiting no fault-tolerance provisions, halved the system throughput when a first fault was injected, and stopped the system functions on the second injection.

Indeed, GLIMP provides us with an example of a class of applications that well exploit the current properties of our first $\mathcal{REL}$ prototype—the class of

- parallel or distributed industrial automation applications,

- with real-time requirements such that violations of the agreed deterministic timing behaviour are acceptable *at error recovery time*,

- written in a procedural language such as C,

- structured as a farmer-worker application,

- with a stateless worker process.

The case study reported so far taught us that, at least for this class, it is possible to make effective use of $\mathcal{REL}$ in order to increase both service throughput and dependability with minimal effort in terms of development costs and time. In particular, $\mathcal{REL}$ can be qualitatively assessed as reaching very good values for properties such as SC and SA. On the contrary, with respect to the structural attributes, all the ALFT structures reviewed in Sect. 3 either do not address this class of applications or they would require a complete rewriting of the application in order to obtain similar results. This happens also because GLIMP is an example of those applications that may be called *company asset codes*—longly-run, longly-tested, large applications typically written in a procedural (non-object-oriented) standard language. These hypotheses forbid the adoption of approaches such as MV, AOP, MOP, and the adoption of ad-hoc programming languages. The lack of systems supporting RMP forbids its use as well.

---

[4]As explained in Sect. 5.2.4.1, the AMS allows to detect a node crash as a lack of a TEIF message within a known deadline.

## 6.2 Enhancement of a TIRAN Dependable Mechanism

As observed by the designers of GUARDS [PABD$^+$99], many embedded systems may benefit from an hardware architecture based on *redundancy* and *consensus* in order to fulfil their dependability requirements. As pointed out in Sect. 1.1.1.3, such a hardware architecture needs to be coupled with some application-level strategy or mechanism in order to guarantee end-to-end tolerance of faults[5]. The software architecture described in this dissertation may support in many ways the class of applications eligible for being embedded in a GUARDS system instance. The main tool for this is the TIRAN DV (described in Sect. 5.2.3). This section describes how it is possible to further increase the dependability of the TIRAN DV by using ARIEL as both a configuration and a recovery language and making use of the TIRAN framework. An assessment of these enhancements is reported in Sect. 7.1.

The system realises a sophisticated $N$-version programming executive that implements the software equivalent of an $N$MR system. Because of this assumption, the user is required to supply four versions of the same software, designed and developed according to the $N$-version approach [Avi85]. This requirement can be avoided if it is possible to assume safely that no design faults resulting in correlate failures affect the software.

To simplify the case study it is assumed that the service provided by any of the versions or instances is *stateless*. When this is not the case, the management of the spare would also call for a forward recovery mechanism—e.g., the spare would need to acquire the current state of one of the non-faulty tasks. Furthermore, when using pipelines of N-version tasks under strict real-time requirements, further techniques (currently not part of the prototype presented in this work) would be required in order to restore the state of the spare with no repercussion on the real-time goals (see [BDGG$^+$98] for an example of these techniques).

The enhanced TIRAN DV described in what follows is representative of the class of applications that are best eligible for being addressed via an ALFT structure such as NVP. This section describes how $\mathcal{R\!E\!L}$ provides to those applications two additional features, namely, support of spares and support of fault-specific reconfiguration, which are not part of plain NVP systems.

A system of four nodes is assumed. Nodes are identified by the symbolic constants NODE1, ..., NODE4, defined in the header file `nodes.h`. Header file `my_definitions.h` contains a number of user definitions, including the unique-id of each version task (VERSION1, ..., VERSION4) and the local identifier of each version task (in this case, on each node the same identifier is used, namely VERSION). In the same file also the time-outs of each version are defined (TIMEOUT_VERSION1,..., TIMEOUT_VERSION4).

Let us consider the following ARIEL script:

```
INCLUDE "nodes.h"
INCLUDE "my_definitions.h"

TASK {VERSION1} IS NODE {NODE1}, TASKID {VERSION}
TASK {VERSION2} IS NODE {NODE2}, TASKID {VERSION}
TASK {VERSION3} IS NODE {NODE3}, TASKID {VERSION}
```

---

[5]This way, also software design faults would be addressed.

```
TASK {VERSION4} IS NODE {NODE4}, TASKID {VERSION}

N-VERSION TASK {TMR_PLUS_ONE_SPARE}
VERSION 1 IS TASK {VERSION1} TIMEOUT {TIMEOUT_VERSION1}ms
VERSION 2 IS TASK {VERSION2} TIMEOUT {TIMEOUT_VERSION2}ms
VERSION 3 IS TASK {VERSION3} TIMEOUT {TIMEOUT_VERSION3}ms
VERSION 4 IS SPARE TASK {VERSION4} TIMEOUT {TIMEOUT_VERSION4}ms
VOTING ALGORITHM IS MAJORITY
METRIC "tmr_cmp"
ON SUCCESS TASK 20
ON ERROR TASK 30
END N-VERSION

IF [ PHASE (T{VERSION1}) == {HAS_FAILED} || FAULTY T{VERSION1} ]
THEN
      STOP T{VERSION1}

      SEND {WAKEUP} T{VERSION4}
      SEND {VERSION1} T{VERSION4}

      SEND {VERSION4} T{VERSION2}
      SEND {VERSION4} T{VERSION3}
FI

IF [ PHASE (T{VERSION2}) == {HAS_FAILED} || FAULTY T{VERSION2} ]
THEN
      STOP T{VERSION2}

      SEND {WAKEUP} T{VERSION4}
      SEND {VERSION2} T{VERSION4}

      SEND {VERSION4} T{VERSION1}
      SEND {VERSION4} T{VERSION3}
FI

IF [ PHASE (T{VERSION3}) == {HAS_FAILED} || FAULTY T{VERSION3} ]
THEN
      STOP T{VERSION3}

      SEND {WAKEUP} T{VERSION4}
      SEND {VERSION3} T{VERSION4}

      SEND {VERSION4} T{VERSION2}
```

```
              SEND {VERSION4} T{VERSION1}
    FI
```

The above script consists of two parts—one for managing the configuration of tasks and tools, the other for describing a recovery strategy. The following two subsections describe the two parts.

## 6.2.1 Configuration

The configuration section of the script just shown defines four tasks, each running on a different node of the system. This decision has been taken to reduce the probability of a common source for multiple failures, in case of a crash of a node. It is worth noting that this design decision—the physical location of the tasks—*is made outside the application code* and can be changed with no repercussion on it, thus allowing location transparency and high SC.

Three of the four tasks are then configured as versions of a TIRAN DV (see Sect. 5.2.3). The fourth task (VERSION4) is configured as a spare. This means that, when task TMR_PLUS_ONE_SPARE is launched, the first three tasks arrange themselves as components of a TIRAN DV, with each task assisted by a local voter as described in the aforementioned section. On the contrary, the fourth task is blocked waiting for the arrival of a special "wakeup" signal.

Feeding the ARIEL translator with the configuration script as described in Sect. 5.3.3.1, a number of source codes are produced, including:

- The basic user tasks of a TIRAN DV consisting of three modules, an example of which can be seen in Table 5.3.

- The spare user task, initially waiting for the arrival of a wakeup message.

- Task TMR_PLUS_ONE_SPARE. This task is in charge of the:

    1. Transparent set up of the TIRAN DV via the TIRAN_CreateTask function of the TIRAN BSL.

    2. Management of the replication of the input value.

    3. Declaration and insertion in the TIRAN TOM of the set of time-outs that represent an upper limit to the duration of the base tasks. This upper limit is set by the user and known by one of the hypotheses of the timed-asynchronous distributed system model defined on p. 50 (all services are timed).

    4. De-multiplexing and delivering the output value.

In the absence of faults, task TMR_PLUS_ONE_SPARE would appear to the user as yet another version providing the same service supplied by tasks VERSION1,...,VERSION4. The only difference would be in terms of a higher reliability (see Eq. 7.1) and a larger execution time, mainly due to the voting overhead (see Sect. 7.2.2.4).

Location transparency in this case is supported by ARIEL, while replication transparency is supported by the TIRAN DV. The degree of code intrusion in the application source code is reduced to the one instruction to spawn task `TMR_PLUS_ONE_SPARE`. No support for the automatic generation of makefiles is provided in the current version of the `ARIEL` translator, so the user needs to properly instruct the compilation of the source files written by the translator.

## 6.2.2   Recovery

The recovery section of the `ARIEL` script on p. 117 defines a recovery strategy for task `TMR_PLUS_ONE_SPARE`. When any error is detected in the system and forwarded to the backbone, the backbone replies to this stimulus as follows:

1. It stores the error notification in its local copy of the DB.

2. It updates the $\alpha$-count variable related to the entity found in error.

3. If the notification is local, i.e., related to a local task, then the local component of the BB forwards the notification to the other components.

4. If the local BB entity plays the role of coordinator, it initiates error recovery by sending a "wakeup" message to the recovery interpreter.

This latter orderly reads and executes the r-codes. Table 6.1 shows a trace of the execution of some of the r-codes produced when translating a simplified version of the `ARIEL` script on p. 117. As already mentioned, RINT implements a virtual machine with a stack architecture. Line 5 starts the scanning of a guard. Line 6 stores on RINT's run-time stack the current value of the phase of task 0 (the value of symbolic constant `VERSION1`). Line 7 compares the top of the run-time stack with integer `HAS_FAILED`. The result, 0 (false) is stored on top of the stack. Line 8 checks the top of the stack for being 0. The condition is fulfilled, so a jump is made to line 18 of the r-code list. That line corresponds to the end of the current IF statement—the guard has been found as false, therefore the corresponding recovery actions have been skipped. Then, on line 19, some internal variables are reset. Line 20 starts the evaluation of the clause of the second IF of the recovery script. The scenario is similar to the one just described, though (on line 22) the phase of task 1 (that is, `VERSION2`) is found to be equal to `HAS_FAILED`. The following conditional jump is therefore skipped, and a stream of recovery actions is then executed: task `VERSION2` is terminated on line 24, then value 10 (`WAKEUP`) is sent to task `VERSION4` by means of r-code `SEND` (which sends a task the top of the run-time stack), and so forth, until line 33, which closes the current IF. A third guard is then evaluated and found as false. Clearing some internal structures closes the execution of RINT, which again "goes to sleep" waiting for a new "wakeup" message to arrive.

Figure 6.6, 6.7, and 6.8 show three different views to a TMR-and-one-spare system at work, as displayed by the TIRAN monitoring tool in a Netscape client [DFDT+98]. The first picture renders the current state and shape of the overall target application. Figure 6.7 summarises the framework-level events related to processing node 0. In particular, a value domain failure is injected on the voter on node 2. This triggers the execution of a recovery script which reconfigures the TMR isolating the voter on node 2 and switching in the spare voter on node 3. The execution trace of the r-codes in this script is displayed in Fig. 6.8.

```
5       IF statement.
6       STORE-PHASE: stored phase of task 0, i.e., 0.
7       COMPARING(9999 vs. 0): Storing 0.
8       Conditional GOTO, fulfilled, 18.
18      FI statement.
19      OA-RENEW.
20      IF statement.
21      STORE-PHASE: stored phase of task 1, i.e., 9999.
22      COMPARING(9999 vs. 9999): Storing 1.
23      Conditional GOTO, unfulfilled, 24.
24      KILLING TASK 1.
25      PUSH(10).
26      SEND MSG 10 to TASK 3.
27      PUSH(1).
28      SEND MSG 1 to TASK 3.
29      PUSH(3).
30      SEND MSG 3 to TASK 2.
31      PUSH(3).
32      SEND MSG 3 to TASK 0.
33      FI statement.
34      OA-RENEW.
35      IF statement.
36      STORE-PHASE: stored phase of task 2, i.e., 0.
37      COMPARING(9999 vs. 0): Storing 0.
38      Conditional GOTO, fulfilled, 48.
48      FI statement.
49      OA-RENEW.
```

**Table 6.1: Trace of execution of the r-codes corresponding to the recovery section of the ARIEL script on p. 117. Number 9999 is the value of constant HAS_FAILED.**

It is worth noting how, modifying the base recovery strategy of the ARIEL script on p. 117 does not require any modification in the application source code—not even recompiling the code, in case the r-codes are read from an external means (e.g., from a file). The two design concerns—that of the fault-tolerance software engineer and that of the application software engineer—are kept apart, with no repercussions on the maintainability of the service overall. To prove this, let us consider the ARIEL excerpt in Table 6.2. Such a strategy is again based on a TMR plus one spare, though now, before switching off the faulty version and in the spare, the current value of the $\alpha$-count filter related to the current version is compared with a threshold supplied by the user in the configuration section. If the value is below the threshold, it is not possible to assess that the fault affecting {VERSION1} is permanent or intermittent. In this case, the faulty task is restarted rather than substituted. In other words, another chance is given to that version, while its "black list" (its $\alpha$-counter) is updated.  As mentioned in Sect. 5.2.4.2, research studies have shown that, whatever the threshold, the $\alpha$-count is bound to exceed it when the fault is permanent or intermittent. In the latter case, therefore, sooner or later the second strategy is to be executed, permanently removing the faulty version. It is worth noting how the adoption of such a scheme, which is no longer purely based on masking the occurrence of a fault, in general implies an execution overhead that may violate the expected real-time behaviour—*during error-recovery*.

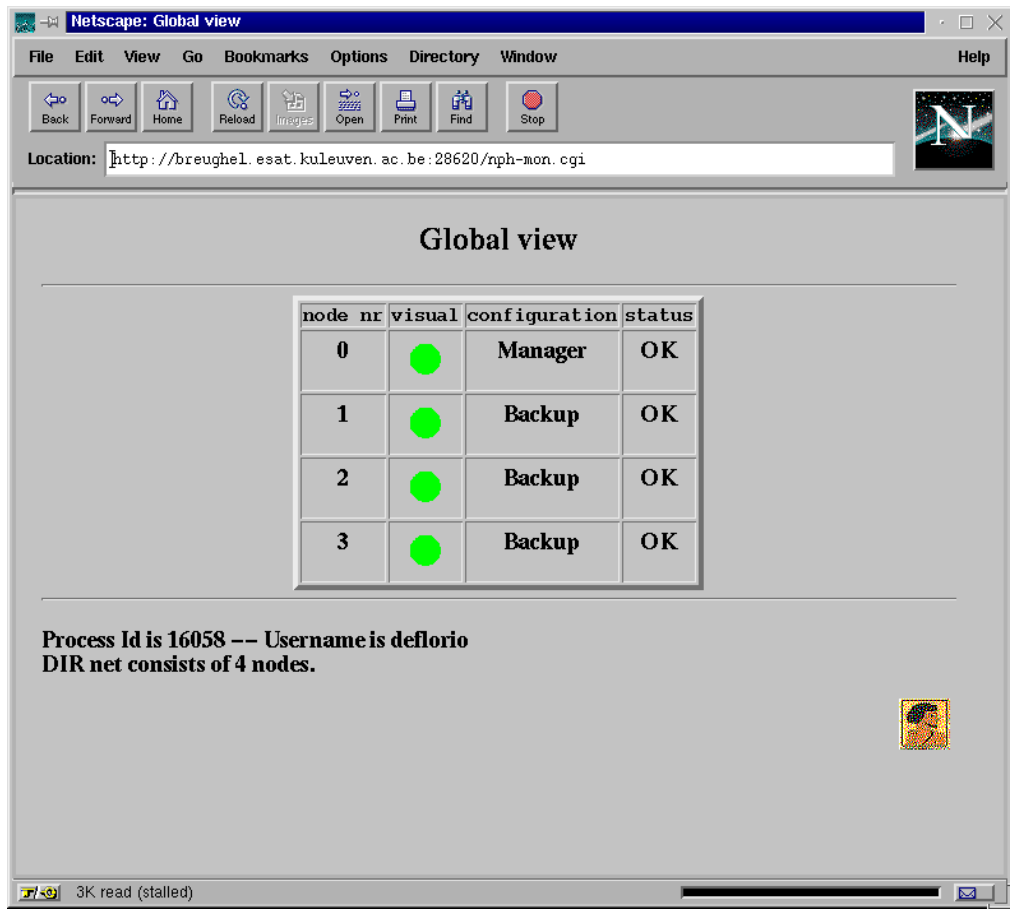

**Figure 6.6:  A global view of the state and shape of the target application, as rendered by a monitoring CGI script. In this case a four processing node system is used. Node 0 hosts the main component of the backbone. The circular icons are hypermedia links (see Fig. 5). The small icon on the bottom right links to the execution trace of the** ARIEL **interpreter.**

Further support towards graceful degradation when spares are exhausted could also be foreseen.

### 6.2.3   Lessons learned

Enhancing the TIRAN DV allowed to include two special services (spare support and fault-identification) into those offered by NVP systems.  For those applications that best match with the MV ALFT approach, and with NVP in particular, those services allow to further increase the dependability without being detrimental to sc—i.e., without intruding code. Eligible applications include coarse-grained, distributed or parallel applications, i.e., those addressed by $\mathcal{REL}$. Furthermore, safety-critical parallel or distributed airborne and spaceborne applications appear to be well addressed by those services. It must be remarked, though, that the safety-critical character of these applications calls for specific framework-level and system-level support, such as the one provided by the GUARDS architecture [PABD$^+$99], a generic, parametrisable hardware and OS architecture for real-time systems, which straightforwardly supports the NVP scheme.

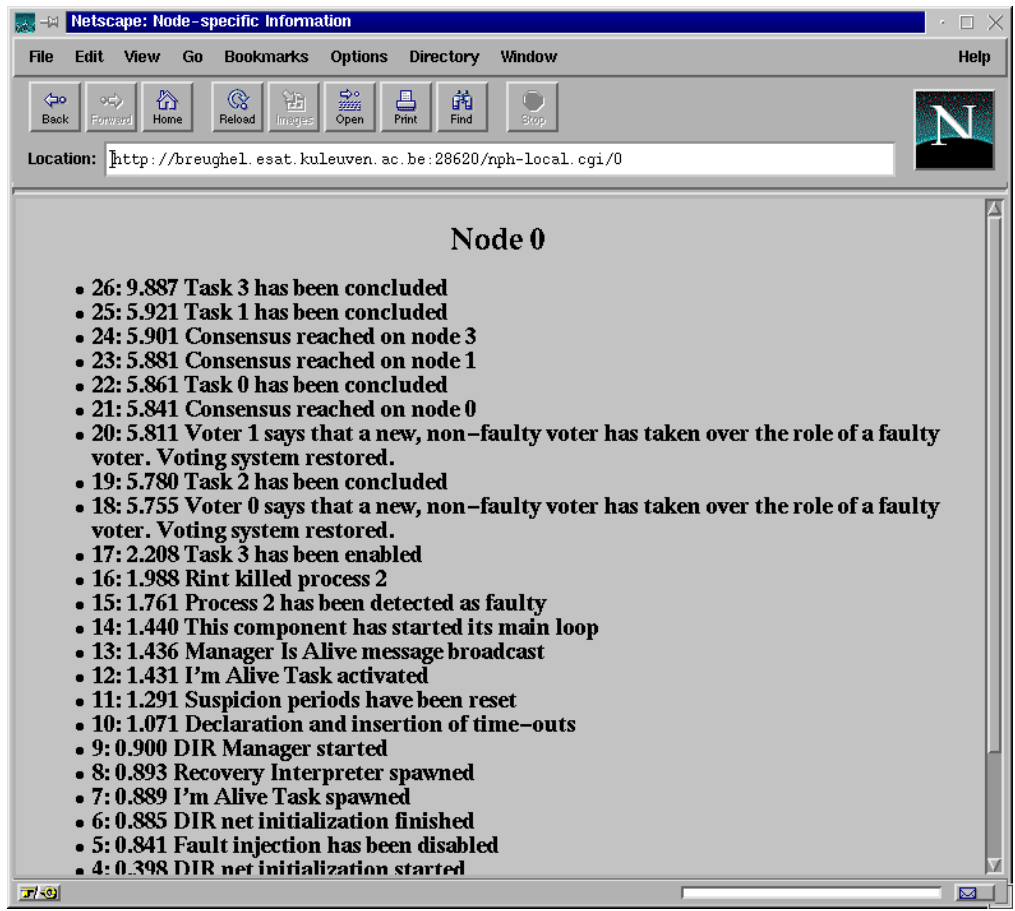

**Figure 6.7:** A view to the events tracked while monitoring processing node 0. This view can be obtained by selecting the top circular icon of Fig. 6.6. Events are ordered decreasingly with respect to their conclusion time. Times are expressed in seconds and measured through the local clock. Going from bottom to top: on line 15 process 2 is detected as faulty. On 16, that process is terminated. On 17, a new process (a spare voter) is awaken. On 18 and 20, respectively voter 0 and 1 acknowledge the reconfiguration. This allows to retry a global consensus that succeeds (see lines 21, 23, and 24). Finally all voters acknowledge their termination.

**Execution trace of the RINT virtual machine:**

```
5      IF statement.
6      STORE-PHASE: stored phase of task 0, i.e., 0.
7      COMPARING(9999 vs. 0): Storing 0.
8      Conditional GOTO, fulfilled, 18.
18     FI statement.
19     OA-RENEW.
20     IF statement.
21     STORE-PHASE: stored phase of task 1, i.e., 0.
22     COMPARING(9999 vs. 0): Storing 0.
23     Conditional GOTO, fulfilled, 33.
33     FI statement.
34     OA-RENEW.
35     IF statement.
36     STORE-PHASE: stored phase of task 2, i.e., 9999.
37     COMPARING(9999 vs. 9999): Storing 1.
38     Conditional GOTO, unfulfilled, 39.
39     KILLING TASK 2.
40     PUSH(10).
41     SEND MSG 10 to TASK 3.
42     PUSH(2).
43     SEND MSG 2 to TASK 3.
44     PUSH(3).
45     SEND MSG 3 to TASK 1.
46     PUSH(3).
47     SEND MSG 3 to TASK 0.
48     FI statement.
49     OA-RENEW.
```

**Figure 6.8: As soon as an error is detected—in this case, a voter has been found in minority—RINT starts executing the r-codes. This picture is the execution trace of the r-codes in Table 5.5 and in Table 5.8. Numbers refer to the code lines in those tables. Note how, at line 37, the state information of task 2 is found to be equal to "9999" (HAS_FAILED). As a consequence, a number of actions are executed in lines 39–47. In particular, task 2 is terminated at line 39.**

## 6.3   The Recovery Language Approach at ENEL

This section reports on the use that is being made of the TIRAN framework and, in particular, of the ARIEL language, at ENEL in the framework of their collaboration with the research centre CESI, a member of the TIRAN consortium. The main reference for this section is [BDPC00].

At the time of writing, no complete prototype of an ARIEL strategy has been coded at ENEL yet, though its overall design has been completed. The level of detail of this design is such that the configuration and the recovery actions to be adopted have been determined. The corresponding scripts will be written at ENEL and demonstrated at the final review meeting of the TIRAN project.

The rest of this section is structured as follows:

- The general context of the case study and its motivations are introduced in Sect. 6.3.1.

```
ALPHACOUNT {VERSION1} IS threshold = 3.0, factor = 0.4 END
IF [ PHASE (T{VERSION1}) == {HAS_FAILED} || FAULTY T{VERSION1} ]
THEN
        IF [ TRANSIENT T{VERSION1} ]
        THEN
            RESTART T{VERSION1}
            SEND {VERSION1} T{VERSION2}
            SEND {VERSION1} T{VERSION3}
        ELSE
            STOP T{VERSION1}

            SEND {WAKEUP} T{VERSION4}
            SEND {VERSION1} T{VERSION4}

            SEND {VERSION4} T{VERSION2}
            SEND {VERSION4} T{VERSION3}
        FI
FI
```

**Table 6.2: An** `ARIEL` **script for a TMR-plus-one-spare system that exploits the $\alpha$-count technique to avoid unnecessary reconfigurations.**

- Next, the application grounding the case study is described in Sect. 6.3.2.

- Section 6.3.3 briefly summarises the dependability requirements of the ENEL pilot application.

- The TIRAN case study is then drawn in Sect. 6.3.4.

- Finally, Sect. 6.3.5 summarises the lessons learned from the reported experience.

## 6.3.1  General Context and Motivations

As mentioned before, ENEL is the main electricity supplier in Italy. ENEL produces, transports, and distributes electricity all over the country. These processes are nowadays largely automated, and huge investments are currently being made at ENEL in order to update the automation systems designed to support these processes.

The ENEL energy distribution system is a meshed network connecting high-voltage lines to the final users. Its nodes of interconnection are:

- Primary substations (PS), connected to high-voltage lines, transform and distribute energy to secondary substations and to medium-voltage customers. The PS consist of a set of electrical component (switches, insulators, transformers, capacitors and so forth) and of an automation system.

- Secondary substations, which transform and distribute energy to low-voltage customers.

As part of this major activity, ENEL is currently renewing their park of so-called primary substation automation systems (PSAS) which serve 2000 PS. Plans are to renew 30–50 PS per year. Another goal at ENEL is to reduce the dependency to dedicated hardware systems, at the same time guaranteeing the current degree of dependability and efficiency and reducing maintenance costs.

The PSAS architecture is controlled by an *automation application*, consisting of a so-called "*local control level*" (LCL) and a number of peripheral units distributed on the plant. Among its functions, the LCL performs local control, supervision of the peripheral units, switching in of protection provisions, and interfacing to remote control systems and to local and remote operators.

## 6.3.2   The ENEL Pilot Application

The case study described in the rest of this section originates in the framework of the above internal activities of ENEL. It is a pilot application from the domain of *energy distribution automation systems*, embedded in their PSAS. The automation application behaves as a cyclic state machine:

- it reads sampled input from the field (i.e., the plant);

- on the basis of this input and of the current state, it produces the next state and provides a new output to the field (that is, it acts as a Moore automaton, see [AH75]).

This state machine is in fact composed by many cooperating automata. The state transitions of these automata are synchronised and lead to a state transition of the whole state machine. The cycle time ($T_{\text{cycle}}$) for the state machine to complete its execution, namely, the time between two subsequent input reads, is a key parameter that depends on the plant constraints and characterises the automation application.

The TIRAN pilot application derives from the just sketched automation application and provides a subset of the functions of the LCL module, namely those related to the protection and the management of the PS and of the plant.

The software architecture for the ENEL pilot application is illustrated in Fig. 6.9. It includes the following modules:

- An application code (APL) that executes control functions.

- The so-called "basic software" (BSW), replicated on the available nodes and responsible for

  – providing the application code with input values and states,

  – receiving the new output value from the APL, and

  – invoking fault-tolerance provisions.

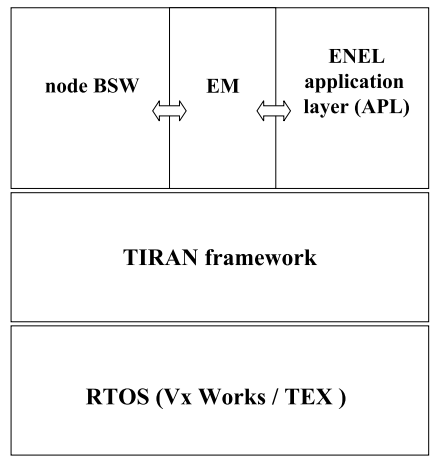

**Figure 6.9: The demonstrator layered architecture.**

- The TIRAN framework, which provides fault-tolerance functionality.

- The TEX/VxWorks kernels, which provides low-level support for distributed communication and other basic services [TXT97, Win99].

Both APL and BSW are distributed applications whose module run on each node of the system. On each node, the APL and the BSW communicate exclusively through a shared memory bank called the "exchange memory" (EM)[6]. The overall system evolves by executing alternatively the BSW and the APL as depicted in Fig. 6.10.

The basic cycle of the APL is as follows:

- At the beginning of the current cycle, the inputs from the field and the current state of the automata are read from the EM.

- During the current cycle a new state and set of outputs are computed.

- At the end of the current cycle, the computed new state and outputs are stored into the EM.

The basic cycle of the BSW is as follows:

- At the beginning of the current cycle, the BSW receives input from the field, timer values from the local clock, and stable states from a stable memory module (SM), and writes all these data into the EM. Furthermore, the BSW activates a set of provisions among those of the TIRAN fault-tolerance framework, including SM.

- It then waits until the end of the current cycle.

---

[6]The EM is virtually structured as a single bank for the whole application; physically, it is partitioned and distributed over the available nodes according to the the strategy of partitioning and allocation of the application. On the start of each new cycle, the blocks of the EM are synchronised and updated by the BSW modules.

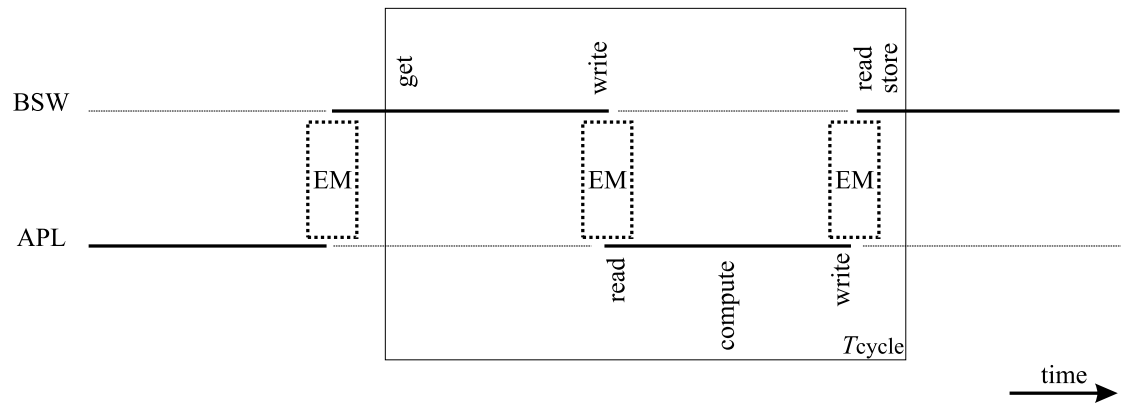

**Figure 6.10: Scheme of execution of the BSW and the APL. Only one of them is in the run state at any time $t$. The ticker line represents the time spent in the run state. Action "get" is executed by the BSW to receive data from the field and stable states from SM. "Write" and "read" act on the EM. "Compute" is carried out by the APL and is the elaboration of the new state and outputs. The BSW executes action "store" to write these values into SM. The rectangle represents the duration of a system cycle ($T_{\text{cycle}}$).**

- At the end of the cycle the BSW reads output and new states from the EM and stores them into the SM.

- The new cycle is then initiated.

### 6.3.3  Dependability Requirements

A number of dependability requirements characterise the TIRAN pilot application [BCDF$^+$99], focusing in particular on availability, integrity and safety. They include:

1. Any first fault affecting either the electric network or the fault-tolerance system must be tolerated.

2. In case of a second fault, the plant must be left in a safe state preserving human operators and electric components, and must prevent wrong output.

3. The automation system must guarantee EMI compatibility compliance [Uni95].

4. It must tolerate transient faults affecting the system behaviour.

### 6.3.4  Description of the Case Study

The software component that constitutes the subject of the case study proposed by ENEL is the BSW. The main reason for this choice lies in the fact that that component is the one managing the fault-tolerance aspects of the overall automation application.

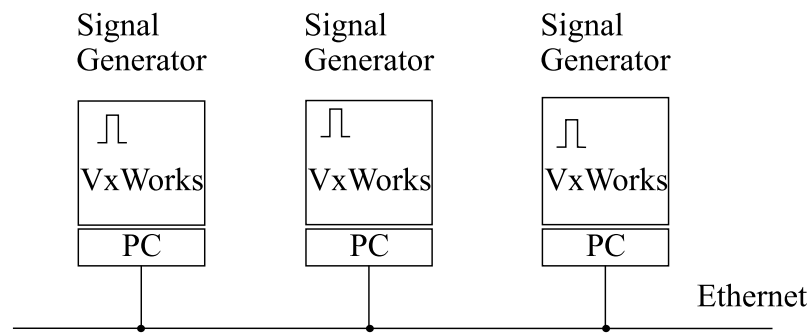

**Figure 6.11: The target system for BSW2.**

Two fault-tolerance strategies have been envisaged at ENEL. The **first strategy** derives from early ENEL automation systems that performed hardware-based fault-tolerance. In these systems, the entire automation system evolved cyclically according to a fixed time duration, at the end of which *an external signal forced a cold restart*. This was done for containing the effects of faults within one cycle. The duration of the cycle, $T_{\text{cycle}}$, is a parameter related to the real-time requirements of the system. The usage of the periodic restart signal translated in an effective strategy, validated by several years of profitable utilisation, though it implies hard constraints for the underlying hardware and OS. In particular, these constraints forbid the adoption of COTS components and translate into huge performance penalties. Further information on this strategy can be found in [DBC+98]. This strategy still focuses on the periodic restart signal as the main fault-tolerance provision. Furthermore, it does not focus on ARIEL, so it will not be described in what follows. Because of the above mentioned constraints, ENEL is currently trying to eliminate this dependency by defining a **second strategy** addressing industrial PC hardware and COTS OS. The rest of this paragraph describes this new strategy, called "BSW2", which is centred on ARIEL.

Figure 6.11 describes the target system for BSW2, based on industrial PC/VxWorks. In this case no external synchronisation and no periodic restart is adopted. Key points with BSW2 are the following ones:

- Detection and isolation of faulty nodes or software components.

- Reconfiguration.

- Substituting the systematic, predetermined periodic restart with an occasional reset of the isolated node, triggered by the detection of errors.

- Re-integration of the recovered nodes and application components.

In this context, ARIEL and the TIRAN backbone play the following roles:

- Initial configurations of tasks, logicals (groups) and nodes are specified in ARIEL.

- Error detection is managed via custom diagnostic functions developed at ENEL and with instances of the TIRAN watchdog or of the "redundant watchdog" component described in Sect. 6.4. Both tools execute the `RaiseEvent` function to notify the backbone of their deductions.

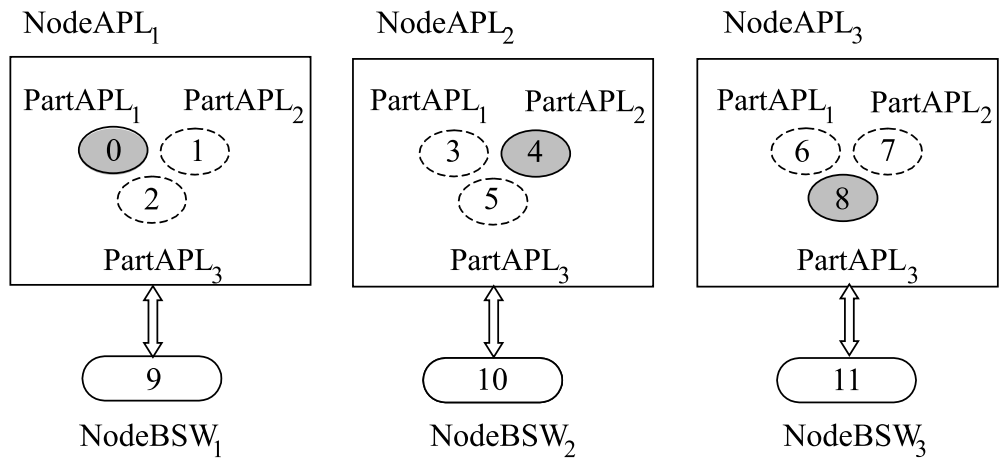

**Figure 6.12: Initial configuration of the APL in a system of 3 nodes (strategy BSW2).**

- ARIEL is in charge for the specification and the management of the complete reconfiguration strategy. Error recovery is managed exploiting linguistic support towards:

  – Low-level support to node isolation.

  – Application reconfiguration.

  – Propagation of the alarm to the operator.

BSW2 exploits the BB services and ARIEL to accomplish error detection and consequent reconfiguration of BSW and APL layers.

It is assumed that the target system consists of $N > 1$ nodes. The APL is present on each of the $N$ nodes with $N$ replicas. On node $n \in [1, N]$, these replicas are collectively denoted as group $PartAPL_n$. Only one of the $N$ replicas available on node $n$ is activated. $NodeAPL_n$ denotes the APL task, belonging to $PartAPL_n$, that is active on node $n$. Also the BSW is present on each of the $N$ nodes, but no replication scheme is adopted. $NodeBSW_n$ represents the BSW component located on node $n$.

The following two paragraphs detail the current state of design for the strategies of configuration and error recovery developed at ENEL for the TIRAN pilot application.

**Configuration.** Figure 6.12 shows the initial configuration of the APL in a system with three nodes. The APL is partitioned in three blocks, of which only one is active per node (the one depicted as a filled circle).

Table 6.3 gives the mapping between portion identifiers and the corresponding TIRAN global and local task identifiers. This configuration step is performed in ARIEL as it has been described in Sect. 5.3.2.1. ARIEL is also used to define the following three logicals:

| Blocks | Unique-id | local task-id, node-id |
|--------|-----------|------------------------|
| NodeAPL$_1$ | 0 | $(t_0, n_1)$ |
| NodeAPL$_2$ | 4 | $(t_1, n_2)$ |
| NodeAPL$_3$ | 8 | $(t_2, n_3)$ |
| NodeBSW$_1$ | 9 | $(t_3, n_1)$ |
| NodeBSW$_2$ | 10 | $(t_3, n_2)$ |
| NodeBSW$_3$ | 11 | $(t_3, n_3)$ |

**Table 6.3: Map of blocks to unique-ids and local task identifiers on each node.**

| Logicals | Unique-id list |
|----------|----------------|
| PartAPL$_1$ | 0,3,6 |
| PartAPL$_2$ | 1,4,7 |
| PartAPL$_3$ | 2,5,8. |

Tasks are addressed by the above logicals—the actual identity of the active task within each partition is kept transparent.

**Reconfiguration.** To date, no complete ARIEL scripts have been written yet by ENEL for their reconfiguration strategy. Nonetheless, some key design elements for that strategy have been chosen and are reported herein:

- Each nodeBSW entity will communicate with the BB by means of function `RaiseEvent`. Messages will hold the current algorithmic phase the application is currently in.

- Some of these event notifications will only be informational and just update the database of the BB without triggering any recovery action. Other notifications will trigger the execution of the recovery actions.

- Recovery actions written in ARIEL, translated into r-codes and statically linked to the application, will use the information in the BB's DB to accomplish recovery by means of the ARIEL constructs such as ERRN and PHASE.

- The $N$ replicas of the APL represent a group consisting of an active component plus $N-1$ inactive spares. Switching off the current active component and electing a new spare as new active component will then be achieved through simple recovery actions composed in terms of ARIEL's directives, such as STOP, START, or RESTART, on tasks, logicals and nodes of the system.

- A TIRAN watchdog or the redundant watchdog will notify the BB when a NodeBSW has not correctly computed the task pertaining to the current cycle. The time of expiration of the adopted watchdog will be equal to the duration of the system cycle: the expiration of the watchdog cycle will be interpreted as the fact that something on some node has caused the cycle to terminate incorrectly. In such a case, the watchdog will signal the event to the backbone via a specific call of function `RaiseEvent`. This notification will trigger recovery. The recovery will operate as a function of the algorithmic phase the distributed application is in in order to accomplish an appropriate reconfiguration.

The exploitation of the $\alpha$-count feature (see Sect. 5.2.4.2) in order "to distinguish between the various natures of the faults [...] and to take different levels of recovery" is currently under consideration at ENEL [BDPC00]. Actions would be reconfiguration for graceful degradation in case of permanent or intermittent faults, and task restart in case of transient faults.

### 6.3.5   Lessons Learned

Preliminary lessons learned from the above reported experience include the fact that $\mathcal{REL}$ can be effectively used for composing fault-tolerance strategies for energy distribution automation systems. The $\mathcal{REL}$ support towards coarse-grained error recovery strategies appear to match well with the requirements of the applications to run on those systems. Similar result may be obtained with an approach such as AOP, though this would require the adoption of custom languages and the design of appropriate aspect programs. Other ALFT approaches do not appear to be eligible in this case: MOPs, for instance, may not provide enough support to the cross-cutting issues of coarse-grained ALFT [LVL00].

## 6.4   The Redundant Watchdog

This section describes how it is possible to compose dependable mechanisms addressing specific needs by means of the $\mathcal{REL}$ architecture devised in TIRAN. The case reported in the rest of this section originated by a requirement of ENEL. In present day's fault-tolerance systems at ENEL, a hardware watchdog timer is available and integrated in several of their automation applications. Such device is a custom (non-COTS) hardware component that is capable of guaranteeing both high availability and high integrity. This allows its use in contexts where safety is the main concern. As mentioned in Sect. 6.3, ENEL has started an internal plan that aims at renewing and enlarging their park of automation systems, also with the goal of improving their efficiency and quality of service. Within this plan, ENEL is investigating the possibility to substitute this hardware device with a software component, exploiting the redundancy of COTS parallel or distributed architectures, in such a way as to guarantee acceptable levels of safety and availability for their applications. While taking part in TIRAN, CESI posed the above problems to the consortium and also proposed a solution based on the $\mathcal{REL}$ approach. This solution is currently being tested at ENEL and will be the basis of a demonstration at the TIRAN final review meeting and at the TIRAN Workshop. It is briefly described in the rest of this section.

### 6.4.1   The Strategy and Its Key Components

The strategy proposed by ENEL exploits the following components:

- The BSL, and specifically its function `TIRAN_Send`, which multicasts a message to a logical, i.e., a group of tasks.

- The configuration support provided by ARIEL.

- The TIRAN BB and its DB.

- The recovery language ARIEL.

- The watchdog BT, i.e., a node-local software component in level 1.1 of the TIRAN architecture (see Fig. 5.1).

The latter, which executes on a single processing node, can not guarantee the required degree of availability when used as a substitute of the ENEL hardware watchdog. This notwithstanding, the adoption of the above components allowed to *compose*—rather than *program*—a new prototypic DM, the so-called Redundant Watchdog (RW). This composition is made in terms of the above elements of the TIRAN framework, with ARIEL playing the role of coordinator.

In order to introduce the strategy, the following scenario is assumed:

- A distributed system is available, consisting of three nodes, identified as $N_1$, $N_2$ and $N_3$.

- On each node of this system, a number of application tasks and an instance of the TIRAN watchdog are running.

The design goal to be reached is enhancing the dependability of this basic scheme by means of a technique that does not overly increase, at the same time, the complexity of the overall software tool.

The adopted strategy is now explained—each step has been tagged with a label describing the main $\mathcal{REL}$ feature being exploited.

**Configuration:** Define and configure the three watchdogs by means of the provisions described in Sect. 5.3.2.3. In particular,

- Assign them the unique-ids $W_1, W_2$, and $W_3$.

- Specify that, on a missed deadline, a notification is to be sent to the TIRAN BB.

- Deploy the watchdogs on different nodes.

**Configuration:** Define logical $L$, consisting of tasks $W_1, W_2$, and $W_3$.

**Recovery:** Define in ARIEL an "*AND-strategy*", that triggers an alarm when each and every watchdog notifies the BB, an "*OR-strategy*", in which the alarm is executed when any of the three watchdog expires, and a "*2-out-of-3 strategy*", in which a majority of the watchdogs needs to notify the BB in order to trigger the alarm. In the current prototype, the alarm is a notification to the task the unique-id of which is $A$.

```
INCLUDE "watchdogs.h"

TASK {W1} IS   NODE {N1}, TASKID {W1}
TASK {W2} IS   NODE {N2}, TASKID {W2}
TASK {W3} IS   NODE {N3}, TASKID {W3}

WATCHDOG {W1}
  HEARTBEATS EVERY {HEARTBEAT} MS
  ON ERROR WARN BACKBONE
END WATCHDOG

WATCHDOG {W2}
  HEARTBEATS EVERY {HEARTBEAT} MS
  ON ERROR WARN BACKBONE
END WATCHDOG

WATCHDOG {W3}
  HEARTBEATS EVERY {HEARTBEAT} MS
  ON ERROR WARN BACKBONE
END WATCHDOG

LOGICAL {L} IS  TASK {W1}, TASK {W2}, TASK {W3}  END LOGICAL
```

**Table 6.4: Configuration of the Redundant Watchdog.**

```
IF [ PHASE (TASK{W1}) == {EXPIRE} AND
     PHASE (TASK{W2}) == {EXPIRE} AND
     PHASE (TASK{W3}) == {EXPIRE} ]
THEN
     SEND {ALARM} TASK{A}
     REMOVE PHASE LOGICAL {L} FROM ERRORLIST
FI
```

**Table 6.5: The AND-strategy of the Redundant Watchdog. Action `REMOVE` resets the phase corresponding to the tasks of logical $L$.**

The configuration step is coded as in Table 6.4. Table 6.5 lists the recovery actions corresponding to the AND-strategy.

When a watched task sends "watchdog $L$" its heartbeats, the BSL relays these messages to the three watchdogs on the three nodes. In absence of faults, the three watchdogs[7] process these message in the same way—each of them in particular resets the internal timer corresponding to the client task that sent the heartbeat. When a heartbeat is missing, the three watchdogs expire and send a notification to the BB, one at the time. The reply of the BB to these notifications is the same: RINT is awoken and the r-codes are interpreted. The difference between the three strategies is then straightforward:

- The OR-strategy triggers the alarm as soon as any of the watchdog expires. This tolerates the case in which up to two watchdogs have crashed, or are faulty, or are unreachable. This intuitively reduces the probability that a missing heartbeat goes undetected, hence can be regarded as a "*safety-first*" strategy. At the same time, the probability of "false alarms" (mistakingly triggered alarms) is increased. Such alarms possibly lead to temporary pauses of the overall system service, and may imply costs.

- The AND-strategy, on the other hand, requires that *all* the watchdogs reach consensus before triggering the system alarm. It decreases the probability of false alarms but at the same time decreases the error detection coverage of the watchdog BT. It may be regarded as an "*availability-first*" strategy.

- Strategy 2-out-of-3 requires that a majority of watchdogs expire before the system alarm is executed. Intuitively, this corresponds to a trade-off between the two above strategies.

---

[7]In this case study, three instances of the same software component have been used. Clearly this does not protect the system from *design faults in the watchdog component itself*. Using NVP when developing the watchdogs may possibly guarantee statistical independence between failures.

More sophisticated strategies, corresponding to other design requirements, may also be composed. Other schemes, such as **meta-watchdogs** (watchdogs watching other watchdogs) can also be straightforwardly set up.

### 6.4.2   Lessons Learned

The reported experience demonstrates how $\mathcal{REL}$ allows to fast-prototype complex strategies by composing a set of building blocks together out of those available in the TIRAN framework, and by building *system-wide, recovery-time coordination strategies* with ARIEL. This allows to set up sophisticated fault-tolerance systems while keeping the management of their complexity outside the user application. The compact size of the ARIEL configuration scripts is one of the argument that can be used as evidence to this claim. Transparency of replication and transparency of location are also reached in this case. No similar support is provided by the ALFT approaches reviewed in Chapter 3.

## 6.5   Conclusions

This chapter has shown how $\mathcal{REL}$ can be used in order to

1. augment a dependable mechanism (the TIRAN distributed voting tool) adding new features to it—in this case, the management of spares and support to fault-specific recovery strategies—as well as to

2. compose and orchestrate new dependable mechanisms using the TIRAN basic tools as building blocks.

The chapter also reported a case study from postal automation and the current state of an experimentation in energy distribution automation. In particular, in the postal automation case study, the application gained the ability to withstand crash failures of nodes and tasks with minimal intrusion in the code (actually, no code intrusion in the GLIMP code and limited error detection code intrusion in the farmer application). In that case, the increase in throughput and dependability was reached with a reduced cost and time penalty, despite the architecture being used was a prototypic one.

# Chapter 7

# Analysis and Simulations

This chapter describes a number of results pertaining to the $\mathcal{R}\mathcal{E}\mathcal{L}$ architecture described in Chapter 5. These results have been obtained via reliability and performance analysis and through measurements resulting from simulations. Three basic properties are estimated: **reliability**, **performance**, and **coding efforts** (influencing both development and maintenance costs). Reliability Markov models have been used to estimate reliability (see Sect. 7.1). Simulations and discrete mathematics have been used in Sect. 7.2 to estimate the performance. Section 7.3 reports on the coding efforts related to fault-tolerance software development and maintenance. All the models, analyses, and simulations in this chapter have been devised, solved, and carried out by the author of this dissertation.

## 7.1   Reliability Analysis

As mentioned in Sect. 6.2, a number of applications are structured in such a way as to be straightforwardly embedded in a fault-tolerance architecture based on redundancy and consensus. Applications belonging to this class are, for instance, parallel airborne and spaceborne applications. The TIRAN DV provides application-level support to these applications. This section analyses the effect on reliability of the enhancements to the TIRAN DV described in Sect. 6.2, that is, management of spares, dealt with in Sect. 7.1.1, and fault-specific recovery strategies supported by the $\alpha$-count feature, analysed in Sect. 7.1.2.

### 7.1.1   Using ARIEL to Manage Spares

This section analyses the influence of one of the features offered by ARIEL—its ability to manage spare modules in $N$-modular redundant systems—that has been introduced in Sect. 5.3.3.1 and Sect. 6.2.

Reliability can be greatly improved by this technique. Let us first consider the following equation:

$$R^{(0)}(t) = 3R(t)^2 - 2R(t)^3, \tag{7.1}$$

i.e., the equation expressing the reliability of a TMR system with no spares, $R(t)$ being the reliability of a single, non-replicated (simplex) component. Equation (7.1) can be derived for instance via Markovian reliability modelling under the assumption of independence between the occurrence of faults [Joh89].

With the same technique and under the same hypothesis it is possible to show that, even in the case of non-perfect error detection coverage, this equation can be considerably improved by adding one spare. This is the equation resulting from the Markov model in Fig. 7.1, expressed as a function of error recovery coverage ($C$, defined as the probability associated with the process of identifying the failed module out of those available and being able to switch in the spare [Joh89]) and time ($t$):

$$R^{(1)}(C,t) = (-3C^2 + 6C) \times [R(t)(1 - R(t))]^2 + R^{(0)}(t). \tag{7.2}$$

Appendix A gives some mathematical details on Eq. (7.2).

Adding more spares obviously implies further improving reliability. In general, for any $N \geq 3$, it is possible to consider a class of monotonically increasing reliability functions,

$$\left( R^{(M)}(C,t) \right)_{M>0}, \tag{7.3}$$

corresponding to systems adopting $N + M$ replicas. Depending on both cost and reliability requirements, the user can choose the most-suited values for $M$ and $N$.

Note how quantity (7.2) is always greater than quantity (7.1) as $R^{(0)}(t)$ and $(-3C^2 + 6C)$ are always positive for $0 < C \leq 1$. Figure 7.2 compares Eq. (7.1) and (7.2) in the general case while Fig. 7.3 covers the case of perfect coverage. In the latter case, the reliability of a single, non-redundant (simplex) system is also portrayed. Note furthermore how the crosspoint between the three-and-one-spare system and the non-redundant system is considerably lower than the crosspoint between the latter and the TMR system—$R(t) \approx 0.2324$ vs. $R(t) = 0.5$.

The reliability of the system can therefore be increased from the one of a pure NMR system to that of $N$-and-$M$-spare systems (see Fig. 7.2 and Fig. 7.3).

## 7.1.2 Using the $\alpha$-count Feature

As it has been already mentioned in Sect. 6.2.2, the `TRANSIENT` clause of `ARIEL`, exploiting the $\alpha$-count fault identification mechanism introduced in Sect. 5.2.4.2, can be adopted for attaching different recovery or reconfiguration strategies to the detection of a permanent or intermittent fault, in general requiring reconfiguration and redundancy exhaustion, and to the detection of a transient fault, that could be tolerated with a recovery technique such as resetting the faulty component. This section describes a simple Markov reliability model of a TMR system discriminating between these two cases. Figure 7.4 shows the model. $T$ is the probability that the current fault is a transient one and $R$ is the probability of successful recovery for a component affected by a transient fault. To simplify the model, recovery is assumed to be instantaneous. The system degrades only when permanent faults occur. Solving the model brings to the following equation:

$$R^\alpha_{\text{TMR}}(t) = 3 \exp^{-2(1-RT)\lambda t} -2 \exp^{-3(1-RT)\lambda t}. \tag{7.4}$$

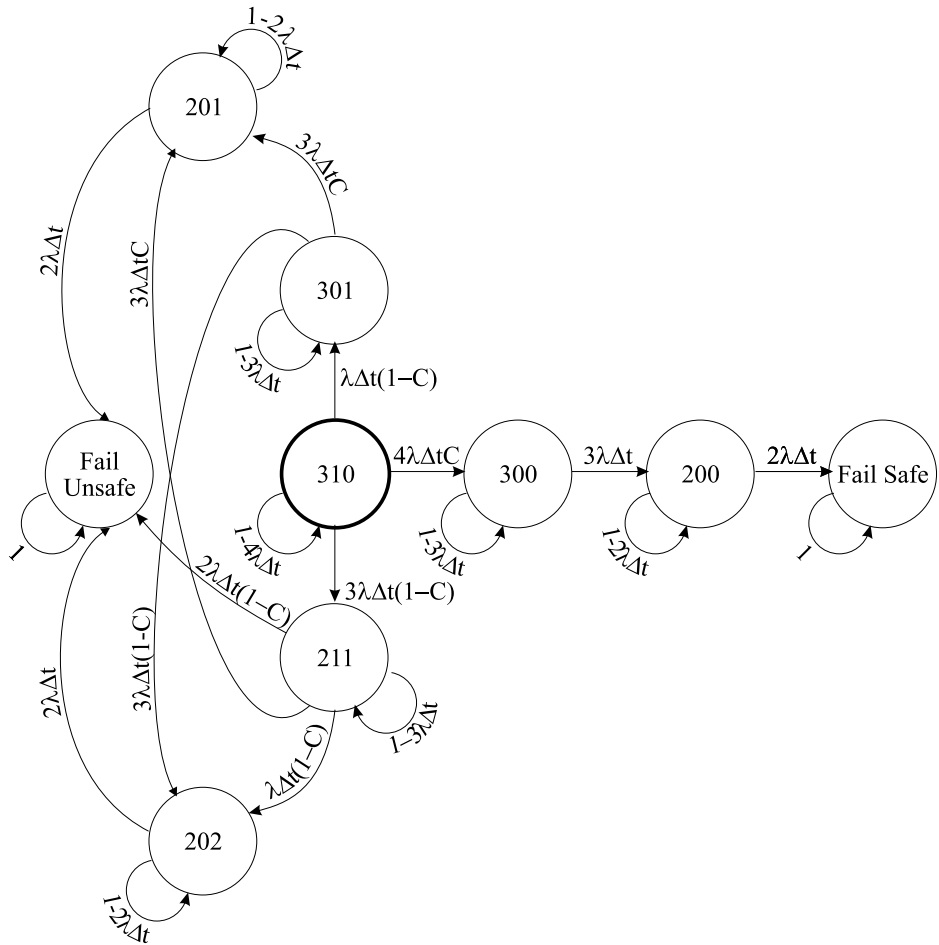

**Figure 7.1:** Markov reliability model for a TMR-and-1-spare system. $\lambda$ is the failure rate, $C$ is the error recovery coverage factor. A "fail safe" state is reached when the system is no more able to correctly perform its function, though the problem has been safely detected and handled properly. In 'Fail unsafe,' on the contrary, the system is incorrect, though the problem has not been handled or detected. Every other state is labeled with three digits, $d_1 d_2 d_3$, such that $d_1$ is the number of non-faulty modules in the TMR system, $d_2$ is the number of non-faulty spares (in this case, 0 or 1), and $d_3$ is the number of undetected, faulty modules. The initial state, 310, has been highlighted. This model is solved by Eq. (7.2).

Eq. 7.4 can be written as

$$R_{\text{TMR}}^{\alpha}(t) = 3R(t)^{2(1-RT)} - 2R(t)^{3(1-RT)},  \tag{7.5}$$

with $R(t)$ the reliability of the simplex system. Note how, in a sense, the introduction of the $\alpha$-count technique results in a modification of the exponents of Eq. 7.1 by a factor equal to $1 - RT$.

In general, the resulting system shows a reliability that is larger than the one of a TMR system. Figure 7.5 compares these reliabilities. As it is evident from that image, the crosspoint between the reliability graphs of the simplex and that of the TMR system extended with the

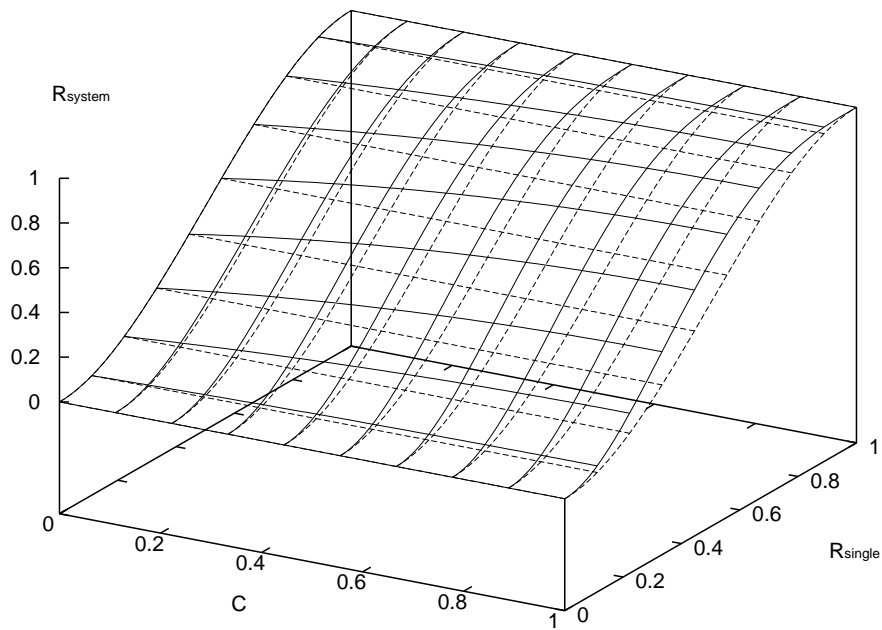

**Figure 7.2:  Graphs of Eq. (7.1) and (7.2) as functions of $C$ and of $R$.  Note how the graph of (7.2) is strictly above the other.**

$\alpha$-count technique is in general lower.  Its exact value is given by the function

$$\text{crosspoint}(R, T) = 0.5^{1/(1 - RT)}$$

that is drawn in Fig. 7.6.

Appendix A.2 provides the mathematical details leading to Eq. (7.4).

## 7.2   Performance Analysis

In this section a number of measurements are reported concerning the performance of different subsystems of the $\mathcal{REL}$ prototype developed within TIRAN. Most of these measurements relate to the development platform version, a Pentium-133 PC running Windows/NT, and to a preliminary version developed on a Parsytec Xplorer, a parallel system based on PowerPC-66 and running EPX, and on a Parsytec CC, also running EPX on PowerPC-133 nodes.

The results of these *simulations* provide the reader with an idea on the order of magnitude of the duration of some algorithms, though they are clearly dependent on "external" variables such as

- the processor architecture and technology (clock speed, bus width, memory ports, cache subsystem, and so forth),

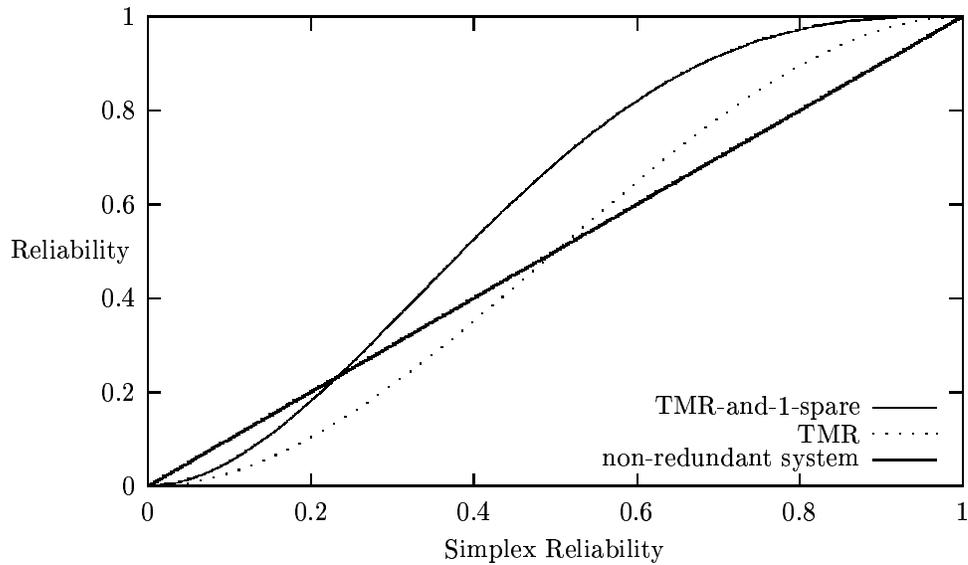

**Figure 7.3: Graphs of Eq. (7.1) and (7.2) when** $C = 1$ **(perfect error detection coverage). The reliability of a single, non-redundant system is also portrayed.**

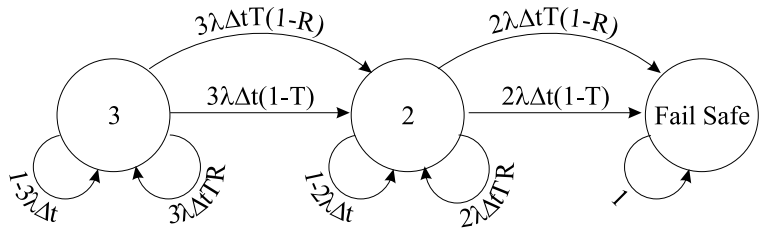

**Figure 7.4: The Markov reliability model of a TMR system that gracefully degrade in the presence of a permanent or intermittent fault and "cleans up" (that is, rejuvenates) the components affected by transient faults.**

- the network architecture and technology (topology, organisation, diameter, number of available independent channels, etc.),

- the characteristics of the underlying operating system (task scheduling policy, virtual memory organisation, etc.)

- the number and type of nodes constituting the system,

as well as on "internal" variables such as

- the algorithms (especially their complexity and scalability),

- the performance of the adopted network protocols (for instance, choosing UDP as a transport protocol implies an overhead that differs from that of TCP),

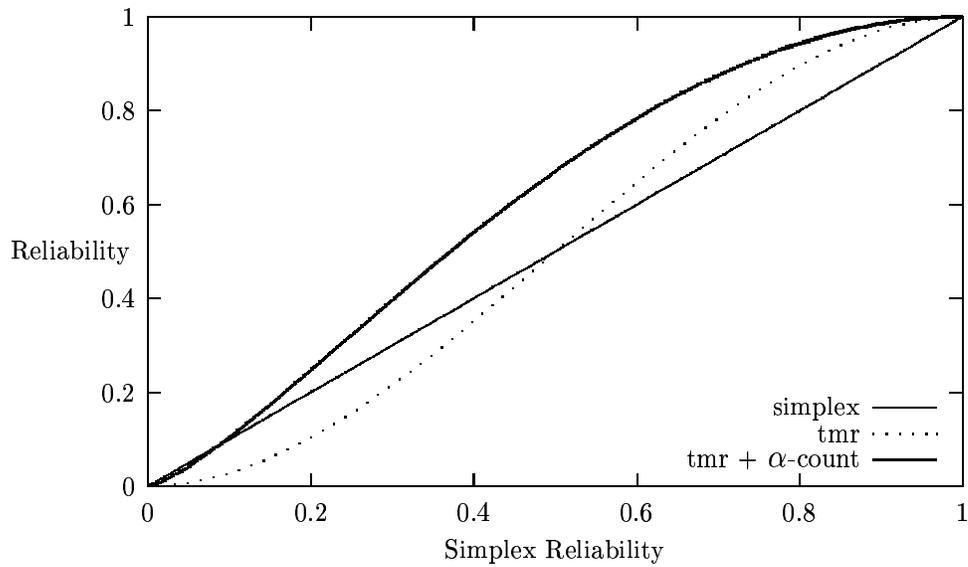

**Figure 7.5: Reliability of a simplex system, of a TMR system, and of a modified TMR system exploiting the $\alpha$-count fault identification technique. In this case $T$ is 50% and $R = 0.6$.**

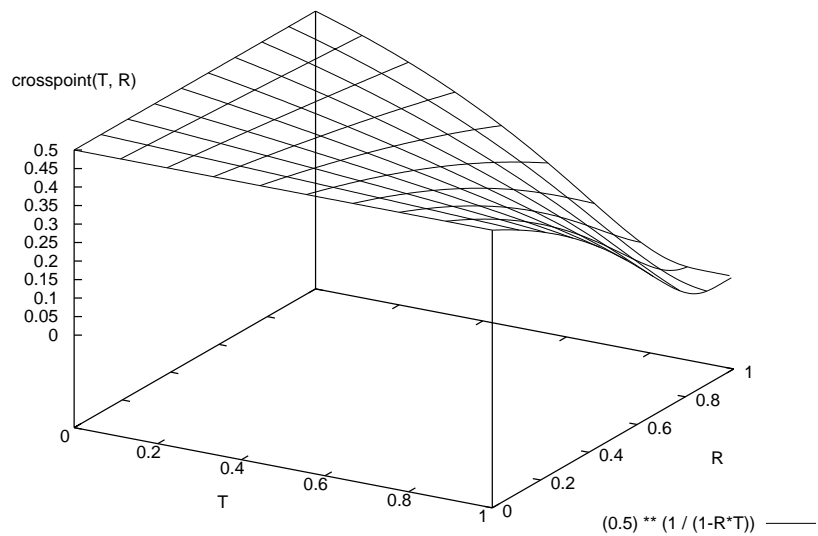

**Figure 7.6: The crosspoint between the graph of the reliability of the simplex system and that of $R_{\text{TMR}}^{\alpha}$ as a function of $R$ and $T$.**

- other design choices affecting the organisation of data structures (dynamic memory allocation vs. static, pre-allocated objects, etc.)

The aim of this section is therefore not only to provide the reader with an idea of the performance of a $\mathcal{REL}$-based tool on a particular system or set of systems—which basically means describing a few points in a wide, complex space. It also *analyses* one of the core internal variables (a base service used in the TIRAN backbone and in one of the dependable mechanisms of the TIRAN library), and solves it in such a way as to identify which external variables would be optimal in order to reach good performance. Section 7.2.1 reports on this analysis [DFDL00c], while Sect. 7.2.2 reports a set of quantitative measurements related to the performance of some of the elements of the $\mathcal{REL}$-prototype developed in TIRAN—namely, the backbone (in Sect. 7.2.2.1), the recovery interpreter (in Sect. 7.2.2.3) and the distributed voting tool (in Sect. 7.2.2.4). Other studies are also reported in [EVM$^+$98].

## 7.2.1 Analysis of the Gossiping Service

A number of distributed applications, such as those based on distributed consensus [LSP82] or those based on the concept of restoring organs ($N$-modular redundancy systems with $N$-replicated voters), require a base service called gossiping [HMS72, HHL88]. Gossiping is also a core algorithm for the distributed voting tool (DV) described in Sect. 5.2.3 and for the TIRAN backbone (BB) described in Sect. 5.2.4. As such, and because of its repercussions on the performance of the $\mathcal{REL}$ architecture described in Sect. 5, the TIRAN gossiping algorithm is analysed in the rest of this section. This algorithm and its analysis are original contributions of the author of this dissertation.

Informally speaking, gossiping is a communication scheme such that every member of a set has to communicate a private value to all the other members. Gossiping is clearly an expensive service, as it requires a large amount of communication. A gossiping service is used in the DV as a mechanism for allowing the voters to exchange their input values such that each of them can vote on the whole input set. The same service is used in the BB as a means to forward each local event notification allowing each component to keep a consistent copy of the system database. In general, implementations of this service can have a great impact on the throughput of its client applications and perform very differently depending on the number of members in the set. Indeed, their basic need—that is, keeping a coherent global view of the event taking place in the distributed system—implies a performance penalty that is mainly due to the gossiping algorithm itself, together with the physical constraints of the communication line and on the number of available independent channels. The analysis carried out in the following shows that, under the hypotheses of discrete time, of constant time for performing a `send` or `receive`, and of a crossbar communication system, and depending on an algorithmic parameter, gossiping can use from O($N^2$) to O($N$) time, with $N + 1$ communicating members. Three cases are briefly investigated. The last and best-performing case, whose activity follows the execution pattern of pipelined hardware processors, is shown to exhibit an efficiency constant with respect to $N$. This translates in unlimited scalability of the corresponding gossiping service. When performing multiple consecutive gossiping sessions, (as in restoring organs or in the TIRAN

BB), the throughput of the system can reach the value of $t/2$, $t$ being the time for sending one value from one member to another. This means that a full gossiping is completed every two basic communication steps.

All results presented in what follows of this section are original contributions of the author of this dissertation [DFDL00c]. For the sake of readability, proofs of propositions have been presented in Appendix A.3.

### 7.2.1.1  Formal Model

The analysis is carried out under the following hypotheses: $N + 1$ processors are interconnected via some communication means that allows them to communicate with each other by means of full-duplex point-to-point communication lines. Communication is synchronous and blocking. Processors are uniquely identified by integer labels in $\{0, \ldots, N\}$. The processors own some local data they need to share (for instance, to execute a voting algorithm or to update the TIRAN DB). In order to share their local data, each processor needs to broadcast its own data to all the others via multiple sending operations, and to receive the $N$ data items owned by the others. A discrete time model is assumed—events occur at discrete time steps, one event at a time per processor. This is a special class of the general family of problems of information dissemination known as *gossiping* [HHL88][1].

Let us assume that the time to send and receive a message is constant. This amount of time be called "*time step*". On a given time step $t$, processor $i$ may be:

1. sending a message to processor $j$, $j \neq i$; using relation $S^t$ this is represented as relation $i\,S^t j$;

2. receiving a message from processor $j$, $j \neq i$; using relation $R^t$ this is represented as $i\,R^t j$;

3. blocked, waiting for messages to be *received* from any processor; where both the identities of the processors and $t$ can be omitted without ambiguity, the symbol "$-$" will be used to represent this case;

4. blocked, waiting for a message to be *sent*, i.e., for another processor to enter the receiving state; under the same assumptions of case 3, the symbol "$\curvearrowright$" will be used.

The above cases are referred to as "the actions" of time step $t$.

A slot is a temporal "window" one time step long, related to a processor. On each given time step there are $N + 1$ available slots within the system. Within that time step, a processor may *use* that slot (if it sends or receives a message during that slot), or it may *waste* it (if it is in one of the remaining two cases). In other words, processor $i$ is said to *use* slot $t$ if and only if

$$U(t, i) = \{\exists j \ (i\,S^t j \lor i\,R^t j)\}$$

---

[1]It is worth mentioning that a relevant difference with respect to the basic Gossip Problem [HHL88] is that, in the case presented herein, on each communication run, members are only allowed to broadcast their own *local* data, while, in general [HMS72], they "pass on to each other as much [information] they know at the time", including data received from the other members in previous communication runs.

is true; on the contrary, processor $i$ is said to *waste* slot $t$ if and only if $\neg U(t, i)$. The following notation,

$$\delta_{i,t} = \begin{cases} 1 & \text{if } U(t, i) \text{ is true,} \\ 0 & \text{otherwise,} \end{cases}$$

will be used to count used slots.

Let us define four state templates for a finite state automaton to be described later on:

$WR$ **state.** A processor is in state $WR_j$ if it is waiting for the arrival of a message from processor $j$. Where the subscript is not important it will be omitted. Once there, a processor stays in state $WR$ for *zero* (if it can start receiving immediately) or more time steps, corresponding to the same number of actions "wait for a message to come".

$S$ **state.** A processor is in state $S_j$ when it is sending a message to processor $j$. Note that, by the above assumptions and definitions, this transition lasts exactly one time step. For each transition to state $S$ there corresponds exactly one "send" action.

$WS$ **state.** A processor that is about to send a message to processor $j$ is said to be in state $WS_j$. In the following, the subscript will be omitted when this does not result in ambiguities. The permanence of a processor in state $WS$ implies *zero* (if the processor can send immediately) or more occurrences in a row of the "wait for sending" action.

$R$ **state.** A processor that is receiving a message from processor $j$ is said to be in state $R_j$. By the above definitions, this state transition also lasts one time step.

For any $i \in [0, N]$, let $\mathcal{P}_1, \ldots, \mathcal{P}_N$ represent a permutation of the $N$ integers $[0, N] - \{i\}$. Then the above state templates can be used to compose $N + 1$ linear finite state automata (FSA) with a structure like the one in Fig. 7.7, which shows the state diagram of the FSA to be executed by processor $i$. The first row represents the condition that has to be reached before processor $i$ is allowed to begin its broadcast: a series of $i$ pairs $(WR, R)$.

Once processor $i$ has successfully received $i$ messages, it gains the right to broadcast, which it does according to the rule expressed in the second row of Fig. 7.7: it orderly sends its message to its fellows, the $j$-th message being sent to processor $\mathcal{P}_j$.

The third row of Fig. 7.7 represents the receipt of the remaining $N - i$ messages, coded as $N - i$ pairs like those in the first row.

**Proposition 1** *For any permutation $\mathcal{P}$, the distributed algorithm described by the state diagram of Fig. 7.7 solves the above mentioned gossiping problem without deadlocks.*

(The proof is given in Sect. A.3.1).

Let us call:

**Run** the collection of slots needed to fully execute the above algorithm on a given system, plus the value of the corresponding actions.

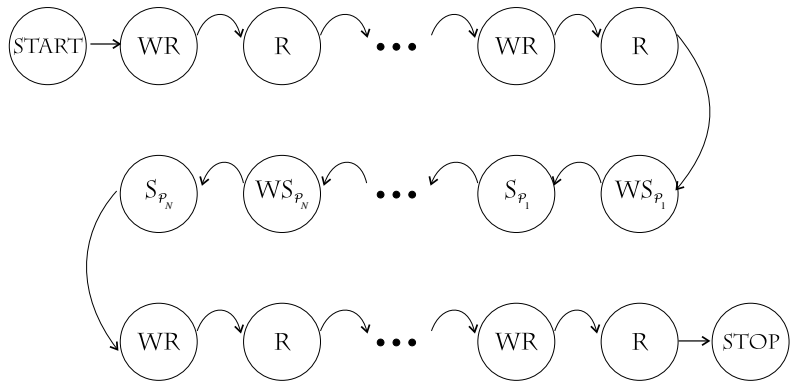

**Figure 7.7: The state diagram of the FSA run by processor $i$. The first row consists of $i$ pairs $(WR, R)$. (For $i = 0$ this row is missing). $(\mathcal{P}_1, \ldots, \mathcal{P}_N)$ represents a permutation of the $N$ integers $[0, N] - \{i\}$. The last row contains $N - i$ pairs $(WR, R)$.**

**Average slot utilisation** The number of slots that have been used in the average within a run. It represents the average degree of parallelism exploited in the system. It will be indicated as $\mu_N$, or simply as $\mu$. It varies between 0 and $N + 1$.

**Efficiency** the percentage of used slots over the total number of slots available during a run. $\varepsilon_N$, or simply $\varepsilon$, will be used to represent efficiency.

**Length** the number of time steps in a run. It represents a measure of the time needed by the distributed algorithm to complete. $\lambda_N$, or simply $\lambda$, will be used for lengths.

**The number of slots** available within a run of $N + 1$ processors be denoted as $\sigma(N) = (N + 1)\lambda_N$.

**The number of used slots** i.e., the number of slots that have been used during time step $t$, be denoted as

$$\nu_t = \sum_{i=0}^{N} \delta_{i,t}.$$

**The utilisation string** is the $\lambda$-tuple $\vec{\nu} = [\nu_1, \nu_2, \ldots, \nu_\lambda]$, representing the number of used slots for each time step, respectively.

### 7.2.1.2  Discussion

In the following, three cases of $\mathcal{P}$ are introduced and discussed. It is shown how varying the structure of $\mathcal{P}$ may develop very different values for $\mu$, $\varepsilon$, and $\lambda$. This fact, coupled with the physical constraints of the communication line and with the number of available independent channels, determines the overall performance of this algorithm and of the services based on it.

| id ↓ / step → | 1 | 2 | 3 | 4 | 5 | 6 | 7 | 8 | 9 | 10 | 11 | 12 | 13 | 14 | 15 | 16 | 17 | 18 |
|---|---|---|---|---|---|---|---|---|---|---|---|---|---|---|---|---|---|---|
| 0 | $S_1$ | $S_2$ | $S_3$ | $S_4$ | $R_1$ | — | $R_2$ | — | — | — | $R_3$ | — | — | — | $R_4$ | — | — | — |
| 1 | $R_0$ | ⌢ | ⌢ | ⌢ | $S_0$ | $S_2$ | $S_3$ | $S_4$ | $R_2$ | — | — | $R_3$ | — | — | — | $R_4$ | — | — |
| 2 | — | $R_0$ | — | — | — | $R_1$ | $S_0$ | ⌢ | $S_1$ | $S_3$ | $S_4$ | — | $R_3$ | — | — | — | $R_4$ | — |
| 3 | — | — | $R_0$ | — | — | — | $R_1$ | — | — | $R_2$ | $S_0$ | $S_1$ | $S_2$ | $S_4$ | — | — | — | $R_4$ |
| 4 | — | — | — | $R_0$ | — | — | — | $R_1$ | — | — | $R_2$ | — | — | $R_3$ | $S_0$ | $S_1$ | $S_2$ | $S_3$ |
| $\vec{v}$ → | 2 | 2 | 2 | 2 | 2 | 2 | 4 | 2 | 2 | 2 | 4 | 2 | 2 | 2 | 2 | 2 | 2 | 2 |

**Table 7.1:** A run ($N = 4$), with $\mathcal{P}$ equal to the identity permutation. The step row represents time steps. Id's identify processors. $\vec{v}$ is the utilisation string (as defined on p. 145). In this case $\mu$, or the average utilisation is 2.22 slots out of 5, with an efficiency $\varepsilon = 44.44\%$ and a length $\lambda = 18$. Note that, if the slot is used, then entry $(i, t) = \mathcal{R}_j$ in this matrix represents relation $i\,\mathcal{R}^t j$.

**First Case: Identity Permutation.** As a first case, let $\mathcal{P}$ be equal to the identity permutation:

$$\begin{pmatrix} 0, \ldots, i-1, i+1, \ldots, N \\ 0, \ldots, i-1, i+1, \ldots, N \end{pmatrix}, \tag{7.6}$$

i.e., in cycle notation [Knu73], $(0) \ldots (i-1)(i+1) \ldots (N)$. Note that in this case only singleton cycles are present.

Using the identity permutation means that, once processor $i$ gains the right to broadcast, it will first send its message to processor 0 (possibly having to wait for processor 0 to become available to receive that message), then it will do the same with processor 1, and so forth up to $N$, obviously skipping itself. This is represented in Table 7.1 for $N = 4$. Let us call this a run-table. Let us also call "run-table $x$" a run-table of a system with $N = x$.

It is possible to characterise precisely the duration of the algorithm that adopts this permutation:

**Proposition 2** *For $N > 0$, $\lambda_N = \frac{3}{4}N^2 + \frac{5}{4}N + \frac{1}{2}\lfloor N/2 \rfloor$.*

(The proof is given in Sect. A.3.2).

**Lemma 1** *The number of columns with 4 used slots inside, for a run with $\mathcal{P}$ equal to the identity permutation and $N + 1$ processors, is*

$$\frac{N^2 - 2N + [N \text{ is odd}]}{4},$$

*where "$[N$ is odd$]$" is 1 when $N$ is odd and 0 otherwise.*

Figure 7.8 shows the typical shape of run-tables in the case of $\mathcal{P}$ being the identity permutation, also locating the 4-used slot clusters.

The following Propositions locate the asymptotic values of $\mu$ and $\varepsilon$:

**Proposition 3** $\lim_{N \to \infty} \varepsilon_N = 0$,

**Proposition 4** $\lim_{N \to \infty} \mu_N = \frac{8}{3}$.

(The proofs are given in Sect. A.3.4 and A.3.5).

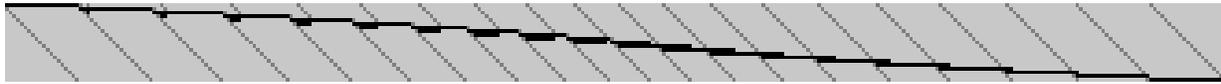

**Figure 7.8: A graphical representation for run-table 20 when $\mathcal{P}$ is the identity permutation. Light gray pixels represent wasted slots, gray pixels represent $R$ actions, black slots are sending actions.**

| id ↓ step → | 1 | 2 | 3 | 4 | 5 | 6 | 7 | 8 | 9 | 10 | 11 | 12 | 13 | 14 | 15 | 16 | 17 | 18 | 19 | 20 | 21 | 22 | 23 | 24 |
|---|---|---|---|---|---|---|---|---|---|---|---|---|---|---|---|---|---|---|---|---|---|---|---|---|
| 0 | $S_5$ | $S_1$ | $S_3$ | $S_2$ | $S_4$ | $R_1$ | $-$ | $-$ | $-$ | $-$ | $R_2$ | $-$ | $-$ | $-$ | $R_3$ | $-$ | $-$ | $-$ | $-$ | $R_4$ | $R_5$ | $-$ | $-$ | $-$ |
| 1 | $-$ | $R_0$ | $S_5$ | $\curvearrowright$ | $\curvearrowright$ | $S_0$ | $S_3$ | $S_2$ | $S_4$ | $R_2$ | $-$ | $-$ | $-$ | $R_3$ | $-$ | $-$ | $-$ | $-$ | $R_4$ | $R_5$ | $-$ | $-$ | $-$ | $-$ |
| 2 | $-$ | $-$ | $-$ | $R_0$ | $-$ | $-$ | $-$ | $R_1$ | $S_5$ | $S_1$ | $S_0$ | $S_3$ | $S_4$ | $-$ | $-$ | $R_3$ | $-$ | $-$ | $-$ | $-$ | $R_4$ | $R_5$ | $-$ | $-$ |
| 3 | $-$ | $-$ | $R_0$ | $-$ | $-$ | $R_1$ | $-$ | $-$ | $-$ | $-$ | $R_2$ | $S_5$ | $S_1$ | $S_0$ | $S_2$ | $S_4$ | $-$ | $-$ | $-$ | $R_4$ | $R_5$ | $-$ | $-$ | $-$ |
| 4 | $-$ | $-$ | $-$ | $-$ | $R_0$ | $-$ | $-$ | $-$ | $R_1$ | $-$ | $-$ | $-$ | $R_2$ | $-$ | $-$ | $-$ | $R_3$ | $S_5$ | $S_1$ | $S_0$ | $S_3$ | $S_2$ | $-$ | $R_5$ |
| 5 | $R_0$ | $-$ | $R_1$ | $-$ | $-$ | $-$ | $-$ | $-$ | $R_2$ | $-$ | $-$ | $-$ | $R_3$ | $-$ | $-$ | $-$ | $-$ | $R_4$ | $\curvearrowright$ | $S_1$ | $S_0$ | $S_3$ | $S_2$ | $S_4$ |
| $\vec{v}$ → | 2 | 2 | 4 | 2 | 2 | 2 | 2 | 2 | 4 | 2 | 2 | 2 | 4 | 2 | 2 | 2 | 2 | 2 | 2 | 2 | 4 | 4 | 4 | 2 | 2 |

**Table 7.2: Run-table 5 when $\mathcal{P}$ is chosen pseudo-randomly. $\mu$ is $2.5$ slots out of 6, which implies an efficiency of 41.67%.**

**Second Case: Pseudo-random Permutation.**   This section covers the case where $\mathcal{P}$ is a pseudo-random[2] permutation of the integers $0, \ldots, i-1, i+1, \ldots, N$.

Figure 7.9 (top picture) shows the values of $\lambda$ using the identity and pseudo-random permutations and draws the parabola which best fits to these values in the case of pseudo-random permutations. Experiments show that the choice of the identity permutation is even "worse" than choosing permutations at random. The same conclusion is suggested by Fig. 7.10 that compares the averages and efficiencies in the above two cases.

Table 7.2 shows run-table 5, and Fig. 7.11 shows the shape of run-table 20 in this case.

**Third Case: Algorithm of Pipelined Broadcast.**   Let $\mathcal{P}$ be the following permutation:

$$\begin{pmatrix} 0, \ldots, i-1, i+1, \ldots, N \\ i+1, \ldots, N, 0, \ldots, i-1 \end{pmatrix}. \tag{7.7}$$

Note how permutation (7.7) is equivalent to $i$ cyclic logical left shifts of the identity permutation. Note also how, in cycle notation, (7.7) is represented as one cycle; for instance,

$$\begin{pmatrix} 0, 1, 2, 4, 5 \\ 4, 5, 0, 1, 2 \end{pmatrix},$$

---

[2]The standard C function "random" [Sun94] has been used—a non-linear additive feedback random number generator returning pseudo-random integers in the range $[0, 2^{31}-1]$ with a period approximately equal to $16(2^{31}-1)$. A truly random integer has been used as a seed. $\mathcal{P}$ is composed by generating with random pseudo-random numbers in domain $[0, N] - X$, with $X$ initially set to $\{i\}$ and then augmented at each step with the integer generated in the previous step.

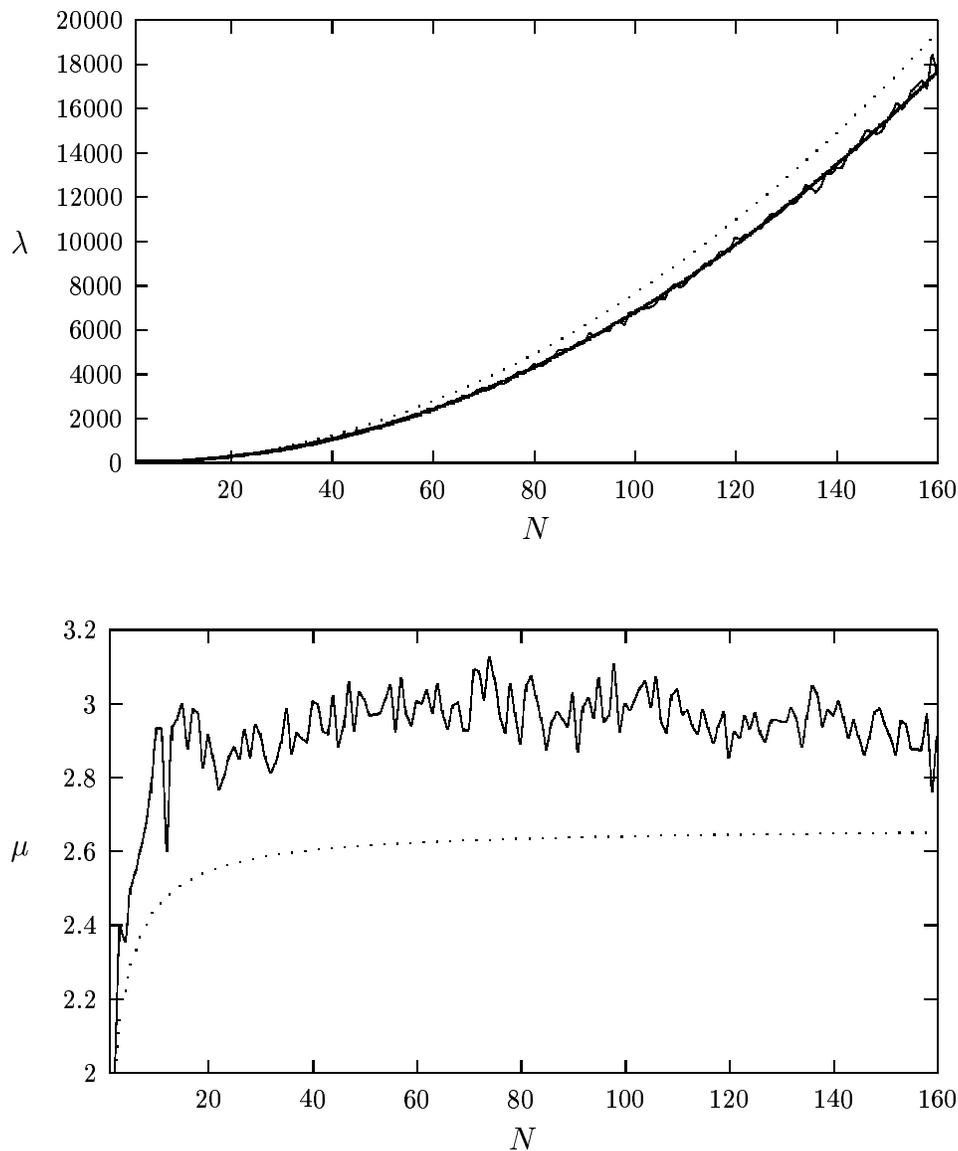

Figure 7.9: **Top picture: comparison between lengths when $\mathcal{P}$ is the identity permutation (dotted parabola) and when it is pseudo-random (piecewise line), $1 \leq N \leq 160$. The lowest curve ($\lambda = 0.71N^2 - 3.88N + 88.91$) is the parabola best fitting to the piecewise line—which suggests a quadratic execution time as in the case of the identity permutation. Bottom picture: comparison between values of $\mu$ when $\mathcal{P}$ is the identity permutation (dotted curve) and when $\mathcal{P}$ is pseudo-random (piecewise line), $1 \leq N \leq 160$. For each run $N$ and for each processor $i \in [0, N]$ with run $N$, $i$ generates a pseudo-random permutation in $[0, N] - \{i\}$ and uses it to structure its broadcast. Note how the piecewise line is strictly over the curve. Note also how $\mu$ tends to a value slightly larger than $2.6$ for the identity permutation, as claimed in Prop. 4.**

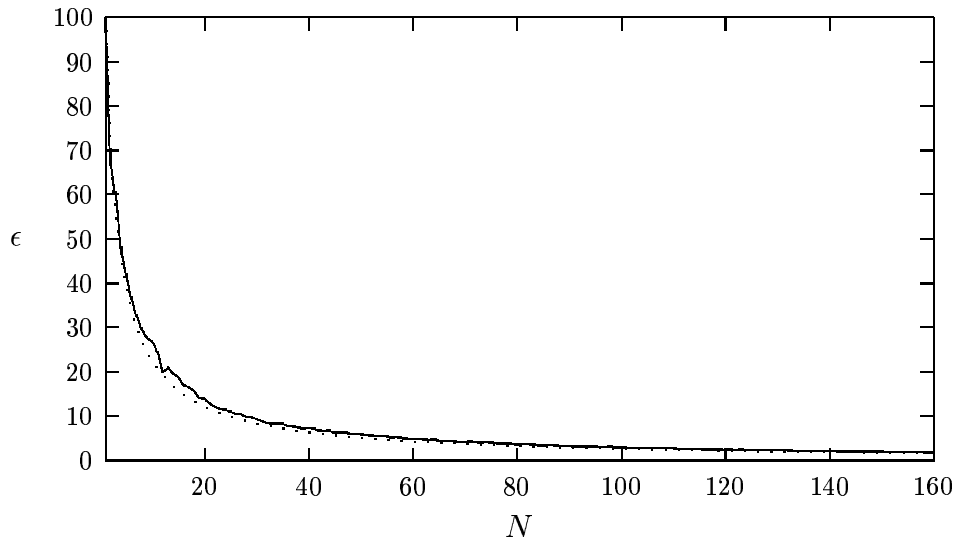

**Figure 7.10:** Comparison between values of $\varepsilon$ in the case of the pseudo-random permutation (piecewise line) and that of the identity permutation (dotted curve), $1 \leq N \leq 160$. Also in this graph the former is strictly over the latter, though they get closer to each other and to zero as $N$ increases, as proven for the identity permutation in Prop. 3.

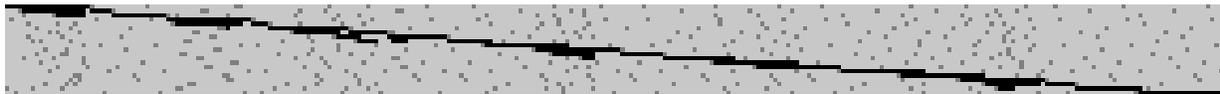

**Figure 7.11:** A graphical representation for run-table 20 when $\mathcal{P}$ is a pseudo-random permutation.

that is, (7.7) for $N = 5$ and $i = 3$, is equivalent to cycle $(0, 4, 1, 5, 2)$.

A value of $\mathcal{P}$ equal to (7.7) means that, once processor $i$ has gained the right to broadcast, it will first send its message to processor $i + 1$ (possibly having to wait for it to become available to receive that message), then it will do the same with processor $i + 2$, and so forth up to $N$, then wrapping around and going from processor 0 to processor $i - 1$. This is represented in Table 7.3 for $N = 9$.

Figures quite similar to Table 7.3 can be found in many classical works on pipelined microprocessors (see e.g. [PH96]). For this and other similarities, the algorithm using permutation (7.7) has been called algorithm of "pipelined broadcast".

Indeed using this permutation leads to better performance. In particular, after a start-up phase (*after filling the pipeline*), sustained performance is close to the maximum—a number of unused slots (*pipeline bubbles*) still exist, even in the sustained region, but here $\mu$ reaches value $N + 1$ half of the times (if $N$ is odd). In the region of decay, starting from time step 19, every new time step a processor fully completes its task. Similar remarks apply to Table 7.4; this is the typical shape of a run-table for $N$ even. This time the state within the sustained region is more steady, though the maximum number of used slots never reaches the number of slots in the system.

| id ↓ / step → | 1 | 2 | 3 | 4 | 5 | 6 | 7 | 8 | 9 | 10 | 11 | 12 | 13 | 14 | 15 | 16 | 17 | 18 | 19 | 20 | 21 | 22 | 23 | 24 | 25 | 26 | 27 |
|---|---|---|---|---|---|---|---|---|---|---|---|---|---|---|---|---|---|---|---|---|---|---|---|---|---|---|---|
| 0 | $S_1$ | $S_2$ | $S_3$ | $S_4$ | $S_5$ | $S_6$ | $S_7$ | $S_8$ | $S_9$ | − | $R_1$ | $R_2$ | $R_3$ | $R_4$ | $R_5$ | $R_6$ | $R_7$ | $R_8$ | $R_9$ | − | − | − | − | − | − | − | − |
| 1 | $R_0$ | ⌢ | $S_2$ | $S_3$ | $S_4$ | $S_5$ | $S_6$ | $S_7$ | $S_8$ | $S_9$ | $S_0$ | − | $R_2$ | $R_3$ | $R_4$ | $R_5$ | $R_6$ | $R_7$ | $R_8$ | $R_9$ | − | − | − | − | − | − | − |
| 2 | − | $R_0$ | $R_1$ | ⌢ | $S_3$ | $S_4$ | $S_5$ | $S_6$ | $S_7$ | $S_8$ | $S_9$ | $S_0$ | $S_1$ | − | $R_3$ | $R_4$ | $R_5$ | $R_6$ | $R_7$ | $R_8$ | $R_9$ | − | − | − | − | − | − |
| 3 | − | − | $R_0$ | $R_1$ | $R_2$ | ⌢ | $S_4$ | $S_5$ | $S_6$ | $S_7$ | $S_8$ | $S_9$ | $S_0$ | $S_1$ | $S_2$ | − | $R_4$ | $R_5$ | $R_6$ | $R_7$ | $R_8$ | $R_9$ | − | − | − | − | − |
| 4 | − | − | − | $R_0$ | $R_1$ | $R_2$ | $R_3$ | ⌢ | $S_5$ | $S_6$ | $S_7$ | $S_8$ | $S_9$ | $S_0$ | $S_1$ | $S_2$ | $S_3$ | − | $R_5$ | $R_6$ | $R_7$ | $R_8$ | $R_9$ | − | − | − | − |
| 5 | − | − | − | − | $R_0$ | $R_1$ | $R_2$ | $R_3$ | $R_4$ | ⌢ | $S_6$ | $S_7$ | $S_8$ | $S_9$ | $S_0$ | $S_1$ | $S_2$ | $S_3$ | $S_4$ | − | $R_6$ | $R_7$ | $R_8$ | $R_9$ | − | − | − |
| 6 | − | − | − | − | − | $R_0$ | $R_1$ | $R_2$ | $R_3$ | $R_4$ | $R_5$ | ⌢ | $S_7$ | $S_8$ | $S_9$ | $S_0$ | $S_1$ | $S_2$ | $S_3$ | $S_4$ | $S_5$ | − | $R_7$ | $R_8$ | $R_9$ | − | − |
| 7 | − | − | − | − | − | − | $R_0$ | $R_1$ | $R_2$ | $R_3$ | $R_4$ | $R_5$ | $R_6$ | ⌢ | $S_8$ | $S_9$ | $S_0$ | $S_1$ | $S_2$ | $S_3$ | $S_4$ | $S_5$ | $S_6$ | − | $R_8$ | $R_9$ | − |
| 8 | − | − | − | − | − | − | − | $R_0$ | $R_1$ | $R_2$ | $R_3$ | $R_4$ | $R_5$ | $R_6$ | $R_7$ | ⌢ | $S_9$ | $S_0$ | $S_1$ | $S_2$ | $S_3$ | $S_4$ | $S_5$ | $S_6$ | $S_7$ | − | $R_9$ |
| 9 | − | − | − | − | − | − | − | − | $R_0$ | $R_1$ | $R_2$ | $R_3$ | $R_4$ | $R_5$ | $R_6$ | $R_7$ | $R_8$ | ⌢ | $S_0$ | $S_1$ | $S_2$ | $S_3$ | $S_4$ | $S_5$ | $S_6$ | $S_7$ | $S_8$ |
| $\vec{\nu}$ → | 2 | 2 | 4 | 4 | 6 | 6 | 8 | 8 | 10 | 8 | 10 | 8 | 10 | 8 | 10 | 8 | 10 | 8 | 10 | 8 | 8 | 6 | 6 | 4 | 4 | 2 | 2 |

**Table 7.3:** Run-table of a run for $N = 9$ using permutation of permutation (7.7). In this case $\mu$, or the average utilisation is 6.67 slots out of 10, with an efficiency $\varepsilon = 66.67\%$ and a length $\lambda = 27$. Note that $\vec{\nu}$ is in this case a palindrome.

| id ↓ / step → | 1 | 2 | 3 | 4 | 5 | 6 | 7 | 8 | 9 | 10 | 11 | 12 | 13 | 14 | 15 | 16 | 17 | 18 | 19 | 20 | 21 | 22 | 23 | 24 |
|---|---|---|---|---|---|---|---|---|---|---|---|---|---|---|---|---|---|---|---|---|---|---|---|---|
| 0 | $S_1$ | $S_2$ | $S_3$ | $S_4$ | $S_5$ | $S_6$ | $S_7$ | $S_8$ | − | $R_1$ | $R_2$ | $R_3$ | $R_4$ | $R_5$ | $R_6$ | $R_7$ | $R_8$ | − | − | − | − | − | − | − |
| 1 | $R_0$ | ⌢ | $S_2$ | $S_3$ | $S_4$ | $S_5$ | $S_6$ | $S_7$ | $S_8$ | $S_0$ | − | $R_2$ | $R_3$ | $R_4$ | $R_5$ | $R_6$ | $R_7$ | $R_8$ | − | − | − | − | − | − |
| 2 | − | $R_0$ | $R_1$ | ⌢ | $S_3$ | $S_4$ | $S_5$ | $S_6$ | $S_7$ | $S_8$ | $S_0$ | $S_1$ | − | $R_3$ | $R_4$ | $R_5$ | $R_6$ | $R_7$ | $R_8$ | − | − | − | − | − |
| 3 | − | − | $R_0$ | $R_1$ | $R_2$ | ⌢ | $S_4$ | $S_5$ | $S_6$ | $S_7$ | $S_8$ | $S_0$ | $S_1$ | $S_2$ | − | $R_4$ | $R_5$ | $R_6$ | $R_7$ | $R_8$ | − | − | − | − |
| 4 | − | − | − | $R_0$ | $R_1$ | $R_2$ | $R_3$ | ⌢ | $S_5$ | $S_6$ | $S_7$ | $S_8$ | $S_0$ | $S_1$ | $S_2$ | $S_3$ | − | $R_5$ | $R_6$ | $R_7$ | $R_8$ | − | − | − |
| 5 | − | − | − | − | $R_0$ | $R_1$ | $R_2$ | $R_3$ | $R_4$ | ⌢ | $S_6$ | $S_7$ | $S_8$ | $S_0$ | $S_1$ | $S_2$ | $S_3$ | $S_4$ | − | $R_6$ | $R_7$ | $R_8$ | − | − |
| 6 | − | − | − | − | − | $R_0$ | $R_1$ | $R_2$ | $R_3$ | $R_4$ | $R_5$ | ⌢ | $S_7$ | $S_8$ | $S_0$ | $S_1$ | $S_2$ | $S_3$ | $S_4$ | $S_5$ | − | $R_7$ | $R_8$ | − |
| 7 | − | − | − | − | − | − | $R_0$ | $R_1$ | $R_2$ | $R_3$ | $R_4$ | $R_5$ | $R_6$ | ⌢ | $S_8$ | $S_0$ | $S_1$ | $S_2$ | $S_3$ | $S_4$ | $S_5$ | $S_6$ | − | $R_8$ |
| 8 | − | − | − | − | − | − | − | $R_0$ | $R_1$ | $R_2$ | $R_3$ | $R_4$ | $R_5$ | $R_6$ | $R_7$ | ⌢ | $S_0$ | $S_1$ | $S_2$ | $S_3$ | $S_4$ | $S_5$ | $S_6$ | $S_7$ |
| $\vec{\nu}$ → | 2 | 2 | 4 | 4 | 6 | 6 | 8 | 8 | 8 | 8 | 8 | 8 | 8 | 8 | 8 | 8 | 8 | 8 | 6 | 6 | 4 | 4 | 2 | 2 |

**Table 7.4:** Run-table of a run for $N = 8$ using permutation (7.7). $\mu$ is equal to 6 slots out of 9, with an efficiency $\varepsilon = 66.67\%$ and a length $\lambda = 24$. Note how $\vec{\nu}$ is a palindrome.

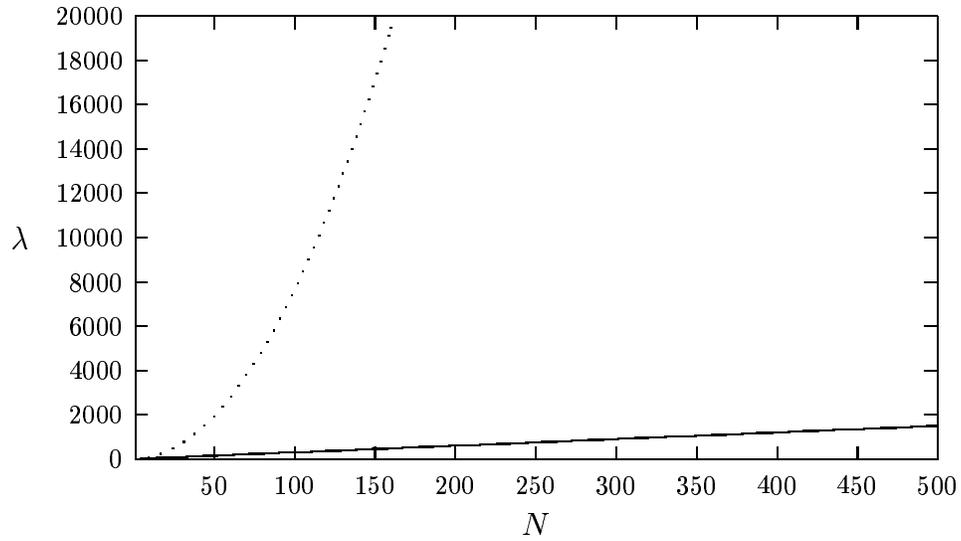

Figure 7.12: Comparison between run lengths resulting from the identity permutation (dotted parabola) and those from permutation (7.7). The former are shown for $1 \le N \le 160$, the latter for $1 \le N \le 500$.

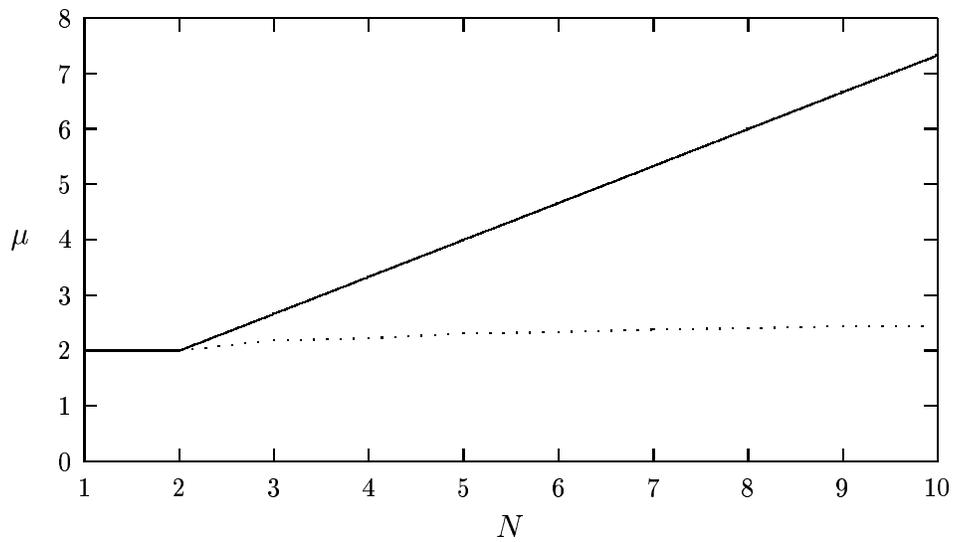

Figure 7.13: Comparison between values of $\mu$ derived from the identity permutation (dotted parabola) and those from permutation (7.7) for $1 \le N \le 10$.

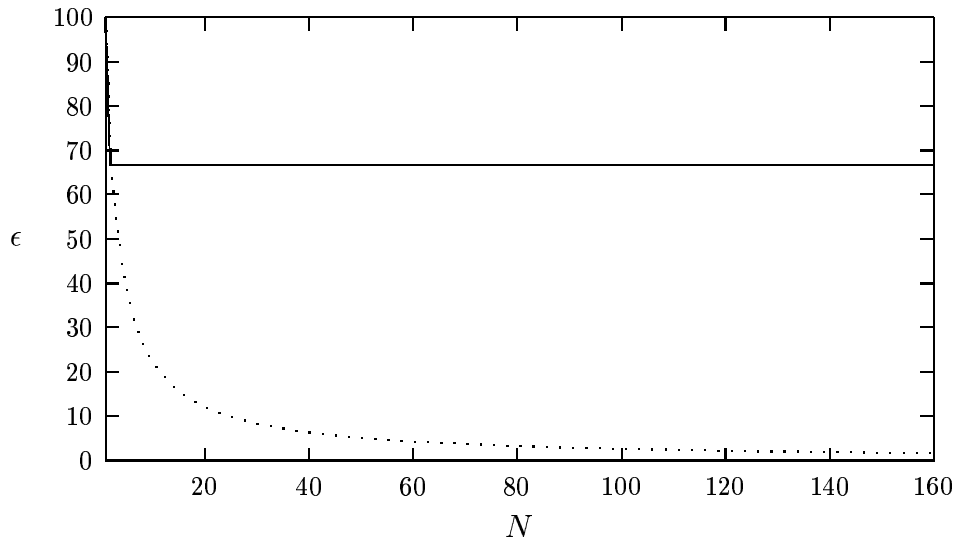

**Figure 7.14:** **Comparison of efficiencies when $\mathcal{P}$ is the identity permutation and in the case of permutation (7.7), for** $1 \leq N \leq 160$.

It is possible to show that the distributed algorithm described in Fig. 7.7, with $\mathcal{P}$ as in (7.7), can be computed in linear time:

**Proposition 5** $\forall N : \lambda_N = 3N$.

(The proof is given in Sect. A.3.6).

The efficiency of the algorithm of pipelined broadcast does not depend on $N$:

**Proposition 6** $\forall N > 0 : \varepsilon_N = 2/3$.

**Proposition 7** $\mu_N = \frac{2}{3}(N + 1)$.

(The proofs are given in Sect. A.3.7 and A.3.8).

Table 7.5 shows how a run-table looks like when multiple broadcasting sessions take place one after the other. As a result, the central area corresponding to the best observable performance is prolonged. In that area, $\varepsilon$ can be counted to be equal to $N/(N+1)$ and the throughput, or the number of gossiping sessions per time step, has been found to be equal to $t/2$, $t$ being the duration of a time step. In other words, within that area a gossiping session is fully completed every two time steps on the average.

### 7.2.1.3  Conclusions

A formal model has been introduced for the sake of characterising the performance of a key sub-service of two TIRAN components, the distributed voting tool and the backbone, namely gossiping. Within this model, a set of parameters has been located such that gossiping can be characterised by unlimited scalability and high-throughput—in case of full-duplex, point-to-point communication. This result could be used, for instance, to devise special-purpose hardware architectures that best support these application components.

|   | 1 | 2 | 3 | 4 | 5 | 6 | 7 | 8 | 9 | 10 | 11 | 12 | 13 | 14 | 15 | 16 | 17 | 18 | 19 | ... |
|---|---|---|---|---|---|---|---|---|---|---|---|---|---|---|---|---|---|---|---|---|
| 0 | $S_1$ | $S_2$ | $S_3$ | $S_4$ | $-$ | $R_1$ | $R_2$ | $R_3$ | $R_4$ | ... | $\curvearrowright$ | $S_1$ | $S_2$ | $S_3$ | $S_4$ | $-$ | $R_1$ | $R_2$ | $R_3$ | $R_4$ ... $\curvearrowright$ $S_1$ $S_2$ $S_3$ $S_4$ $-$ $R_1$ $R_2$ $R_3$ $R_4$ $-$ $-$ $-$ |
| 1 | $R_0$ | $\curvearrowright$ | $S_2$ | $S_3$ | $S_4$ | $S_0$ | $-$ | $R_2$ | $R_3$ | $R_4$ | $R_0$ | $\curvearrowright$ | $S_2$ | $S_3$ | $S_4$ | $S_0$ | $-$ | $R_2$ | $R_3$ | ... $R_4$ $R_0$ $\curvearrowright$ $S_2$ $S_3$ $S_4$ $S_0$ $-$ $R_2$ $R_3$ $R_4$ $-$ $-$ |
| 2 | $-$ | $R_0$ | $R_1$ | $\curvearrowright$ | $S_3$ | $S_4$ | $S_0$ | $S_1$ | $-$ | $R_3$ | $R_4$ | $R_0$ | $R_1$ | $\curvearrowright$ | $S_3$ | $S_4$ | $S_0$ | $S_1$ | $-$ | ... $R_3$ $R_4$ $R_0$ $R_1$ $\curvearrowright$ $S_3$ $S_4$ $S_0$ $S_1$ $-$ $R_3$ $R_4$ $-$ |
| 3 | $-$ | $-$ | $R_0$ | $R_1$ | $R_2$ | $\curvearrowright$ | $S_4$ | $S_0$ | $S_1$ | $S_2$ | $-$ | $R_4$ | $R_0$ | $R_1$ | $R_2$ | $\curvearrowright$ | $S_4$ | $S_0$ | $S_1$ | ... $S_2$ $-$ $R_4$ $R_0$ $R_1$ $R_2$ $\curvearrowright$ $S_4$ $S_0$ $S_1$ $S_2$ $-$ $R_4$ |
| 4 | $-$ | $-$ | $-$ | $R_0$ | $R_1$ | $R_2$ | $R_3$ | $\curvearrowright$ | $S_0$ | $S_1$ | $S_2$ | $S_3$ | $-$ | $R_0$ | $R_1$ | $R_2$ | $R_3$ | $\curvearrowright$ | $S_0$ | ... $S_1$ $S_2$ $S_3$ $-$ $R_0$ $R_1$ $R_2$ $R_3$ $\curvearrowright$ $S_0$ $S_1$ $S_2$ $S_3$ |
|   | 2 | 2 | 4 | 4 | 4 | 4 | 4 | 4 | 4 | 4 | 4 | 4 | 4 | 4 | 4 | 4 | 4 | 4 | 4 | 4 ... 4 4 4 4 4 4 4 4 4 4 4 2 2 |

**Table 7.5:** The algorithm is modified so that multiple gossiping sessions take place. As a consequence, the central, best performing area is prolonged. Therein $\varepsilon$ is equal to $N/(N+1)$. Note how, within that area, there are consecutive "zones" of ten columns each, within which five gossiping sessions reach their conclusion. For instance, such a zone is the region between columns 7 and 16: therein, at entries $(4, 7)$, $(0, 9)$, $(1, 10)$, $(2, 11)$, and $(3, 12)$, a processor gets the last value of a broadcast and can perform some work on a full set of values. This brings to a throughput of $t/2$, where $t$ is the duration of a slot.

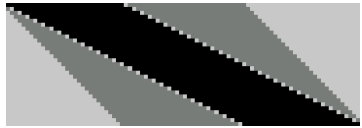

**Figure 7.15:** A graphical representation for run-table 30 when $\mathcal{P}$ is permutation (7.7).

### 7.2.2 Performance of the TIRAN tools

In the rest of this section, the results obtained from some simulations with the $\mathcal{R\varepsilon L}$ prototypes described in Chapter 5 are reported.

#### 7.2.2.1 Performance of the TIRAN BB

The performance of the backbone, as mentioned in Sect. 7.2.1, highly depends on its base gossiping service, already analysed in the cited section. It is worth reporting now on the performance of the BB when faults occur. How much time is in general required for the BB to assess that an assistant is faulty and to isolate it? This question is clearly very system-dependent, especially with respect to the communication network. Indeed, the algorithm of mutual suspicion (see Sect. 5.2.4.1) is based on the values of the deadlines of a number of time-outs—namely, the MIA (manager is alive), TAIA (this assistant is alive), and TEIF (this entity is faulty) time-outs. Let us consider the case of a crashed node hosting the coordinator. The error detection latency is in this case a contribution of the time required to detect that a MIA message is missing (equal to the deadline of a MIA time-out) plus the time required to detect that a TEIF message is missing, plus some delay due to communication and execution overheads. Figure 7.16 collects the durations of the intervals between the starting time in which the node of the manager becomes suspected by an assistant and that in which that node is assumed to have crashed and election of a new manager is started.

Performance highly depends on the communication network—the faster the network is, and the larger its throughput, the shorter can be the deadlines of the above mentioned time-outs be without congesting the network. Clearly the choice of which deadlines to adopt is an important one:

- Too short deadlines may congest the network and bring many periods of instability. These periods are characterised by system partitioning, in which correct nodes are disconnected from the rest of the system for the entire duration of the period (see the hypotheses of the timed asynchronous distributed system model on p. 50). This calls for further overhead for managing the following partition merging (see Sect. 5.2.4.2). Clearly this activity could be represented as a dynamic system, and studied as it has been done for instance in [DF95]. Further reducing the deadlines would exacerbate the instability and result in chaotic behaviours.

- On the other hand, too long deadlines translate in less messages, which asks less resources to the communication network, though also imply a larger error detection latency, and consequently a larger delay for the initiation of recovery.

In general, a proper trade-off between communication overhead and detection latency is required. Both for the prototype realized on a Parsytec Xplorer and for that running on Windows NT the same values are being used:

- A deadline of 1 second for the time-outs for sending a MIA or TAIA message.

- A deadline of 2 seconds for the time-outs corresponding to declaring a missing MIA or TAIA message, thus initiating a so-called "suspicion period".

- A deadline of 0.9 second for declaring a missing TEIF message and thus initiating the graceful degradation of the backbone.

Results are shown in Fig. 7.16 for a version running on two Windows NT nodes.

### 7.2.2.2   Performance of the TIRAN TOM

Another key point in the TIRAN strategy and its overall performance is given by its time-out manager TOM (described in Sect. 5.2.5). Indeed, the TIRAN implementation of the core algorithm of the backbone, the AMS (see Sect. 5.2.4.1), is based on the services provided by TOM, which translate time-related events into delivery of messages: for instance, when cyclic time-out `IAT_SET_TIMEOUT` elapses, an alarm is called, which sends the local component of the backbone message `IAT_SET`. When the backbone gets this message, it sets the local "" flag to state that it is active.

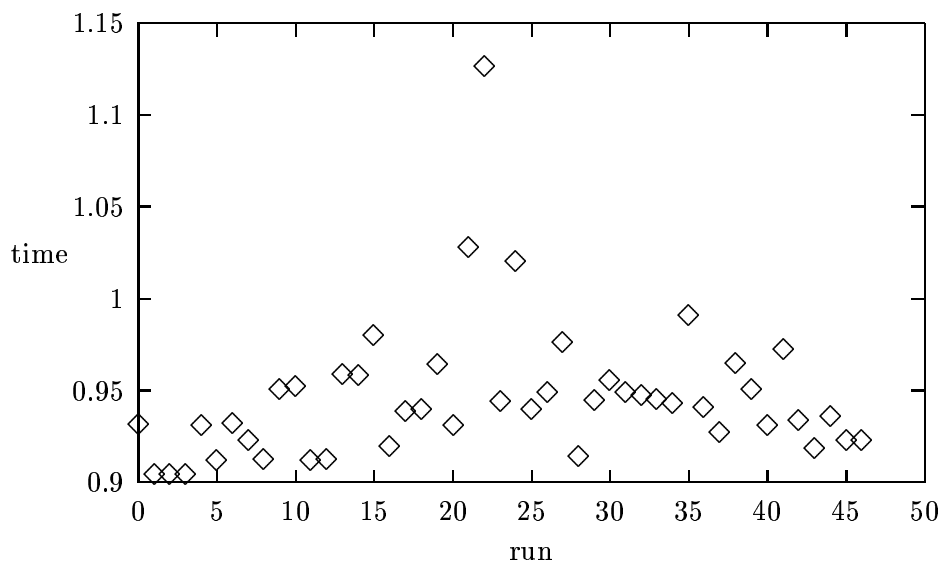

**Figure 7.16:** **Time to detect that an assistant is faulty and to remove it from the TIRAN BB. Times are expressed in seconds. 47 runs are reported. The deadline of the TEIF time-out is 0.9 second in this case. Communication and execution overheads bring to a maximum of about 1.2 seconds. Values are related to a backbone running on two Windows NT workstations connected via a 10Mbps Ethernet and both based on a Pentium processor at 133MHz.**

Clearly this scheme works under the assumption that the delay between the expiration of a time-out and the delivery of the associated alarm is negligible. This may not always be the case, of course, especially because the duration of the alarm (the time required for TOM to execute the alarm function, and thus for sending the message) may be relevant. This time has been found to be indeed negligible for the TIRAN prototypes, mainly because the function to send an alarm is non-blocking and terminates after copying its argument (a buffer of 20 bytes) into the outgoing mailbox buffer—an operation lasting approximately $50\mu s$ on a DEC Alpha board using the TEX nanokernel [TXT97].

The author of this dissertation has performed an analysis on the effects of congestion to the execution of alarms competing for the same set of resources—for instance, the communication subsystem [DFDL00b]. A novel architecture for TOM has been proposed in the cited paper and realized by the author. This system exploits multiple alarm execution threads in order to reduce the congestion due to concurrent alarm execution and the consequent run-time violations. This new TOM is not being used in TIRAN because of the characteristics of the TIRAN alarms and of its non-blocking communication functions–the additional complexity would provide no benefit in this case. Appendix B describes the new mechanism and reports on some experiments that show in which case the adoption of multiple alarm execution threads may provide control over alarm execution congestion.

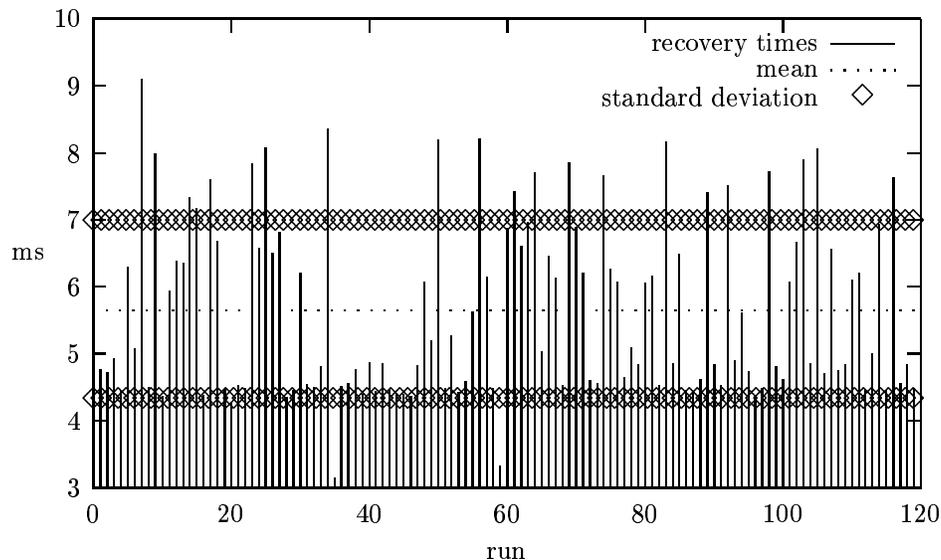

**Figure 7.17: Recovery times.** $X$ **values represent consecutive runs of the application,** $Y$ **values are times for RINT to interpret the r-codes in Fig. 6.8 on a Parsytec PowerXplorer multicomputer.**

### 7.2.2.3 Performance of the TIRAN Recovery Interpreter

In order to estimate the time required for error recovery, including the overhead of the RINT recovery interpreter, the application described in Sect. 6.2.2 has been run for 120 times in a row, injecting a value fault on one of the versions (version 2). The system replied executing the recovery script described in Sect. 6.2, which switched version 2 off and version 4 in, thus restoring an errorless TMR. Figure 6.6 and 6.7 display images taken in a similar experiment.

Figure 6.8 shows the actual list of r-codes executed by RINT. Figure 7.17 displays the time required for executing them. These times include the additional overhead for writing the elapsed time of each run into a file. Times, in milliseconds, are related to the prototype running on the Parsytec Xplorer (based on PowerPC 601 processors with a clock frequency of 66MHz).

### 7.2.2.4 Performance of the Distributed Voting Tool

This section focuses on estimating the voting delay of the TIRAN distributed voting tool via some experiments. Overheads are expressed in terms of system resources (threads, memory occupation and so forth).

**Time and Resources Overheads of the Distributed Voting Tool.** All measurements have been performed running a restoring organ consisting of $N$ processing nodes, $N = 1, \ldots, 4$. The executable file has been obtained on a Parsytec CC system with the "ancc" C compiler using the "-O" optimisation flag. During the experiments, the CC system was fully dedicated to the execution of that application.

The application has been executed in four runs, each of which has been repeated fifty times, increasing the number of voters from 1 to 4, one voter per node. Wall-clock times have been collected. Averages and standard deviations are shown in Table 7.6.

| number of nodes | average | standard deviation |
|---|---|---|
| 1 | 0.0615 | 0.0006 |
| 2 | 0.1684 | 0.0022 |
| 3 | 0.2224 | 0.0035 |
| 4 | 0.3502 | 0.0144 |

**Table 7.6: Time overhead of the voting farm for one to four node systems (one voter per node). The unit is milliseconds.**

The system requires $N$ threads to be spawned, and $N$ mailboxes are needed for the communication between each user module and its local voter. The network of voters calls for another $N \times (N - 1)/2$ mailboxes.

## 7.3   Coding Effort

Coding efforts can be measured by the amount of fault-tolerance code intruded in the user application. When few coding effort is required by an ALFT structure, then the SA and SC characterising that structure may be assessed as satisfactory. As it has been shown in Chapter 6, with $\mathcal{R}\mathcal{E}\mathcal{L}$ typically the sole code intrusion is given by the deployment of a set of error detection provisions, a limited amount of calls to the `RaiseEvent` function, and possibly by application-specific replies to messages from the RINT module, aiming, for instance at executing functions to roll back to a previously checkpointed state. In the case described in Sect. 6.1 the code intrusion is limited to the deployment of a watchdog. In the case of Sect. 6.2, code intrusion is absent, and the user is only requested to supply the base versions of the functional service. Configuration and recovery are supplied as an ARIEL script, hence they do not intrude in the functional code. A few lines need to be added to the application "makefile" so to instruct the compilation of the ancillary source codes produced by the ARIEL translator. Also in the ENEL demonstrator being developed at ENEL and described in Sect. 6.3, code intrusion is expected to be limited to a few `RaiseEvent`'s in their diagnostic functions (see on p. 129) and in a function call to a configured instance of the TIRAN watchdog—this configuration being carried out in ARIEL. This translates in low development costs related to the dependability design goals. Also what maintainability is concerned, the use of ARIEL as a configuration language allows to modify even the API to the TIRAN functions with no effect on the application code. In the redundant watchdog described in Sect. 6.4, the diverse strategies are coded outside the functional application layer, and the tool results from a compositional process supported by the configuration language. Hence, also in this case the coding effort is very limited.

## 7.4 Conclusions

This aim of this chapter has been the evaluation of some key properties of $\mathcal{REL}$—namely, reliability, performance and coding efforts. The effect of spares and fault-specific strategies on reliability has been observed. An analysis of the performance of some key algorithmic components revealed their range of complexity and hinted at possible improvements. Other analyses and simulations have been reported, concerning the performance of some key architectural components of the TIRAN prototype—its backbone, the time-out manager, the recovery interpreter, and a basic tool. Finally, a few considerations on coding efforts related to $\mathcal{REL}$ and their effect on development and maintenance costs have been drawn.

# Chapter 8

# Open Challenges and Conclusions

This chapter concludes the dissertation and draws a few elements for further research. In particular, a number of open challenges are described in Sect. 8.1 while an overall conclusion with a summary of the contributions of the author is drawn in Sect. 8.2.

## 8.1 Open Challenges and Research Problems

This section casts the basis for further investigations on the fault-tolerance linguistic structure provided by $\mathcal{REL}$.

First, a straightforward extension for high adaptability (structural attribute A) is described in Sect. 8.1.1. Next, an open challenging research path is stated in Sect. 8.1.2: extending $\mathcal{REL}$ to unlimited distributed systems, such as Internet-based wide-area metacomputing systems or electronic-commerce applications. Section 8.1.3 briefly mentions a few other open research paths.

### 8.1.1 A $\mathcal{REL}$ Architecture for High Adaptability

As already mentioned, in ARIEL the fault-tolerance part of a computer program is separated from the functional part, to the point that one could change the former with no impact on the latter—not even recompiling the application code. This happens because, while the target application code is a "plain" code, compiled with a conventional programming language such as, e.g., C or C++, the fault-tolerance code is available in the form of a binary pseudo-code to be interpreted at run-time by the r-code virtual machine. This property allows to create an infrastructure such that this fault-tolerance code is not *statically* read from a file or from a data segment in the running application—as it happens in the TIRAN prototype—but *dynamically* negotiated, at run-time, from an external r-code broker, i.e., a server specific of the given environment. This strategy, depicted in Fig. 8.1, would translate into the following properties:

- Moving a code to a new environment simply means that that code receives "recovery applets" from a different r-code broker.

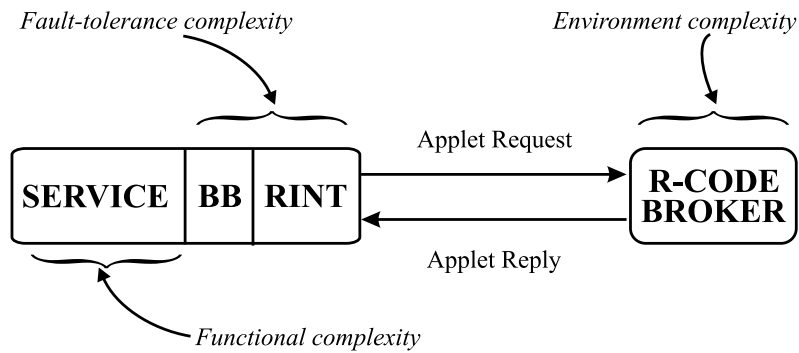

**Figure 8.1: The process of negotiating the r-codes: on the left, the user application runs together with the backbone and the recovery interpreter. The actual set of r-codes to be interpreted (called "recovery applet") is requested from an external component, called the r-code broker. This scheme partitions the complexity into three separate blocks—functional (service) complexity, related to the functional aspects and dealt with in the application code; fault-tolerance complexity, i.e., the complexity pertaining to the management of the fault-tolerance provisions, which is dealt with in the $\mathcal{REL}$ system and in the r-codes; and environment complexity, defined as the complexity required to cope in an optimal way with the varying environmental conditions, which is charged to the r-code broker.**

- Updating the recovery strategies of a set of client codes only requires updating a library of recovery applets managed by the local r-code broker.

- Dynamic management of the applets in this r-code broker allows the whole set of clients to adapt dynamically to a changing environment. This allows transparent, immediate adaptation to a new environment or to new environmental conditions, in order to guarantee the degree of quality of service agreed in the service specifications despite any changes in the environment.

- The burden of facing a changing environment (called "*environment complexity*" in Fig. 8.1) is not charged to each application component. On the contrary, this complexity is shifted to a single specialised external component, the r-code broker.

It is worth remarking how, in such an architecture, the complexity required to manage the fault-tolerance provisions, as well as the one required to adjust the fault-tolerance provisions to the current environmental conditions, are no more *exclusively in the application*. As drawn in Fig. 8.1, this complexity can be decomposed into its three components and tackled in different contexts—possibly by different teams of experts, possibly at different times. According to the *divide et impera* design principle, this may turn into an effective scheme for controlling the complexity in a fault-tolerance distributed application. Note also how, in the above mentioned scheme, the only general-purpose task is purely the service. Indeed, the system for the management of the fault-tolerance provisions and the r-code broker are *special-purpose tasks*, and as such they are more eligible for being cost-effectively optimised with respect to both performance and dependability. The maturity of hardware-based and OS-based fault-tolerance, with respect to the complex scenario of application-level fault-tolerance described in this dissertation, may be used as an argument to bring evidence to the above statement.

## 8.1.2 Extending $\mathcal{REL}$ to Unlimited Distributed Systems

Another challenging research path would be to devise a $\mathcal{REL}$ approach suited for the adoption in unlimited distributed systems, such as Internet-based wide-area metacomputing applications [Buy99] or WAP-based [WAP00] electronic-commerce applications. The peculiar characteristics of $\mathcal{REL}$, and especially its optimal adaptability, may prove to match well with the requirements of these classes of applications. The rest of this paragraph deals with the latter class.

Nowadays WAP-based [WAP00] electronic-commerce applications allow to perform commercial transactions through hand-held devices. Clearly, the novel possibilities opened by this new scenario inevitably let in new problems to be investigated and solved. For instance, client applications written for being embedded into those hand-held devices, as well as the applications remotely serving those clients, *must be designed in such a way as to cope with the typical scenarios those applications may run into*. Indeed, the mobility of these clients inplies error-proneness and *communication autonomy* [VT00], that is, devices are not always reachable through the network and they are often disconnected. In other words, the environment in which the clients execute may be unpredictably subjected to sudden, drastic changes, which could in turn result, for instance, in temporary disconnections. The $\mathcal{REL}$ approach described in this dissertation could be extended as described in Sect. 8.1.1 so to bring *adaptable fault-tolerance* into this context. Fine-tuning the degree of redundancy to the *current conditions* of the channel allows to minimise the cost and time overheads. A $\mathcal{REL}$ architecture may provide this solution without introducing large amounts of complexity into the applications. Minimising costs and information redundancy matches well with the characteristics of nowadays' hand-held electronic-business devices, i.e., scarce and and expensive limited bandwidth [VT00].

## 8.1.3 Other Research Paths

Other future research paths should concern the conception of resilient strategies for the error recovery process managed by RINT—an important requisite for developing an effective architecture compliant to $\mathcal{REL}$. Furthermore, the dynamic behaviour of systems such as the TIRAN backbone or the time-out management system may be worth investigating. Typical questions to be investigated would be, what happens to a dynamic system that models the activity of the backbone when one varies the deadlines of its time-outs? What external behaviour shows up when shrinking those deadlines and hence increasing the frequency of the periods of instability? Similar investigations were carried out, e.g., in [DF95], by the author of this dissertation. Another path would be that of investigating on the capability of $\mathcal{REL}$ as a system structure for the expression of aspects such as security. Other investigation may concern the automatic translation of a recovery language program into a set of logic clauses with whom to fed some verification tool. Furthermore, experimenting structures other than $\mathcal{REL}$'s guarded actions may further increase the syntactical adequacy. Another important research path should concern the extension of the approach towards hard real-time. The integration of the application-level fault-tolerance architecture of ARIEL with a system-level fault-tolerance architecture such as the one of GUARDS [PABD$^+$99], which also addresses hard real-time, could provide an adequate end-to-end architecture addressing both dependability and real-time at all levels.

## 8.2 Conclusions

The thesis of this dissertation ("*it is possible to catch up satisfactory values for the structural attributes, at least in the domain of soft real-time, distributed and parallel applications, by means of a fault-tolerance linguistic structure that realises a distinct, secondary, programming and processing context, directly addressable from the application layer*") has been justified and validated throughout this text by means of a number of arguments.

These arguments are recalled in Sect. 8.2.1, which discusses the key lessons learned. Finally, Sect. 8.2.2 summarises the main contributions of this work.

### 8.2.1 Key Lessons Learned

One of the **lessons learned** while doing this work has been that an effective structure for application fault-tolerance can take the form of a couple

$$(\text{application services}, \text{fault-tolerance services}),$$

with limited dependencies between its two components. Such a separation allows to have independent design, development and maintenance processes for both the application and the fault-tolerance services. This turns into optimal values of the structural attribute **separation of design concerns** (SC).

Moreover, the $\mathcal{R}\mathcal{L}$ model allows the use of

- a widespread language such as C for the functional aspects, and

- an ad hoc linguistic structure for the fault-tolerance aspects,

thus reaching both the benefits of the evolutionary approaches (better portability, compliancy to well-accepted and known standards, . . . ) and those of the revolutionary ones (less syntactical constraints, more elegance and efficiency, explicitness of the fault-tolerance design requirements, possibility to extend or add new features freely, therefore enhancing expressiveness and allowing to reach optimal values of **syntactical adequacy** (SA).

One can conclude that the $\mathcal{R}\mathcal{L}$ approach and the linguistic structure of the configuration and recovery language ARIEL allow to prove the truth of conjecture "*the coexistence of two separate though cooperating layers for the functional and the fault-tolerance aspects allows to reach both the benefits inherent to the evolutionary approaches and those of the revolutionary solutions*" (see on p. 45) as well as of conjecture "*it can be possible to address, within one efficacious linguistic structure, a wide set of fault-tolerance provisions*" (see on p. 47).

Another important **lesson learned** has been that in ARIEL, the separation of concerns is such that *even the corresponding executable codes are separable*[1]. This fact can be exploited as suggested in Sect. 8.1.1 in order to design a distributed architecture in which the dynamic management of r-code applets both provides high **adaptability** (A) with respect to current environmental conditions and at the same time does not charge the application with the burden of managing large amounts of non-functional complexity. This happens because complexity is partitioned into three distinct and separated components. This brings evidence to our conjecture (on p. 23) that argued that *a satisfactory solution to the design problem of the management of the fault-tolerance code can translate into an optimal management of the fault-tolerance provisions with respect to varying environmental conditions*.

## 8.2.2 Contributions

This dissertation provides a number of original contributions to the state-of-the-art of the structures to express fault-tolerance in the application layer of a computer system. These contributions are now briefly recalled.

The first contribution has been the definition in **Chapter 1** of the *structural attributes*, a sort of base of variables with which application-level fault-tolerance structures can be qualitatively assessed and compared with each other and with respect to above mentioned need.

**Chapter 2** provides a summary of the Laprie's model of dependability. The relation between the management of the fault-tolerance code, the efficient management of redundancy, and complexity control, has been also introduced.

**Chapter 3** is an in-depth survey of state-of-the-art of fault-tolerance structure. A qualitative assessment is supplied for each class with respect to the structural attributes defined in Chapter 1. Two conjectures have been drawn on the characteristics of an optimal class. In particular, the coexistence of two separate layers for the functional and the non-functional aspects has been isolated as a key factor for a linguistic structure for fault-tolerance.

**Chapter 4** introduces a novel linguistic structure for the fault-tolerance of distributed applications—the recovery language approach. System, application and fault models are proposed. A workflow diagram is supplied. The overall design of the approach is given and an investigation of possible technical foundations is carried out. Other contributions include an analysis of the specific differences between the new structure and the ALFT approaches reviewed in Chapter 3, and an analysis of the specific limitations of the recovery language approach.

**Chapter 5** describes the prototype system compliant to the recovery language approach that has been developed in project TIRAN. Contributions of the author include: the definition of the TIRAN architecture; the concept of recovery and configuration language; the design of the configuration and recovery language ARIEL; the design of a recovery pseudo-code (the r-code); the design and development of the ARIEL-to-r-code translator; the design and development of the ARIEL run-time executive; the design and development of the TIRAN backbone, including its database; the conception of the algorithm of mutual suspicion; the design and development of the distributed voting tool; the design and development of the TIRAN time-out manager; and the design decision to adopt the $\alpha$-count fault-identification mechanism.

---

[1]When one considers an entire executable image as one *code word*, the corresponding code is said to be *separable* [Joh89, p. 84] with respect to the fault-tolerance and the functional aspects.

**Chapter 6** provides several case studies. In particular, the GLIMP software and the ENEL redundant watchdog have been co-authored by the author, while the enhancements to the TIRAN distributed voting tool (adding spares to an $N$-version programming executive, or using recovery strategies that exploit the identification of the nature of each fault) have been designed and implemented by the author.

**Chapter 7** focuses on the assessment of the recovery language approach. A number of results have been obtained via reliability and performance analysis and via measurements obtained through simulations. Key properties being assessed are reliability, performance, and coding efforts—the latter affecting both development and maintenance costs. It is shown that the approach and algorithms described in this work may reach, under certain circumstances, high performance and throughput, unlimited scalability, high reliability, and very limited usage costs. Main contributions of the author in this chapter include: the definition of a class of gossiping algorithms as a base service of the distributed voting tool; the conception of a formal model for those algorithms and their analysis; the algorithm of pipelined broadcast and related theorems; the Markovian reliability models for the analysis of the effect of spares and of $\alpha$-count in the distributed voting tool; the design and development of the system for congestion control in the time-out manager; and the analysis of this system.

**Chapter 8** then casts the elements for future research proposing possible evolutions and improvements, including an architecture for the adaptable management of the fault-tolerance code.

As an overall conclusion, this dissertation has addressed the lack of a structuring technique capable to offer its users a satisfactory solution to the demands of dependable-software engineering. While a generally applicable solution does not exist, a particular one has been found, analysed, and discussed, for the domain of soft real-time, distributed and parallel applications. In this domain, the solution and its prototype exhibit satisfactory values of the structural attributes. Experimental results suggest that systems based on the novel solution are able to fulfil dependability and performance design goals.

# Appendix A

# Mathematical Details

## A.1   Mathematical Details Related to Eq. 7.2

The basic steps leading to Eq. 7.2, i.e.,

$$R^{(1)}(C,t) = (-3C^2 + 6C) \times [R(t)(1 - R(t))]^2 + R^{(0)}(t),$$

are described in what follows.

The Markov reliability model of Fig. 7.1 brings to the following set of equations:

$$
\begin{cases}
p_{310}(t + \Delta t) &= p_{310}(t)(1 - 4\lambda\Delta t) \\
p_{300}(t + \Delta t) &= p_{300}(t)(1 - 3\lambda\Delta t) + p_{310}(t)4\lambda\Delta t C \\
p_{200}(t + \Delta t) &= p_{200}(t)(1 - 2\lambda\Delta t) + p_{300}(t)3\lambda\Delta t \\
p_{\text{FS}}(t + \Delta t) &= p_{\text{FS}}(t) + p_{200}(t)2\lambda\Delta t \\
p_{211}(t + \Delta t) &= p_{211}(t)(1 - 3\lambda\Delta t) + p_{310}(t)3\lambda\Delta t(1 - C) \\
p_{301}(t + \Delta t) &= p_{301}(t)(1 - 3\lambda\Delta t) + p_{310}(t)\lambda\Delta t(1 - C) \\
p_{201}(t + \Delta t) &= p_{201}(t)(1 - 2\lambda\Delta t) + p_{301}(t)3\lambda\Delta t C + \\
&\quad p_{211}(t)3\lambda\Delta t C \\
p_{202}(t + \Delta t) &= p_{202}(t)(1 - 2\lambda\Delta t) + p_{301}(t)3\lambda\Delta t(1 - C) + \\
&\quad p_{211}(t)\lambda\Delta t(1 - C) \\
p_{\text{FU}}(t + \Delta t) &= p_{\text{FU}}(t) + p_{201}(t)2\lambda\Delta t + \\
&\quad p_{211}(t)2\lambda\Delta t(1 - C) + p_{202}(t)2\lambda\Delta t.
\end{cases}
$$

The above equations can be written as follows:

$$
\begin{cases}
\frac{dp_{310}(t+\Delta t)}{dt} & = & -4\lambda p_{310}(t) \\
\frac{dp_{300}(t+\Delta t)}{dt} & = & -3\lambda p_{300}(t) + 4\lambda C p_{310} \\
\frac{dp_{200}(t+\Delta t)}{dt} & = & -2\lambda p_{200}(t) + 3\lambda p_{300}(t) \\
\frac{dp_{\mathrm{FS}}(t+\Delta t)}{dt} & = & 2\lambda p_{200}(t) \\
\frac{dp_{211}(t+\Delta t)}{dt} & = & -3\lambda p_{211}(t) + 3\lambda(1-C) p_{310}(t) \\
\frac{dp_{301}(t+\Delta t)}{dt} & = & -3\lambda p_{301}(t) + \lambda(1-C) p_{310}(t) \\
\frac{dp_{201}(t+\Delta t)}{dt} & = & -2\lambda p_{201}(t) + 3\lambda C p_{301}(t) + 3\lambda C p_{211}(t) \\
\frac{dp_{202}(t+\Delta t)}{dt} & = & -2\lambda p_{202}(t) + \lambda(1-C) p_{211}(t) + 3\lambda(1-C) p_{301}(t) \\
\frac{dp_{\mathrm{FU}}(t+\Delta t)}{dt} & = & 2\lambda p_{202}(t) + 2\lambda p_{201}(t) + 2\lambda(1-C) p_{211}(t).
\end{cases}
$$

For any state $s$, let us now call $L_s = L(p_s(t))$, where $L$ is the Laplace transform. Furthermore, as $(310)$ is the initial state, it is reasonable to assume that $p_{310}(0) = 1$ and $\forall s \neq (310) : p_s(0) = 0$. Then taking the limit of the above equations as $\Delta t$ goes to zero and taking the Laplace transform brings to

$$
\begin{cases}
L_{310} & = & \frac{1}{s+4\lambda} \\
L_{300} & = & \frac{4C}{s+3\lambda} - \frac{4C}{s+4\lambda} \\
L_{200} & = & \frac{6C}{s+4\lambda} - \frac{12C}{s+3\lambda} + \frac{6C}{s+2\lambda} \\
L_{\mathrm{FS}} & = & \frac{C}{s} - \frac{3C}{s+4\lambda} + \frac{8C}{s+3\lambda} - \frac{6C}{s+2\lambda} \\
L_{211} & = & \frac{3(1-C)}{s+3\lambda} - \frac{(3(1-C)}{s+4\lambda} \\
L_{301} & = & \frac{1-C}{s+3\lambda} - \frac{1-C}{s+4\lambda} \\
L_{201} & = & 6C(1-C)(\frac{1}{s+4\lambda} - \frac{2}{s+3\lambda} + \frac{1}{s+2\lambda}) \\
L_{202} & = & 3(1-C)^2(\frac{1}{s+4\lambda} - \frac{2}{s+3\lambda} + \frac{1}{s+2\lambda}).
\end{cases}
$$

Inverting the Laplace transform, the following probabilities can be found:

$$
\begin{cases}
p_{310}(t) & = & \exp^{-4\lambda t} \\
p_{300}(t) & = & 4C \exp^{-3\lambda t} - 4C \exp^{-4\lambda t} \\
p_{200}(t) & = & 6C \exp^{-4\lambda t} - 12C \exp^{-3\lambda t} + 6C \exp^{-2\lambda t} \\
p_{211}(t) & = & 3(1-C) \exp^{-3\lambda t} - 3(1-C) \exp^{-4\lambda t} \\
p_{301}(t) & = & (1-C) \exp^{-3\lambda t} - (1-C) \exp^{-4\lambda t} \\
p_{201}(t) & = & 6C(1-C)(\exp^{-4\lambda t} - 2\exp^{-3\lambda t} + \exp^{-2\lambda t}) \\
p_{202}(t) & = & 3(1-C)^2(\exp^{-4\lambda t} - 2\exp^{-3\lambda t} + \exp^{-2\lambda t})
\end{cases}
$$

(only useful states have been computed).

Let us denote with $R$ the reliability of the basic component of the system, i.e., $\exp^{-\lambda t}$, and $R_{\mathrm{TMR}}$ as the reliability of the TMR system based on the same component. The reliability of the three and one spare system, $R^{(1)}(C, t)$, is given by the sum of the above probabilities:

$$
\begin{aligned}
R^{(1)}(C,t) &= R^4(-3C^2 + 6C) + R^3(6c^2 - 12C - 2) + R^2(-3C^2 + 6C + 3) \\
&= (-3C^2 + 6C)(R(1-R))^2 + (3R^2 - 2R^3) \\
&= (-3C^2 + 6C)(R(1-R))^2 + R_{\mathrm{TMR}},
\end{aligned}
$$

which proves Eq. (7.2).

## A.2   Mathematical Details Related to Eq. 7.4

This section describes the basic steps leading to Eq. 7.4:

$$
R^\alpha_{\mathrm{TMR}}(t) = 3\exp^{-2(1-RT)\lambda t} - 2\exp^{-3(1-RT)\lambda t}.
$$

The Markov reliability model of Fig. 7.4 leads to the following set of equations:

$$
\left\{
\begin{aligned}
p_3(t+\Delta t) &= p_3(t)[1 - 3\lambda\Delta t(1-RT)] \\
p_2(t+\Delta t) &= p_2(t)[1 - 2\lambda\Delta t(1-RT)] + p_3(t)[3\lambda\Delta t(1-RT)] \\
p_F(t+\Delta t) &= p_F(t) + p_2(t)[2\lambda\Delta t(1-RT)].
\end{aligned}
\right.
$$

The above equations can be written as follows:

$$
\left\{
\begin{aligned}
\frac{dp_3(t+\Delta t)}{dt} &= -3\lambda(1-RT)p_3(t) \\
\frac{dp_2(t+\Delta t)}{dt} &= -2\lambda p_2(t)(1-RT) + 3\lambda p_3(t)(1-RT) \\
\frac{dp_F(t+\Delta t)}{dt} &= 2\lambda p_2(t)(1-RT).
\end{aligned}
\right.
$$

For any state $s$, let us now again call $L_s = L(p_s(t))$, where $L$ is the Laplace transform. Furthermore, as (3) is the initial state, it is assumed that $p_3(0) = 1$ and $\forall s \neq (3) : p_s(0) = 0$. Then taking the limit of the above equations as $\Delta t$ goes to zero and taking the Laplace transform brings to

$$
\left\{
\begin{aligned}
L_3 &= \frac{1}{s+3\lambda(1-RT)} \\
L_2 &= \frac{3}{s+2\lambda(1-RT)} - \frac{3}{s+3\lambda(1-RT)} \\
L_F &= \frac{1}{s} - \frac{2}{s+3\lambda(1-RT)} - \frac{3}{s+2\lambda(1-RT)}.
\end{aligned}
\right.
$$

Inverting the Laplace transform, the following probabilities can be found:

$$
\left\{
\begin{aligned}
p_3(t) &= \exp^{-3(1-RT)\lambda t} \\
p_2(t) &= 3\exp^{-2(1-RT)\lambda t} - 3\exp^{-3(1-RT)\lambda t} \\
p_F(t) &= 1 + 2\exp^{-3(1-RT)\lambda t} - 3\exp^{-2(1-RT)\lambda t}.
\end{aligned}
\right.
$$

Let us denote as $R^\alpha_{\mathrm{TMR}}$ the reliability of a TMR system exploiting the $\alpha$-count mechanism. Indeed $R^\alpha_{\mathrm{TMR}} = p_3(t) + p_2(t)$ is Eq. (7.4).

# A.3    Proofs of Propositions of Sect. 7.2.1

## A.3.1    Proof of Proposition 1

For any permutation $\mathcal{P}$, the distributed algorithm described by the state diagram of Fig. 7.7 solves the gossiping problem of Sect. 7.2.1 without deadlocks.

**PROOF 1** *The algorithm trivially implements gossiping through $N$ broadcast sessions, i.e., $N$ sequences of $N-1$ sends and $N-1$ corresponding receipts. The state diagram of Fig. 7.7 is a* linear *sequence of $4N+2$ states—hence, the corresponding algorithm has no selective or recurrence statements. It is now shown how no "send-send" deadlock situations are possible, i.e., no two processors $p_i$ and $p_j$ can be found such that, at the same time, $p_i$ is in state $WS_j$ and $p_j$ is in state $WS_i$.*

*Let us first observe how, by construction, for all $i \in [0, N]$, if processor $p_i$ has acquired $j$ input values, then, for all $k \le j$, processor $p_k$ has already passed through states $WS_i$ and $S_i$ (shown the second row of Fig. 7.7).*

*If, ab absurdo, $\exists i, j$ such that processor $p_j$ is in state $WS_i$ and processor $p_i$ is in state $WS_j$ (i.e., the system is in deadlock), then either $i < j$ or the other way around. If $i < j$ then $p_j$ has already passed through states $WS_i$ and $S_i$, which contradicts the hypothesis. The same contradiction is reached if $j < i$.*  $\square$

## A.3.2    Proof of Proposition 2

$$\lambda_N = \tfrac{3}{4}N^2 + \tfrac{5}{4}N + \tfrac{1}{2}\lfloor N/2 \rfloor.$$

**PROOF 2** (by induction). *The property holds for $N = 1$. Let us consider run-table $N + 1$. Let us strip off its last row; then wipe out the $\lfloor (N+1)/2 \rfloor - 1$ leftmost columns which contain the element $S_{N+1}$. Let us also cut out the whole right part of the table starting at the column containing the last occurrence of $S_{N+1}$. Finally, let us rename all the remaining $S_{N+1}$'s as "$-$".*

*Our first goal is showing that what remains is run-table $N$. To this end, let us first point out how the only actions that affect the content of other cells in a run-table are the $S$ actions. Their range of action is given by their subscript: an $S_{N+1}$ for instance only affects an entry in row $N + 1$.*

*Now let us consider what happens when processor $i - 1$ sends its message to processor $i$ and the latter gains the right to broadcast as well: at this point, processor $i$ starts sending to processors in the range $\{0, \ldots, i - 2\}$ i.e., those "above"; as soon as it tries to reach processor $i - 1$, in the case the latter has not finished its own broadcast, $i$ enters state $WS$ and blocks.*

*This means that:*

1. *processors "below" processor $i$ will not be allowed to start their broadcast, and*

2. *for processor $i$ and those "above", $\mu$, or the degree of parallelism, is always equal to 2 or 4—no other value is possible. This is shown for instance in Table 7.1, row "$\vec{v}$ ".*

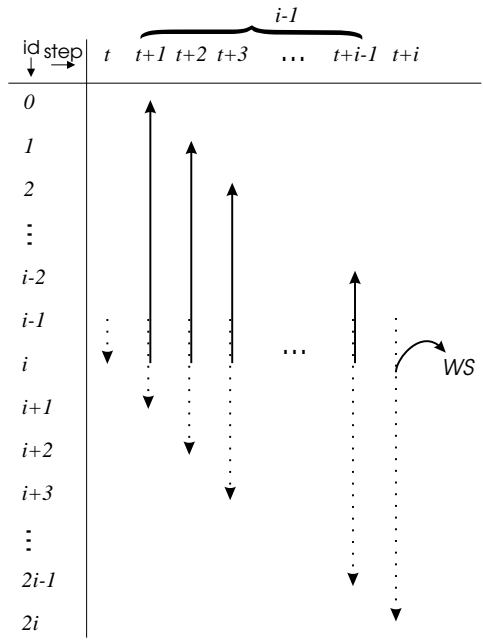

**Figure A.1:** **Processor** $i - 1$ **blocks processor** $i$ **only if** $2i - 1 < N$. **A transmission i.e., two used slots, is represented by an arrow. In dotted arrows the sender is processor** $i - 1$, **for normal arrows it is processor** $i$. **Note the cluster of** $i - 1$ **columns with two concurrent transmissions (adding up to 4 used slots) in each of them.**

As depicted in Fig. A.1, processor $i$ gets blocked only if it tries to send to processor $i - 1$ while the latter is still broadcasting, which happens when $i < \frac{N+1}{2}$. This condition is true for any processor $j \in \{0, \ldots, \lfloor \frac{N+1}{2} \rfloor - 1\}$. Note how a "cluster" appears, consisting of $j - 1$ columns with 4 used slots inside (Table A.1 can be used to verify the above when $N$ is 7). Removing the first $\lfloor \frac{N+1}{2} \rfloor$ occurrences of $S_{N+1}$ (from row 0 to row $\lfloor \frac{N+1}{2} \rfloor - 1$) therefore simply shortens of one time step the stay of each processor in their current waiting states. All the remaining columns containing that element cannot be removed—these occurrences simply vanish by substituting them with a "$-$" action.

Finally, the removal of the last occurrence of $S_{N+1}$ from the series of sending actions which constitute the broadcast of processor $N$ allows the removal of the whole right sub-table starting at that point. The obtained table contains all and only the actions of run $N$; the coherence of these actions is not affected; and all broadcast sessions are managed according to the rule of the identity permutation. In other words, this is run-table $N$.

Now let us consider $\sigma(N + 1)$: according to the above argument, this is equal to:

1. the number of slots available in an $N$-run i.e., $\sigma(N) = (N + 1)\lambda_N$,

2. plus $N + 1$ slots from each of the columns that experience a delay i.e.,

$$\lfloor (N + 1)/2 \rfloor \cdot (N + 1),$$

| id \ step → | 1 | 2 | 3 | 4 | 5 | 6 | 7 | 8 | 9 | 10 | 11 | 12 | 13 | 14 | 15 | 16 | 17 | 18 | 19 | 20 | 21 | 22 | 23 | 24 |
|---|---|---|---|---|---|---|---|---|---|---|---|---|---|---|---|---|---|---|---|---|---|---|---|---|
| 0 | $S_1$ | $S_2$ | $S_3$ | $S_4$ | $S_5$ | $S_6$ | $S_7$ | $R_1$ | – | $R_2$ | – | – | – | – | – | – | $R_3$ | – | – | – | – | – | $R_4$ | – |
| 1 | $R_0$ | ⌢ | ⌢ | ⌢ | ⌢ | ⌢ | ⌢ | $S_0$ | $S_2$ | $S_3$ | $S_4$ | $S_5$ | $S_6$ | $S_7$ | $R_2$ | – | – | $R_3$ | – | – | – | – | – | $R_4$ |
| 2 | – | $R_0$ | – | – | – | – | – | $R_1$ | $S_0$ | ⌢ | ⌢ | ⌢ | ⌢ | $S_1$ | $S_3$ | $S_4$ | $S_5$ | $S_6$ | $S_7$ | $R_3$ | – | – | – | – |
| 3 | – | – | $R_0$ | – | – | – | – | – | $R_1$ | – | – | – | – | $R_2$ | $S_0$ | $S_1$ | ⌢ | ⌢ | $S_2$ | $S_4$ | $S_5$ | $S_6$ | – | – |
| 4 | – | – | – | $R_0$ | – | – | – | – | – | $R_1$ | – | – | – | – | $R_2$ | – | – | – | – | $R_3$ | $S_0$ | $S_1$ | – | – |
| 5 | – | – | – | – | $R_0$ | – | – | – | – | – | $R_1$ | – | – | – | – | – | $R_2$ | – | – | – | – | $R_3$ | – | – |
| 6 | – | – | – | – | – | $R_0$ | – | – | – | – | – | $R_1$ | – | – | – | – | – | $R_2$ | – | – | – | – | – | $R_3$ |
| 7 | – | – | – | – | – | – | $R_0$ | – | – | – | – | – | $R_1$ | – | – | – | – | – | $R_2$ | – | – | – | – | – |
| $\vec{v}$ → | 2 | 2 | 2 | 2 | 2 | 2 | 2 | 2 | 2 | 4 | 2 | 2 | 2 | 2 | 2 | 2 | 4 | 2 | 2 | 2 | 2 | 2 | 4 | 4 |

| id \ step → | 25 | 26 | 27 | 28 | 29 | 30 | 31 | 32 | 33 | 34 | 35 | 36 | 37 | 38 | 39 | 40 | 41 | 42 | 43 | 44 | 45 | 46 | 47 |
|---|---|---|---|---|---|---|---|---|---|---|---|---|---|---|---|---|---|---|---|---|---|---|---|
| 0 | – | – | – | $R_5$ | – | – | – | – | – | $R_6$ | – | – | – | – | – | $R_7$ | – | – | – | – | – | – | – |
| 1 | – | – | – | – | $R_5$ | – | – | – | – | – | $R_6$ | – | – | – | – | – | $R_7$ | – | – | – | – | – | – |
| 2 | $R_4$ | – | – | – | $R_5$ | – | – | – | – | – | $R_6$ | – | – | – | – | – | $R_7$ | – | – | – | – | – | – |
| 3 | $S_7$ | $R_4$ | – | – | – | $R_5$ | – | – | – | – | – | $R_6$ | – | – | – | – | – | $R_7$ | – | – | – | – | – |
| 4 | $S_2$ | $S_3$ | $S_5$ | $S_6$ | $S_7$ | – | – | $R_5$ | – | – | – | – | – | $R_6$ | – | – | – | – | – | $R_7$ | – | – | – |
| 5 | – | – | $R_4$ | $S_0$ | $S_1$ | $S_2$ | $S_3$ | $S_4$ | $S_6$ | $S_7$ | – | – | $R_5$ | – | – | – | – | – | $R_6$ | – | – | – | $R_7$ |
| 6 | – | – | – | $R_4$ | – | – | – | – | $R_5$ | $S_0$ | $S_1$ | $S_2$ | $S_3$ | $S_4$ | $S_5$ | $S_7$ | – | – | – | – | – | – | $R_7$ |
| 7 | $R_3$ | – | – | – | $R_4$ | – | – | – | – | $R_5$ | – | – | – | – | $R_6$ | $S_0$ | $S_1$ | $S_2$ | $S_3$ | $S_4$ | $S_5$ | $S_6$ | – |
| $\vec{v}$ → | 4 | 2 | 2 | 2 | 4 | 2 | 2 | 2 | 2 | 2 | 4 | 2 | 2 | 2 | 2 | 2 | 2 | 2 | 2 | 2 | 2 | 2 | 2 |

**Table A.1: Run-table 7 for $\mathcal{P}$ equal to the identity permutation. Average utilisation is 2.38 slots out of 8, or an efficiency of 29.79%.**

3. *plus the slots in the right sub-matrix, not counting the last row i.e.,*

$$(N+1)(N+2),$$

4. *plus an additional row.*

In other words, $\sigma(N+1)$ can be expressed as the sum of the above first three items multiplied by a factor equal to $\frac{N+2}{N+1}$. This can be written as an equation as

$$\sigma(k+1) = \left(\sigma(k) + \lfloor \tfrac{k+1}{2}\rfloor(k+1) + (k+1)(k+2)\right)\frac{k+2}{k+1}. \qquad (A.9)$$

By the definition of $\sigma$ (see on p. 146), this leads to the following recursive relation:

$$\ell(k+1) = \ell(k) + \lfloor \tfrac{k+1}{2}\rfloor + k + 2. \qquad (A.10)$$

Furthermore, the following is true by the induction hypothesis:

$$\lambda_N = \ell(N) = \frac{3}{4}N^2 + \frac{5}{4}N + \frac{1}{2}\lfloor N/2\rfloor. \qquad (A.11)$$

*Our goal is to show that Eq. (A.10) and Eq. (A.11) together imply that $\ell(N+1) = \lambda_{N+1}$, the latter being*

$$\lambda_{N+1} = \frac{3}{4}(N+1)^2 + \frac{5}{4}(N+1) + \frac{1}{2}\lfloor\frac{N+1}{2}\rfloor \tag{A.12}$$

*Let us call quantity $\frac{1}{2}\lfloor\frac{N}{2}\rfloor + \lfloor\frac{N+1}{2}\rfloor$ as $\Delta$.*
*Let us consider now*

$$\begin{aligned}
\ell(N+1) &= \ell(N) + \lfloor\frac{N+1}{2}\rfloor + N + 2 \\
&= \lambda_N + \frac{3}{2}N + 2 \\
&= \frac{3}{4}N^2 + \frac{5}{4}N + \Delta + N + 2 \\
&= \frac{3}{4}(N+1)^2 - \frac{6}{4}N - \frac{3}{4} + \frac{9}{4}N + 2 + \Delta \\
&= \frac{3}{4}(N+1)^2 + \frac{5}{4}(N+1) - \frac{N}{2} + \Delta.
\end{aligned}$$

*The thesis is proved by showing that*

$$\frac{1}{2}\lfloor\frac{N+1}{2}\rfloor = \Delta - \frac{N}{2}. \tag{A.13}$$

*Indeed, Eq. (A.13) can be written as*

$$\begin{aligned}
\frac{1}{2}\lfloor\frac{N+1}{2}\rfloor &= \frac{1}{2}\lfloor\frac{N}{2}\rfloor + \lfloor\frac{N+1}{2}\rfloor - \frac{N}{2} \\
\lfloor\frac{N+1}{2}\rfloor &= \lfloor\frac{N}{2}\rfloor + 2\lfloor\frac{N+1}{2}\rfloor - N \\
N &= \lfloor\frac{N+1}{2}\rfloor + \lfloor\frac{N}{2}\rfloor.
\end{aligned} \tag{A.14}$$

*The truth of Eq. (A.14) proves our claim.*

## A.3.3  Proof of Lemma 1

The number of columns with 4 used slots inside, for a run with $\mathcal{P}$ equal to the identity permutation and $N + 1$ processors, is

$$\frac{N^2 - 2N + [N \text{ is odd}]}{4}.$$

**PROOF 3** *Figure A.1 shows also how, for any processor $1 \leq i \leq \lfloor(N+1)/2\rfloor$, there exists only one cluster of $i - 1$ columns such that each column contains exactly 4 used slots. Moreover Fig. A.2 shows that, for any processor $\lfloor(N+1)/2\rfloor + 1 \leq i \leq N$, there exists only one cluster of $N - i$ columns with that same property.*

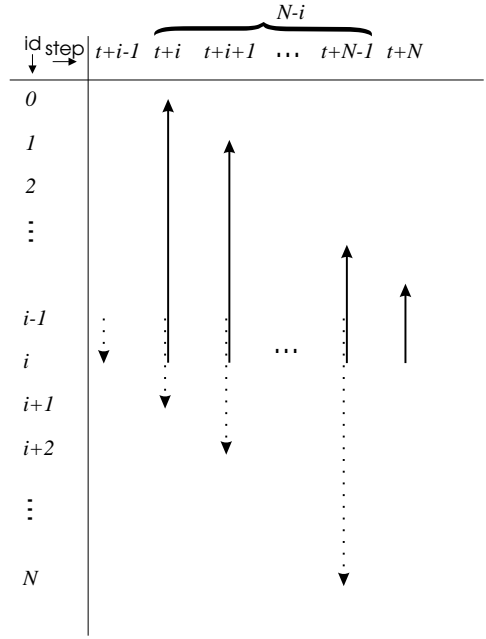

**Figure A.2:** **For any processor** $i > \lfloor (N+1)/2 \rfloor$**, there exists only one cluster of** $N - i$ **columns with 4 used slots inside.**

*Let us call $u_4$ this number and count such columns:*

$$u_4 = \sum_{i=1}^{\lfloor (N+1)/2 \rfloor} (i-1) + \sum_{i=\lfloor (N+1)/2 \rfloor + 1}^{N} (N-i).$$  (A.15)

*Two well-known algebraic transformations on sums (see e.g., in [GK86, PjB85]) bring to*

$$u_4 = \sum_{j=0}^{\lfloor (N+1)/2 \rfloor - 1} j + \sum_{j=0}^{N - \lfloor (N+1)/2 \rfloor - 1} (N - \lfloor \frac{N+1}{2} \rfloor - 1 - j).$$  (A.16)

*Note that, with $m = N - \lfloor (N+1)/2 \rfloor - 1$, the second sum can be written as $\sum_{j=0}^{m}(m-j)$, which is equal to $\frac{m(m+1)}{2}$. Hence,*

$$
\begin{aligned}
u_4 &= \frac{(\lfloor \frac{N+1}{2} \rfloor - 1)(\lfloor \frac{N+1}{2} \rfloor)}{2} + \frac{(N - \lfloor \frac{N+1}{2} \rfloor - 1)(N - \lfloor \frac{N+1}{2} \rfloor)}{2} \\
&= \frac{1}{2}(N^2 + 2(\lfloor \frac{N+1}{2} \rfloor)^2 - 2N\lfloor \frac{N+1}{2} \rfloor - N) \\
&= \frac{N^2 - 2N + [N \text{ is odd}]}{4},
\end{aligned}
$$  (A.17)

*where "$[N$ is odd]" is 1 when $N$ is odd and 0 otherwise.*   □

### A.3.4  Proof of Proposition 3

$$\lim_{N \to \infty} \varepsilon_N = 0.$$

**PROOF 4** *Let us call $U(N)$ the number of used slots in a run of $N$ processors. As a consequence of Lemma 1, the number of used slots in a run is*

$$
\begin{aligned}
U(N) &= 2\frac{N^2 - 2N + [N \text{ is odd}]}{4} + 2\lambda_N \\
&= \frac{N^2 - 2N + [N \text{ is odd}]}{2} + 2(\frac{3}{4}N^2 + \frac{5}{4}N + \frac{1}{2}\lfloor N/2 \rfloor).
\end{aligned}
\tag{A.18}
$$

*The definition of $\sigma$ (see on p. 146) implies that*

$$\varepsilon_N = \frac{U(N)}{(N+1)\lambda_N}. \tag{A.19}$$

*Clearly $\deg[U(N)] = 2$, while from Prop. 2 one knows that $\deg[(N+1) \cdot \lambda_N] = 3$. As a consequence, $\varepsilon_N$ tends to zero as $N$ tends to infinity.* □

### A.3.5  Proof of Proposition 4

$$\lim_{N \to \infty} \mu_N = \frac{8}{3}.$$

**PROOF 5** *From*

$$\mu_N = \frac{U(N)}{\lambda_N}, \tag{A.20}$$

*one can derive that*

$$
\begin{aligned}
\mu_N &= \frac{2\frac{N^2}{4} + 2\frac{3}{4}N^2 + \ldots \text{some 1st degree elements}}{\frac{3}{4}N^2 + \ldots \text{some 1st degree elements}} \\
&= \frac{2N^2 + \ldots \text{some 1st degree elements}}{\frac{3}{4}N^2 + \ldots \text{some 1st degree elements}},
\end{aligned}
\tag{A.21}
$$

*which tends to $\frac{8}{3}$, or $2.\overline{6}$, when $N$ goes to infinity.* □

### A.3.6  Proof of Proposition 5

$$\lambda_N = 3N.$$

**PROOF 6** *Let us consider run-table $N + 1$. Let us strip off its last row; then remove each occurrence of $S_{N+1}$, shifting each row left over one position. Remove also each occurrence of $R_{N+1}$. Finally, remove the last column, now empty because of the previous rules.*

*Our first goal is showing that what remains is run-table $N$. To this end, let us remind the reader that each occurrence of an $S_{N+1}$ action only affects row $N + 1$, which has been cut out.*

*Furthermore, each occurrence of $R_{N+1}$ comes from an $S$ action in row $N + 1$. Finally, due to the structure of the permutation, the last action in row $N+1$ has to be an $S_N$—as a consequence, row $N$ shall contain an $R_{N+1}$, and the remaining rows shall contain the action "$-$". Removing the $R_{N+1}$ allows to remove the last column as well, with no coherency violation and no redundant steps. This proves our first claim.*

*With a reasoning similar to the one followed for Prop. 2 it is possible to show that*

$$\sigma(N + 1) = \big(\sigma(N) + 3(N + 1)\big)\frac{N + 2}{N + 1}. \tag{A.22}$$

*that is, by Definition 7.2.1.1,*

$$\lambda(N + 1) = \lambda(N) + 3. \tag{A.23}$$

*Recursive relation (A.23) represents the first (or forward) difference of $\lambda(N)$ (see for instance [PW79]). The solution of the above is $\lambda(N) = 3N$.* □

## A.3.7   Proof of Proposition 6

$$\forall N > 0 : \varepsilon_N = 2/3.$$

**PROOF 7** *Again, let $U(N)$ be the number of used slots in a run of $N$ processors. From Prop. 5, run-table $N$ differs from run-table $N + 1$ only for $N + 1$ "S" actions, $N + 1$ "R" actions, and the last row consisting of another $N + 1$ pairs of useful actions plus some non-useful actions. Then*

$$U(N + 1) = U(N) + 4(N + 1). \tag{A.24}$$

*Via, e.g., the method of trial solutions for constant coefficient difference equations introduced in [PW79, p. 16], it is possible to get to $U(N) = 2N(N + 1)$, which clearly satisfies recursive relation (A.24) being $2N(N + 1) + 4(N + 1) = 2(N + 1)(N + 2)$.*

*Hence*

$$\varepsilon_N = \frac{U(N)}{\sigma(N)} = \frac{2N(N + 1)}{\lambda_N(N + 1)} = \frac{2}{3}. \tag{A.25}$$

□

## A.3.8   Proof of Proposition 7

$$\forall N > 0 : \mu_N = \frac{2}{3}(N + 1).$$

**PROOF 8** *The proof follows immediately from*

$$\mu_N = U(N)/\lambda_N = 2N(N + 1)/(3N).$$

□

# Appendix B

# The TIRAN Time-Out Manager

This chapter describes the architecture, the application programmer's interface (i.e., the client-side view), as well as the server-side protocol of an enhanced version of the TIRAN TOM. TOM can be regarded as a client-server application such that the client issues requests according to a well-defined protocol, while a server thread fulfils these requests: registering, updating, modifying, purging entries of the time-out list, and executing the corresponding alarms. This server side is totally transparent to the user module (the client).

## B.1   The Architecture of the TOM System

Figure B.1 shows TOM's architecture. In

(1) the client process sends requests to the time-out list manager via the API to be described in Sect. B.2; in

(2) the time-out list manager accordingly updates the time-out list with the server-side protocol to be described in Sect. B.3;

(3) each time a time-out reaches its deadline, a request for execution of the corresponding alarm is sent to a task called alarm scheduler ($\mathcal{AS}$);

(4) the latter allocates an alarm request to the first available thread out of those in a circular list of alarm threads ($\mathcal{AT}$'s), possibly waiting until such a thread becomes available.

The availability of $\mathcal{AS}$ and $\mathcal{AT}$'s can have positive consequences on fulfilling real-time requirements. These aspects are dealt with in Sect. B.4.

## B.2   Client-Side Protocol

The TOM class appears to the user as a couple of new types and some function calls. The first of these types comes into action the moment the user starts using the time-out management service.

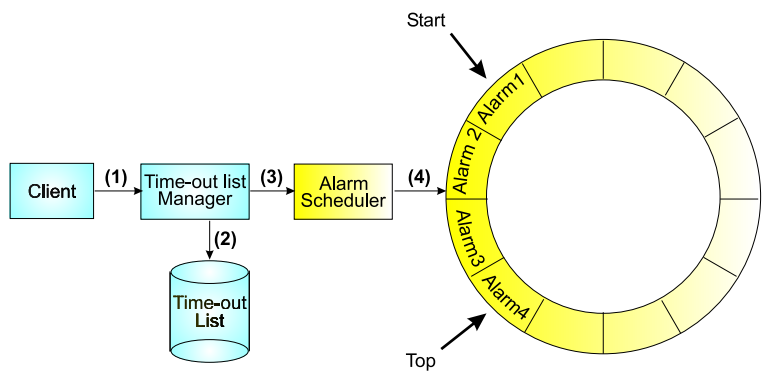

**Figure B.1: Architecture of the time-out management system.**

Specifically, to declare a time-out manager, the user needs to define a pointer to a `TOM` object.
This is more or less like defining a `FILE` pointer in standard C: `TOM *tom`.

Within the standard C class "FILE", the function `fopen` attaches an object to that pointer
and opens a connection with the file system; likewise, the function `tom_init` defines a time-out
manager object, and sets up a connection with TOM's server-side:

$$\texttt{int alarm(TOM *); tom = tom\_init( alarm );}$$

(the function `alarm` is the default function to be called when a time-out expires. A class-specific
variable stores this function pointer. This default value can be changed afterwards by setting
another, object-specific, variable). Function `tom_init` is node-specific; that is, the first time it
is called on a node, a custom thread is spawned on that node. That thread is the actual time-out
manager, that is, TOM's server-side.

At this point the user is allowed to define time-outs. This is done via type `timeout_t` and
function `tom_declare`; an example follows:

$$\texttt{timeout\_t t;}$$
$$\texttt{tom\_declare(\&t,TOM\_CYCLIC, TOM\_SET\_ENABLE, TID, DEADLINE).}$$

In the above, time-out `t` is defined as:

- a cyclic time-out (renewed on expiration; as opposed to `TOM_NON_CYCLIC`, i.e., purged
  on expiration),

- enabled (only enabled time-outs "fire", i.e., call their alarm on expiration; an alarm is
  disabled with `TOM_SET_DISABLE`),

- with a deadline of `DEADLINE` clock ticks before expiration.

Furthermore, time-out `t` is identified by the integer `TID`.

Once defined, a time-out can be submitted to the time-out manager for insertion in its running
list of time-outs—this will be explained in more detail in Sect. B.3. From the user viewpoint,
this is managed by calling the function `tom_insert`.

```
1.  /* declarations */
    TOM *tom; timeout_t t1, t2;
    int my_alarm(TOM*), another_alarm(TOM*);
2.  /* definitions */
    tom ← tom_init(my_alarm);
    tom_declare(&t1, TOM_NON_CYCLIC, TOM_SET_ENABLE,
                     TIMEOUT1, DEADLINE1);
    tom_declare(&t2, TOM_CYCLIC, TOM_SET_DISABLE,
                     TIMEOUT2, DEADLINE2);
    tom_set_action(&t2, another_alarm);
3.  /* insertion */
    tom_insert(tom, &t1),
    tom_insert(tom, &t2);
4.  /* control */
    tom_enable(tom, &t2);
    tom_set_deadline(&t2, NEW_DEADLINE2);
    tom_renew(tom, &t2);
    tom_delete(tom, &t1);
5.  /* deactivation */
    tom_close(tom);
```

**Table B.1: An example of usage of the TOM class.**

After successful insertion (to be tested via the return value of `tom_insert`), and in case no further control is specified, an enabled time-out will trigger the call of the default alarm function after the specified deadline. This behaviour would be cyclic if the original time-out is defined as such. Further control is still possible though. For instance, a time-out can be temporarily suspended while in the time-out list via function `tom_disable` and (re)-enabled via function `tom_enable`.

Furthermore, via similar functions, the user is allowed to specify a function other than the default one to be called upon expiration of the deadline (function `tom_set_action`), as well as to specify new deadline values (function `tom_set_deadline`). TOM also allows to delete a time-out from the list (`tom_delete`) as well as to *renew* a time-out—which means removing a time-out and inserting it with the original values set again (`tom_renew`). Function `tom_close` terminates the time-out management service also performing garbage collection. Table B.1 summarises the above client-side protocol with an example.

# B.3   Server-Side Protocol

The server-side protocol is run by a component called time-out list manager (TOLM). TOLM basically checks every `TOM_CYCLE` whether

- there are any incoming requests for manipulating the time-out list; if so, it deals with those requests;

- there are time-outs whose deadlines are reached.  If so, it manages the execution of the corresponding alarm.

This section describes the server-side protocol of TOLM, i.e., the way TOLM manages the list of time-outs.

Each time-out `t` is characterised by its *deadline*, namely `t.deadline`, a positive integer representing the number of clock ticks that must separate the time of insertion or renewal from the scheduled time of alarm. Functions `tom_declare` and `tom_set_deadline` set this field. Each time-out `t` holds also a field, `t.running`, initially set to `t.deadline`.

Each time-out list object, say `tom`, hosts a variable representing the origin of the time axis. This variable, `tom.starting_time`, is related in particular to the time-out at the top of the time-out list—the idea is that the top of the list is the only entry whose `running` field needs to be compared with current time in order to verify the occurrence of the time-out-expired event. For the time-outs behind the top one, that field represents relative values, i.e., distances from expiration time of the closest, preceding time-out. In other words, the overall time-out list management aims at isolating a "closest to expiration" time-out, or head time-out, that is the one and only time-out to be tracked for expiration, and at preserving the coherence of a list of "relative time-outs".

Let us call `TimeNow` the system function returning the current value of the clock register. In an ordered, coherent time-out list, residual time *for the head time-out* `t`, is given by

$$\texttt{t.running} - (\texttt{TimeNow} - \texttt{tom.starting\_time}), \tag{B.1}$$

that is, residual time minus time already passed by.  Let us call quantity (B.1) as $r_1$, or head residual.  For time-out $n$, $n > 1$, that is for the time-out located $n - 1$ entries "after" the top block, let us define

$$r_n = r_1 + \sum_{i=2}^{n} \texttt{t}_i.\texttt{running} \tag{B.2}$$

as the $n$-th residual, or residual time for time-out at entry $n$. If there are $m$ entries in the time-out list, let us define $r_j = 0$ for any $j > m$.

It is now possible to formally define the fundamental operations on a time-out list: insertion and deletion of an entry.

## B.3.1   Insertion

There are three possible insertions, namely on top, in the middle, and at the end of the list.

### B.3.1.1   Insertion on Top

In this case a new time-out object, say $t$, has to be inserted on top, such that $t.\mathtt{deadline} < r_1$, that is, the deadline of which is less than the head residual. Let us call $u$ the current top of the list. Then the following operations need to be carried out:

$$\begin{cases} t.\mathtt{running} & \leftarrow & t.\mathtt{deadline} + \mathtt{TimeNow} - \\ & & \mathtt{tom.starting\_time} \\ u.\mathtt{running} & \leftarrow & r_1 - t.\mathtt{deadline}. \end{cases}$$

Note that the first operation is needed in order to verify relation

$$\begin{aligned} t.\mathtt{running} & - & (\mathtt{TimeNow} - \mathtt{tom.starting\_time}) \\ & = & t.\mathtt{deadline}, \end{aligned}$$

while the second operation aims at turning the absolute value kept in the $\mathtt{running}$ field of the "old" head of the list into a value relative to the one stored in the corresponding field of the "new" top of the list.

### B.3.1.2   Insertion in the Middle

In this case a time-out $t$ such that, for some $j$,

$$r_j \leq t.\mathtt{deadline} < r_{j+1},$$

is to be inserted. Let us call $u$ time-out $j + 1$. (Note that both $t$ and $u$ exist by hypothesis). Then the following operations need to be carried out:

$$\begin{cases} t.\mathtt{running} & \leftarrow t.\mathtt{deadline} - r_j \\ u.\mathtt{running} & \leftarrow u.\mathtt{running} - t.\mathtt{running}. \end{cases}$$

Note how, both in the case of insertion on top and in that of insertion in the middle of the list, the time interval $[0, r_m]$ has not changed its length. It has, however, been further subdivided, and is now to be referred to as $[0, r_{m+1}]$.

### B.3.1.3   Insertion at the End

Let us suppose the time-out list consists of $m > 0$ items, and that a time-out $t$ has to be inserted, with $t.\mathtt{deadline} \geq r_m$. In this case the item is simply appended and initialised so that

$$t.\mathtt{running} \leftarrow t.\mathtt{deadline} - r_m.$$

Note how insertion at the end of the list is the only way to prolong the range of action from a certain $[0, r_m]$ to a larger period $[0, r_{m+1}]$.

### B.3.2    Deletion

The other basic management operation on the time-out list is deletion. Three types of deletions can be found: deletion from the top, from the middle, and from the end of the list.

#### B.3.2.1    Deletion from the Top

A singleton list is a trivial case. Let us suppose there are at least two items in the list. Let us call $t$ the top of the list and $u$ the next element, the one that will be promoted to top of the list. From its definition it is known that

$$
\begin{aligned}
r_2 &= u.\texttt{running} + r_1 \\
&= u.\texttt{running} + t.\texttt{running} - \\
&\quad (\texttt{TimeNow} - \texttt{tom.starting\_time}).
\end{aligned}
\tag{B.3}
$$

By (B.1), the bracketed quantity is elapsed time, so the amount of absolute time units that separate current time from the expiration time is given by $u.\texttt{running} + t.\texttt{running}$. In order to "behead" the list $t$ needs to be updated as follows:

$$
u.\texttt{running} \leftarrow u.\texttt{running} + t.\texttt{running}.
$$

#### B.3.2.2    Deletion from the Middle

Let us suppose that there are two consecutive time-outs in the list, $t$ followed by $u$, such that $t$ is not the top of the list. Again, before actually removing $t$ from the list, the following step is required:

$$
u.\texttt{running} \leftarrow u.\texttt{running} + t.\texttt{running}.
$$

#### B.3.2.3    Deletion from the End

Deletion from the end means deleting an entry which is not referenced by any further item in the list. Physical deletion can be performed with no need for any updating. Only the interval of action is shortened.

The variable `tom.starting_time` is never touched when deleting from or inserting entries into a time-out list, except when inserting the first element, when it is set to the current value returned by `TimeNow`.

## B.4    Real-Time Support

What has been reported so far applies in the ideal situation such that

- all management operations occur instantaneously, that is, inducing no real-time delay,

- time-out detection delay is negligible,

- other threads/processes running on the same processing node have negligible influence on the performance of the manager,

- alarm execution delay is negligible.

This is in general far from being true. Management tasks do impose a non-negligible overhead, time-out checks occur once every `TOM_CYCLE` clock ticks (one may reduce this value, though this is detrimental to performance, because this increases the frequency of checking, which implies more context switches and more CPU cycles assigned to TOM), overloading the process with threads results in a smaller time slice per thread, and, last but not least, the alarm function is a big source of non-determinism due to the fact that it is defined by the user. In our experience, for instance, the alarm function is often related to some communication task—typically the alarm triggers the transfer of a message, possibly across the network. Frequently this happens through synchronous links and results in blocking the sender until the receiver is ready to receive and the transfer has been made.

In conclusion, there is a non negligible delay between the time at which $r_1$ becomes zero and the time that event is managed; such delay can be expressed as

$$(\texttt{TimeNow} - \texttt{tom.starting\_time}) - t.\texttt{running}.$$

This quantity must be immediately propagated to those entries who follow the top of the list. This is done by determining the integer $j \geq 1$ such that

$$r_j < 0 \quad \wedge \quad r_{j+1} \geq 0, \tag{B.4}$$

where this time $r_1$ can also be negative. The time-out management thread needs therefore first to check whether $r_1$ is less than zero; if so, it must calculate index $j$ such that (B.4) is verified; and finally command the execution of all the corresponding alarm functions.

Finally, if the list is not empty, that thread must adjust the corresponding `running` field as follows: let $t$ be time-out $j + 1$; then

$$t.\texttt{running} \leftarrow t.\texttt{running} + r_j.$$

Clearly the above mechanism only works well if there is a way to keep under control the congestion that is due to alarm execution. The rest of this section describes a mechanism with this aim.

Alarm execution is managed via the mechanism shown in Fig. B.1, points (3) and (4), i.e., through the $\mathcal{AS}$, which gathers all alarm execution requests and forwards them to the next entry of a circular list of threads. As already said, alarms often imply communications, therefore using a list of concurrent threads might in principle result in better performance, should the underlying system offer means for managing I/O in parallel. In general, if the alarms do not compete "too much" for the same resources at the same times, this scheme should allow a better exploitation of the available resources as well as a higher probability of controlling alarm congestion. The present section deals with the estimation of the capability of this mechanism to control alarm congestion under different levels of congestion and in two opposite cases of alarm interference, i.e., no competition and full competition.

In order to estimate the average run-time delay imposed by TOM, the following experiment has been performed: 1000 non-cyclic time-outs have been generated pseudo-randomly with a deadline uniformly distributed in time interval $]0, T]$. Each time a time-out triggered an alarm, the difference between expected and real-time of execution of the alarm has been computed. Let us call $\delta$ this value. TOM's period, i.e., TOM_CYCLE was set to 50000 clock ticks, 1 clock tick being $1\mu$s on the target machines. The experiment was performed on a single node of a Parsytec PowerXplorer, using a PowerPC 601 at 66MHz, and has been repeated on a Parsytec CC system, with a PowerPC 604 at 133MHz. More or less the same results have been observed on both machines (the clock frequency only influences alarm list management times, which is a small percentage of the processing time).

In order to measure the alarm congestion capability of TOM, the experiment has been repeated configuring TOM with no threads in the circular list of $\mathcal{AT}$'s (alarm execution managed by TOLM) and later adding more and more threads in that list. Let us call $\tau$ the number of threads in the list.

Furthermore, the following overheads have been artificially imposed on the alarm functions: nearly no overhead (just that of calling a function and copying a 20-byte message—this is the typical delay used in most of our applications, lasting approximately $50\mu$s on a DEC Alpha board using the TEX nanokernel [TXT97],—10ms, 100ms, and so on. This overhead has been imposed either by loading the alarms with some purely computation-oriented tasks, so that they compete for the same resource (in this case, the one CPU in the system), or by executing a function (`TimeWait`) that puts the calling thread in the wait state for the specified amount of time—this way alarms do not compete at all with each other.

Figure B.2 summarises the results of the first experiment. The experiments show that minimal (effective) overhead is imposed on the alarms of 1000 time-outs, the deadline of which is generated pseudo-randomly in $]0, 100s]$. TOLM directly executes alarms in this case ($\tau = 0$).

## B.4.1 Best Case Scenario

Another experiment aimed at evaluating what happens when $\delta$ is increased while keeping $\tau$ equal to zero and there is no competition among the alarms. The results have been also summarised in Table B.2 which reports the number of items that resulted in a real-time violation (alarm delay greater than TOM_CYCLE clock ticks) as well as other statistics. Particularly meaningful is the largest violation, which, in the case of a delay of 20ms, was equal to 124535 clock ticks, or 2.49 times TOM_CYCLE.

In particular, when the duration of the alarm became larger than TOM_CYCLE (the checking quantum of the time-out list manager), an ever increasing violation of real-time requirements due to alarm execution congestion has been observed.

It has been experimentally found that the presence of the $\mathcal{AS}$ and of the circular list of threads has a great consequence on alarm congestion control: Table B.3 for instance summarises the results of increasing $\tau$ from 0 to 5 when $\delta$ is set to $20\mu$s. In particular when $\tau$ is equal to 2, the number of "run-time violations", i.e., the number of time-outs for which the true time of execution of the alarm minus the expected time of execution of the alarm is greater than TOM_CYCLE, drops from 264 to 46 items, while with $\tau = 3$ the worst case violation was equal to 53934 clock ticks, or just 1.07868 times TOM_CYCLE—a violation of something less than 8%. Table B.4 shows a similar behaviour when $\delta$ is larger than TOM_CYCLE.

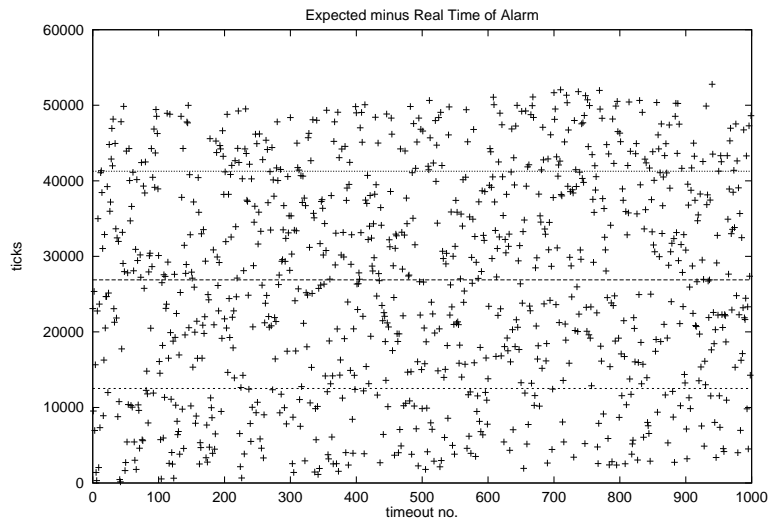

**Figure B.2: 1000 time-outs are uniformly generated in $]0, 100]$s. The duration of the alarms is in this case $\approx 50\mu$s. Only one alarm thread is adopted in this case. The maximum value is 52787 "clock ticks", minimum is 157 (1 tick=1 $\mu$s). The average is 26892.74, standard deviation is 14376.89 ticks; 20 of the 1000 time-outs exceeded TOM_CYCLE.**

| delay | population | average | stdev | max |
|---|---|---|---|---|
| $\approx 50\mu$s | 20 | 50967.10 | 762.07 | 52782 |
| 1ms | 34 | 51282.32 | 975.85 | 53263 |
| 5ms | 77 | 53077.68 | 2774.48 | 64328 |
| 10ms | 132 | 57605.80 | 6917.56 | 88183 |
| 20ms | 238 | 65639.74 | 14571.78 | 124535 |

**Table B.2: The table reports on experiments aiming at measuring alarm execution congestion. The first column represents the duration of each alarm. 50$\mu$s is approximately the average time required by the alarm() function used in the TIRAN BB: a context switch time plus the time to store a 20-byte message into the input mailbox of the receiver process (measurements pertaining the TEX operating system [DEC97]). Values of the average, standard deviation and maximum are in clock ticks (microseconds).**

## B.4.2 Worst Case Scenario

The worst case is when all $\mathcal{AT}$'s compete for the same set of resources at the same time. An easy way to accomplish this has been to let each alarm function perform a pure integer computation task so that each thread competes with all the others for the single CPU of our systems. Given a fixed $\delta = 20$ms, $\tau$ has been increased and the corresponding real-time violations have been measured. The results have been summarised in Table B.5. Adding threads did not produce any useful result in this case. Note how, adopting a parallel system, one could automatically and transparently schedule the parallel execution of alarms on the available processors to further enhance the ability of TOM to fight alarm execution congestion.

| $\tau$ | $\mu$ | $\sigma$ | $\gamma$ | max | $\mu'$ | $\sigma'$ |
|---|---|---|---|---|---|---|
| 0 | 35962.85 | 22683.49 | 264 | 130993 | 65483.29 | 14160.47 |
| 1 | 30200.69 | 18447.08 | 108 | 108991 | 64676.18 | 15647.9 |
| 2 | 27556.86 | 14471.61 | 46 | 63659 | 52228.13 | 2926.68 |
| 3 | 27286.77 | 14370 | 45 | 53934 | 51280.73 | 1014.65 |
| 4 | 27493.73 | 14058.12 | 47 | 53372 | 51526.17 | 921.99 |
| 5 | 27422.84 | 14079.80 | 46 | 54077 | 51357.2 | 1079.07 |

**Table B.3: A run with $\delta = 20$ms and $0 \leq \tau \leq 5$.** $\mu$ and $\sigma$ are resp. the average and std. deviation of the 1000 outcomes. $\gamma$ is the number of time-outs for which real time of alarm minus expected time of alarm was $>$ TOM_CYCLE (50000 clock ticks). $\mu'$ and $\sigma'$ are resp. the average and std. deviation of the $\gamma$ time-outs. "max" represents the largest outcome, i.e., the worst run-time violation.

| $\tau$ | $\mu$ | $\sigma$ | $\gamma$ | max | $\mu'$ | $\sigma'$ |
|---|---|---|---|---|---|---|
| 0 | 3692158.86 | 2966039.23 | 974 | 12277216 | 3790057.80 | 2943375.88 |
| 1 | 33674.46 | 27688.17 | 140 | 264882 | 83244.94 | 38052.06 |
| 2 | 28177.01 | 16801.97 | 54 | 146684 | 66737.87 | 20705.76 |
| 3 | 27621.63 | 14276.78 | 45 | 100260 | 52871.73 | 7358.90 |
| 4 | 27435.40 | 14087.22 | 44 | 53712 | 51375.93 | 1001.36 |
| 5 | 27475.84 | 14107.32 | 44 | 53992 | 51443.95 | 1008.77 |

**Table B.4: A run with $\delta = 100$ms and $0 \leq \tau \leq 5$.**

| $\tau$ | $\mu$ | $\sigma$ | $\gamma$ | max | $\mu'$ | $\sigma'$ |
|---|---|---|---|---|---|---|
| 0 | 35251.93 | 21667.08 | 226 | 126931 | 66223.70 | 13713.40 |
| 1 | 35533.45 | 21584.35 | 248 | 127403 | 64377.88 | 13803.45 |
| 2 | 35735.19 | 21606.70 | 250 | 125910 | 64681.21 | 13305.83 |

**Table B.5: A run with $\delta = 20$ms and $0 \leq \tau \leq 2$. This time $\mathcal{AT}$'s do compete for a unique system resource, so adding threads does not improve the behaviour.**

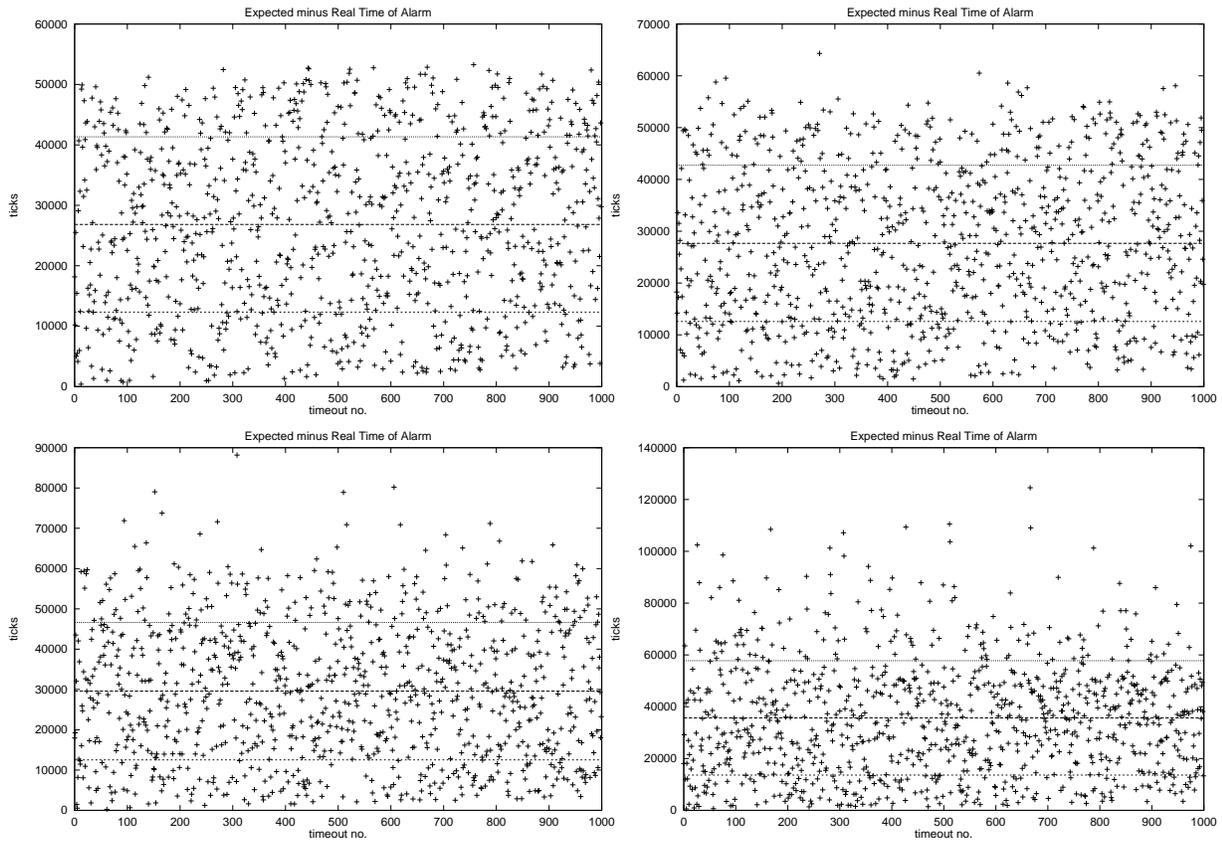

**Figure B.3: Real-time minus expected time of execution of the alarm when the alarms last 1ms (upper left picture), 5ms (upper right picture), 10ms (lower, left picture), and 20ms (lower, right picture). 1000 pseudo-random time-outs with deadlines uniformly distributed in $]0, 100s]$, $\tau = 0$. In each picture, the central straight line represents the average ($y = a$), while the others are $y = a \pm d$, $d$ being the standard deviation.**

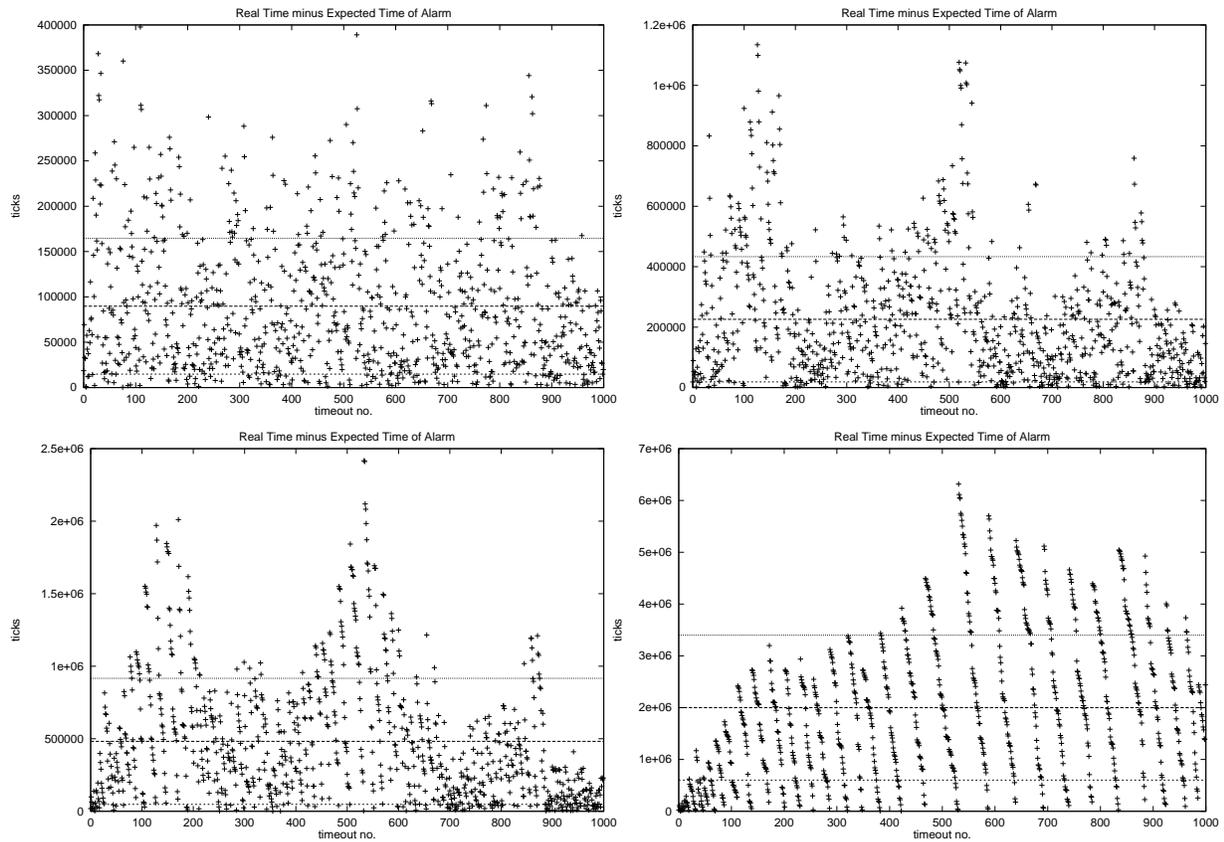

**Figure B.4: The experiment is repeated with alarms lasting respectively 60ms, 80ms, 90ms, and 100ms. Run-time violation grows without bounds.**

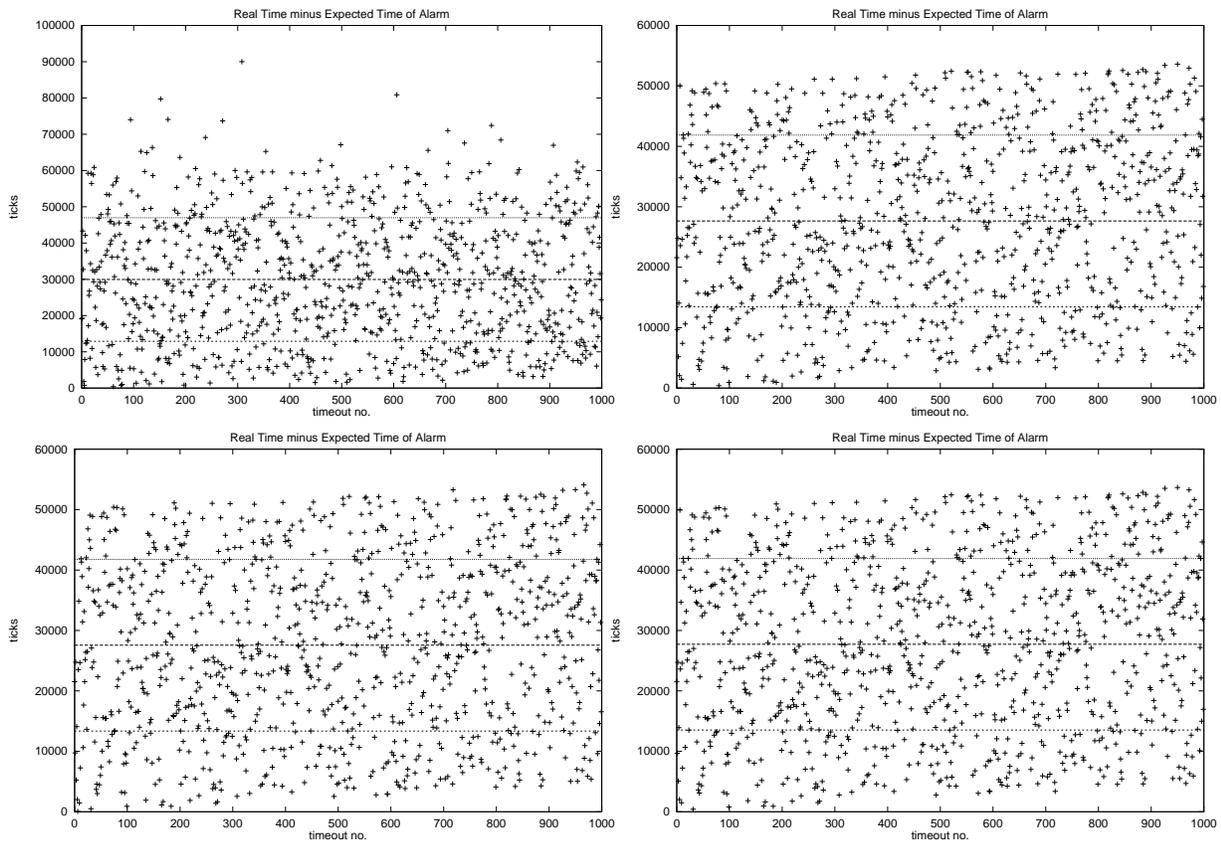

**Figure B.5: For $\delta$=10ms, $\tau$ is varied from 0 to 3.**

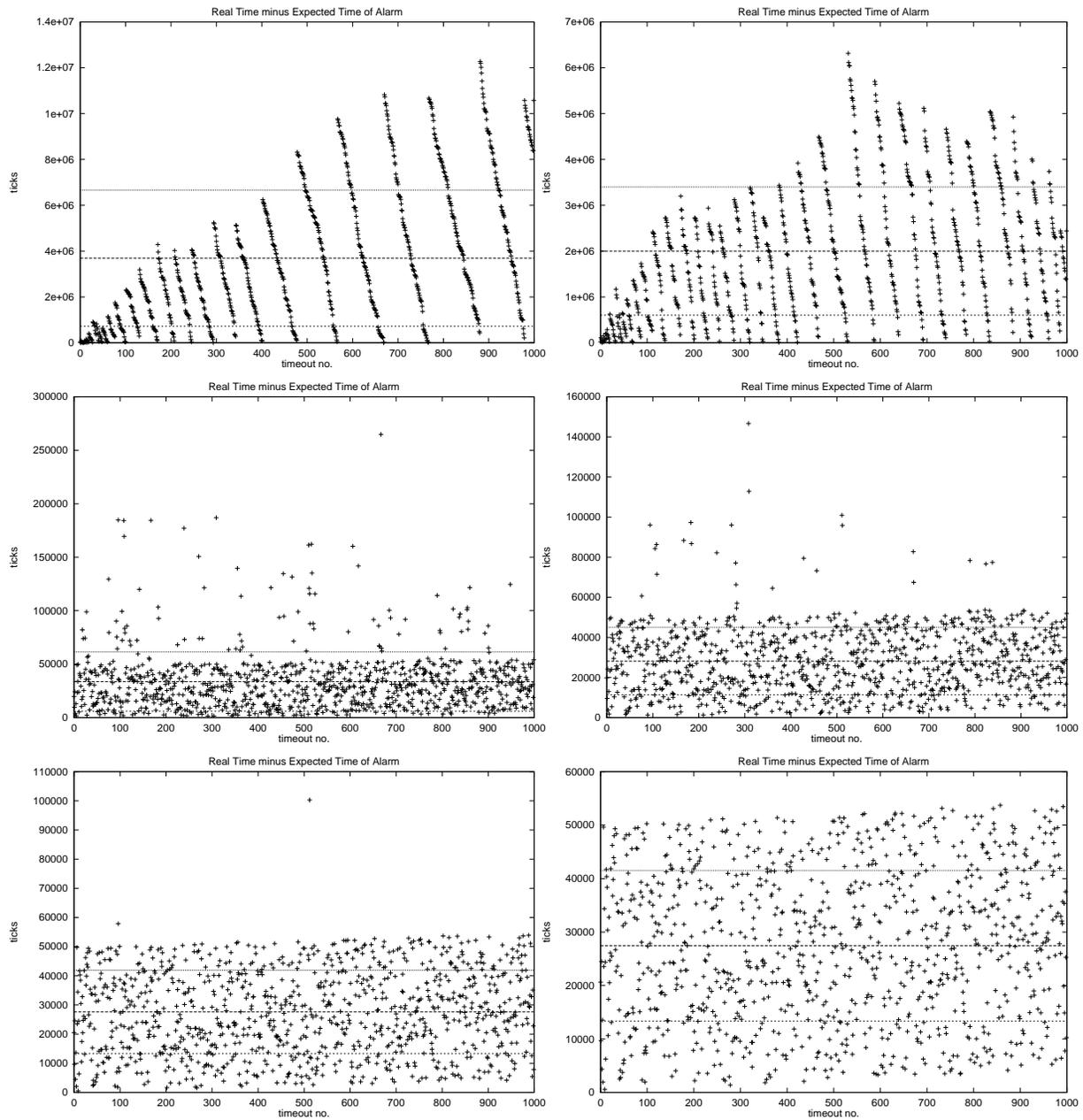

**Figure B.6:** **The pictures show how it is possible to control alarm execution congestion, even in the scenario of the last picture of Fig. B.4 ($\delta = 100\mu$s), by increasing the number of $\mathcal{AT}$'s. From left to right and up to down, picture represents the cases of $\tau$ equal to $0, \ldots, 5$.**

# Appendix C

# The Grammar of ARIEL

The grammar of the ARIEL language is provided in this chapter in the form of an abridged source file witten with the yacc [Joh75] parser generator. The C code that implements the semantics of ARIEL has been omitted for the sake of readability. For the same reason, the terminal symbols and their definitions, in the form of a lex [LMB92] lexical analyser, have not been included.

```
rlstats:
      | rlstats rlstat
      | error '\n'
      ;

rlstat:   '\n'
      | definition '\n'
      | default_action '\n'
      | number_of_nodes '\n'
      | section '\n'
      | include '\n'
      | timeouts '\n'
      | definitions '\n'
      | identifiers '\n'
      | alphacounts '\n'
      | aliases '\n'
      | logicals '\n'
      | watchdog '\n'
      | nversiontask '\n'
      | injection '\n'
      ;

aliases   KEYW_TASK '[' NUMBER ',' NUMBER ']' KEYW_IS KEYW_MBOX
          '[' NUMBER ',' NUMBER ']' ',' KEYW_ALIAS
          '[' NUMBER ',' NUMBER ']'
```

```
        |       KEYW_TASK NUMBER  KEYW_IS KEYW_MBOX NUMBER ','
                KEYW_ALIAS NUMBER
        ;

tasklist:  KEYW_TASK NUMBER tasklist
        |       ',' KEYW_TASK NUMBER tasklist
        |       ',' NUMBER tasklist
        |
        ;

logicals:  KEYW_LOGICAL NUMBER '=' STRING
    KEYW_IS tasklist KEYW_ENDLOGICAL
        ;

nversiontask: nversion_start nversion_args nversion_end
        ;

nversion_start: NVERSION KEYW_TASK NUMBER '\n'
        ;

nversion_args:
        | nv_version    nversion_args
        | nv_voting     nversion_args
        | nv_metric     nversion_args
        | nv_on_error   nversion_args
        | nv_on_success nversion_args
        ;

spare:
        |   KEYW_SPARE
        ;

nv_version:  KEYW_VERSION NUMBER KEYW_IS spare
        KEYW_TASK NUMBER '\n'
        |       KEYW_VERSION NUMBER KEYW_IS spare
        KEYW_TASK NUMBER
        TIMEOUT NUMBER seconds '\n'
        ;

nv_voting: VOTING ALGORITHM KEYW_IS MAJORITY '\n'
        ;
```

```
nv_metric: METRIC STRING '\n'
    ;

nv_on_error: ON ERROR KEYW_TASK NUMBER '\n'
    ;

nv_on_success: ON SUCCESS KEYW_TASK NUMBER '\n'
    ;

nversion_end: KEYW_ENDNVERSION '\n'
    ;

watchdog: watchdog_start watchdog_args watchdog_end
    ;

watchdog_args:
    | on_error watchdog_args
    | heartbeats watchdog_args
    | w_alphacount watchdog_args
    ;

watchdog_start: WATCHDOG KEYW_TASK NUMBER WATCHES
                KEYW_TASK NUMBER '\n'
    |           WATCHDOG KEYW_TASK NUMBER '\n'
    ;

w_alphacount:   ALPHACOUNT KEYW_IS THRESHOLD '=' REAL
                ',' FACTOR '=' REAL KEYW_ENDALPHA '\n'
    ;

seconds : MILLISEC | MICROSEC
    ;

heartbeats: HEARTBEATS EVERY NUMBER seconds '\n'
    ;

on_error: ON ERROR WARN KEYW_TASK NUMBER '\n'
    |     ON ERROR WARN BACKBONE '\n'
    |     ON ERROR REBOOT '\n'
    |     ON ERROR RESTART '\n'
    ;
```

```
watchdog_end: KEYW_ENDWATCHDOG '\n'
        ;

alphacounts:    ALPHACOUNT NUMBER KEYW_IS THRESHOLD '=' REAL
                ',' FACTOR '=' REAL KEYW_ENDALPHA
        ;

identifiers:    KEYW_TASK NUMBER '=' STRING KEYW_IS
                KEYW_NODE NUMBER ',' KEYW_TASKID NUMBER
        |
                KEYW_TASK '[' NUMBER ',' NUMBER ']' '=' STRING
                KEYW_IS KEYW_NODE NUMBER ',' KEYW_TASKID
                '[' NUMBER ',' NUMBER ']'
        ;

definitions:    NUMTASKS NUMBER '=' NUMBER
        ;

timeouts:    MIA_TIMEOUT '=' NUMBER
        |    TAIA_TIMEOUT '=' NUMBER
        |    ALIVE_TIMEOUT '=' NUMBER
        |    MIA_TIMEOUT_B '=' NUMBER
        |    TAIA_TIMEOUT_B '=' NUMBER
        |    ALIVE_TIMEOUT_B '=' NUMBER
        |    TEIF_TIMEOUT '=' NUMBER
        |    REQUEST_DB_TIMEOUT '=' NUMBER
        |    REPLY_DB_TIMEOUT '=' NUMBER
        |    MID_TIMEOUT '=' NUMBER
        ;

fault:  BFAULT | MFAULT
        ;

what:   NODE | COMPONENT
        ;

ticks:
        | TICKS
        ;

injection:  INJECT fault ON what NUMBER AFTER NUMBER ticks
        ;
```

```
definition: DEF list '=' ROLE
     |          DEF interval '=' ROLE
     ;

list:      NUMBER
     |      list ',' NUMBER
     ;

interval: NUMBER '-' NUMBER
     ;

number_of_nodes: NPROCS '=' NUMBER
     ;

separator:   '\n'  |  ';'
     ;

those   :         AT | TILDE
     ;

expr:      status entity
     | status those
     | '(' expr ')'
     | expr AND expr
     | expr OR expr
     | NOT expr
     | ERRN     '(' entity ')'  compare   NUMBER
     | ERRT     '(' entity ')'  compare   NUMBER
     | PHASE    '(' entity ')'  compare   NUMBER
     | DEADLOCKED entity entity
     ;

compare:  EQ | NEQ | GT | GE | LT | LE
     ;

include:  INCLUDE STRING
     ;

section:  if elif else fi
     ;

if:    IF '[' expr  ']'     separator
```

```
        THEN separator actions
    ;

elif:   |
ELIF '[' expr ']'    separator
THEN separator actions
elif
    ;

else:   |
ELSE
separator actions
    ;

fi:    FI
    ;

status: FAULTY | RUNNING | REBOOTED | STARTED |
        ISOLATED | RESTARTED | TRANSIENT | REINTEGRATED
    ;

entity:    GID |NID |TID
    ;

actions:
    | actions action
    ;

action: '\n'
    |
section       separator
    |
recovery_action    separator
    ;

recovery_action:  STOP    entity
    | STOP those
    | ERR NUMBER entity WARN entity
    | ERR NUMBER entity WARN entity '(' list ')'
    | ERR NUMBER entity WARN those
    | ERR NUMBER entity WARN those '(' list ')'
    | recovery_action AND WARN entity
    | recovery_action AND WARN entity '(' list ')'
```

```
|   recovery_action AND WARN those
|   recovery_action AND WARN those '(' list ')'
|   RESTART TID
|   RESTART GID
|   RESTART NID
|   RESTART those
|   START   TID
|   START   GID
|   START   NID
|   START those
|   REMOVE PHASE entity FROM ERRORLIST
|   REMOVE ANY entity FROM ERRORLIST
|   SEND NUMBER TID
|   SEND FAULTY TID
|   ENABLE TID
|   ENABLE GID
|   ENABLE  NID
|   ENABLE those
|   CALL NUMBER
|   CALL NUMBER '(' list ')'
|   PAUSE NUMBER
;
```